\documentclass[aps,prb,10pt,twocolumn,superscriptaddress,longbibliography]{revtex4-2}

\usepackage[version=4]{mhchem}
\usepackage{graphicx} % Required for inserting images

\usepackage{amssymb}
\usepackage{amsmath,mathtools}

\usepackage[table,xcdraw,svgnames]{xcolor}
\usepackage[ruled,vlined]{algorithm2e}

\usepackage[english]{babel}
\usepackage{multirow}

\usepackage{bibunits}

\usepackage{hyperref}
\usepackage{url}

\hypersetup{
    citecolor=blue,
    colorlinks=true,
    linkcolor=blue,
    filecolor=magenta,      
    urlcolor=cyan,
    }

\usepackage[capitalize]{cleveref}

\crefname{figure}{Figure}{Figures}
\crefname{section}{Section}{Sections} 
\crefname{table}{Table}{Tables} 

\usepackage{float}
\crefname{enumi}{Criterion}{Criterion}
\usepackage{booktabs}

\newcommand{\beginsupplement}{%

        \setcounter{table}{0}
        \renewcommand{\thetable}{S\arabic{table}}%
        \setcounter{figure}{0}
        \renewcommand{\thefigure}{\arabic{figure}}
        \setcounter{equation}{0}
        \renewcommand{\theequation}{S\arabic{equation}}%
        \setcounter{page}{0}
        \renewcommand{\thepage}{S\arabic{page}} 
        \setcounter{section}{0}
        \renewcommand{\thesection}{S\arabic{section}} 
        
        \setlength{\parskip}{5pt}
        \everypar{\setlength{\parindent}{24pt}}
}

\usepackage[T1]{fontenc}
\usepackage{newtxtext,newtxmath}

\usepackage{siunitx}
\renewcommand{\AA}{\text{\r{A}}}   
\DeclareSIUnit{\angstrom}{\text{\AA}}

\usepackage{enumitem}
\usepackage[normalem]{ulem}

\makeatletter
\@ifundefined{contentsname}{%
}{}
\newcommand{\sitablesofcontents}{%
  \begingroup
    \renewcommand{\addcontentsline}[3]{}%
    \section*{Table of Contents}%
  \endgroup
  \begingroup
    \hypersetup{linkcolor=black}%
    \@starttoc{stoc}%
  \endgroup
}
\newcommand{\usestocfile}{%
  \let\si@saved@addcontentsline\addcontentsline
  \renewcommand{\addcontentsline}[3]{%
    \def\si@tocfile{##1}%
    \def\si@tocname{toc}%
    \ifx\si@tocfile\si@tocname
      \si@saved@addcontentsline{stoc}{##2}{##3}%
    \else
      \si@saved@addcontentsline{##1}{##2}{##3}%
    \fi
  }%
}
\makeatother

\makeatletter
\renewcommand{\l@section}{\@dottedtocline{1}{0em}{2.8em}}
\renewcommand{\l@subsection}{\@dottedtocline{2}{2.2em}{3.8em}}
\renewcommand{\l@subsubsection}{\@dottedtocline{3}{4.5em}{4.8em}}

\renewcommand{\@dotsep}{2}  % Reduce dot separation (default is 4.5)
\makeatother

\begin{document}

\title{
Vacancy-enhanced Li solid-solution anodes for solid-state batteries: 
computational screening enabling high critical current density
}

% Vacancy-enhanced Li solid-solution anodes with high critical current density for solid-state batteries: a computational screening

% Vacancy-enhanced Li solid solutions for solid-state batteries: a DFT screen for high critical current density

% Vacancy-enhanced Li solid-solution anodes for solid-state batteries: a computational screening for high critical current density

\author{Arseniy S. Burov}
\altaffiliation{Corresponding author: arseniy.burov@skoltech.ru}
\affiliation{Skolkovo Institute of Science and Technology, 121205 Moscow, Russian Federation}

\author{Nadezhda A. Ashikhmina}
\affiliation{Moscow Institute of Physics and Technology, 141701 Dolgoprudny, Russian Federation}

\author{Nurlan Zhanabayev}
\affiliation{Skolkovo Institute of Science and Technology, 121205 Moscow, Russian Federation}

\author{Anton O. Boev}
\affiliation{Skolkovo Institute of Science and Technology, 121205 Moscow, Russian Federation}

\author{Aleksandra D. Radina}
\affiliation{Skolkovo Institute of Science and Technology, 121205 Moscow, Russian Federation}

\author{Ilya V. Chepkasov}
\affiliation{Skolkovo Institute of Science and Technology, 121205 Moscow, Russian Federation}

\author{Alexander G. Kvashnin}
\affiliation{Skolkovo Institute of Science and Technology, 121205 Moscow, Russian Federation}

\author{Dmitry A. Aksyonov}
\affiliation{Skolkovo Institute of Science and Technology, 121205 Moscow, Russian Federation}

\date{\today}

\begin{abstract}

    Sluggish vacancy-mediated self-diffusion in lithium metal limits the critical current density (CCD) of Li anodes in solid-state batteries by promoting interfacial void formation during stripping. To enhance the effective vacancy concentration while preserving the bcc Li host, we screen 50 substitutional solutes using density functional theory (DFT) within a funnel based on automatic phase diagram construction, solution energies, solute--vacancy binding energies, and experimental solubility limits. Using an adaptation of the Lomer model for the dilute limit,, we show that only Ag, Ge, and Cd solid solutions are predicted to reach $1~\mathrm{mA~cm^{-2}}$ within their solubility limits. The thermodynamically estimated CCD of Ag at $\sim 9$ at.\% reaches $\sim 4~\mathrm{mA~cm^{-2}}$ at a theoretical capacity of $1522~\mathrm{mAh~g^{-1}}$. Migration barrier calculations indicate that the enhancement may be partially offset by vacancy trapping, necessitating further validation via molecular dynamics simulations. Notably, Hume--Rothery rules alone prove insufficient as prefilters, whereas machine-learning models trained on the DFT data can eliminate most unsuitable solutes early in the funnel and extend the workflow to related solid-solution searches.
    
\end{abstract}

\maketitle

\begin{bibunit}[apsrev4-2]

\section{Introduction}

The pursuit of high ionic conductivity in solid electrolytes (SEs) has often come at the expense of addressing the equally critical challenge of achieving a high critical current density (CCD). For solid-state batteries to become a practical reality, several interface-related issues between the lithium metal anode and the SE must be overcome~\cite{sarkar2022critical}. Among these, the sluggish self-diffusion of Li in the metal anode represents a primary bottleneck. During stripping, the rate of lithium replenishment from the bulk to the interface is insufficient to match the rate of electrochemical removal~\cite{krauskopf2019toward,krauskopf2020fast,yang2021interfacial_mech}. 
This imbalance initiates a Kirkendall-type void formation process~\cite{smigelskas1947zinc} at the Li/SE interface, resulting in contact loss, localized current hotspots, and ultimately cell degradation~\cite{krauskopf2019toward,Banerjee_2020,wang2024suppressing}.

Despite significant advances, state-of-the-art sulfide-based (e.g., \ce{Li_{10}GeP_2S_{12}}, LGPS) and oxide-based (e.g., \ce{Li7La3Zr2O12}, LLZO) electrolytes currently achieve CCDs of only 1--10 mA cm$^{-2}$, typically under applied pressure or elevated temperature~\cite{lu2021critical}. Under ambient conditions, the challenge is even more pronounced; for example, the natural CCD of a Li|LLZO|Li cell is limited to just 280~$\mu$A~cm$^{-2}$~\cite{flatscher2020natural}. This disparity arises because the kinetics of Li stripping are inherently slower than those of plating, making the CCD fundamentally stripping-limited~\cite{kasemchainan2019critical}. We propose that enhancing the vacancy diffusion rate within the Li anode is key to synchronizing the rates of stripping and plating, thereby increasing the practical CCD of solid-state batteries.

Lithium self-diffusion in Li metal is generally understood to follow a vacancy mechanism, with experimentally measured diffusion coefficients of $10^{-10}$--$10^{-11}$~cm$^2$ s$^{-1}$ at room temperature, indicative of intrinsically sluggish Li transport in the metal~\cite{mali19886li,frank1996first,messer1975nuclear}.
The vacancy mechanism is further supported by first-principles calculations. In Li metal, the calculated formation energy of a vacancy is 0.5 eV, lower than the 0.6 eV formation energy predicted for a self-interstitial along the $[111]$ direction~\cite{yang2021interfacial}, indicating that vacancies are energetically more favorable as intrinsic point defects. Although the equilibrium vacancy concentration is difficult to quantify experimentally using positron annihilation spectroscopy~\cite{hautojarvi1979positrons}, first-principles calculations estimate it to be on the order of $2\times10^{-7}$ at room temperature~\cite{frank1996first}. 
The role of vacancies in Li transport is further supported by the close agreement between experimentally measured vacancy diffusion coefficients, $\sim3\times10^{-4}$~cm$^2$~s$^{-1}$~\cite{mali19886li,messer1975nuclear,lodding1970isotope_li_diffusion}, and values obtained from molecular dynamics simulations~\cite{yang2021interfacial,sergeev2023anomalous,burov2026machinelearningassistedmolecular}. Importantly, the low equilibrium vacancy concentration limits the rate at which Li can be transported through the metal. Crucially, increasing the vacancy concentration in the anode enhances vacancy transport away from the Li/SE interface, thereby facilitating Li replenishment at the interface during stripping. This enhanced transport can help prevent the accumulation of vacancies and suppress the formation of the voids and pores commonly observed during cycling of Li|SE|Li cells~\cite{krauskopf2019toward,krauskopf2020fast,yang2021interfacial_mech}.

One of the most widely explored strategies for enhancing Li transport in the anode is alloying Li with a second element. Promising Li-based alloy anodes have been identified using a range of criteria, including energy density, with reported screening thresholds above 500~Wh~kg$^{-1}$~\cite{bi2022anode_towards}, specific capacity and average operating potential~\cite{lewis2022promise}, and overall electrochemical performance~\cite{jeong2024electrochemical}. 
More recently, machine-learning approaches applied to existing materials databases have enabled the rapid screening of large chemical spaces for promising alloy candidates. However, such data-driven approaches may overlook unconventional compositions; for example, Li--Ag alloys have recently emerged as promising anode materials despite not being consistently identified in earlier screening efforts~\cite{shi2025machine,huang2024unlocking}.

Despite these advances, many Li-based alloy systems require substantial concentrations of the solute element, typically $\gtrsim 20$~at.\%, which can significantly reduce the specific capacity of the anode. Moreover, alloying often alters the underlying crystal structure and may adversely affect Li-ion transport. An alternative approach is to introduce dilute solute atoms that form a solid solution while retaining the crystal structure of Li. Such solutes can modify the local vacancy thermodynamics and increase the effective vacancy concentration, potentially enhancing Li transport without substantially compromising the capacity of the Li host. To date, however, relatively few studies have explored this strategy, with most efforts focused on a limited number of elements, such as Ag and Zn~\cite{jin2020solid,ye2023solid}. Although these solid-solution anodes have demonstrated CCDs exceeding 3~mA~cm$^{-2}$, the atomistic and thermodynamic origins of their enhanced performance remain poorly understood.

In this work, we address this gap through a systematic first-principles investigation of Li-based solid solutions designed to enhance the effective vacancy concentration relative to pristine Li. We develop a simple yet robust descriptor based on solution energies and solute--vacancy binding energies to quantify the effect of different solute elements on vacancy thermodynamics. By combining DFT calculations with automated data processing and machine-learning techniques, and by leveraging existing materials databases such as the Materials Project~\cite{jain2013commentary_mpdatabase}, we systematically screen a broad chemical space of potential Li solid solutions. We further examine the roles of lattice relaxation and next-nearest-neighbor interactions in determining the calculated energetics. Finally, we assess the applicability of conventional empirical guidelines, including the Hume--Rothery rules, to this design problem and demonstrate why a more quantitative, energetically informed framework is required for the rational design of high-diffusivity Li alloys.

\section{Methods and computational setup}

\begin{figure*}[!htbp]
\centering
\includegraphics[width=0.99\textwidth]{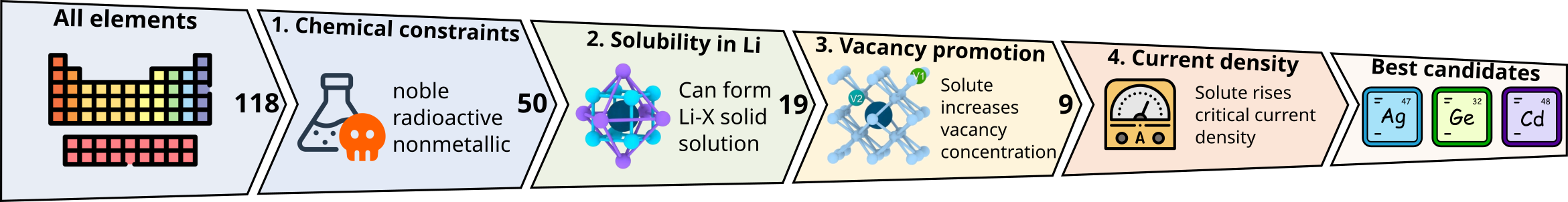}
\caption{Schematic of the multi-step screening workflow used to identify promising solute elements for Li-based solid solutions. The screening consists of four sequential filters: (1) exclusion of noble, radioactive, and non-metallic elements; (2) removal of elements with negligible solid solubility in Li; (3) selection of solutes that increase the equilibrium vacancy concentration; and (4) identification of candidates capable of increasing the critical current density (CCD) to a practically relevant threshold of 1~mA~cm$^{-2}$.}
\label{fig:scheme_main}
\end{figure*}

The solute screening workflow is illustrated in~\cref{fig:scheme_main}. The screening was performed in four sequential steps: (i) exclusion of noble, radioactive, and non-metallic elements; (ii) removal of elements with negligible solid solubility in Li; (iii) selection of solutes predicted to increase the equilibrium vacancy concentration; and (iv) identification of candidates capable of increasing the critical current density (CCD) to a practically relevant threshold of 1~mA~cm$^{-2}$.

\subsection{Density functional theory}

Density functional theory (DFT) calculations were performed using the projector augmented wave (PAW) method as implemented in the VASP code~\cite{Kresse:1996,Kresse:1999}. The generalized gradient approximation (GGA) in the Perdew--Burke--Ernzerhof (PBE) form~\cite{Perdew:PRL:1996,Perdew:PRL:1997} was used for the exchange-correlation (XC) functional. Electron-ion interactions were described using PAW potentials from version 54, with the same valence electron configurations  as in the Materials Project database~\cite{jain2013commentary_mpdatabase}. For the most promising solute candidates, lattice parameters were optimized by calculating the total energy for 11 unit-cell volumes spanning both contracted and expanded structures. The volume corresponding to the minimum total energy was then used in subsequent calculations to reduce the influence of Pulay stresses.

Nudged elastic band (NEB) calculations~\cite{jonsson1998nudged} were performed in VASP. Initial images were generated using image-dependent pair potential (IDPP) interpolation~\cite{henkelman2000_neb_improved} with seven intermediate images per path.

For structural relaxations, the total-energy convergence criterion was set to $10^{-5}$~eV, while the maximum residual force was required to be below 20~meV~\AA$^{-1}$. Brillouin-zone integration was performed using second-order Methfessel--Paxton smearing with a smearing width of 0.1~eV.
Four computational setups were employed: \textit{CoarseSet} --- used for screening all solutes, \textit{MPSet} --- used for phase-diagram calculations, \textit{TightSet} --- used for high-accuracy calculations on the most promising solutes, and \textit{NEBSet} --- used for all NEB calculations. 
The parameters for these sets were chosen based on convergence tests of total energies and relative energies such as the vacancy formation energy for \textit{CoarseSet}. 
The corresponding parameters and convergence results are provided in~\cref{sec:convergence_tests}.  For \textit{CoarseSet}, three valence electrons ($1s^2 2s^1$) were included for Li, the plane-wave cutoff energy ($E_{\mathrm{cut}}$) was set to 400~eV, and a $\Gamma$-centered $k$-point mesh with a spacing of less than 0.23~\AA$^{-1}$ was used. These parameters yielded estimated numerical uncertainties of approximately 10~meV per atom for total energies and 25~meV per atom for relative energies. The remaining computational setups employ higher cutoff energies and denser $k$-point meshes and are therefore expected to provide smaller numerical errors.

Structure generation and calculations were performed using the SIMAN~\cite{AKSYONOV2018449_siman}, Pymatgen~\cite{ong2013python_pymatgen}, and ASE~\cite{larsen2017atomic_ase} Python packages. Structural visualization was performed using VESTA~\cite{momma2011vesta}.

\subsection{Thermodynamic stability}

Only substitutional solutes were considered in this work.
The solution energy of a substitutional solute ($X$) was calculated as follows:
\begin{equation}
        E_{\mathrm{s}} = E(N-1 + X) - [(N-1)/N]E(N) - \mu_{X}, 
    \label{eq:sol_en}
\end{equation}
where $E(N-1+X)$ is the total energy of the supercell containing a substitutional  $X$ atom, $E(N)$ is the energy of the perfect $N$-atom supercell, and $\mu_{\mathit{X}}$ is the chemical potential of the solute element $X$.

The solute--vacancy binding energy ($E_{\mathrm{b}}$) was calculated as:

\begin{equation}
    E_{\mathrm{b}} = E(N-2+X+\mathrm{Vac, \, sep}) - E(N-2+X+\mathrm{Vac, \, NN}),
    \label{eq:e_bind}
\end{equation}
where $E(N-2+X+\mathrm{Vac, \, sep})$ and $E(N-2+X+\mathrm{Vac, \, NN})$ are the total energies of $4 \times 4 \times 4$ supercells (with a lattice constant of 13.757~\AA\ for pristine Li) containing a substitutional solute atom ($X$) and a vacancy at the maximum separation available in the supercell (11.91~\AA) and at a nearest-neighbor (NN) site (2.98~\AA), respectively. With this definition, a positive $E_{\mathrm{b}}$ indicates an attractive interaction between the solute and the vacancy, whereas a negative value indicates repulsion.

\subsection{Crystal orbital Hamilton population analysis}
The crystal orbital Hamilton population (COHP) method was employed to analyze chemical bonding in the Li-solute systems. Bonding information was extracted from the electronic density of states (DOS) via the crystal orbital overlap population (COOP) analysis, which quantifies the population of mutually overlapping crystal orbitals by multiplying DOS by the number of overlaps. This approach separates the total electron density into bonding and anti-bonding states. The sign of COHP is opposite to that of COOP, therefore $-$COHP diagrams are always plotted for clarity. The mathematical form of COHP is similar to that of a projected DOS equation, but the density-of-state matrix is weighted by the Hamiltonian matrix elements.

For a periodic crystal, COHP can be defined as follows~\cite{ruggiero2015origins}:

\begin{equation}
\mathrm{COHP}_{AB}(\epsilon)=
\sum_{\nu \in A}\sum_{\mu \in B} \mathrm{COHP}_{\mu\nu}(\epsilon)
\label{eq:cohp_total}
\end{equation}

\begin{equation}
\begin{split}
\mathrm{COHP}_{\mu\nu}(\epsilon) = \frac{2}{V_{BZ}} \int_{BZ} H_{\mu\nu}(k) & \\
\times \left\{ \sum_j c^*_{\mu j}(k) c_{\nu j}(k) \, \delta\bigl(\epsilon - \epsilon_j(k)\bigr) \right\} & \, dk,
\end{split}
\label{eq:cohp_integrated}
\end{equation}
where $H_{\mu\nu}(k)$ is the Hamiltonian matrix in reciprocal space, $c_{\mu j}(k)$ are the expansion coefficients of the $j$-th occupied crystal orbital with the energy $\epsilon_j(k)$ in terms of Bloch functions, and integration over $k$ is performed in the Brillouin zone ($BZ$) with the volume $V_{BZ}$. COHP calculations and analysis were carried out using the LOBSTER algorithm~\cite{deringer2011crystal,Nelson2020}.

\subsection{Machine-learning calculations}

A suite of machine-learning (ML) models was employed, including linear regularized regressors (Ridge, Lasso, and ElasticNet), support vector regression (SVR), ensemble methods (Random Forest and GradientBoosting), and gradient-boosting frameworks (XGBoost and CatBoost). Implementations were based on Scikit-learn~\cite{pedregosa2011scikit} and the respective native libraries~\cite{chen2016xgboost,prokhorenkova2018catboost}. Feature representation was constructed using numerical descriptor vectors generated via MatMiner~\cite{ward2018matminer}, supplemented by Hume--Rothery rule-based features and materials data obtained from the Materials Project API~\cite{ONG2015209_API}.

Models were trained on the 19 optimized solute candidates using an opt-only hold-out split (9 train / 10 test). Model selection used 3-fold cross-validation RMSE on the training fold, and hyperparameters were optimized via randomized search. All code, parameters, and results are available in the accompanying GitHub repository~\href{https://github.com/purphy42/solid_solutions_li}{github.com/purphy42/solid\_solutions\_li}.

\section{Results}

\subsection{Criteria for the solute candidates}

The solute candidates were screened according to the following criteria:

\begin{enumerate}
\item \label{eq:funnel_rule_1} Noble gases, radioactive elements, and nonmetallic elements are excluded from the screening.
\item \label{eq:funnel_rule_2} The solution energy in bcc Li was required to satisfy $E_{\mathrm{s}} < 1.0$~eV.
\item \label{eq:funnel_rule_3} The best candidates were selected according to the effective vacancy formation energy defined as:
\begin{equation}
E_{\mathrm{eff}} =
\begin{cases}
E_{\mathrm{s}} - E_{\mathrm{b}} < 0, & \text{if } E_{\mathrm{s}} > 0 \\[4pt]
-E_{\mathrm{b}} < 0, & \text{if } E_{\mathrm{s}} < 0
\end{cases}
\label{eq:vac_eff_energy}
\end{equation}
% Candidates are accepted if $E_{\mathrm{eff}} < 0$.
and candidates were retained if $E_{\mathrm{eff}} < 0$.
\item \label{eq:funnel_rule_4} The estimated critical current density was calculated as
\begin{equation}
J_{\mathrm{crit}}^{\mathrm{max}}\simeq
\left[1+c_{\max}\exp\!\left(E_{\mathrm{b}}/k_{\mathrm{B}}T\right) \right]\times CCD^{\mathrm{natural}}_{\mathrm{SE}},
\label{eq:ccd_at_solubility}
\end{equation}
where $c_{\mathrm{max}}$ is the atomic fraction of the solute element at its solubility limit (see~\cref{tab:solubility}), and $CCD^{\mathrm{natural}}_{\mathrm{SE}}$ is the natural critical current density of the anode|SE interface (in $\mathrm{mA~cm^{-2}}$). Candidates are accepted if $J_{\mathrm{crit}}^{\mathrm{max}} \ge 1~\mathrm{mA~cm^{-2}}$. This equation is derived from the Lomer model~\cite{lomer1958equilibrium} at dilute solute limit (see~\cref{sec:eff_vac_increase}).
\end{enumerate}

The second criterion limits the solution energy to identify solutes with appreciable solubility in bcc Li. The third criterion selects solutes that can lower the effective vacancy formation energy and thereby increase the equilibrium vacancy concentration; its derivation is provided in~\cref{sec:vac_conditions}. Finally, the fourth criterion ensures that the predicted enhancement in vacancy concentration is sufficient to achieve a critical current density (CCD) of at least 1~mA~cm\(^{-2}\). This threshold represents a practically relevant intermediate target toward the industrially desirable CCD of approximately 4~mA~cm\(^{-2}\)~\cite{lu2021critical}.

\subsection{Phase diagrams and chemical potentials}
The chemical potential of the solute is a key input for evaluating the thermodynamic stability of substitutional solutes in Li.  To obtain physically meaningful chemical potentials under Li-rich conditions, we identified the stable Li--$X$ phases that are in equilibrium with Li metal. Binary phase diagrams were constructed following the methodology described in~\cref{sec:calculated_phase_diagrams}, using all available information about stable phases for each Li--$X$ system. The most stable Li-rich phase identified from the obtained phase diagrams was selected as the end-member for the corresponding solute. 
The energy of this phase was then recalculated using our computational setup (\textit{MPSet}), to obtain more accurate solute chemical potentials via~\cref{eq:chem_potentials}. The resulting absolute and relative chemical potentials of Li and the 50 alloying elements ($X$)  are summarized in~\cref{tab:chemical_potentials_all}.
For most solutes, the end-member phases identified by the automated workflow are consistent with the experimentally reported phase diagrams. Four exceptions were identified: Cd, Sn, Pb, and Hg. For Cd, the calculated end-member is \ce{LiCd}, while the closest experimentally reported phase is \ce{Li3Cd}, which is listed as a non-experimental (theoretical) phase in the Materials Project database (see~\cref{fig:phase_diagram_cd}). For Sn, the calculated end-member is \ce{Li17Sn4}, whereas the closest experimentally reported phase is \ce{Li22Sn5} (see~\cref{fig:phase_diagram_sn}). For Pb, the workflow identifies \ce{Li17Pb4}, compared with the experimental phase \ce{Li4Pb} (see~\cref{fig:phase_diagram_pb}). For Hg, \ce{Li3Hg} (mp-1646) was selected as the end-member, whereas the experimentally reported Li-rich phase is \ce{Li6Hg} (see~\cref{fig:phase_diagram_hg}). The latter could not be included directly because its crystal structure has not been reported and no corresponding structure is available in the Materials Project database. The discrepancies for Sn and Pb are unlikely to substantially affect the calculated chemical potentials, since the Li atomic fractions of the selected and experimentally reported phases differ by less than 1\% (e.g., for Sn, $\frac{22}{27}-\frac{17}{21} \simeq 0.5\%$). We therefore expect the resulting chemical potentials to closely approximate those associated with the experimental end-member phases. Importantly, we retained the phases selected by the automated workflow throughout the screening procedure rather than introducing manual corrections. This choice ensures that the subsequent solute screening remains fully automated and internally consistent across all Li--$X$ systems.

\subsection{Thermodynamic properties}

\subsubsection{All candidates.} 
\begin{figure}[!htbp]
\centering
\includegraphics[width=0.90\columnwidth]{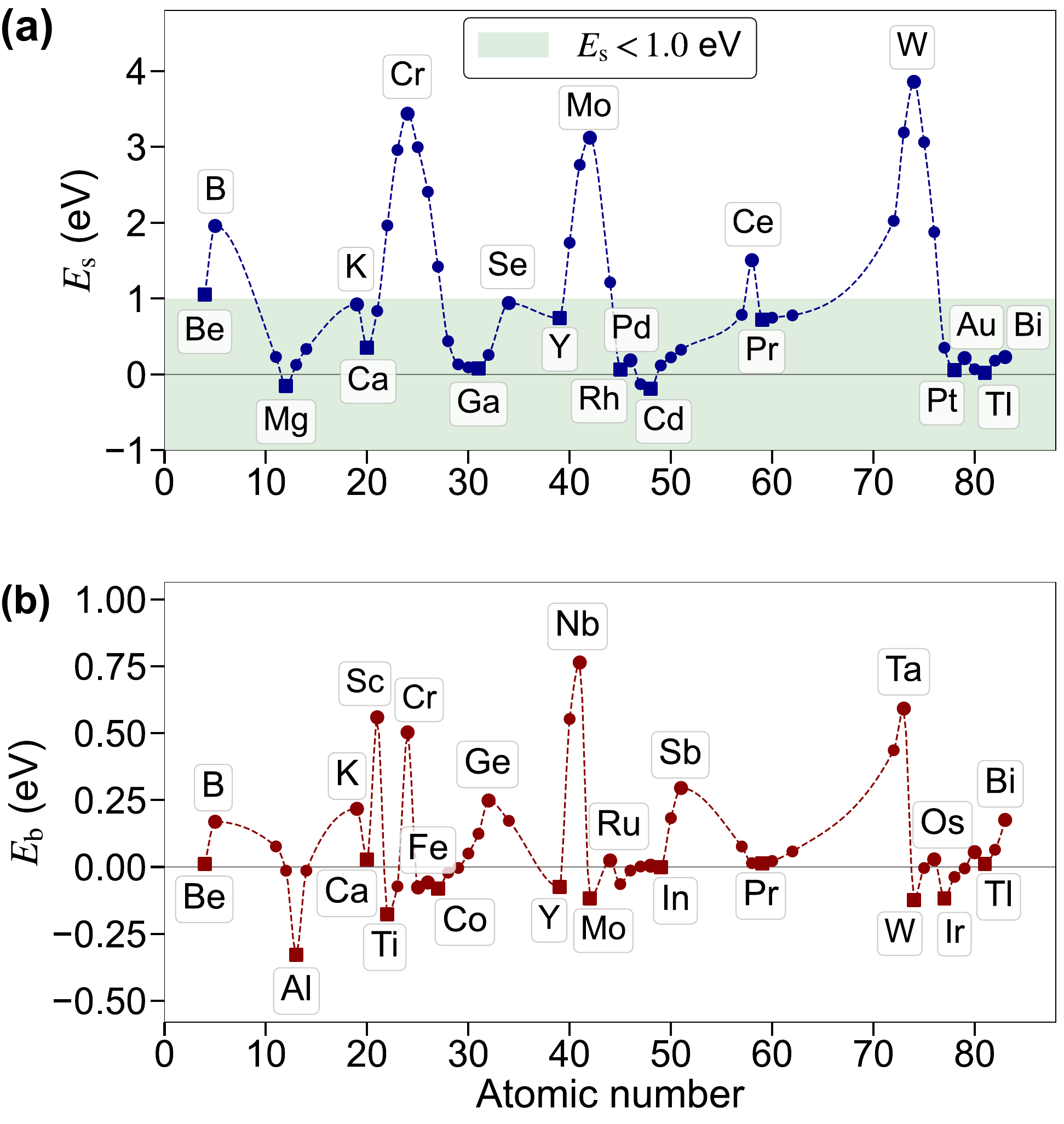}
\caption{(a) Solution energies ($E_{\mathrm{s}}$) for all 50 solutes, calculated using~\cref{eq:sol_en}. The green shaded region marks solutes satisfying the solubility criterion $E_{\mathrm{s}} < 1.0$~eV (\cref{eq:funnel_rule_2}). (b) Solute--vacancy binding energies ($E_{\mathrm{b}}$) for the same set of solutes, calculated using~\cref{eq:e_bind}. Local extrema are labeled in both panels to highlight the trends.} 
\label{fig:sol_bind_num}
\end{figure}

The solution energies ($E_{\mathrm{s}}$) were calculated using~\cref{eq:sol_en} and the \textit{CoarseSet} setup, with the results visualized in~\cref{fig:sol_bind_num}.a. The profile exhibits strong periodic trends, with a local minimum at Cd ($-0.19$~eV) and pronounced peaks for $d$-block (Ti, V, Cr, Mn, Fe, Co, Zr, Nb, Mo, Ru, Hf, Ta, W) and $f$-block metals (Ce, Pr, Nd, Sm). This behavior reflects the tendency of Cd to form stable Li-rich intermetallic phases ($\ce{Li_{3}Cd}$), which lowers the energy cost of substituting a Li atom. The elements with the lowest solution energies are Cd, Mg, and Ag, at $-0.19$, $-0.15$, and $-0.13$~eV, respectively. Consequently, they also exhibit the highest solubility limits, of 18, 76, and 9\%, respectively (see~\cref{tab:solubility}). In contrast, transition metals with partially filled $d$ shells experience larger lattice strain and electronic repulsion in the Li matrix, leading to positive solution energies (W, Cr, Ta, Mo, and others). The solubility threshold in~\cref{eq:funnel_rule_2} ($E_{\mathrm{s}} < 1.0$~eV) successfully narrowed the candidate pool to 19 candidates. However, it is important to note that a small $E_{\mathrm{s}}$ ensures that solute incorporation is thermodynamically feasible, but does not guarantee an increase in the effective vacancy concentration.

Solute--vacancy binding energies were calculated using~\cref{eq:e_bind} and are shown in~\cref{fig:sol_bind_num}.b. In contrast to the solution energies, the binding energies display no clear periodic trend with atomic number, suggesting that local electronic and strain effects dominate over simple atomic number or group dependencies.

In~\cref{sec:latt_opt_effect}, we assessed the effect of lattice optimization on the solution energies $E_{\mathrm{s}}$, vacancy binding energies $E_{\mathrm{b}}$, and chemical potentials for the best-performing solutes. We also examined the influence of including vacancies in the second-nearest-neighbor (2NN, 3.41~\AA\ distance) coordination shell. The resulting max absolute errors (MaxAE), mean errors (ME) and mean absolute errors (MAE) are summarized in~\cref{tb:effect_of_latt_opt}. The MaxAEs are 49~meV for $E_{\mathrm{s}}$ and 156~meV for $E_{\mathrm{b}}$. To avoid excluding viable solutes, we therefore set the tolerance for~\cref{eq:vac_eff_energy} to the sum of these absolute errors as 0.21~eV.

\cref{tab:data_therm_all} summarizes all computed thermodynamic quantities, including solute solution energies ($E_{\mathrm{s}}$), vacancy binding energies ($E_{\mathrm{b}}$), effective energies ($E_{\mathrm{eff}}$), and magnetic moments of solutes in the Li host. Screening the 50 initial solutes using~\cref{eq:funnel_rule_2} and~\cref{eq:funnel_rule_3} yielded 19 candidates that proceeded to the next funnel stage. Importantly, all solutes that passed the screening exhibited zero magnetic moments in their reference states. Therefore, spin-polarized calculations were omitted in the subsequent \textit{TightSet} setup, reducing computational cost without compromising accuracy.

\subsubsection{Best candidates.} Next, the 19 best solutes were recalculated with higher accuracy using the \textit{TightSet} setup. All corresponding end-members and three supercell structures (pristine, solute-containing, and vacancy-containing) were also recalculated at this level of theory. For each solute, we considered a vacancy located on both the first (1NN) and second (2NN) coordination spheres relative to the solute site. The solute--vacancy binding energy was then defined as the maximum energy gain (i.e., the lowest-energy supercell configuration containing both a solute and a vacancy) across these two vacancy positions. The results of this refined analysis are summarized in~\cref{tb:best_solutes}.

\begin{figure}[!htbp]
\centering
\includegraphics[width=0.99\columnwidth]{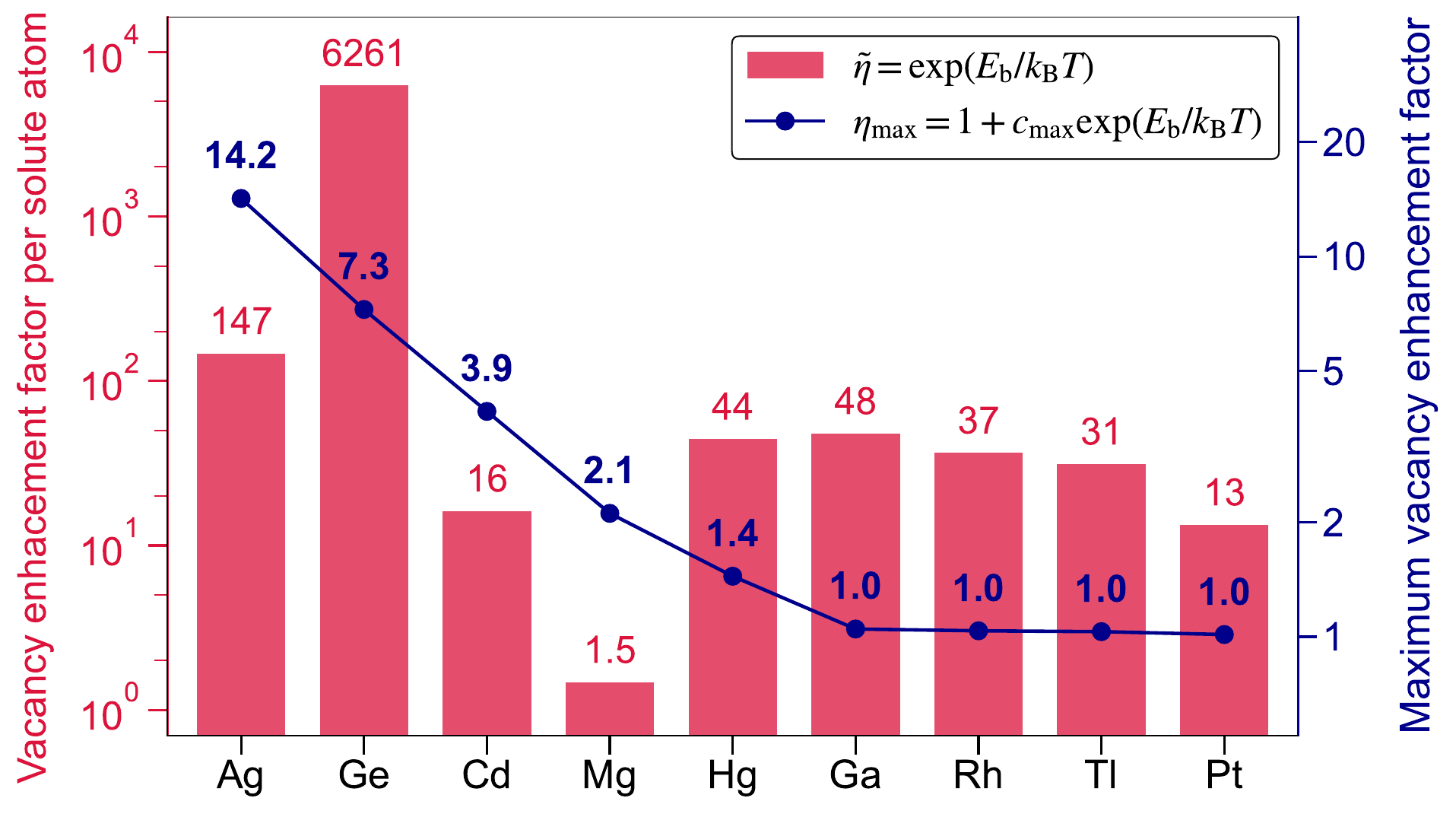}
\caption{Vacancy capture and enhancement metrics for the most promising solute candidates at room temperature ($300\text{ K}$). Red bars (left axis, logarithmic scale) show the Boltzmann factor $\tilde{\eta} = \exp(E_{\mathrm{b}}/k_{\mathrm{B}}T)$, representing the vacancy enhancement per solute atom, where $E_{\mathrm{b}}$ is the vacancy binding energy. The blue curve (right axis, logarithmic scale) shows the corresponding maximum vacancy enhancement factor $\eta_{\mathrm{max}} = 1 + c_{\mathrm{max}} \exp(E_{\mathrm{b}}/k_{\mathrm{B}}T)$, evaluated at the maximum solubility limit $c_{\mathrm{max}}$ of each element (see \cref{tab:solubility}).}

\label{fig:final_plot_eff_en}
\end{figure}

A solute is considered suitable for increasing the effective vacancy concentration in bcc Li when its calculated effective energy ($E_{\mathrm{eff}}$) is negative. The promising solutes satisfying this criterion are listed in~\cref{tb:vac_conc_gain} and visualized in~\cref{fig:final_plot_eff_en}. Among these, Ge and Ag provide the highest enhancement per solute atom ($\tilde{\eta} = \exp(E_{\mathrm{b}}/k_{\mathrm{B}}T)$, see \cref{tb:vac_conc_gain} in \cref{sec:eff_vac_increase}), increasing the vacancy concentration by factors of 6261 and 147, respectively. In contrast, Rh, Pt, Tl, Hg, and Ga yield almost negligible changes in the effective vacancy concentration at the solubility limit (see~\cref{sec:eff_vac_increase}).

\subsection{Minimum critical current density threshold}

\begin{figure}[!htbp]
\centering
\includegraphics[width=0.90\columnwidth]{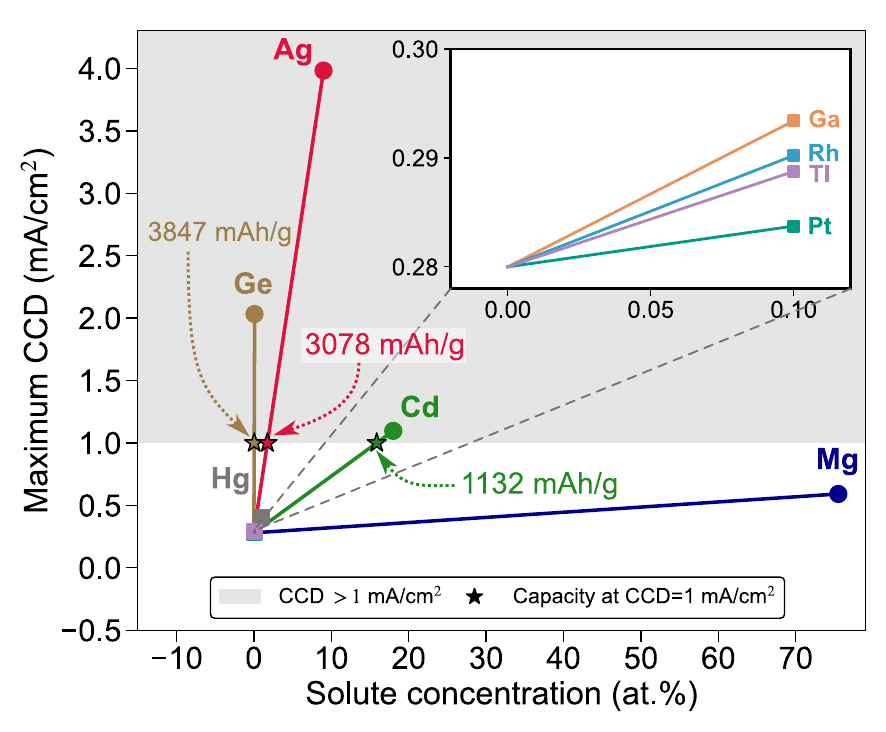}
\caption{Critical current density (CCD) as a function of the molar concentration of solute elements (see~\cref{sec:ccd_estimations_si}). The gray-shaded area denotes the baseline performance threshold of $1\ \text{mA cm}^{-2}$. The main plot (circles) and inset (squares) display the maximum CCD gain achieved at the solubility limit for each element (see~\cref{tab:solubility}). Stars indicate the minimum solute concentration required to theoretically meet the $1\ \text{mA cm}^{-2}$ threshold, alongside the corresponding theoretical capacity. }
\label{fig:ccd_gain_final}
\end{figure}

Next, we estimated the critical current density (CCD) of the best solute candidates for alloys, as described in~\cref{sec:ccd_estimations_si}. The natural CCD limit is achieved by void formation at the Li/SE interface, rapid interface resistance increase and further contact loss, as was shown by Krauskopf~\textit{et al.} for Li/LLZO interface~\cite{krauskopf2019toward, krauskopf2020fast}. This finding is consistent with slower stripping and lower stripping current than plating~\cite{kasemchainan2019critical, lu2021critical}. Assuming an optimistic scenario, an $N$-fold increase in the effective vacancy concentration in Li metal would translate to a proportional $N$-fold increase in the CCD limit. The full methodological and theoretical framework underlying this estimation is presented in~\cref{sec:ccd_estimations_si}, where we estimated both the vacancy enhancement factor per atomic fraction of solute ($\tilde{\eta}$) and that at the solubility limit ($\eta_{\mathrm{max}}$). Additionally, we adopt the optimistic premise that the maximum solute solubility achieved at elevated temperatures can be quenched and retained as a metastable state at room temperature upon alloying. The resulting CCD predictions are summarized in~\cref{tab:ccd_gain_all} and illustrated in~\cref{fig:ccd_gain_final}.

In addition, we loosened the CCD threshold to 1~mA~cm$^{-2}$, below the industrial benchmark of 4~mA~cm$^{-2}$~\cite{lu2021critical}, to account for the substantial changes in vacancy diffusivity that can occur upon alloying, as previously found in solid electrolytes~\cite{li2025superionic}. Even with this reduced requirement, only Ag, Ge, and Cd reach the threshold, at reasonable concentrations of 9, 0.1, and 18 at.\%, respectively. Among the other candidates, Mg is still promising for future studies due to its high solubility limit ($\approx 75.5$~at.\%), which is expected to significantly alter vacancy migration kinetics in Li--Mg solid solutions. The remaining elements do not meet the criterion, primarily because of their low solubility limits.

\subsection{DFT-NEB calculations}

\subsubsection{Percolation pathways}

\begin{figure*}[!htp]
\centering
\includegraphics[width=0.9\textwidth]{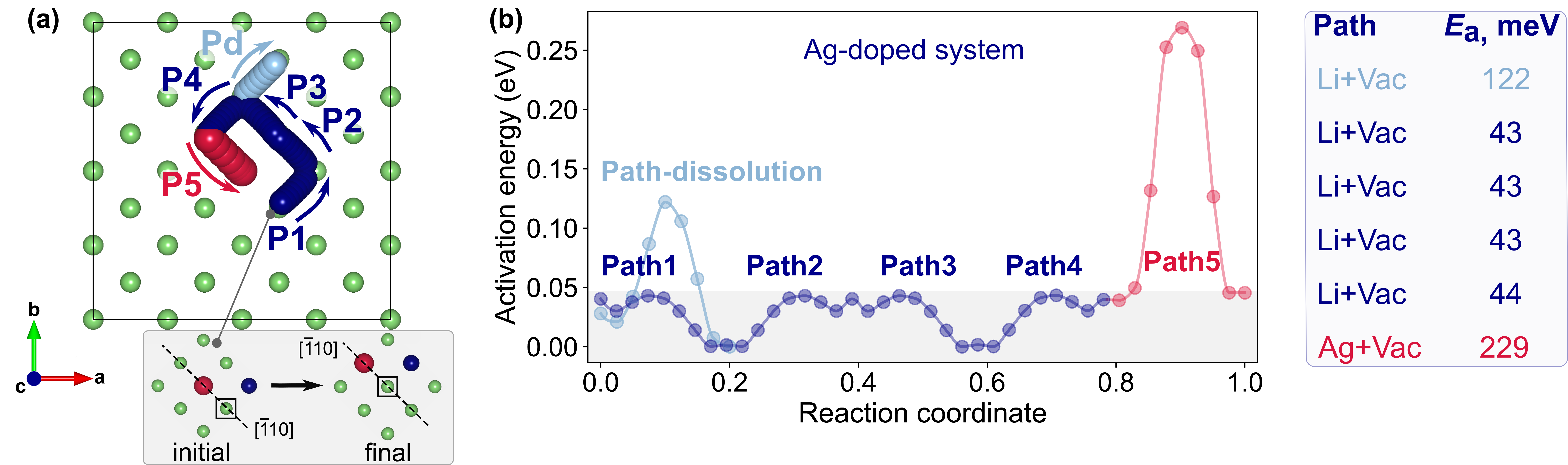}
\caption{DFT-NEB migration barriers for Li vacancy-mediated diffusion in solid-solution Li electrodes. (a) Crystal structure of the Ag-alloyed system (\ce{Li_{127}Ag}), showing the percolation pathway along the $[\overline{1}10]$ direction.  Blue arrows indicate nearest-neighbor (NN) Li--vacancy exchanges (paths 1--4), the red arrow indicates the NN Ag--vacancy exchange (path 5), and path D represents the dissolution of the Ag--vacancy complex via the NN hop. The inset depicts the initial and final configurations of the migrating Li, Ag, and vacancy species. (b) Corresponding energy profiles along the migration coordinate. The inset table lists the extracted activation energies $E_{\mathrm{a}}$: 122~meV (sky-blue) for Ag--Li dissolution, 43--44~meV (blue) for Li--vacancy exchanges, and 229~meV (red) for Ag--vacancy exchange.}
\label{fig:neb_panel_main}
\end{figure*}

First, we investigated how the presence of solutes (Ag and Mg) affects the migration of a nearby vacancy. Ag was selected as the solute with the highest maximum effective vacancy enhancement factor, while Mg was chosen as the solute with the lowest migration barrier. To this end, we constructed a five-stage percolation process for the solute--vacancy complex, consisting of four nearest-neighbor Li migrations and one nearest-neighbor solute migration. This process provides a percolation pathway for Li--solute migration along the $[\overline{1}10]$ direction. The results are visualized for Ag in~\cref{fig:neb_panel_main}.a--b and for Mg in~\cref{fig:neb_all_mg}. In both cases, the migration barriers for Li hops remain almost unchanged compared to the pristine vacancy-mediated Li migration, where a single vacancy exists at the site into which the migrating Li atom moves. Thus, the presence of a solute does not significantly influence the migration energetics of Li vacancies.

However, the Ag migration barrier is substantially higher than that for Li (0.22 eV vs. 0.04 eV, see \cref{fig:solutes_NN_NNN}.a), indicating that Ag diffuses much more slowly than Li within the single-hop vacancy model. Furthermore, the positive Ag--vacancy binding energy (0.13 eV) indicates that Ag acts as a trapping center for vacancies, effectively hampering their long-range diffusion. To quantitatively estimate the net effect of Ag alloying on the vacancy diffusion coefficient, we calculated the kinetic barrier required to separate the Ag--vacancy complex (shown in sky blue in \cref{fig:neb_panel_main}). The calculated escape barrier is $\sim$0.12 eV, which is significantly higher than the pristine migration barrier of 0.04 eV.

To capture the net effect, accounting for the enhanced vacancy concentration, the kinetic barrier for vacancy escape from the Ag environment, and the limited coordination volume around the solute, we performed a comprehensive analysis (\cref{sec:ag_vacancy_trapping}). This analysis reveals that the combined kinetic and volumetric effects reduce the net enhancement of vacancy diffusion from the purely thermodynamic estimate of $\sim$146.9 to approximately 1.5 at dilute alloying limit (1 at.\% Ag). However, a more accurate assessment requires large-scale MD simulations that naturally include lattice vibrations, finite-temperature effects, cooperative migration, and alternative mechanisms such as interstitial hopping, which are not captured in the present static DFT-NEB framework.

\subsubsection{Cooperative multi-ion diffusion effects}

The cooperative migration of several atoms can reduce the activation energy of diffusion, as previously shown by He~\textit{et al.}~\cite{He2017}. To model the simplest cooperative process, we considered two-stage jumps of the Li--solute pair: solute$\xleftrightarrow{\text{Path 1}}$Vac and solute$\xleftrightarrow{\text{Path 2}}$Vac, as well as the corresponding process in which both jumps occur simultaneously. In this case, Li migration requires a NNN jump, as no such simple two-stage mechanism exists for NN jumps. The results for Ag are visualized in~\cref{fig:neb_coop_ag} and for Cd in~\cref{fig:neb_coop_cd}.

As a result, the trajectories of Li and Cd migration become slightly curved, and the activation energy of this cooperative process is reduced from 281~meV (for the single-ion process) to 209~meV. Similarly, for Ag, the Li NNN barrier decreases from 183~meV (see~\cref{fig:neb_li_both}) to 103~meV along path 1 in~\cref{fig:neb_coop_ag}, accompanied by a 0.7~\AA~displacement of the Ag atom toward the vacancy. Collectively, these observations suggest that cooperative effects can substantially lower the effective solute--vacancy migration barrier. While the NN migration mechanism was not explicitly modeled due to the computational expense of the DFT-NEB method and the combinatorial explosion of possible configurations, our results strongly underscore the need for large-scale molecular dynamics simulations to capture such multi-ion cooperative effects and derive reliable vacancy diffusion coefficients in solute-containing systems.

\subsection{Hume--Rothery rules validation}

In this work, we did not use any empirical rules to filter the first batch of candidates. However, several such rules do exist, for instance, the Hume--Rothery (H--R) rules, which govern the conditions under which an element can dissolve in a metal to form a solid solution~\cite{hume1934freezing}. The four basic rules are as follows:

\begin{enumerate}
    \item \label{eq:hr_rules_1} The atomic radius of the solute and solvent atoms must differ by no more than 15\%:
    $$
    \Delta r_{i} = \left( \frac{r_{\mathrm{solute}} - r_{\mathrm{solvent}}}{r_{\mathrm{solvent}}} \right) \times 100\% \leq 15\%
    $$
    \item \label{eq:hr_rules_2} The crystal structures of solute and solvent must be similar.
    \item \label{eq:hr_rules_3} Complete solubility occurs when the solvent and solute have the same valency. A metal is more likely to dissolve a metal of higher valency than vice versa.
    \item \label{eq:hr_rules_4} The solute and solvent should have similar electronegativity. If the electronegativity difference is too great, the metals tend to form intermetallic compounds instead of solid solutions:
    $$
    \Delta \chi_{i} = \left( \frac{\chi_{\mathrm{solute}} - \chi_{\mathrm{solvent}}}{\chi_{\mathrm{solvent}}} \right) \times 100\% \leq \chi_{\mathrm{opt}}
    $$
\end{enumerate}

First, we divided the solutes into four categories: ``extensive solubility'', when the solid-solution region exceeds 1\%; ``limited solubility'', when it is approximately 1\%; ``minor solubility'', when it is less than 1\%; and ``insoluble'', when the solute does not form a solid solution with Li.

For the first Hume--Rothery rule validation (\cref{eq:hr_rules_1}), we used metallic radii for elements that exist in such phases and covalent radii otherwise. The results are visualized in~\cref{fig:hr_rule_1}. The original threshold for the radii difference, which was defined for atomic radii, is 15\%. We demonstrate that this criterion should be relaxed by a factor of three (to 44\%) to correctly predict all solid solutions with Li.

We also verified the crystal structure (\cref{eq:hr_rules_2},~\cref{fig:hr_rule_2}) and valency (\cref{eq:hr_rules_3},~\cref{fig:hr_rule_3}) criteria. Both were found to be unsuitable for Li-based solid solutions, as they correctly predict only 14\% and 19\% of solid solutions, respectively, while discarding all others. Consequently, these rules cannot serve as pre-conditions in our screening, as doing so would eliminate many potential solute candidates.

\begin{figure}[!htbp]
\centering
\includegraphics[width=0.8\columnwidth]{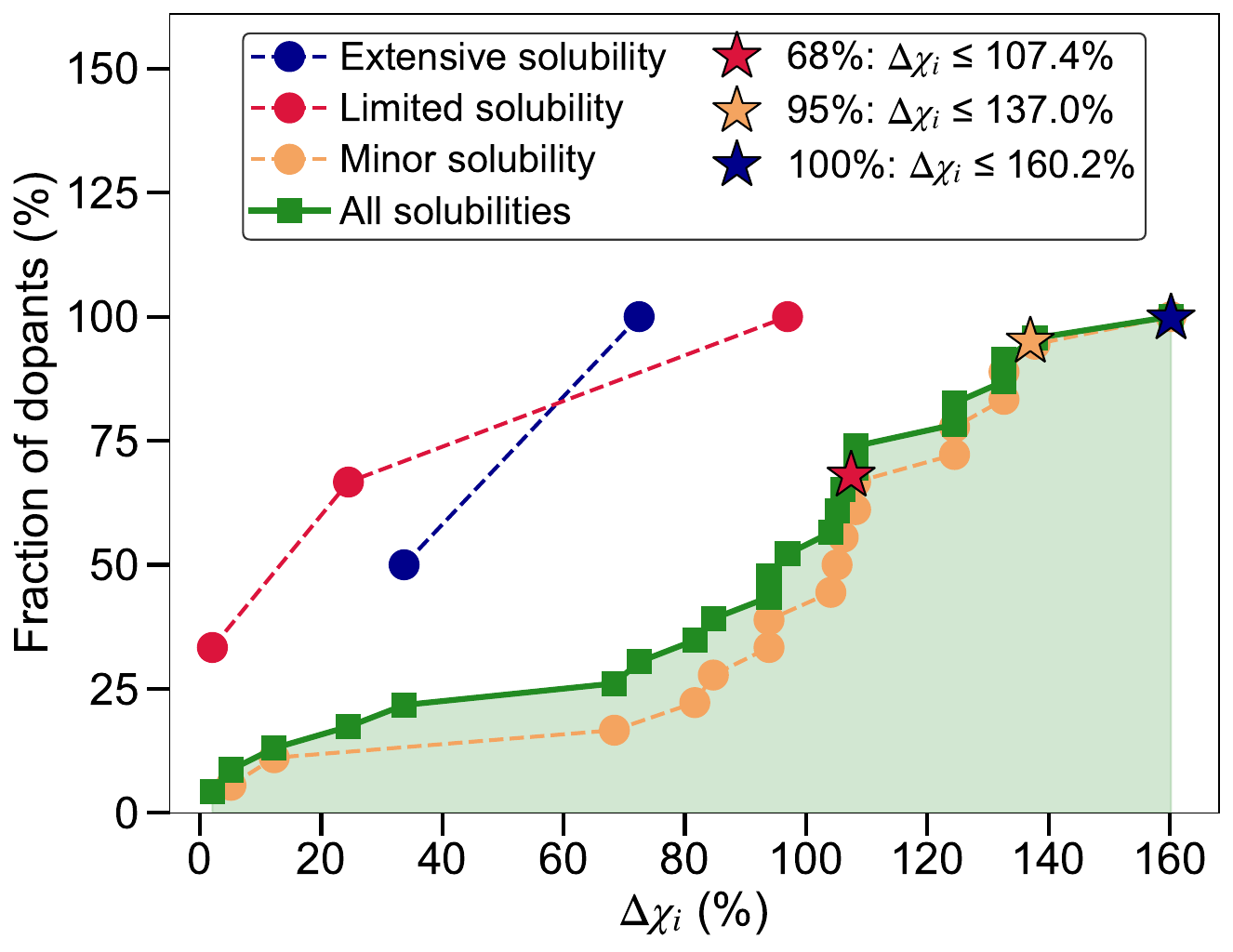}
\caption{Validation of the fourth Hume--Rothery rule, related to the difference in host--solute electronegativities. Blue denotes extensive solubility ($>1\%$), red denotes limited solubility ($\sim 1\%$), orange denotes minor solubility ($<1\%$), and green denotes the combined region. Star symbols mark $\sigma$ (65\% precision), $2\sigma$ (95\% precision), and the ideal prediction (100\% precision).}
\label{fig:hr_rule_4}
\end{figure}

Finally, we validated the electronegativity criterion (\cref{eq:hr_rules_4}), with results shown in~\cref{fig:hr_rule_4}. Owing to the low electronegativity of Li (0.98 on the Pauling scale~\cite{pauling1932nature}), the threshold for the relative difference in Li--solute electronegativities must be increased to 160\% to correctly predict all solid solutions with Li. As a result, solutes with electronegativities up to 2.54 should be considered, and this rule does not allow us to discard any potential solutes during our screening funnel.

\subsection{Crystal orbital Hamilton population analysis}

Crystal orbital Hamilton population (COHP) analysis was used to interpret the solute--vacancy interaction, complementing the binding energy $E_{\mathrm{b}}$, which quantifies the strength of the interaction but not its chemical origin. Whereas $E_{\mathrm{b}}$ is a single scalar, COHP resolves the interaction into bonding and antibonding contributions on individual Li--solute bonds, providing a local, chemically interpretable picture of how a vacancy modifies the surrounding bonding. Negative COHP values denote bonding states, whereas positive values denote antibonding states. The applicability of COHP to metallic systems was previously demonstrated by Deringer~\textit{et al.}~\cite{deringer2011crystal} and further validated by our COHP and electron charge-difference tests in~\cref{sec:cohp_justified}.

\begin{figure*}[!htbp]
\centering
\includegraphics[width=0.99\textwidth]{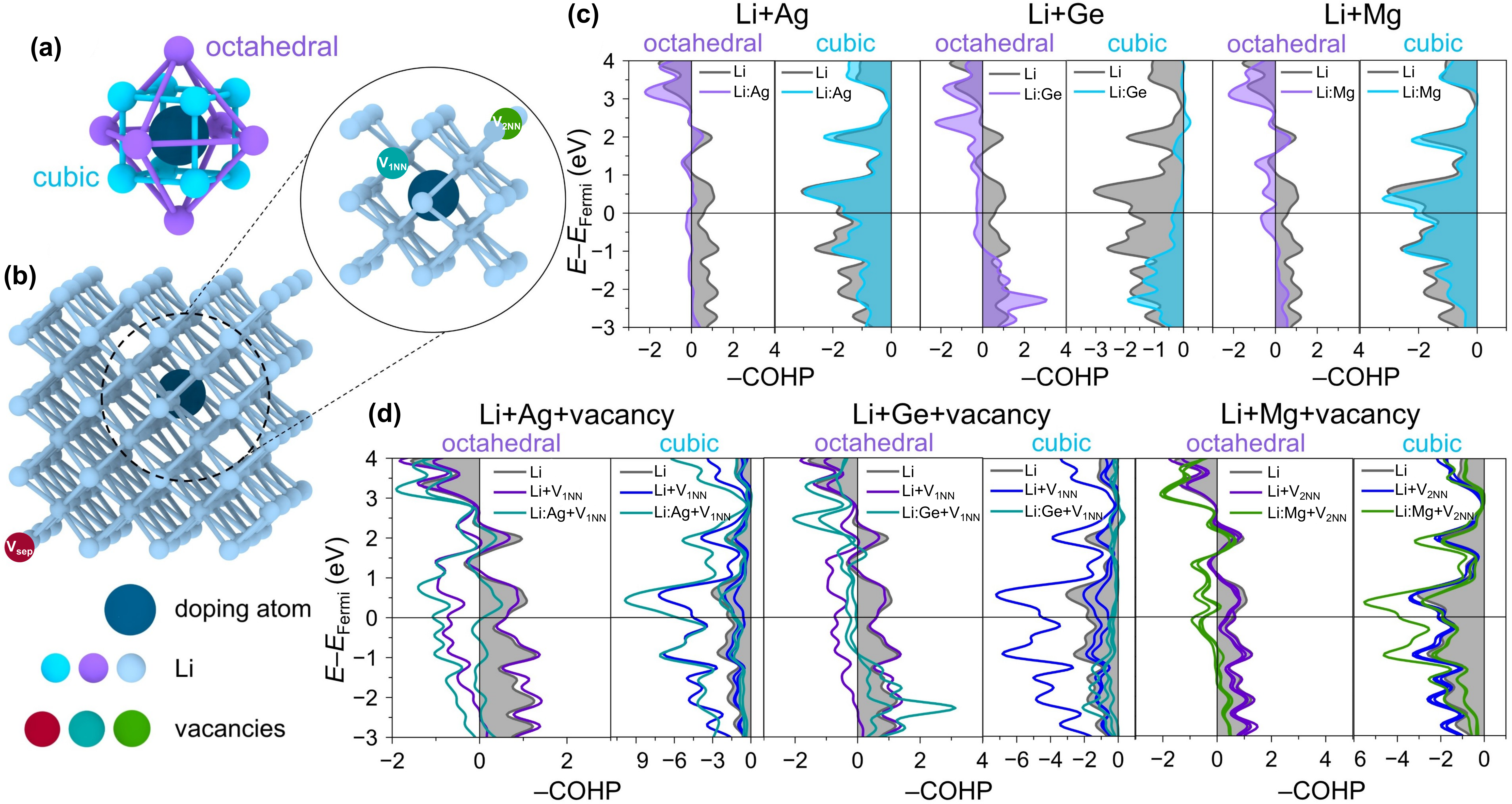}
\caption{(a) Cubic (blue) and octahedral (violet) coordination environments of the solute atom. (b) Crystal structure of the alloyed Li supercell, showing the solute atom and the vacancy sites ($\mathrm{V_{1NN}}$ and $\mathrm{V_{2NN}}$). (c) Crystal orbital Hamilton population (COHP) plots for Li--solute bonds in the cubic (blue) and octahedral (violet) environments for supercells in which one Li atom is replaced by a solute element: Ag, Ge, and Mg. (d) COHPs for these supercells with the most stable vacancy among the $\mathrm{V_{1NN}}$ and $\mathrm{V_{2NN}}$ positions (see~\cref{tb:best_solutes}). For reference, --COHP curves for pristine Li in both the cubic and octahedral environments are shown in gray.}
\label{fig:cohp_main}
\end{figure*}

A negative $\Delta\mathrm{ICOHP}$ therefore indicates that the selected bonds become stronger after the vacancy is introduced. Since COHP cannot be calculated for a vacancy itself, we evaluated it within cubic and octahedral environments (see~\cref{fig:cohp_main}.a--b), which allows us to track the change in different bond lengths arising from the different most stable vacancy positions (nearest-neighbor, $\mathrm{V_{1NN}}$, or next-nearest-neighbor, $\mathrm{V_{2NN}}$) of different solutes. Furthermore, since a vacancy alters different bond lengths differently, we distinguished these cases in the COHP figures in~\cref{fig:cohp_main} (see~\cref{tab:icohp_ag_li_var}). Curves for all solute elements, including those for the separated vacancy ($\mathrm{V_{sep}}$), are shown in~\cref{fig:llli} in~\cref{sec:cohp_calculated}.

A vacancy in pristine bcc Li weakens the surrounding Li--Li bonds relative to vacancy-free Li. The Li--Li ICOHP changes from $-1.85$~eV per bond to $-1.24$~eV per bond ($\Delta\mathrm{ICOHP} = +0.61$~eV, see~\cref{tab:icohp}). This vacancy-containing Li--Li bond was used as a reference for the alloyed structures.

Replacing one Li atom with a solute element, with no vacancy present, does not shift the local bonding in a single direction. Relative to the Li--Li ICOHP of $-1.85$~eV, several solute elements yield more negative Li--solute ICOHPs, including Ag ($-3.61$~eV) and Rh ($-5.79$~eV). Others are positive, including Ge ($+0.68$~eV), Ga ($+0.28$~eV), Si ($+0.60$~eV), and Al ($+0.42$~eV). However, these absolute values cannot be used to compare solute elements, since the presence of a solute simultaneously changes the hybridization, the local potential, and the bond length, as well as the electron count.

Therefore, we used $\Delta\mathrm{ICOHP}$ to quantify the effect of a vacancy adjacent to the solute element and how it alters the Li--solute bonding. This quantity is calculated as follows:

\begin{equation}
\Delta\mathrm{ICOHP} = \mathrm{ICOHP}_{X+\mathrm{vac}} - \mathrm{ICOHP}_{X},
\label{eq:delta_icohp}
\end{equation}
where $\mathrm{ICOHP}_{X+\mathrm{vac}}$ and $\mathrm{ICOHP}_{X}$ are the ICOHPs of the supercells with the solute element $X$ with and without a vacancy near it (1NN or 2NN), respectively.

This quantity describes the selected local bonds. In~\cref{tab:icohp}, Ge exhibits the most negative $\Delta\mathrm{ICOHP}$ ($-0.65$~eV), followed by Ga and Bi, whereas the change for Ag is only $-0.02$~eV. Ge also has the largest solute--vacancy binding energy and therefore the largest vacancy enhancement per atomic fraction of solute. This quantity makes it possible to determine whether the vacancy is attracted to or repelled from the solute, since the largest $E_{\mathrm{b}}$ coincides with the lowest $\Delta\mathrm{ICOHP}$.

However, the constructed linear regressions of $E_{\mathrm{b}}$ on $\Delta\mathrm{ICOHP}$ do not render the two quantities equivalent and reveal only a minor correlation between ICOHP and solute--vacancy binding energies: the best single-descriptor model reaches only $R^{2} = 0.24$ and $r=0.49$ (Pearson correlation coefficient; see~\cref{fig:icohp_delta_vs_energies} and~\cref{tab:icohp-prediction-metrics} in~\cref{sec:icohp_regression}). A similar regression for solution energies likewise shows no correlation, with $R^{2} = 0.13$ and $r=0.36$. A linear combination of different ICOHPs does not provide better predictive metrics for $E_{\mathrm{s}}$, with $R^{2} = 0.13$ and $r=0.37$ (see~\cref{fig:icohp_energy_best_correlation} and~\cref{tab:icohp-prediction-metrics}). Similarly, the prediction of $E_{\mathrm{b}}$ remains almost unchanged, with $R^{2} = 0.25$ and $r=0.49$, which also indicates only a minor correlation.

The binding energy $E_{\mathrm{b}}$ therefore determines whether a solute and a vacancy attract each other, whereas COHP analysis reveals whether that attraction manifests as a redistribution of bonding and antibonding states on the nearby Li--solute bonds. Because ICOHP is evaluated only within a finite environment that captures the most prominent change in electron density, it neglects long-range interactions that also contribute to $E_{\mathrm{b}}$ (see~\cref{fig:charge_diff}). Consequently, COHP cannot serve as a standalone replacement for binding energy calculations; its role is instead to provide a quantitative, bond-resolved metric that rationalizes why a given solute strengthens or weakens the surrounding Li lattice. That information is not contained in $E_{\mathrm{b}}$ alone.

\subsection{Machine-learning surrogate models}

\begin{figure*}[!htbp]
\centering
\includegraphics[width=0.99\textwidth]{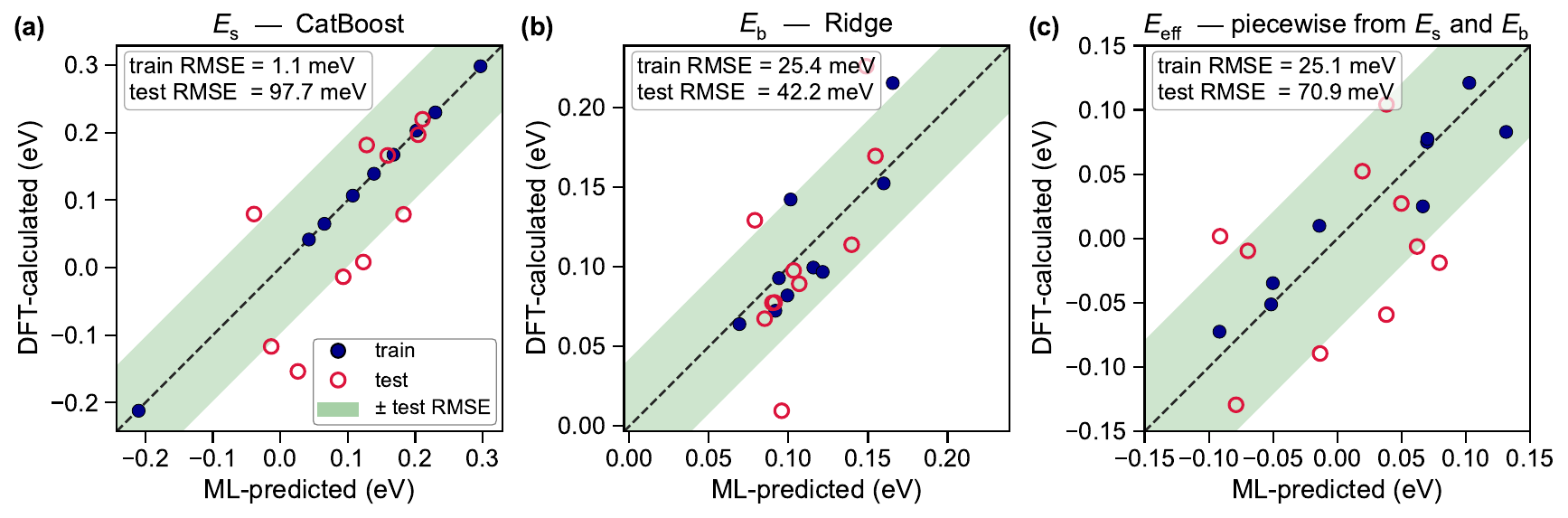}
\caption{Performance of optimal machine-learning models on the test set, evaluated using the root-mean-square error (RMSE). Results are shown for (a) the solution energy $E_{\mathrm{s}}$ predicted by CatBoost, (b) the solute--vacancy binding energy $E_{\mathrm{b}}$ predicted by Ridge regression, and (c) the effective vacancy energy $E_{\mathrm{eff}}$, derived from the best $E_{\mathrm{s}}$ and $E_{\mathrm{b}}$ models via the piecewise relation: $E_{\mathrm{eff}} = E_{\mathrm{s}} - E_{\mathrm{b}}$ if $E_{\mathrm{s}} > 0$, and $E_{\mathrm{eff}} = -E_{\mathrm{b}}$ otherwise. The green shaded region indicates the $\pm$ RMSE interval around the parity line.}
\label{fig:ml_part_main}
\end{figure*}

Surrogate models were trained to predict $E_{\mathrm{s}}$, $E_{\mathrm{b}}$, and $E_{\mathrm{eff}}$ from descriptors that require no additional DFT calculations: the Hume--Rothery quantities (\cref{eq:hr_rules_1,eq:hr_rules_2,eq:hr_rules_3,eq:hr_rules_4}) and MatMiner features of the end-member and room-temperature phase of the solute (\cref{sec:feature_significance}); see \cref{fig:ml_part_main}. The training set is limited to the 19 solutes that were recalculated with a relaxed cell and with both 1NN and 2NN vacancies, because those two choices change all three target energies. Fitting the same models to all 50 solutes, without that refinement, performs substantially worse; those tests are in the accompanying repository (\href{https://github.com/purphy42/solid_solutions_li}{github.com/purphy42/solid\_solutions\_li}).

End-member features rank above room-temperature ones for every target. The Hume--Rothery descriptors place no higher than fifth, behind MatMiner quantities such as the packing fraction and the number of $d$ valence electrons, consistent with the high solution energies of the $d$-block elements (\cref{fig:sol_bind_num}.a). No single feature tracks all three energies. The maximum Mendeleev number, for example, correlates with $E_{\mathrm{b}}$ ($r = 0.39$), barely with $E_{\mathrm{s}}$ ($r = 0.06$), and anticorrelates with $E_{\mathrm{eff}}$ ($r = -0.11$; \cref{fig:pearson_correlation_all}). Each target therefore keeps its own set. After significance and correlation filtering, 17 features remain per target.

For each target, eight models were trained on a 9/10 hold-out split of the 19 solutes (9 for training, 10 for testing). Errors for every model are reported in \cref{fig:ml_models_all_solution,fig:ml_models_all_binding,fig:ml_models_all_effective} and in \cref{tab:ml_solution,tab:ml_binding,tab:ml_solution_minus_binding}. The best cross-validated models are CatBoost for $E_{\mathrm{s}}$ (test RMSE 98~meV, MaxAE 180~meV) and Ridge regression for $E_{\mathrm{b}}$ (test RMSE 42~meV, MaxAE 86~meV); see \cref{fig:ml_part_main} and \cref{tab:ml_best_models_metrics}. Predicting $E_{\mathrm{eff}}$ directly is less accurate than reconstructing it from these two outputs using the piecewise definition in \cref{eq:vac_eff_energy}: the combined route yields a test MaxAE of 98~meV, against 151~meV for the best direct model (SVR; \cref{fig:ml_best_eff_direct}). Applied to the original 50 solutes, the same predictions eliminate 31 candidates that fail the $E_{\mathrm{s}}$, $E_{\mathrm{b}}$, and $E_{\mathrm{eff}}$ screening criteria before those energies are computed with DFT.

\section{General discussion}

\subsection{Strengths and limitations of the screening methodology}\label{sec:strengths_and_limitations}

The proposed funnel is transferable to other solid-solution systems thanks to three robust choices. First, \textit{automatic end-members} drawn from the Materials Project successfully recover the Li-rich reference for 46 of the 50 solutes, despite the phase diagrams being constructed at $T = 0$~K; the four minor mismatches (Cd, Sn, Pb, and Hg) do not alter which elements pass to the next stage. Second, \textit{thermodynamic energies} provide a reliable shortlist where classical Hume--Rothery rules fall short: the conventional radius and electronegativity cuts miss known Li solid solutions, while the structure and valency rules both accept and reject the wrong solutes. Third, \textit{machine-learning surrogates} trained on these energies safely discard 31 of the 50 solutes before the costly steps, making a repeated search on another host practical. \textit{Lattice relaxation} is also essential: after inserting a solute and a vacancy, relaxing the supercell shifts $E_{\mathrm{s}}$ by up to 49~meV and $E_{\mathrm{b}}$ by up to 156~meV. Their combined uncertainty of 0.21~eV defines the acceptance tolerance, meaning any solute falling within this window of the threshold is uncertain.

The predicted $E_{\mathrm{eff}}$ values do not directly yield diffusion coefficients of alloyed samples, since the latter require molecular dynamics (MD) or kinetic Monte Carlo simulations combined with the equilibrium vacancy concentrations derived in this work. \textit{Nearest-neighbor hops} of a vacancy change only marginally in the presence of a nearby solute, which is precisely why the predicted benefit manifests primarily as a higher vacancy concentration; however, cooperative hops can lower some barriers by roughly 25\% (as seen in the Cd example), and capturing these pathways requires MD. In particular, the DFT-NEB calculations consider only the simplest cooperative pathways; multi-atom concerted mechanisms cannot be adequately captured within this approach. Consequently, $E_{\mathrm{eff}}$ should not yet be used as a direct metric for comparison with experimental results.

This gap is illustrated by Zn, which our funnel overlooks despite experimental evidence that \ce{Li_{20}Zn} enables a CCD of 1~mA~cm$^{-2}$ in full solid-state cells~\cite{jin2020solid, ye2023solid}. The mechanism underlying the superior performance of Zn over pristine Li may differ from the hypothesis proposed here, and this possibility warrants further investigation. To this end, other elements with solubility limits near 1~at.\% (Ca~\cite{jia2019dual} and Au~\cite{leenheer2016phase,chu2026solubility}) should also be examined in MD simulations, despite their vacancy enhancement factors being less than unity. Ultimately, a reliable determination of vacancy diffusion coefficients will require the combined application of our thermodynamic framework and explicit diffusion modeling. Further DFT studies should also incorporate additional critical parameters, including operating voltage ranges, thermal expansion coefficients, and mechanical properties, which are addressed in related computational works~\cite{huang2024unlocking,shi2025machine}.

\subsection{Best solute element candidates}

Most of the identified candidates (Hg, Ga, Rh, Pt, and Tl) are unsuitable because of their low solubility limits. Their thermodynamic estimates, as well as the natural CCD limits with a Li anode, fall far short of the industrial threshold of $>4$~mA~cm$^{-2}$~\cite{lu2021critical}. The most promising solute candidates are Ag and Cd, owing to their positive vacancy-enhancement factors and high solubility limits (9--18~at.\%), as well as Ge, whose exceptionally high solubility--vacancy enhancement coefficient compensates for its low solubility limit of 0.1\%.

\textbf{Li--Ag}. Li--Ag alloys are known to enhance interface stability~\cite{lu2026thermodynamics,ko2025mechanism} by facilitating fast Li diffusion and stabilizing the phase assemblage. In particular, \ce{Li4Ag} and \ce{Li9Ag4} alloys demonstrate uniform nucleation and deposition, alongside reduced capacity fading and polarization in symmetric \ce{Li6PS5Cl} cells~\cite{huang2024unlocking}. An all-solid-state lithium cell assembled with the same solid electrolyte (SE), NMC811 (\ce{LiNi_{0.78}Co_{0.10}Mn_{0.12}O_2}), and a Li--Ag alloy achieved a CCD of 1.6~mA~cm$^{-2}$~\cite{park_2024}. Despite these advantages, the specific capacity of these alloys does not exceed 696~mAh~g$^{-1}$. By contrast, our calculations suggest that solid solutions with $\sim 9$~at.\% Ag offer a theoretical capacity of 1522~mAh~g$^{-1}$ and support a thermodynamically estimated CCD of 4~mA~cm$^{-2}$. In contrast, a similar specific capacity (1660~mAh~g$^{-1}$) can be achieved with thin \ce{Li_{20}Ag} metal foils~\cite{jin2020solid}; however, the achieved current density is limited to only 1~mA~cm$^{-2}$. Notably, while spin-lattice relaxation (SLR) measurements have successfully quantified enhanced Ag diffusion in Li-rich solid solutions compared to pristine Li~\cite{korblein1985diffusion}, a direct comparison of Li diffusion coefficients in solid solutions versus pure Li remains unavailable. Furthermore, the SLR study suggests that Ag diffuses faster than Li~\cite{korblein1985diffusion}, which appears to contradict our DFT-NEB results, where the migration barrier for Li (0.04~eV) is significantly lower than that for Ag (0.22~eV). This discrepancy indicates that the vacancy-mediated mechanism considered in our calculations may be insufficient to describe Ag diffusion. Indeed, previous studies have proposed that Ag primarily migrates via interstitial sites~\cite{korblein1985diffusion} and multi-hop mechanisms, which can reduce the activation energy for Ag hops by approximately twofold~\cite{thomas2024thermodynamic}. These alternative pathways should be considered in future calculations to resolve the apparent contradiction between experimental observations and our theoretical results. Although computational studies indicate that the activation barriers for vacancy migration from Ag to Li sites are low (29~meV), which is proposed to promote uniform plating~\cite{lu2026thermodynamics}, such approaches inherently neglect finite-temperature and size effects that can only be captured by large-scale molecular dynamics (MD) simulations. These controversial findings preclude a definitive conclusion regarding the feasibility of manufacturing a Li--Ag anode with both high specific capacity ($>1500$~mAh~g$^{-1}$) and high CCD ($>2$~mA~cm$^{-2}$). Specifically, it remains unclear whether the current limitations arise from imperfect synthesis and cell assembly or from a fundamental physical constraint.

\textbf{Li--Ge}. Li--Ge alloys combine high theoretical capacity with comparatively fast Li transport relative to Si-based analogues. Full lithiation toward the Li-rich phases \ce{Li15Ge4} and \ce{Li22Ge5} corresponds to theoretical capacities of approximately 1385 and 1625~mAh~g$^{-1}$, respectively~\cite{baggetto2009lithium}. Experimentally, nanocrystalline and thin-film Ge electrodes have delivered reversible capacities of up to $\sim$1400--1700~mAh~g$^{-1}$ with improved rate capability attributed to short diffusion lengths~\cite{graetz2004nanocrystalline}. Consistent with this kinetic advantage, the diffusivity of Li in Ge is reported to be roughly an order of magnitude higher than in Si at elevated temperature, and the extrapolated room-temperature disparity can be even larger~\cite{graetz2004nanocrystalline}. In all-solid-state architectures, thin Ge interlayers have been used to improve Li wetting of garnet electrolytes: a $\sim$20~nm Ge coating reduced the garnet/Li interfacial resistance from $\sim$900 to $\sim$115~$\Omega$~cm$^{2}$ through Li--Ge alloying and enabled stable room-temperature cycling of a $\ce{Li|LLZO|LiFePO4}$ cell~\cite{luo2017reducing}. However, deep lithiation of Ge is accompanied by large volume changes (typically $>250\%$), which promote pulverization and contact loss in thick electrodes~\cite{graetz2004nanocrystalline,baggetto2009lithium}. First-principles surveys of the Li--Ge phase diagram further indicate a rich landscape of stable and metastable germanides whose Li mobilities vary strongly with stoichiometry~\cite{morris2014thermodynamically,iosimovska2024thermodynamic}. By analogy with Li--Ag and Li--Mg, dilute Li--Ge solid solutions are therefore more relevant than stoichiometric germanides if the goal is to retain a Li-metal-like specific capacity while using Ge as a kinetic solute. At present, however, direct comparisons of Li diffusion coefficients in dilute Li--Ge solid solutions versus pristine Li, and of the corresponding CCDs in SE-based cells, remain scarce. As for the other solutes discussed above, resolving whether Ge enhances the effective vacancy-mediated Li diffusivity, or primarily acts through interfacial wetting and alloying layers, will require finite-temperature MD together with carefully controlled solid-solution anode experiments.

\textbf{Li--Mg}. In the range of 2--11~at.\% Mg in solid solutions, both Mg and Li are found to diffuse more slowly than in pristine Li~\cite{korblein1985diffusion}. An experimental study of the $\beta$-phase \ce{Li_{9}Mg_{1}} solid solution, which exhibits a high specific capacity of 2820~mAh~g$^{-1}$, showed no contact loss at room temperature; however, it did not improve the critical current density (CCD) in an LLZO-based cell, which remained at 300~$\mu$A~cm$^{-2}$~\cite{krauskopf2019diffusion}. Notably, even a modest stack pressure of 15~MPa increases the CCD and prevents contact loss. Overall, these findings indicate that at room temperature, planar bulk Li--Mg alloy anodes, characterized by an average Li chemical diffusion coefficient of $D_{\mathrm{Li}} = 3 \times 10^{-11}$~cm$^{2}$~s$^{-1}$, compared to $\sim 6 \times 10^{-11}$~cm$^{2}$~s$^{-1}$ in pristine Li, cannot sustain the fast charging rates required for practical solid-state batteries. This conclusion aligns with our results, which yield a low vacancy enhancement factor of $\tilde{\eta}=1.5$ for Mg (see~\cref{fig:final_plot_eff_en}). Nevertheless, further investigation of the Mg--Li system is warranted to decouple the specific influence of Mg solute atoms on the Li diffusion coefficient.

\textbf{Li--Cd}. The \ce{Li3Cd} alloy has been considered as a potential metal anode material, offering a specific capacity of 604~mAh~g$^{-1}$ and exhibiting a volume change of 268\%~\cite{wang1986investigations,fauteux1993rechargeable}. However, the toxicity of cadmium severely limits its practical application as an anode material. Nevertheless, MD simulations of Li--Cd solid-solutions and their underlying Li diffusion mechanisms remain valuable from a theoretical standpoint, as they may help elucidate the origins of the increased effective vacancy concentration and, consequently, the potentially enhanced vacancy diffusion coefficient.

\section{Conclusions}

In summary, the combination of thermodynamic screening, encompassing solution energy and vacancy binding energy, alongside kinetic analyses, including both single-ion and cooperative DFT-NEB calculations, offers a multifaceted understanding of solute effects in bcc Li. Our funnel-based screening identifies Ag, Ge, and Cd as promising candidates capable of achieving a critical current density of 1~mA~cm$^{-2}$ within their solubility limits. Furthermore, Mg, Zn, Ca and Au merit further investigation via molecular dynamics simulations, as they may enhance vacancy flux at the anode side through increased vacancy diffusion coefficients, rather than through changes in vacancy concentration.

Looking ahead, the proposed extended screening funnel, which integrates optimized Hume--Rothery rules together with machine-learning predictions, can not only accelerate the screening process but also be readily extended to other solid-solution systems. While cooperative diffusion can significantly reduce activation barriers, it remains challenging to describe within conventional vacancy-mediated frameworks. Large-scale, long-timescale molecular dynamics simulations will therefore be essential for accurately determining vacancy diffusivity in Li solid solutions, thereby enabling reliable quantification of vacancy diffusion coefficients.

\section{Data availability}
The data on validation of the Hume--Rothery rules and machine-learning calculations are available at \href{https://github.com/purphy42/solid_solutions_li}{github.com/purphy42/solid\_solutions\_li}.

\section{Acknowledgments}
The authors acknowledge the financial support of Russian Science Foundation project No. 24-73-10204.

\clearpage
\def\bibsection{\section*{References}} 
\putbib[lib]
\end{bibunit}

\clearpage
\begin{bibunit}[apsrev4-2]

\clearpage

\onecolumngrid 
\beginsupplement
\vspace*{-2em} % Adjust to pull it up if needed
\begin{center}
    {\Large \bfseries SUPPLEMENTARY MATERIALS FOR}\label{sec:si}\par
\end{center}
\vspace{1em}
\thispagestyle{empty}

\begin{center}
{\large\bfseries Vacancy-enhanced Li solid-solution anodes for solid-state batteries:\\ 
computational screening enabling high critical current density\par}
\vspace{1.5em}

{\normalsize Arseniy S. Burov$^{1,*}$, Nadezhda A. Ashikhmina$^{2}$, Nurlan Zhanabayev$^{1}$, Anton O. Boev$^{1}$, Aleksandra D. Radina$^{1}$, \\ Ilya V. Chepkasov$^{1}$, Alexander G. Kvashnin$^{1}$, Dmitry A. Aksyonov$^{1}$\par}
\vspace{0.8em}

{\small $^{1}$Skolkovo Institute of Science and Technology, 121205 Moscow, Russian Federation\par}
{\small $^{2}$Moscow Institute of Physics and Technology, 141701 Dolgoprudny, Russian Federation\par}
\vspace{0.5em}
{\small $^{*}$Corresponding author: arseniy.burov@skoltech.ru\par}
\end{center}

\vspace{1.5em}
\date{\today}

\clearpage

\setcounter{page}{1}
\setcounter{figure}{0}
\renewcommand{\thefigure}{S\arabic{figure}}%
\setcounter{table}{0}
\renewcommand{\thetable}{S\arabic{table}}%

\setcounter{section}{0}  
\setcounter{subsection}{0}
\setcounter{subsubsection}{0}
% Define numbering hierarchy cleanly
\makeatletter
\renewcommand{\thesection}{S\arabic{section}}
\renewcommand{\thesubsection}{\thesection.\arabic{subsection}}
\renewcommand{\thesubsubsection}{\thesubsection.\arabic{subsubsection}}
\setcounter{secnumdepth}{3}
\makeatother

\makeatletter
\renewcommand{\p@subsection}{}      % removes the automatic section prefix for subsections
\renewcommand{\p@subsubsection}{}   % likewise for subsubsections
\makeatother

\sitablesofcontents
\usestocfile
\clearpage

\section{Effective vacancy concentration}\label{sec:vac_conditions}

The effective Li diffusion coefficient is proportional to the total vacancy concentration, assuming that the diffusivity of free vacancies and solute--vacancy complexes is comparable:

\begin{equation}
    D_{\mathrm{Li}}^{\mathrm{eff}} \propto c_{\mathrm{V}}^{\mathrm{total}}
    \label{eq:vac_diffusion}
\end{equation}

The total concentration of vacancies can be expressed as:

\begin{equation}
    c_{\mathrm{V}}^{\mathrm{total}} = c_{\mathrm{V}}^{\mathrm{int}} + c_{\mathrm{V}}^{\mathrm{bound}}
    = c_{\mathrm{V}}^{\mathrm{int}} \left[1 + z \exp{\left(-\frac{E_{\mathrm{s}} - E_{\mathrm{b}}}{k_{\mathrm{B}}T} \right)} \right],
    \label{eq:vac_concentration_total}
\end{equation}
where $c_{\mathrm{V}}^{\mathrm{int}}$ is the intrinsic vacancy concentration in pristine bcc Li, $c_{\mathrm{V}}^{\mathrm{bound}}$ is the concentration of solute--vacancy complexes, $z$ is the coordination number (number of equivalent binding sites around a solute), $E_{\mathrm{s}}$ is the solution energy, $E_{\mathrm{b}}$ is the solute--vacancy binding energy, $k_{\mathrm{B}}$ is the Boltzmann constant, and $T$ is the absolute temperature.

The enhancement factor for vacancy concentration due to alloying is:

\begin{equation}
    \frac{c_{\mathrm{V}}^{\mathrm{total}}}{c_{\mathrm{V}}^{\mathrm{int}}} = 1 + z \exp{\left(-\frac{E_{\mathrm{s}} - E_{\mathrm{b}}}{k_{\mathrm{B}}T} \right)}
    \label{eq:enhancement_factor}
\end{equation}

Therefore, a solute increases the effective vacancy concentration when the bound vacancy contribution exceeds the intrinsic one:

\begin{equation*}
    z \exp{\left(-\frac{E_{\mathrm{s}} - E_{\mathrm{b}}}{k_{\mathrm{B}}T} \right)} > 1
    \quad \Rightarrow \quad
    E_{\mathrm{eff}} = E_{\mathrm{s}} - E_{\mathrm{b}} < k_{\mathrm{B}}T \ln{z}
\end{equation*}

At room temperature ($k_{\mathrm{B}}T \approx 26$~meV) and for the 1NN coordination shell in bcc Li ($z = 8$, $\ln{8} \approx 2.08$), the threshold is approximately $52$~meV. As a conservative approximation, we adopt the simplified condition:

\begin{equation*}
    E_{\mathrm{eff}}  < 0
\end{equation*}

Solutes satisfying this criterion (\cref{eq:funnel_rule_3} in the main text) are expected to increase the total vacancy concentration and, under the assumption of comparable diffusivities, enhance vacancy-mediated diffusion.

However, the criterion $E_{\mathrm{eff}} < 0$ becomes invalid when $E_{\mathrm{s}}$ is  negative, as the Boltzmann factor $\exp{\left(-E_{\mathrm{s}}/k_{\mathrm{B}}T\right)}$ then exceeds unity, signaling a phase transition to a pristine solute phase and violating the dilute-solution approximation. In this regime, the enhancement of effective vacancy concentration is primarily governed by the binding energy. Therefore, we adopt the simpler condition $E_{\mathrm{b}} > 0$ (or more precisely $E_{\mathrm{b}} \gg k_{\mathrm{B}}T$) to identify solutes whose complexes significantly increase the vacancy population.

\section{Convergence tests}\label{sec:convergence_tests}

\subsection{Total energies}

\begin{figure}[H]
\centering
\includegraphics[width=0.54\columnwidth]{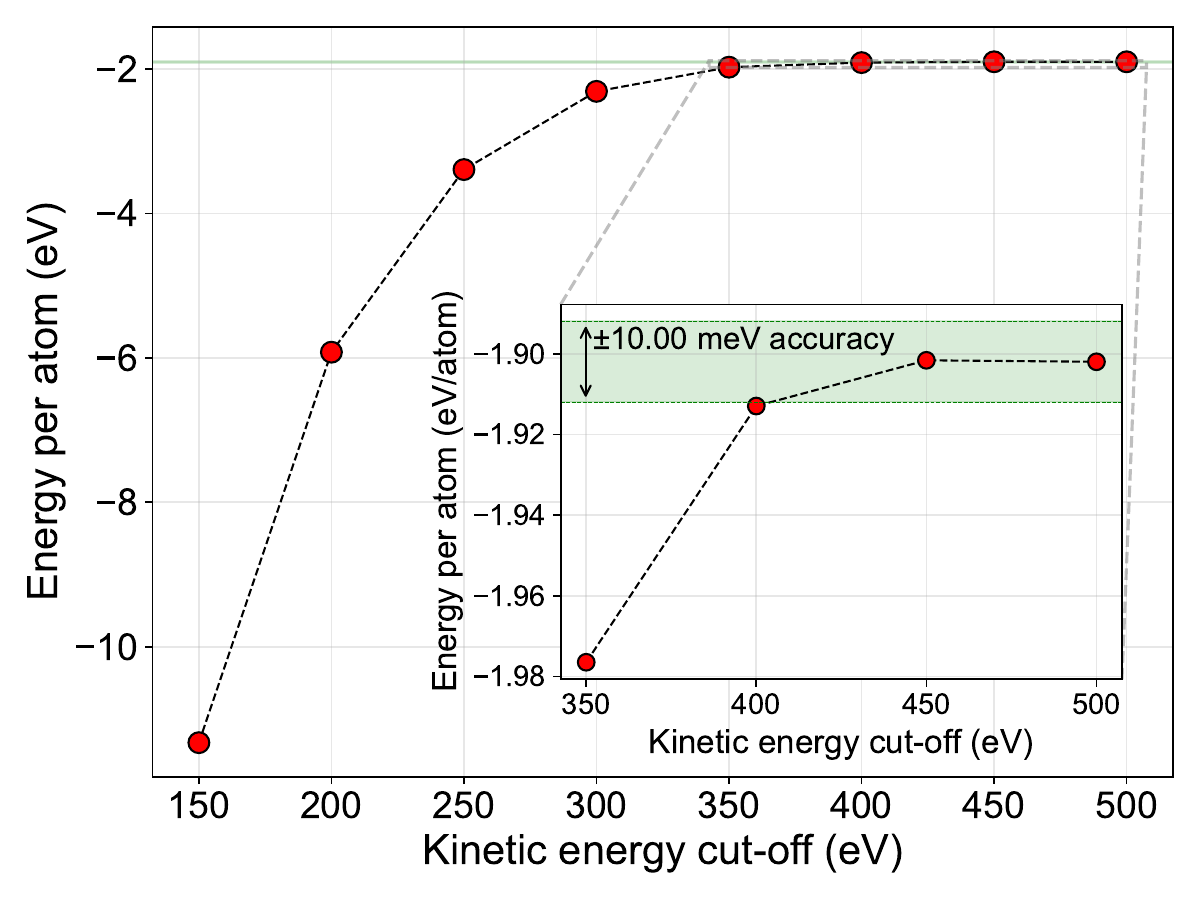}
\caption{Total energy per atom of $\ce{Li_{127}Mg}$ as a function of the plane-wave kinetic energy cut-off. The green shaded area indicates the range of values that lie within this accuracy threshold of 10~meV. }
\label{fig:conv_total_ecut}
\end{figure}

\begin{figure}[H]
\centering
\includegraphics[width=0.54\columnwidth]{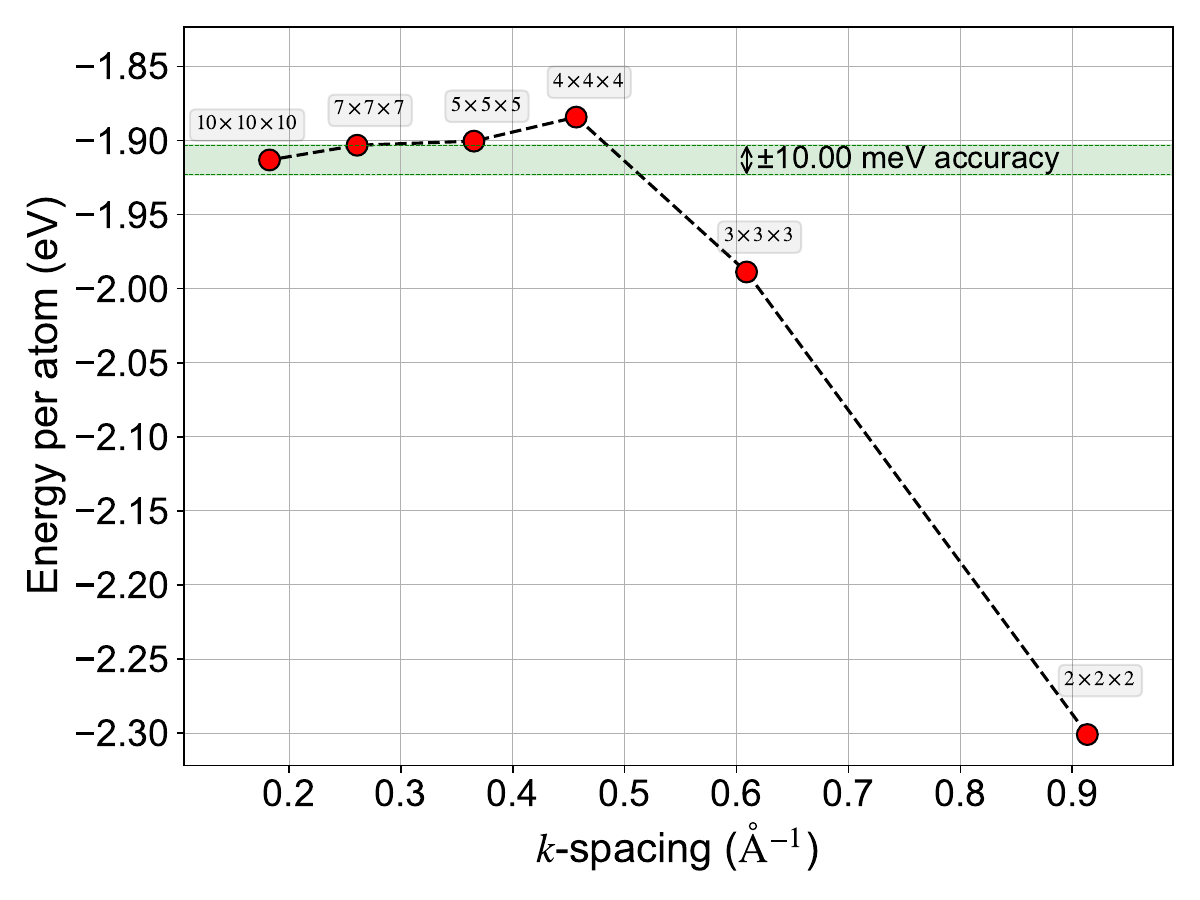}
\caption{Total energy per atom of $\ce{Li_{127}Mg}$ as a function of the $k$-spacing. The green shaded area indicates the range of values that lie within the accuracy threshold of 10~meV per atom. The labels in gray boxes denote the number of $k$-points in each direction for the simulation cell.}
\label{fig:conv_total_ksp}
\end{figure}

\begin{figure}[H]
\centering
\includegraphics[width=0.55\columnwidth]{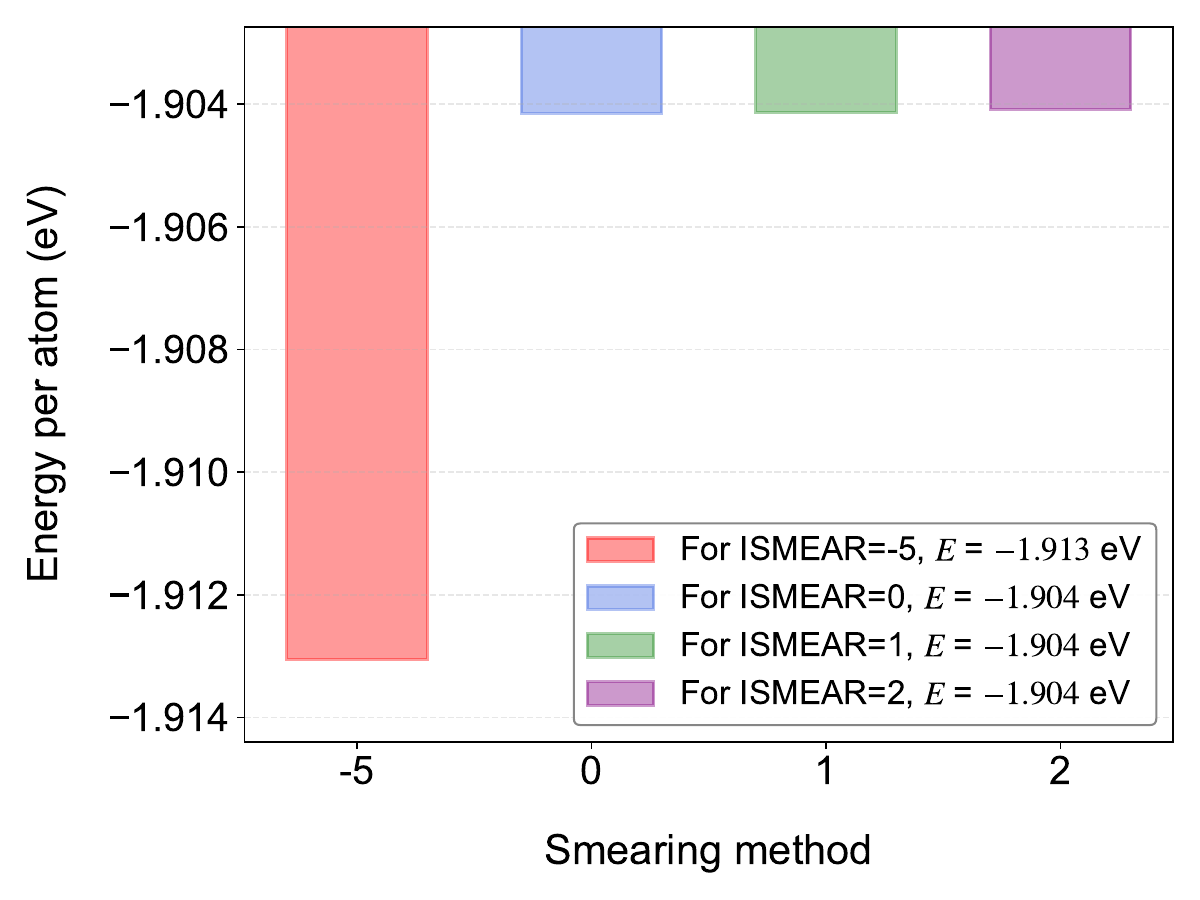}
\caption{Total energy per atom of $\ce{Li_{127}Mg}$ as a function of the Brillouin-zone smearing method. Here, ``$-5$'' denotes the tetrahedron method with Bl\"{o}chl corrections without smearing, ``$0$'' denotes Gaussian smearing, ``$1$'' denotes first-order Methfessel--Paxton smearing, and ``$2$'' denotes second-order Methfessel--Paxton smearing. In all cases the smearing width was set to 0.1~eV. Hereinafter, we employ the second-order Methfessel--Paxton smearing scheme in all further calculations. }
\label{fig:conv_smear}
\end{figure}

\begin{figure}[H]
\centering
\includegraphics[width=0.55\columnwidth]{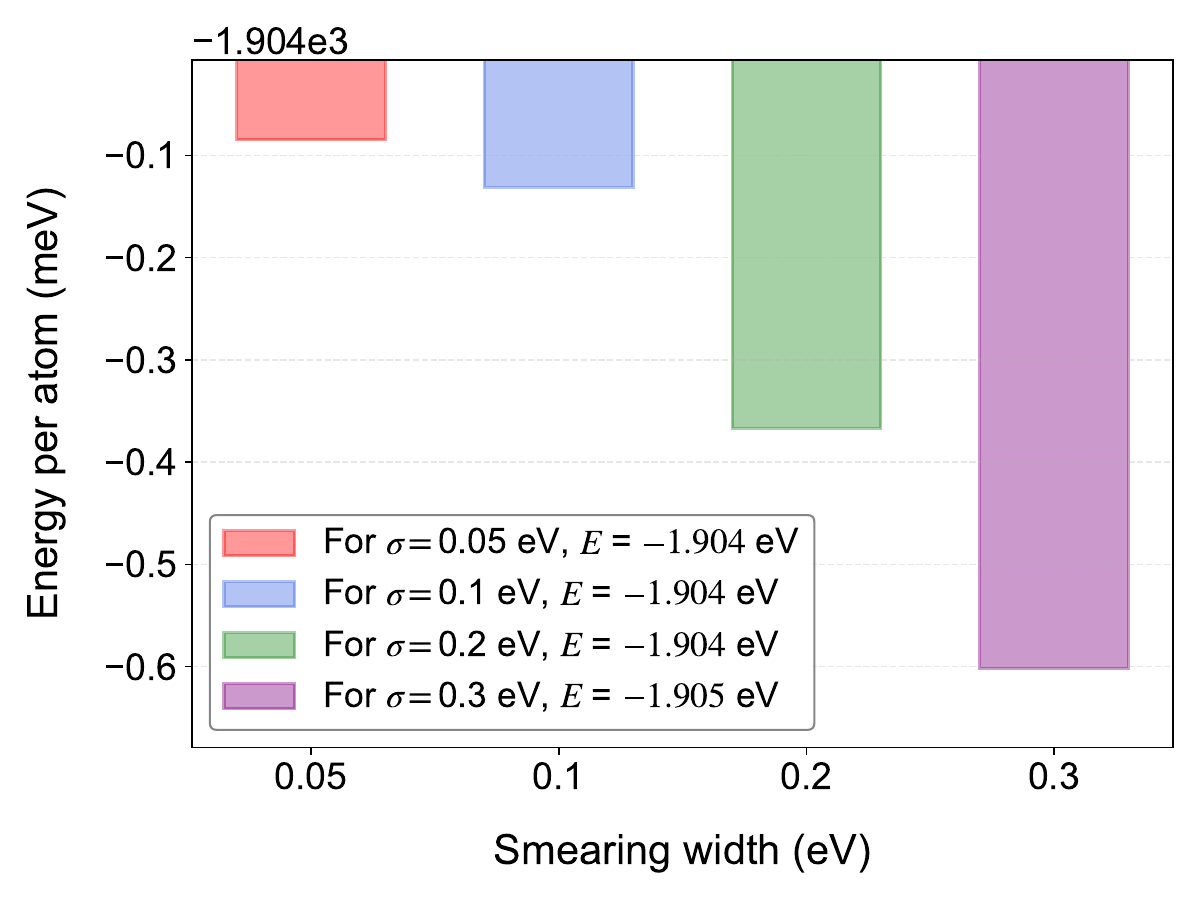}
\caption{Total energy per atom of $\ce{Li_{127}Mg}$ as a function of the smearing width $\sigma$. In all cases, the second-order Methfessel--Paxton smearing scheme was employed. Henceforth, we use a smearing width of 0.1~eV in all further calculations. }
\label{fig:conv_sigma}
\end{figure}

To ensure high accuracy of the DFT calculations, we performed convergence tests on the $\ce{Li_{127}Mg}$ system using a $4 \times 4 \times 4$ supercell with a Li pseudopotential containing one valence electron ($2s^1$). The Mg pseudopotential was chosen as the reference case because it requires the largest minimum cut-off energy of 303~eV. The convergence of the total energy with respect to the plane-wave cut-off energy is shown in~\cref{fig:conv_total_ecut} and with respect to the $k$-point spacing in~\cref{fig:conv_total_ksp}. A cut-off energy of 400~eV and a $k$-spacing of 0.26~\AA$^{-1}$ achieves an accuracy of 10~meV per atom. We then tested different smearing schemes (see~\cref{fig:conv_smear}) and smearing widths $\sigma$ (see~\cref{fig:conv_sigma}). The second-order Methfessel--Paxton smearing with $\sigma = 0.1$~eV yields an energy convergence better than 0.05~meV per atom.

\subsection{Relative energies}

\begin{figure}[H]
\centering
\includegraphics[width=0.5\columnwidth]{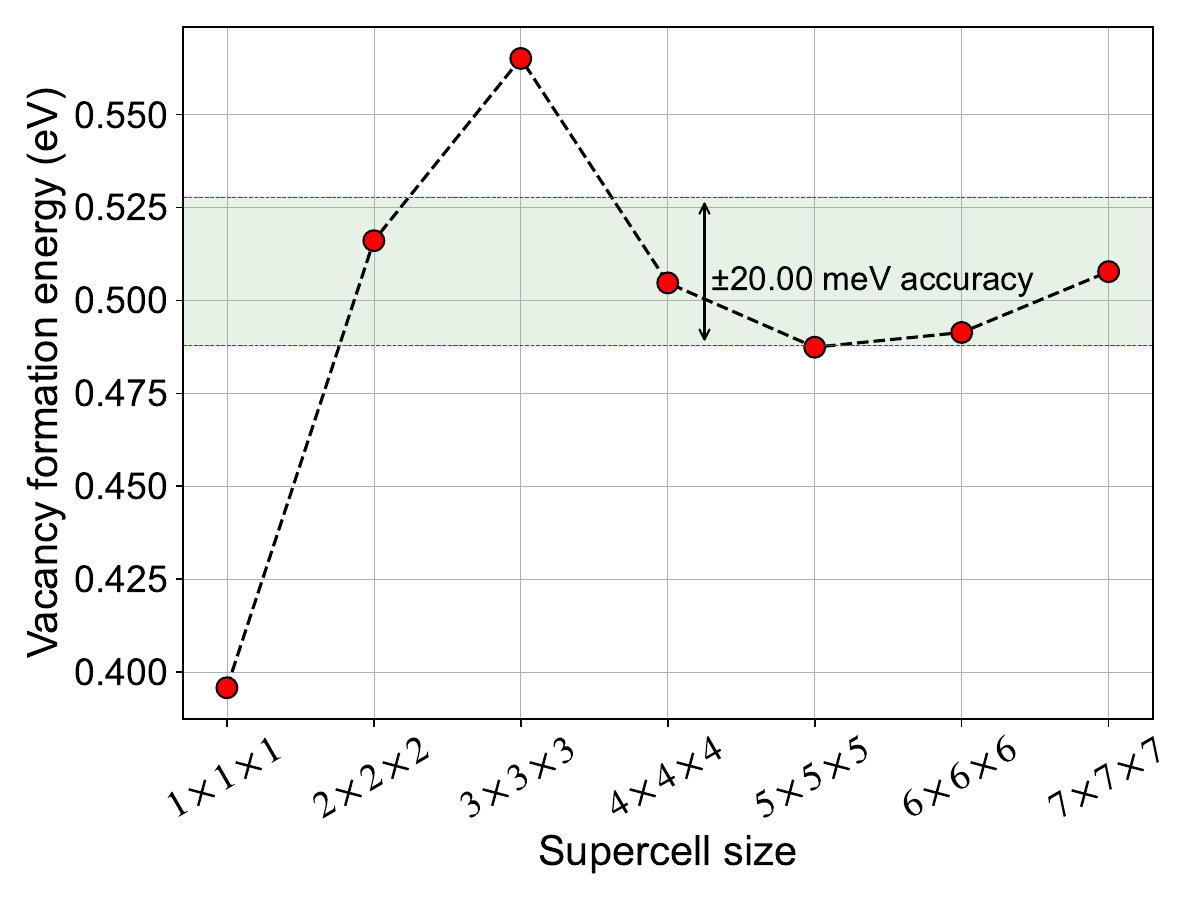}
\caption{Li vacancy formation energy in $\ce{Li127Mg}$ as a function of supercell size.}
\label{fig:conv_li_vac}
\end{figure}

\begin{figure}[H]
\centering
\includegraphics[width=0.5\columnwidth]{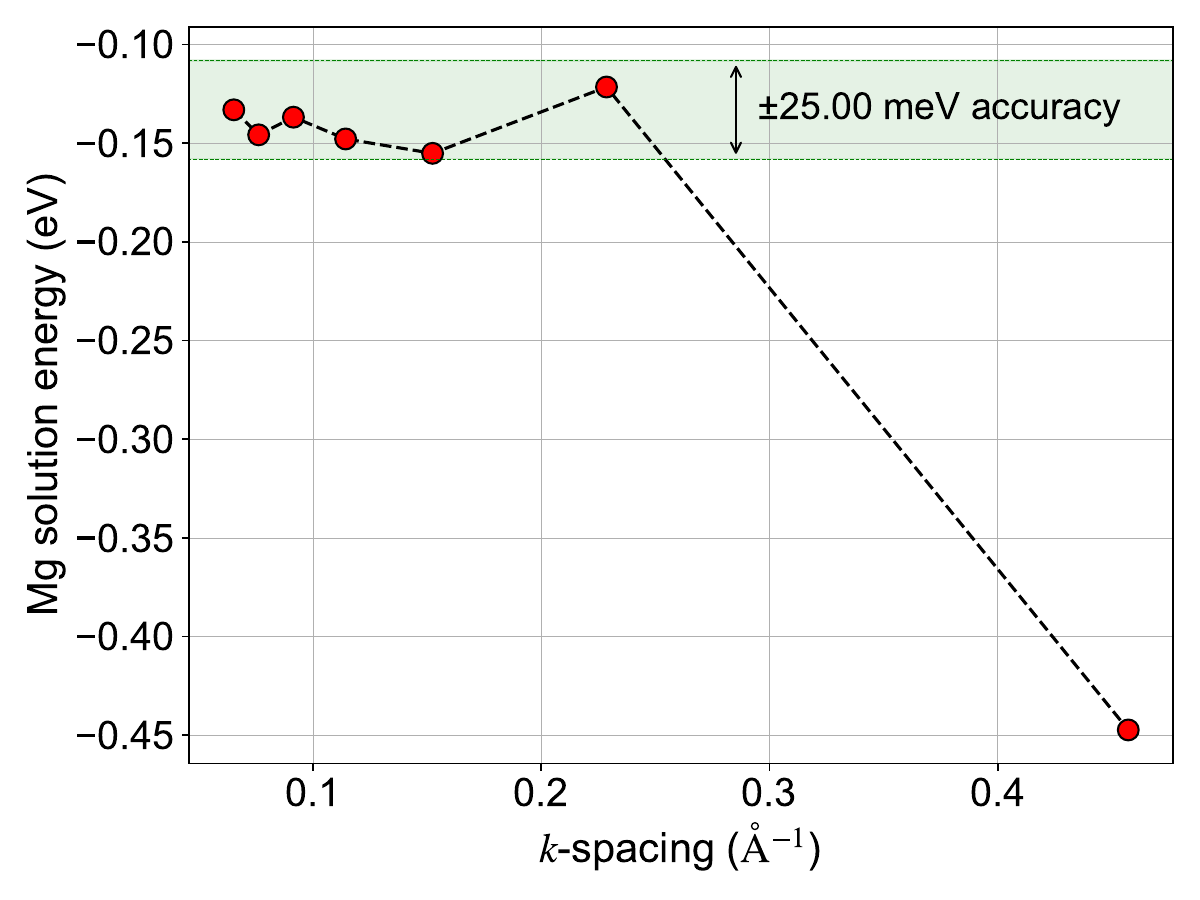}
\caption{Mg solution energy in $\ce{Li_{127}Mg}$ as a function of supercell size. The chemical potential of Mg metal is utilized and was recalculated using the same $k$-spacing values as those employed for the corresponding supercell. }
\label{fig:conv_mg_sol_energy}
\end{figure}

The same setup ensures an accuracy of 25~meV/atom for the relative energy quantities, such as the Li vacancy formation energy (\cref{fig:conv_li_vac}) and the Mg solution energy (\cref{fig:conv_mg_sol_energy}). The supercell size of $4 \times 4 \times 4$ with two $k$-points in each direction gives the $k$-spacing of 0.22~\AA$^{-1}$, which is less than that of 0.26~\AA$^{-1}$ in~\cref{fig:conv_total_ksp}. Hence, this supercell size falls within the specified accuracy for the selected computational setup.

The calculated Li vacancy formation energy is 504~meV (\cref{fig:conv_li_vac}), which agrees well with the previously reported value of 497~meV by Yang~\textit{et al.}~\cite{yang2021interfacial_mech}.

\subsection{DFT-NEB calculations}

\begin{figure}[H]
\centering
\includegraphics[width=0.99\columnwidth]{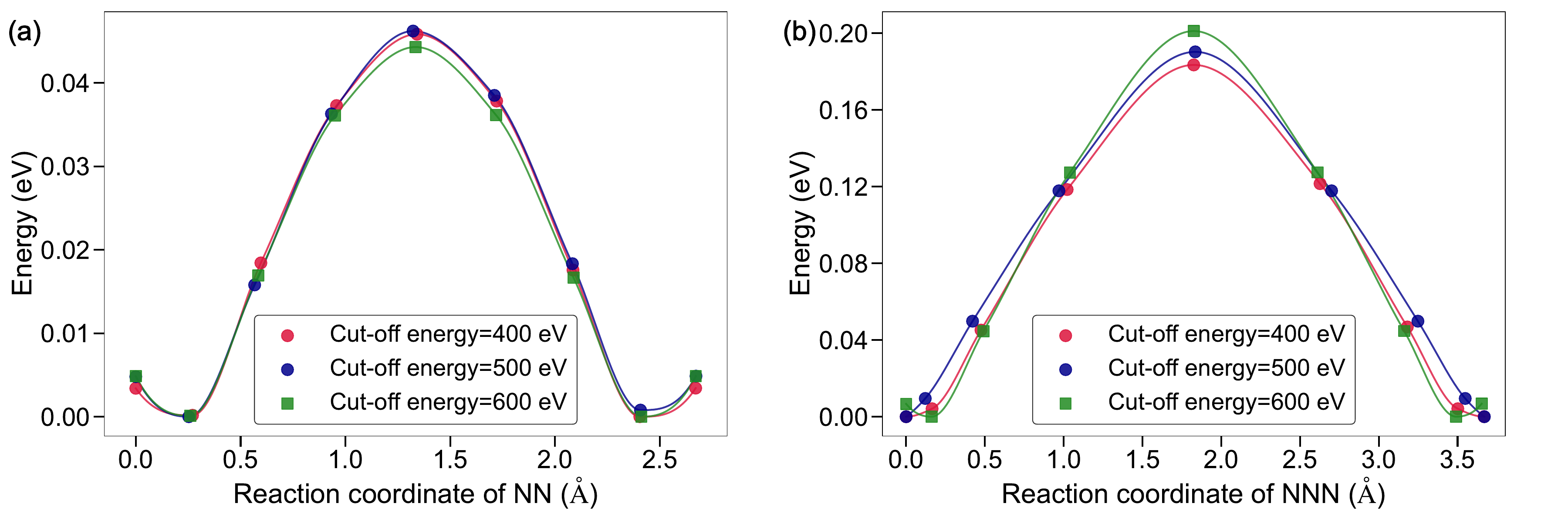}
\caption{DFT-NEB activation energies in a bcc Li $4 \times 4 \times 4$ supercell for different cut-off energies: 400~eV (red), 500~eV (blue), and 600~eV (green). The energy profiles are shown for (a) nearest-neighbor (NN) and (b) next nearest-neighbor (NNN) jumps.}
\label{fig:conv_neb_ecut}
\end{figure}

\begin{figure}[H]
\centering
\includegraphics[width=0.99\columnwidth]{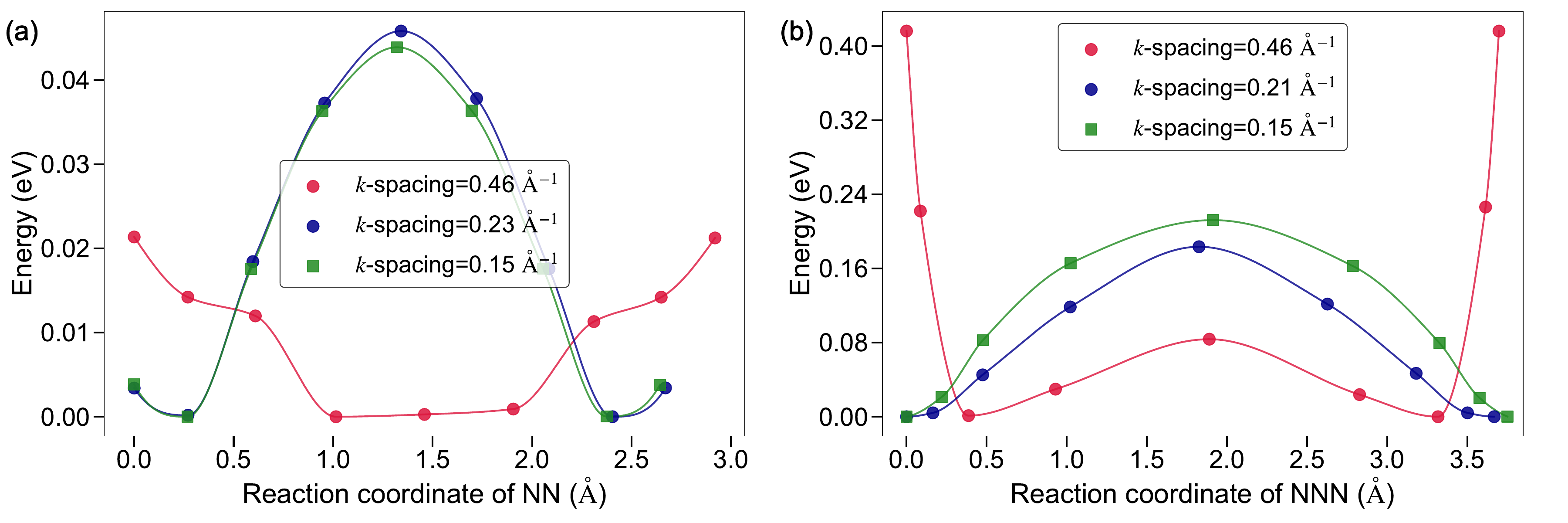}
\caption{DFT-NEB activation energies in a bcc Li $4 \times 4 \times 4$ supercell for different $k$-spacings : $0.46$~\AA$^{-1}$ (red), $0.23$~\AA$^{-1}$ (blue), and $0.15$~\AA$^{-1}$ (green). The energy profiles are shown for (a) nearest-neighbor (NN) and (b) next nearest-neighbor (NNN) jumps. }
\label{fig:conv_neb_ksp}
\end{figure}

The convergence tests for DFT-NEB energies for nearest-neighbor (NN) and next nearest-neighbor (NNN) jumps are visualized in~\cref{fig:conv_neb_ecut} (cutoff energy) and in~\cref{fig:conv_neb_ksp} ($k$-spacing). Our computational setup (\textit{NEBSet}) with a cutoff energy of 400~eV ensures an accuracy of 1~meV for the NN migration barrier and 17~meV for the NNN migration barrier with respect to cutoff energy. The selected $k$-spacing of $0.23$~\AA$^{-1}$ guarantees an accuracy of 2~meV for the NN barrier and 26~meV for the NNN barrier. Overall, our computational parameters yield an accuracy of approximately 3~meV for the NN migration barrier.

\subsection{Technical parameters for computational sets}

The computational parameters for the employed sets are summarized in~\cref{tb:sets_criteria}. Since convergence was achieved with the coarsest setup, the more refined setups are expected to provide even higher accuracy. The examples of input setting for the VASP calculations (INCAR files) are specified in~\cref{sec:example_incars}.

\begin{table}[H]
\setlength{\tabcolsep}{10pt}
\renewcommand{\arraystretch}{1.25}
\centering
\caption{Computational setups for the datasets used in this work: \textit{CoarseSet} --- used for screening all solutes, \textit{MPSet} --- used for phase-diagram calculations, utilizing Materials Project database entries~\cite{jain2013commentary_mpdatabase}, \textit{TightSet} --- used for high-accuracy calculations on the most promising solutes, and \textit{NEBSet} --- for migration barrier calculations using the nudged elastic band (NEB) method. The table lists cut-off energies, $k$-spacing, spin-polarization settings, lattice constant optimization via equation of state (EOS-optimization), and the number of valence electrons for the Li pseudopotential.}
\label{tb:sets_criteria}
\begin{tabular}{l l l l l l }
\hline \hline 
Computational setup & Cut-off energy,~eV & $k$-spacing, \AA$^{-1}$ & Spin-polarization & EOS-optimization & Li valence electrons \\ \hline
\textit{CoarseSet}  & 400  & 0.23  & yes & no & $1s^2 \, 2s^1$ (3)              \\
\textit{MPSet}    & 520  & 0.16   & yes & no  & $1s^2 \, 2s^1$ (3) \\
\textit{TightSet}  & 520   & 0.16   & no & yes & $1s^2 \, 2s^1$ (3)         \\
\textit{NEBSet}  & 400  & 0.23     & no  & yes & $2s^1$ (1) \\ \hline\hline          
\end{tabular}
\end{table}

\clearpage
\section{Example of input files}\label{sec:example_incars}

Below, we provide example input files (INCAR settings) for three setups: all candidates, the best candidates and NEB calculations.

Example 1. INCAR file containing the algorithm and parameter settings used for VASP calculations of all candidates (\textit{CoarseSet}). In this case, we employed the Li\_sv pseudopotential, which accounts for three valence electrons ($1s^2 \, 2s^1$).
\noindent
\begin{verbatim}
SYSTEM = CoarseSet for all dopants 
ALGO = Normal
EDIFF = 1e-05
EDIFFG = -0.02
ENAUG = 600
ENCUT = 400
GGA_COMPAT = .FALSE.
IBRION = 1
ISIF = 4
ISMEAR = 2
ISPIN = 2
MAGMOM = 0.6 0.6 0.6 0.6 0.6 0.6 0.6 0.6 0.6 0.6 0.6 0.6 0.6 0.6 0.6 0.6 0.6 0.6 0.6 0.6 
 0.6 0.6 0.6 0.6 0.6 0.6 0.6 0.6 0.6 0.6 0.6 0.6 0.6 0.6 0.6 0.6 0.6 0.6 0.6 0.6 0.6 0.6 
 0.6 0.6 0.6 0.6 0.6 0.6 0.6 0.6 0.6 0.6 0.6 0.6 0.6 0.6 0.6 0.6 0.6 0.6 0.6 0.6 0.6 0.6
 0.6 0.6 0.6 0.6 0.6 0.6 0.6 0.6 0.6 0.6 0.6 0.6 0.6 0.6 0.6 0.6 0.6 0.6 0.6 0.6 0.6 0.6 
 0.6 0.6 0.6 0.6 0.6 0.6 0.6 0.6 0.6 0.6 0.6 0.6 0.6 0.6 0.6 0.6 0.6 0.6 0.6 0.6 0.6 0.6
 0.6 0.6 0.6 0.6 0.6 0.6 0.6 0.6 0.6 0.6 0.6 0.6 0.6 0.6 0.6 0.6 0.6 0.6 5.0
ISTART = 0
KGAMMA = .TRUE.
LASPH = .TRUE.
LORBIT = 11
LPLANE = .TRUE.
LREAL = Auto
LWAVE = .FALSE.
NBANDS = 250
NELM = 100
NPAR = 1
NSW = 50
POTIM = 0.2
PREC = Accurate
SIGMA = 0.1
KSPACING = 0.23
\end{verbatim}

\clearpage
Example 2. INCAR file containing the algorithm and parameter settings used for VASP calculations used to construct the phase diagram within the Materials Project setup (\textit{MPSet}). In this case, we employed the Li\_sv pseudopotential, which accounts for three valence electrons ($1s^2 \, 2s^1$).
\noindent
\begin{verbatim}
ALGO = MPSet for the phase diagrams construction and reference phases
EDIFF = 1e-05
EDIFFG = -0.05
ENAUG = 780
ENCUT = 520
IBRION = 1
ISIF = 2
ISMEAR = 0
ISPIN = 2
ISTART = 1
KGAMMA = .TRUE.
LAECHG = .TRUE.
LMIXTAU = .TRUE.
LORBIT = 11
LREAL = Auto
LWAVE = .FALSE.
NBANDS = 235
NELM = 100
NSW = 99
POTIM = 0.2
PREC = Accurate
SIGMA = 0.1
KSPACING = 0.16
\end{verbatim}

\clearpage
Example 3. INCAR file containing the algorithm and parameter settings used for VASP calculations of the best candidates (\textit{TightSet}). In this case, we employed the Li\_sv pseudopotential, which accounts for three valence electrons ($1s^2 \, 2s^1$).
\noindent
\begin{verbatim}
SYSTEM = TightSet for the best dopants 
ALGO = Normal
EDIFF = 1e-05
EDIFFG = -0.02
ENAUG = 780
ENCUT = 520
GGA_COMPAT = .FALSE.
IBRION = 1
ISIF = 2
ISMEAR = 2
ISPIN = 1
ISTART = 0
ISYM = 0
KGAMMA = .TRUE.
LASPH = .TRUE.
LCHARG = .False.
LORBIT = 11
LPLANE = .TRUE.
LREAL = Auto
LWAVE = .FALSE.
NBANDS = 110
NELM = 100
NELMIN = 4
NPAR = 1
NSW = 100
POTIM = 0.15
PREC = Accurate
SIGMA = 0.1
SYMPREC = 1e-10
KSPACING = 0.16
\end{verbatim}

\clearpage
Example 4. INCAR file containing the algorithm and parameter settings used for VASP calculations using the nudged elastic band (NEB) method (\textit{NEBSet}). In this case, we employed the Li pseudopotential, which accounts for one valence electron ($2s^1$).
\noindent
\begin{verbatim}
SYSTEM = NEBSet for the DFT-NEB calculations 
ALGO = Normal
EDIFF = 1e-05
EDIFFG = -0.02
ENAUG = 600
ENCUT = 400
GGA_COMPAT = .FALSE.
IBRION = 1
IMAGES = 7
ISIF = 2
ISMEAR = 2
ISPIN = 1
ISTART = 0
ISYM = 0
KGAMMA = .TRUE.
LASPH = .TRUE.
LCHARG = .False.
LORBIT = 11
LPLANE = .TRUE.
LREAL = Auto
LWAVE = .FALSE.
NBANDS = 110
NELM = 100
NELMIN = 4
NPAR = 1
NSW = 100
POTIM = 0.15
PREC = Accurate
SIGMA = 0.1
SYMPREC = 1e-10
KSPACING = 0.23
\end{verbatim}

\clearpage

\section{Calculated phase diagrams and chemical potentials}\label{sec:calculated_phase_diagrams}

To obtain chemical potentials for each solute under realistic conditions, we identified the corresponding end-member phases using the \textit{PhaseDiagram()} module of the \textit{Pymatgen} package, based on calculations performed within the MPSet setup. This module employs an application programming interface (API) to retrieve data from the Materials Project database. Here, we constrained Li, Na, and K to the bcc phases (mp-135, mp-127, and mp-58, respectively), which are stable at room temperature, even though the trigonal $\overline{R}3m$ (mp-1018134), hexagonal $P6_3/mmc$ (mp-10172), and orthorhombic $Pnma$ (mp-1545455) phases are the most stable at 0~K.

In addition, when constructing the phase diagram, we considered only experimentally observed stable and metastable phases (energy above the hull $<5$~meV), as selected by the \textit{theoretical\,=\,False} tag in Pymatgen. This restriction improves the agreement between the calculated and experimental phase diagrams, even though our DFT calculations are performed at 0~K.

\begin{figure}[H]
\centering
\includegraphics[width=0.75\columnwidth]{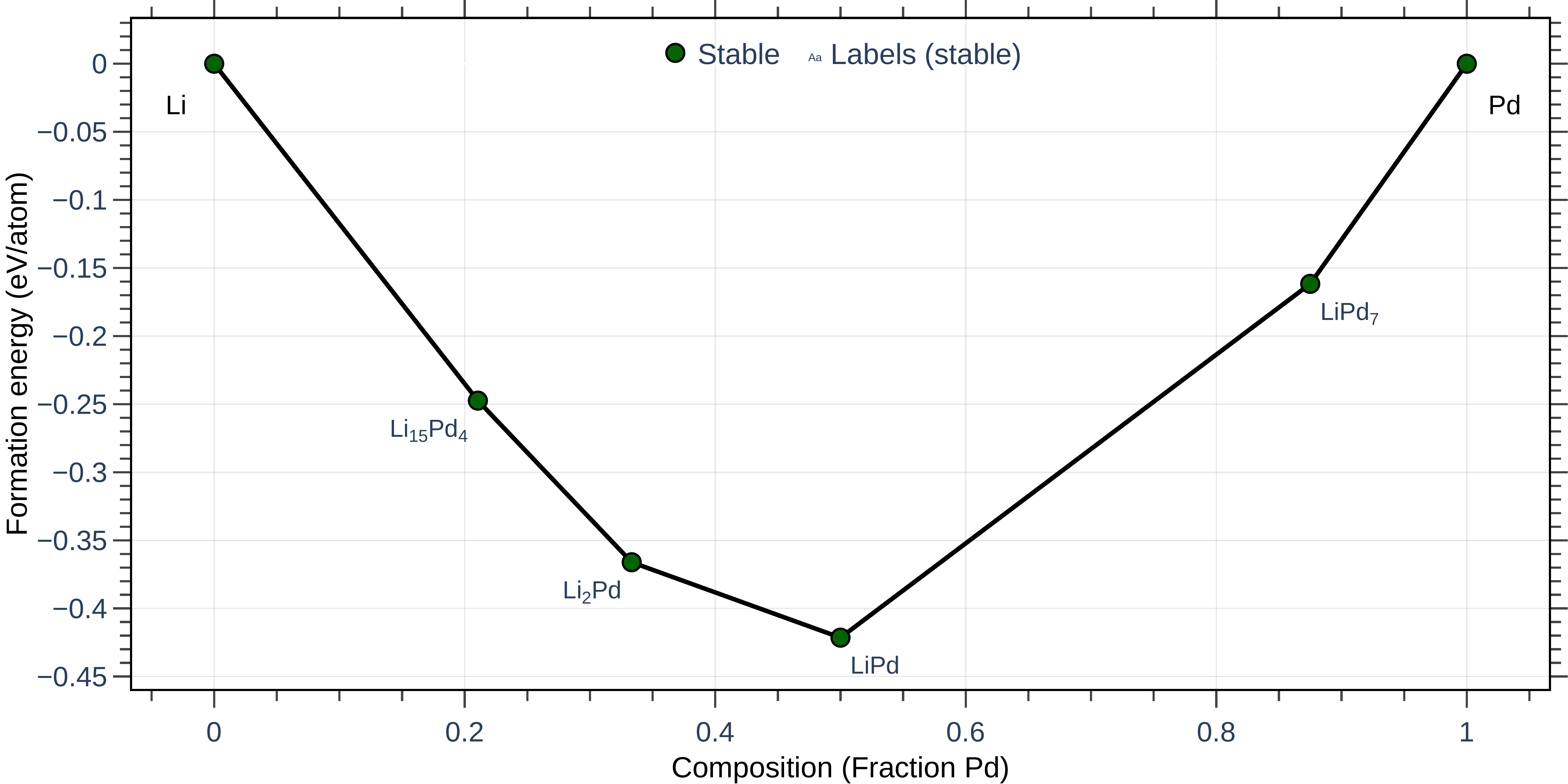}
\caption{Calculated phase diagram of the Li--Pd system obtained via the Materials Project API using the Pymatgen package. Only experimentally observed stable and metastable phases (energy above the hull $<5$~meV) are shown. The end-member is defined as the Li-rich phase closest to Li in composition space.}
\label{fig:calc_pd_lipd}
\end{figure}

Since all our calculations were performed within the GGA-PBE functional, we enforced the use of phases with the same computational setup whenever possible. Otherwise, we selected available data with the r$^2$SCAN functional and applied mixing~\cite{kingsbury2022flexible_mixing} to align these data with GGA-PBE calculations. The Li--Pd phase diagram, shown in~\cref{fig:calc_pd_lipd}, serves as an example of the calculated results. The Materials Project was accessed on 26 December 2026. As it is under active development, some entries or attributes may have changed since then.

We then computed the decomposition energies for all phases and identified the phase with the minimum value as the end-member. We successfully identified the nearest end-member for each solute, except for Cd, Sn, Pb, and Hg.

\begin{figure}[H]
\centering
\includegraphics[width=0.6\columnwidth]{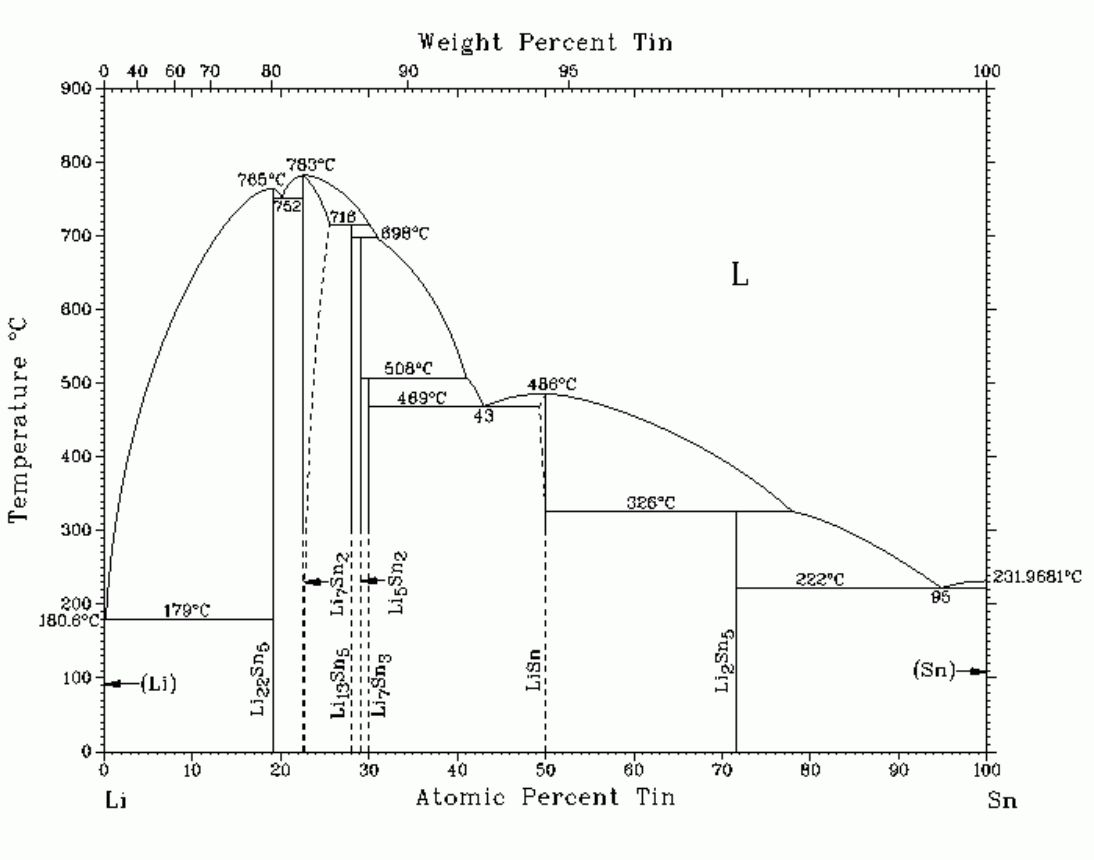}
\caption{Li--Sn phase diagram. Here, Sn forms a solid solution with Li in a narrow region of concentrations. Our reference phase is $\ce{Li17Sn4}$ (mp-573471). Data is from the work by Okamoto~\textit{et al.}~\cite{okamoto2016_sn}.}
\label{fig:phase_diagram_sn}
\end{figure}

\begin{figure}[H]
\centering
\includegraphics[width=0.7\columnwidth]{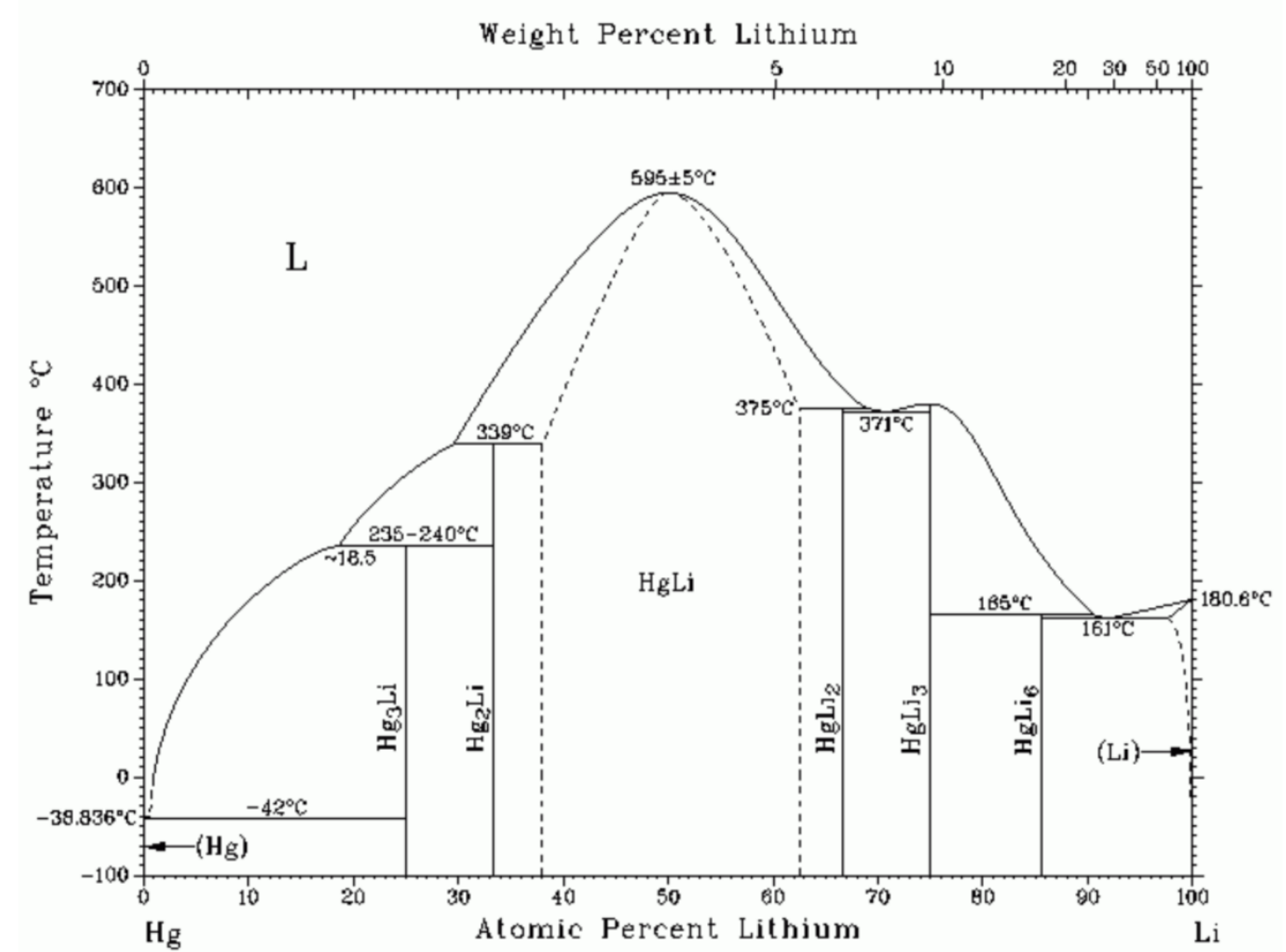}
\caption{Li--Hg phase diagram. Here, Hg forms a solid solution with Li in a narrow range of concentrations. Our reference phase is $\ce{Li3Hg}$ (mp-1646). Data is from the work by Predel~\textit{et al.}~\cite{predel1997hg}.}
\label{fig:phase_diagram_hg}
\end{figure}

\begin{figure}[H]
\centering
\includegraphics[width=0.7\columnwidth]{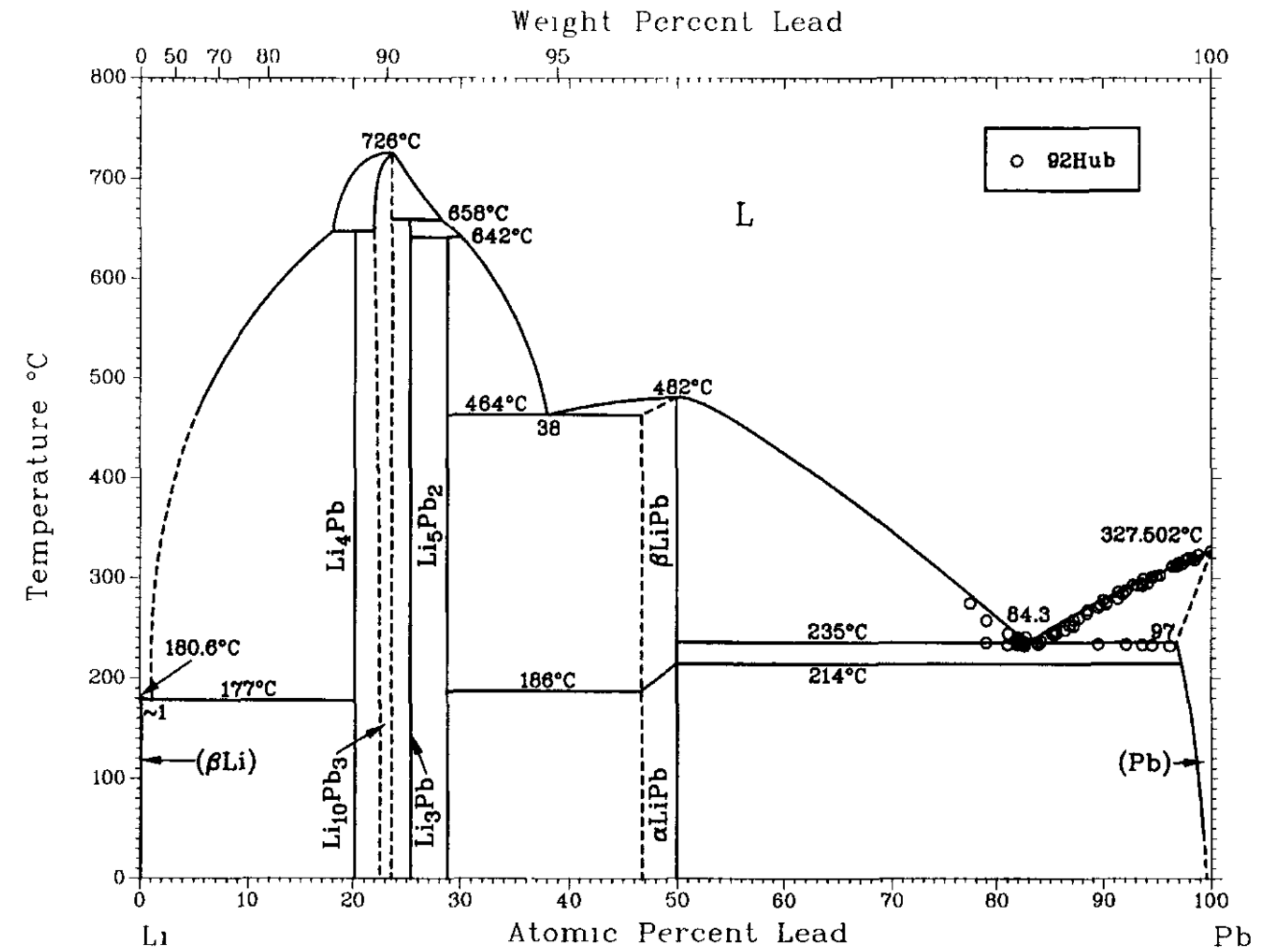}
\caption{Li--Pb phase diagram. Here, Pb forms a solid solution with Li in a narrow range of concentrations. Our reference phase is $\ce{Li17Pb4}$ (mp-574275). Data is from the work by Okamoto~\textit{et al.}~\cite{okamoto1993pb}.}
\label{fig:phase_diagram_pb}
\end{figure}

\begin{itemize}[noitemsep, topsep=6pt]
    \item For Sn, our chosen end-member is $\ce{Li_{17}Sn4}$, whereas the closest phase on the experimental phase diagram is $\ce{Li_{22}Sn5}$ (see~\cref{fig:phase_diagram_sn}).
    \item For Pb, our end-member is $\ce{Li_{17}Pb4}$, whereas the closest experimental phase is $\ce{Li4Pb}$ (see~\cref{fig:phase_diagram_pb}).
    \item For Hg, our end-member is $\ce{Li3Hg}$ (mp-1646), whereas the closest experimental phase is $\ce{Li6Hg}$ (see~\cref{fig:phase_diagram_hg}). However, no literature data exist on the crystal structure of $\ce{Li6Hg}$, and there is no corresponding entry in the MP database. 
\end{itemize}

Nevertheless, the difference in Li fraction between the experimental phases and our chosen phases for Sn and Pb is less than 1\%, so the associated chemical potentials should be close to those of the true phases. In this work, we deliberately did not correct the end-member phases to maintain the automation of our screening funnel.

Next, using the end-members derived from the phase diagram, we recalculated them with our \textit{MPSet} computational setup. Additionally, we performed the same procedure for the room-temperature stable and metastable phases (energy above the hull $<5$~meV) metallic phases of the solutes. To eliminate basis dependence of the chemical potentials, we calculated reference chemical potentials of the solutes as follows:
\begin{equation}
\Delta \mu_{i} = \mu_{i} - \mu_{i}^{0},
\label{eq:chem_potentials}
\end{equation}
where $\mu_{i}$ is the chemical potential of element $i$ from the phase diagram and $\mu_{i}^{0}$ is the chemical potential of the metallic phase of element $i$.

The end-members for each solute, along with their calculated chemical potentials, are summarized in~\cref{tab:chemical_potentials_all}.

\vspace*{-0.2cm}

\begin{table}[H]
\caption{List of end-member phases selected for each solute, along with their corresponding Materials Project IDs. Absolute chemical potentials ($\mu_{i}$) and potentials referenced to the metallic phase ($\Delta \mu_{i}$) are provided for each entry.}
\label{tab:chemical_potentials_all}
\centering
\setlength{\tabcolsep}{12pt}
\renewcommand{\arraystretch}{1.0}
\begin{tabular*}{0.75\columnwidth}{llllrr}
\hline \hline
Element & End-member & Space group  &  Mp-id & $\mu_i$,~eV & $\Delta\mu_i$,~eV \\
\hline
Be       & Be               & $P6_3/mmc$        & mp-87       & $-3.741$  & $0$       \\
B        & LiB              & $P6_3/mmc$        & mp-1001835  & $-7.049$  & $-0.370$  \\
Li       & Li               & $Im\overline{3}m$ & mp-135      & $-1.907$  & $0$       \\
Na       & Na               & $Im\overline{3}m$ & mp-127      & $-1.314$  & $0$       \\
Mg       & Mg               & $P6_3/mmc$        & mp-153      & $-1.488$  & $0$       \\
Al       & Li$_9$Al$_4$     & $C2/m$            & mp-568404   & $-4.298$  & $-0.545$  \\
Si       & Li$_{21}$Si$_5$  & $F\overline{4}3m$ & mp-29720    & $-6.669$  & $-1.245$  \\
K        & K                & $Im\overline{3}m$ & mp-58       & $-1.090$  & $0$       \\
Ca       & Li$_2$Ca         & $Fd\overline{3}m$ & mp-11644    & $-2.064$  & $-0.066$  \\
Sc       & Sc               & $P6_3/mmc$        & mp-67       & $-6.332$  & $0$       \\
Ti       & Ti               & $P6/mmm$        & mp-72       & $-7.898$  & $0$       \\
V        & V                & $Im\overline{3}m$ & mp-146      & $-9.085$  & $0$       \\
Cr       & Cr               & $Im\overline{3}m$ & mp-90       & $-9.624$  & $0$       \\
Mn       & Mn               & $I\overline{4}3m$ & mp-35       & $-9.160$  & $0$       \\
Fe       & Fe               & $Im\overline{3}m$ & mp-13       & $-8.472$  & $0$       \\
Co       & Co               & $P6_3/mmc$        & mp-54       & $-7.109$  & $0$       \\
Ni       & Ni               & $Fm\overline{3}m$ & mp-23       & $-5.780$  & $0$       \\
Cu       & Cu               & $Fm\overline{3}m$ & mp-30       & $-4.104$  & $0$       \\
Zn       & LiZn             & $Fd\overline{3}m$ & mp-1934     & $-1.722$  & $-0.456$  \\
Ga       & Li$_2$Ga         & $Cmcm$            & mp-29210    & $-3.975$  & $-0.946$  \\
Ge       & Li$_{15}$Ge$_4$  & $I\overline{4}3d$ & mp-1777     & $-6.156$  & $-1.641$  \\
Se       & Li$_2$Se         & $Fm\overline{3}m$ & mp-2286     & $-7.361$  & $-3.866$  \\
Y        & Y                & $P6_3/mmc$        & mp-112      & $-6.464$  & $0$       \\
Zr       & Zr               & $P6_3/mmc$        & mp-131      & $-8.548$  & $0$       \\
Nb       & Nb               & $Im\overline{3}m$ & mp-75       & $-10.105$ & $0$       \\
Mo       & Mo               & $Im\overline{3}m$ & mp-129      & $-10.849$ & $0$       \\
Ru       & Ru               & $P6_3/mmc$        & mp-33       & $-9.278$  & $0$       \\
Rh       & LiRh             & $P\overline{6}m2$ & mp-600561   & $-7.770$  & $-0.425$  \\
Pd       & Li$_{15}$Pd$_4$  & $I\overline{4}3d$ & mp-1197547  & $-6.493$  & $-1.307$  \\
Ag       & LiAg             & $Pm\overline{3}m$ & mp-2426     & $-3.288$  & $-0.467$  \\
Cd       & LiCd             & $Fd\overline{3}m$ & mp-1437     & $-1.455$  & $-0.557$  \\
In       & Li$_{13}$In$_3$  & $Fd\overline{3}m$ & mp-510430   & $-3.942$  & $-1.203$  \\
Sn       & Li$_{17}$Sn$_4$  & $F\overline{4}3m$ & mp-573471   & $-5.955$  & $-1.951$  \\
Sb       & Li$_3$Sb         & $Fm\overline{3}m$ & mp-2074     & $-6.766$  & $-2.642$  \\
La       & La               & $P6_3/mmc$        & mp-26       & $-4.926$  & $0$       \\
Pr       & Pr               & $Fm\overline{3}m$ & mp-97       & $-4.775$  & $0$       \\
Ce       & Ce               & $C2/m$            & mp-567332   & $-5.932$  & $0$       \\
Nd       & Nd               & $P6_3/mmc$        & mp-123      & $-4.762$  & $0$       \\
Sm       & Sm               & $P6_3/mmc$        & mp-69       & $-4.646$  & $0$       \\
Hf       & Hf               & $P6_3/mmc$        & mp-103      & $-9.953$  & $0$       \\
Ta       & Ta               & $Im\overline{3}m$ & mp-50       & $-11.843$ & $0$       \\
W        & W                & $Im\overline{3}m$ & mp-91       & $-12.975$ & $0$       \\
Re       & Re               & $P6_3/mmc$        & mp-8        & $-12.446$ & $0$       \\
Os       & Os               & $P6_3/mmc$        & mp-49       & $-11.229$ & $0$       \\
Ir       & IrLi             & $P\overline{6}m2$ & mp-279      & $-9.362$  & $-0.510$  \\
Pt       & Li$_2$Pt         & $P6/mmm$          & mp-2170     & $-7.733$  & $-1.685$  \\
Au       & Li$_{15}$Au$_4$  & $I\overline{4}3d$ & mp-567395   & $-4.995$  & $-1.731$  \\
Hg       & \ce{Li3Hg}               & $P6/mmm$          & mp-1646    & $-1.472$  & $-1.169$  \\
Tl       & Li$_3$Tl         & $Fm\overline{3}m$ & mp-7396     & $-3.151$  & $-0.916$  \\
Pb       & Li$_{17}$Pb$_4$  & $F\overline{4}3m$ & mp-574275   & $-5.324$  & $-1.622$  \\
Bi       & Li$_3$Bi         & $Fm\overline{3}m$ & mp-23222    & $-6.111$  & $-2.224$  \\
\hline \hline
\end{tabular*}
\end{table}

\clearpage

\section{Thermodynamic properties of all candidates}

\vspace*{-0.45cm}

Thermodynamic properties for all candidates were calculated using the \textit{CoarseSet} setup. The results are listed in~\cref{tab:data_therm_all}. Solute solution energies were obtained from~\cref{eq:sol_en} using the chemical potentials in~\cref{tab:chemical_potentials_all}. Solute–vacancy binding energies were computed via~\cref{eq:e_bind} for the first nearest-neighbor (NN) coordination shell.

\vspace*{-0.5cm}
\begin{table}[H]
\centering
\begingroup
\caption{Summary of all phases considered in this work, including their end-members, Materials Project IDs, and key calculated properties. For each solute, the table provides the magnetic moment of solute element ($m_{\mathrm{solute}}$), solution energy ($E_{\mathrm{s}}$), vacancy binding energy in the nearest-neighbor coordination shell ($E_{\mathrm{b}}$), the vacancy effective energy $E_{\mathrm{eff}}$, and the exclusion criterion (see~\cref{eq:funnel_rule_1,eq:funnel_rule_2,eq:funnel_rule_3}) if applicable.}
\label{tab:data_therm_all}
\setlength{\tabcolsep}{10pt}
\begin{tabular*}{1.0\columnwidth}{l l l l r r r r l}
\hline \hline
Element & End-member & Space group & Mp-id & $\mathrm{m}_{\mathrm{solute}}$ ($\mu_{\mathrm{B}}$) & $E_{\mathrm{s}}$ (eV) & $E_{\mathrm{b}}$ (eV) & $E_{\mathrm{eff}}$ (eV) & Rule violated \\ \hline
Be       & Be        & $P6_3/mmc$        & mp-87        & 0.000 & 1.05  & 0.01  & 1.04  & Rule 2 \\
B        & LiB         & $P6_3/mmc$        & mp-1001835   & 0.000 & 1.96  & 0.17  & 1.79  & Rule 2 \\
Na       & Na        & $Im\overline{3}m$ & mp-127       & 0.000 & 0.23  & 0.08  & 0.15  & Proceeds \\
Mg       & Mg        & $P6_3/mmc$        & mp-153       & 0.000 & --0.15 & --0.01 & 0.01  & Proceeds \\
Al       & Li$_9$Al$_4$ & $C2/m$        & mp-568404   & 0.000 & 0.13  & --0.33 & 0.46  & Rule 3 \\
Si       & Li$_{21}$Si$_5$ & $F\overline{4}3m$ & mp-29720 & 0.000 & 0.33  & --0.01 & 0.35  & Rule 3 \\
K        & K         & $Im\overline{3}m$ & mp-58        & 0.000 & 0.92  & 0.22  & 0.71  & Rule 3 \\
Ca       & Li$_2$Ca  & $Fd\overline{3}m$ & mp-11644     & 0.000 & 0.35  & 0.03  & 0.32  & Rule 3 \\
Sc       & Sc        & $P6_3/mmc$        & mp-67        & 0.003 & 0.84  & --0.17 & 1.01  & Rule 3 \\
Ti       & Ti        & $P6/mmm$          & mp-72        & 0.014 & 1.96  & --0.18 & 2.14  & Rule 2 \\
V        & V         & $Im\overline{3}m$ & mp-146       & 2.947 & 2.96  & --0.07 & 3.03  & Rule 2 \\
Cr       & Cr        & $Im\overline{3}m$ & mp-90        & 3.976 & 3.44  & --0.54 & 3.98  & Rule 2 \\
Mn       & Mn        & $I\overline{4}3m$ & mp-35        & 4.038 & 3.00  & --0.08 & 3.08  & Rule 2 \\
Fe       & Fe        & $Im\overline{3}m$ & mp-13        & 2.853 & 2.41  & --0.88 & 3.28  & Rule 2 \\
Co       & Co        & $P6_3/mmc$        & mp-54        & 1.456 & 1.42  & --0.96 & 2.38  & Rule 2 \\
Ni       & Ni        & $Fm\overline{3}m$ & mp-23        & 0.000 & 0.44  & --0.02 & 0.46  & Rule 3 \\
Cu       & Cu        & $Fm\overline{3}m$ & mp-30        & 0.000 & 0.13  & 0.00  & 0.14  & Proceeds \\
Zn       & LiZn      & $Fd\overline{3}m$ & mp-1934      & 0.000 & 0.09  & 0.05  & 0.04  & Proceeds \\
Ga       & Li$_2$Ga  & $Cmcm$            & mp-29210     & 0.000 & 0.08  & 0.12  & --0.05 & Proceeds \\
Ge       & Li$_{15}$Ge$_4$ & $I\overline{4}3d$ & mp-1777 & 0.000 & 0.26  & 0.25  & 0.01  & Proceeds \\
Se       & Li$_2$Se  & $Fm\overline{3}m$ & mp-2286      & 0.000 & 0.94  & 0.17  & 0.77  & Rule 3 \\
Y        & Y         & $P6_3/mmc$        & mp-112       & 0.000 & 0.71  & --0.07 & 0.82  & Rule 3 \\
Zr       & Zr        & $P6_3/mmc$        & mp-131       & 1.045 & 1.74  & --0.24 & 1.97  & Rule 2 \\
Nb       & Nb        & $Im\overline{3}m$ & mp-75        & 2.302 & 2.76  & --0.23 & 3.00  & Rule 2 \\
Mo       & Mo        & $Im\overline{3}m$ & mp-129       & 2.719 & 3.12  & --0.13 & 3.25  & Rule 2 \\
Ru       & Ru        & $P6_3/mmc$        & mp-33        & 0.000 & 1.21  & 0.02  & 1.19  & Rule 2 \\
Rh       & LiRh      & $P\overline{6}m2$ & mp-600561    & 0.000 & 0.06  & --0.06 & 0.13  & Proceeds \\
Pd       & Li$_{15}$Pd$_4$ & $I\overline{4}3d$ & mp-1197547 & 0.000 & 0.19  & --0.01 & 0.20  & Proceeds \\
Ag       & LiAg      & $Pm\overline{3}m$ & mp-2426      & 0.000 & -0.12 & 0.00  & --0.13 & Proceeds \\
Cd       & LiCd      & $Fd\overline{3}m$ & mp-1437      & 0.000 & -0.19 & 0.00  & --0.19 & Proceeds \\
In       & Li$_{13}$In$_3$ & $Fd\overline{3}m$ & mp-510430 & 0.000 & 0.12  & 0.00  & 0.12  & Proceeds \\
Sn       & Li$_{17}$Sn$_4$ & $F\overline{4}3m$ & mp-573471 & 0.000 & 0.22  & 0.18  & 0.04  & Proceeds \\
Sb       & Li$_3$Sb  & $Fm\overline{3}m$ & mp-2074      & 0.000 & 0.32  & 0.29  & 0.03  & Proceeds \\
La       & La        & $P6_3/mmc$        & mp-26        & 0.000 & 0.79  & 0.08  & 0.71  & Rule 3 \\
Pr       & Pr        & $Fm\overline{3}m$ & mp-97        & 0.000 & 0.72  & 0.01  & 0.71  & Rule 3 \\
Ce       & Ce        & $C2/m$            & mp-567332    & 0.000 & 1.51  & 0.01  & 1.49  & Rule 2 \\
Nd       & Nd        & $P6_3/mmc$        & mp-123       & 0.000 & 0.75  & 0.02  & 0.73  & Rule 3 \\
Sm       & Sm        & $P6_3/mmc$        & mp-69        & 0.000 & 0.78  & 0.06  & 0.72  & Rule 3 \\
Hf       & Hf        & $P6_3/mmc$        & mp-103       & 0.617 & 2.02  & --0.25 & 2.28  & Rule 2 \\
Ta       & Ta        & $Im\overline{3}m$ & mp-50        & 1.566 & 3.19  & --0.23 & 3.42  & Rule 2 \\
W        & W         & $Im\overline{3}m$ & mp-91        & 1.987 & 3.86  & --0.13 & 3.99  & Rule 2 \\
Re       & Re        & $P6_3/mmc$        & mp-8         & 0.003 & 3.06  & 0.00  & 3.06  & Rule 2 \\
Os       & Os        & $P6_3/mmc$        & mp-49        & 0.000 & 1.88  & 0.03  & 1.85  & Rule 2 \\
Ir       & IrLi      & $P\overline{6}m2$ & mp-279       & 0.000 & 0.35  & --0.12 & 0.47  & Rule 3 \\
Pt       & Li$_2$Pt  & $P6/mmm$          & mp-2170      & 0.000 & 0.06  & --0.04 & 0.09  & Proceeds \\
Au       & Li$_{15}$Au$_4$ & $I\overline{4}3d$ & mp-567395 & 0.000 & 0.22  & --0.01 & 0.22  & Proceeds \\
Hg       & Li3Hg        & $P6/mmm$          & mp-1646     & 0.000 & 0.07  & 0.06  & 0.01  & Proceeds \\
Tl       & Li$_3$Tl  & $Fm\overline{3}m$ & mp-7396      & 0.000 & 0.02  & 0.01  & 0.01  & Proceeds \\
Pb       & Li$_{17}$Pb$_4$ & $F\overline{4}3m$ & mp-574275 & 0.000 & 0.18  & 0.06  & 0.12  & Proceeds \\
Bi       & Li$_3$Bi  & $Fm\overline{3}m$ & mp-23222     & 0.000 & 0.23  & 0.18  & 0.05  & Proceeds \\
\hline \hline
\end{tabular*}
\endgroup
\end{table}

\section{Lattice optimization effect}\label{sec:latt_opt_effect}

\begin{figure}[H]
\centering
\includegraphics[width=0.9\columnwidth]{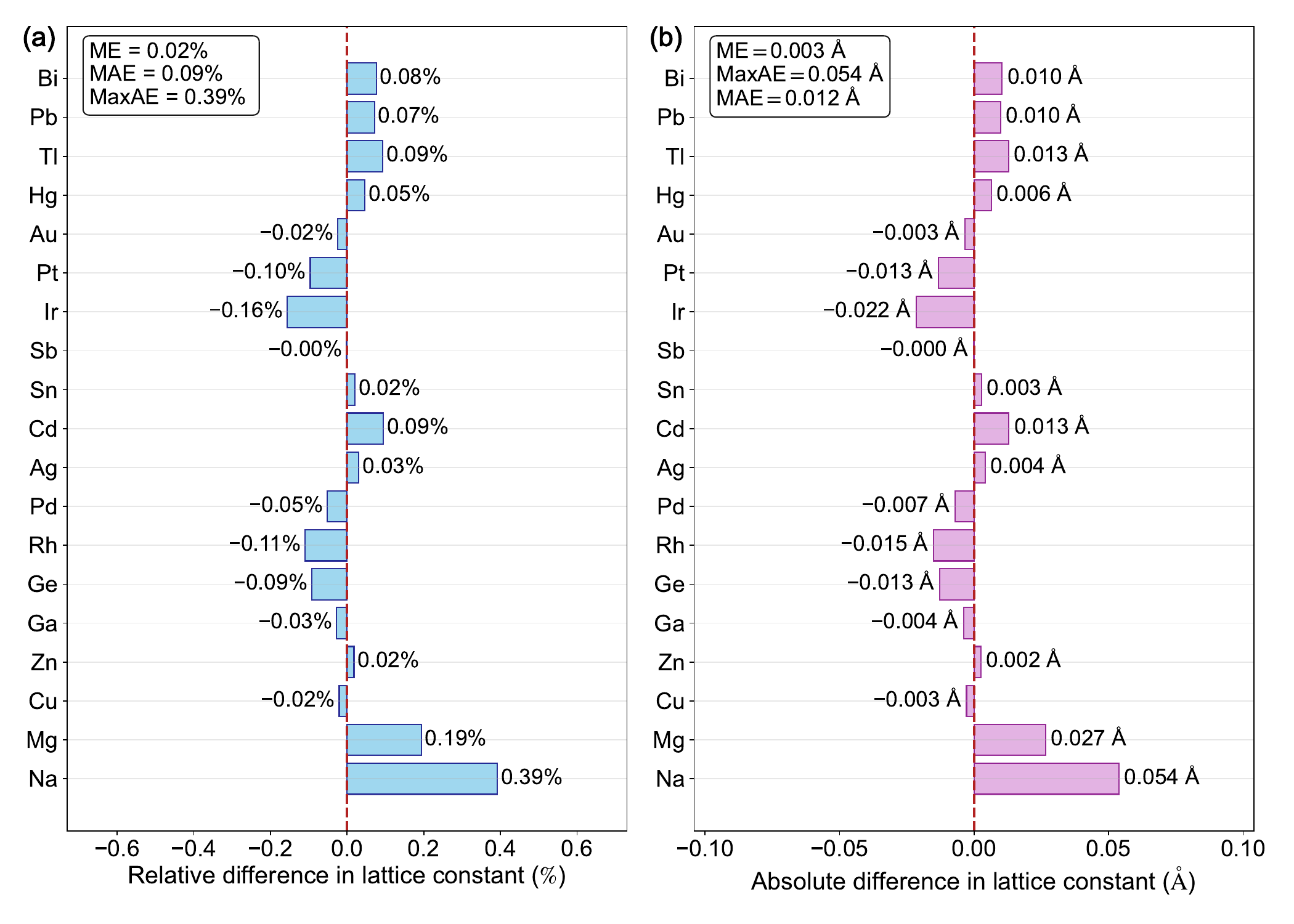}
\caption{Effect of lattice optimization, performed via an equation of state, on the equilibrium lattice constants of supercells alloyed with the best solute candidates ($\ce{Li_{127}\it{X}}$, with $X$ being the solute). Results are presented in (a) relative units (\%) and (b) absolute units (\AA).  }
\label{fig:opt_latt_const}
\end{figure}

Next, we optimized the lattice constants for both solutes and end-members; the results are shown in~\cref{fig:opt_latt_const}. Lattice constants changed by up to 0.075~\AA, with a mean absolute error (MAE) of 0.013~\AA.

\begin{figure}[H]
\centering
\includegraphics[width=0.9\columnwidth]{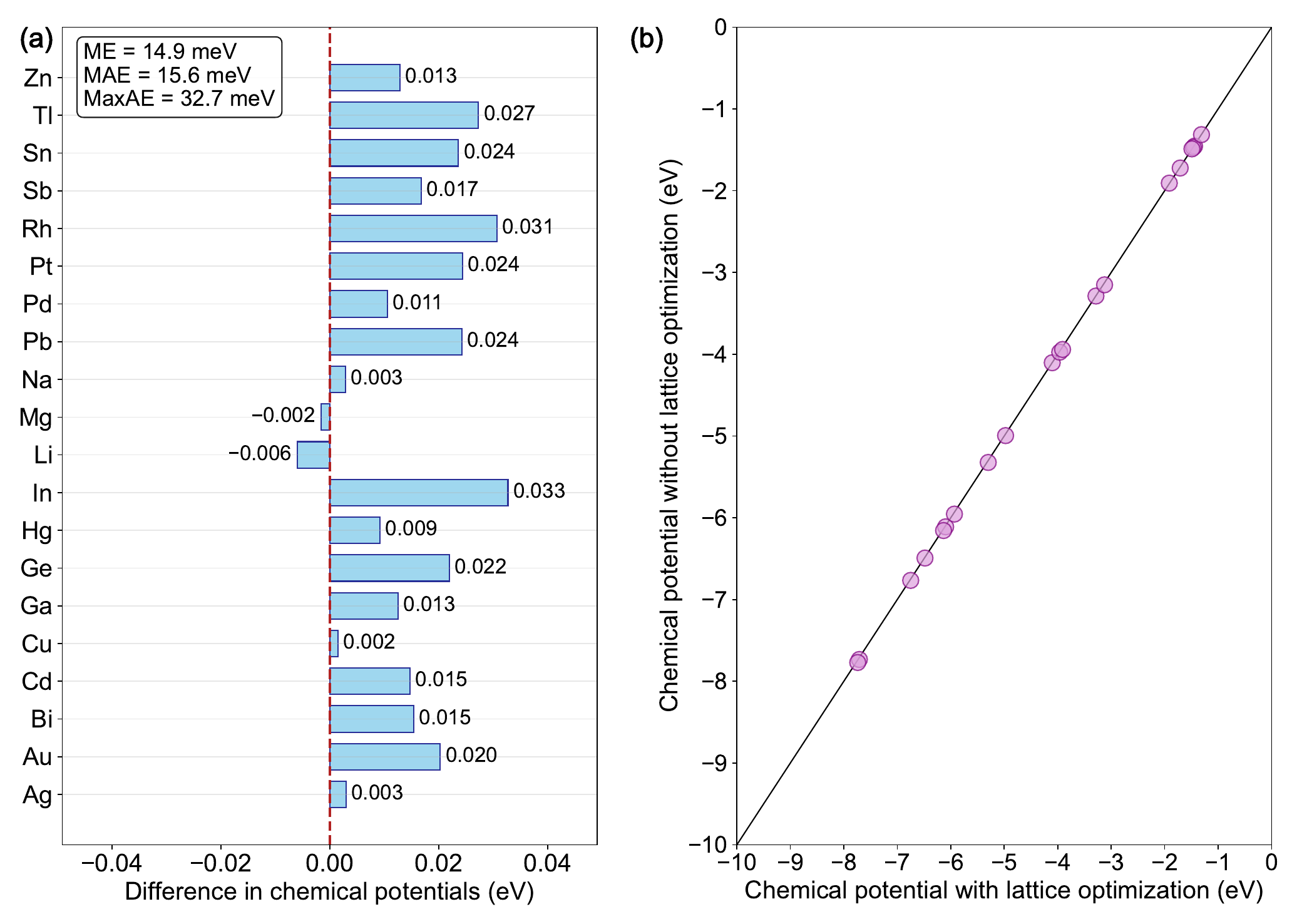}
\caption{Effect of lattice constant optimization on solute chemical potentials ($\mu_i$) from~\cref{tab:chemical_potentials_all}: (a) change in chemical potential for each considered solute; (b) scatter plot of chemical potentials computed with and without lattice optimization.}
\label{fig:opt_chem_pots}
\end{figure}

\begin{figure}[H]
\centering
\includegraphics[width=0.9\columnwidth]{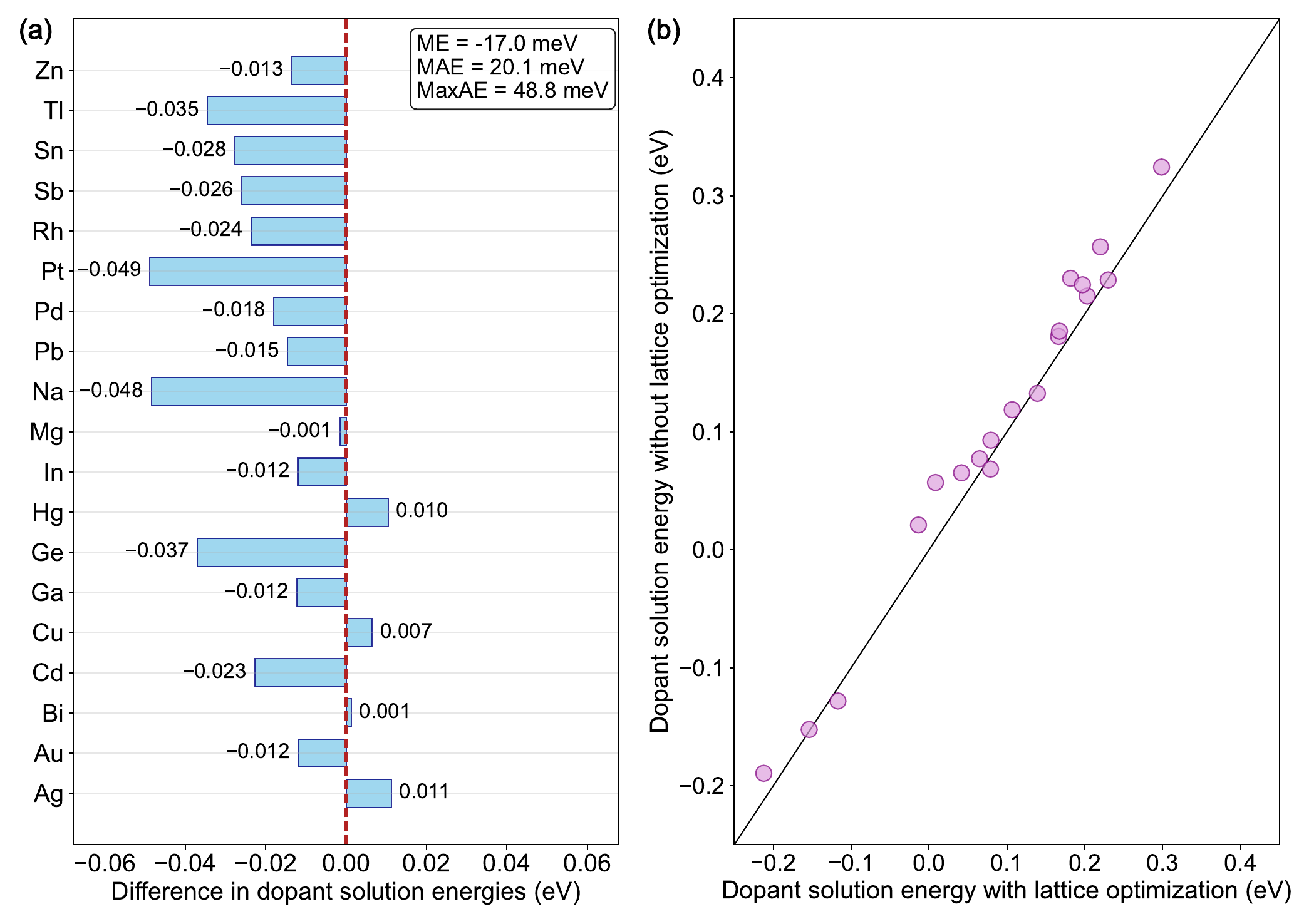}
\caption{Effect of lattice constant optimization on the solute solution energies ($E_\mathrm{s}$) of supercells with the best solute candidates ($\ce{Li_{127}\it{X}}$, where $X$ is the solute): (a) change in solution energy for each considered solute; (b) scatter plot of solution energies computed with and without lattice optimization.}
\label{fig:opt_sol_energy}
\end{figure}

\begin{figure}[H]
\centering
\includegraphics[width=0.9\columnwidth]{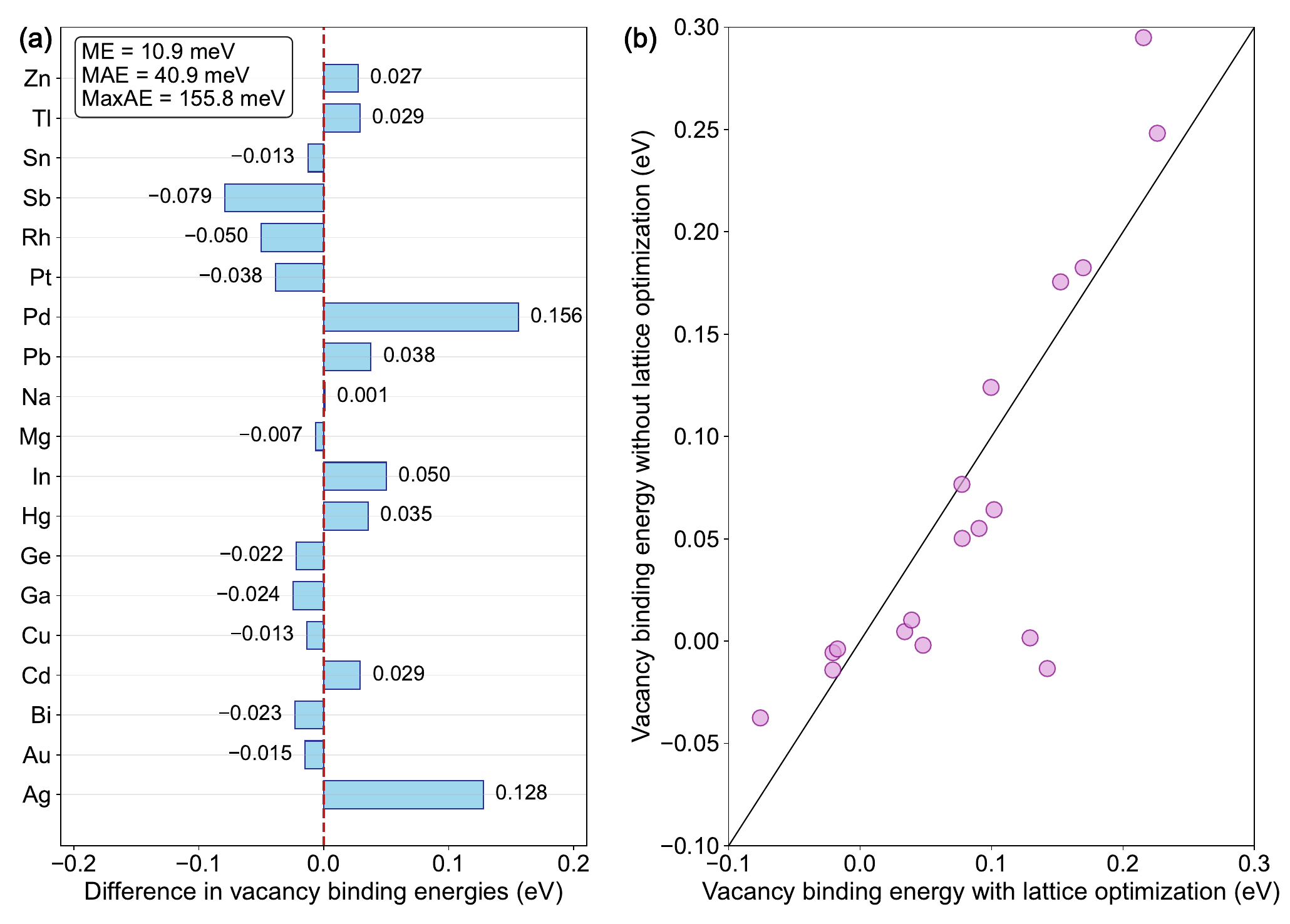}
\caption{Effect of lattice constant optimization on the solute--vacancy binding energies ($E_\mathrm{b}$) of supercells with the best solute candidates ($\ce{Li_{127}\it{X}}$, where $X$ is the solute): (a) change in solution energy for each considered solute; (b) scatter plot of solution energies computed with and without lattice optimization.}
\label{fig:opt_bind_energy}
\end{figure}

\begin{figure}[H]
\centering
\includegraphics[width=0.95\columnwidth]{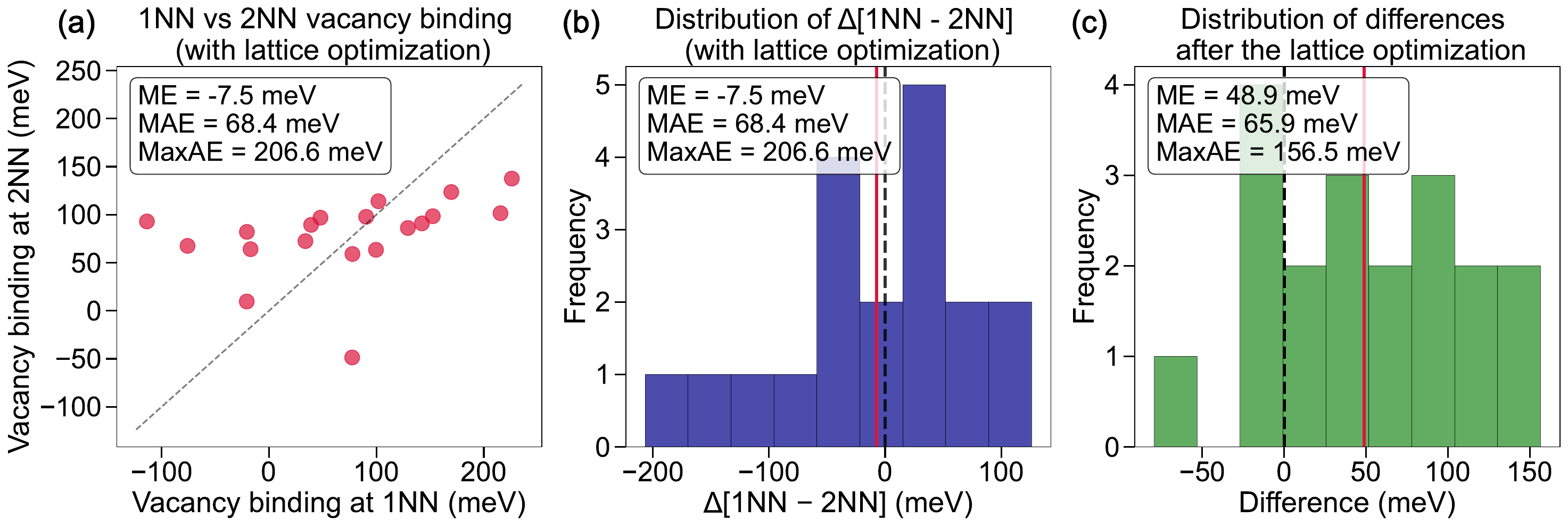}
\caption{Effect of including vacancies in the second-nearest-neighbor (2NN) coordination shell on the binding energies of the most promising solutes. (a) Solute--vacancy binding energies calculated for the first (1NN, 2.98~\AA\ distance) and second (2NN, 3.41~\AA\ distance) coordination shells following lattice constant optimization. (b) Distribution of errors in the solute--vacancy binding energies for the 1NN and 2NN shells after lattice optimization. (c) Error distribution of solute--vacancy binding energies across different calculation stages, comparing the 1NN model without lattice optimization against the maximum binding energy from the 1NN and 2NN shells with lattice optimization, for all solute elements and for the best-performing solutes. The solid red vertical line marks the mean error, and the dashed black line represents the zero-error reference.   }
\label{fig:opt_bind_energy_env}
\end{figure}

Using the optimized lattice constants, we recalculated the chemical potentials $\mu_i$, the solute solution energies $E_{\mathrm{s}}$, and the vacancy binding energies $E_{\mathrm{b}}$. Lattice optimization caused only minor changes in the chemical potentials, as shown in \cref{fig:opt_chem_pots}. As summarized in~\cref{fig:opt_sol_energy}, lattice optimization shifted $\mu_i$ by up to 33~meV, with a mean absolute error (MAE) of 16~meV, and $E_{\mathrm{s}}$ by up to 49~meV, with MAE of 20~meV. A more pronounced effect was observed for the binding energies: optimization changed $E_{\mathrm{b}}$ by as much as 156~meV (MAE=41~meV), as depicted in~\cref{fig:opt_bind_energy}. Additionally, we examined the influence of including vacancies in the second-nearest-neighbor (2NN) shell, which altered the binding energies by up to 157~meV (MAE=66~meV) (see~\cref{fig:opt_bind_energy_env}).

The effect of lattice constant optimization is summarized in~\cref{tb:effect_of_latt_opt}. To avoid excluding potentially suitable solutes when applying the $E_{\mathrm{eff}}<0$ criterion, the possible error introduced should be taken as the sum of the maximum absolute errors for each quantity, which amounts to 0.21~eV.

\begin{table}[H]
\centering
\caption{Effect of lattice optimization on the calculated properties: maximum absolute error (MaxAE), mean error (ME), and mean absolute error (MAE) for lattice constants, chemical potentials $\mu_i$, solute solution energies $E_{\mathrm{s}}$, and solute--vacancy binding energies $E_{\mathrm{b}}$. The effect of including vacancies in the second-nearest-neighbor (2NN) coordination shell on $E_{\mathrm{b}}$ is also reported.}
\label{tb:effect_of_latt_opt}
\setlength{\tabcolsep}{15pt}
\renewcommand{\arraystretch}{1.15} 
\begin{tabular*}{0.9\columnwidth}{llrrr}
\hline \hline
Property and units of measurement                       & MaxAE & ME & MAE \\ \hline
Lattice constant, pm         &   5.4    &  0.3  &  1.2    \\
Chemical potential, meV     &    32.7   & 14.9   &  15.6    \\
Solution energy, meV        &    48.8   &  --17.0  &  20.1   \\
Solute--vacancy binding energy (lattice constant optimization), meV &    155.8   &  10.9  &  40.9  \\ 
Solute--vacancy binding energy (1NN and 2NN effect), meV &    156.5   &  48.9  &  65.9  \\ 

\hline \hline
\end{tabular*}
\end{table}

\clearpage
\section{Thermodynamic properties of the best candidates}

For the best candidates, we performed lattice constant optimization and also calculated the solute--vacancy binding energy at the 2NN coordination sphere. The binding energy was then taken as the more favorable (lower) value between the 1NN and 2NN coordination spheres:

\begin{equation}
     E_{\mathrm{b}}^{\mathrm{refined}} = \max\left( E_{\mathrm{b}}^{\mathrm{1NN}}, \, E_{\mathrm{b}}^{\mathrm{2NN}} \right),
    \label{eq:eff_for_2nn}
\end{equation}
where $E_{\mathrm{b}}^{\mathrm{1NN}}$ and $E_{\mathrm{b}}^{\mathrm{2NN}}$ are the solute--vacancy binding energies when the vacancy is located at the 1NN and 2NN coordination spheres of the solute, respectively.

The compiled data for the best solute candidates after lattice constant optimization is provided in~\cref{tb:best_solutes}. In total, nine candidates have negative $E_{\mathrm{eff}}$ and proceed to further molecular dynamics simulations and experimental validation.

\begin{table}[H]
\caption{Best candidates, end-members on the phase diagram at $T=0$~K, substitutional energy, and vacancy binding energy at 1NN and 2NN coordination spheres.}
\label{tb:best_solutes}
\centering
\setlength{\tabcolsep}{6pt}
\renewcommand{\arraystretch}{1.1} 
\begin{tabular*}{0.99\columnwidth}{@{\extracolsep{\fill}} l l l l r r r r l }
\hline \hline
Element & End-member & Space group & Mp-id & $E_{\mathrm{s}}$ (eV) & $E_{\mathrm{b}}^{\mathrm{1NN}}$ (eV) & $E_{\mathrm{b}}^{\mathrm{2NN}}$ (eV) & $E_{\mathrm{eff}}$ (eV) & Solute \\ \hline
Na  & Na        & $Im\overline{3}m$ & mp-127       & 0.18 & 0.08 & --0.05 & 0.10 & no \\
Mg  & Mg        & $P6_3/mmc$        & mp-153       & --0.15 & --0.02 & 0.01 & --0.01 & yes \\
Cu  & Cu        & $Fm\overline{3}m$ & mp-30        & 0.14 & --0.02 & 0.06 & 0.08 & no \\
Zn  & LiZn      & $Fd\overline{3}m$ & mp-1934      & 0.08 & 0.08 & 0.06 & 0.00 & no \\
Ga  & Li$_2$Ga  & $Cmcm$            & mp-29210     & 0.06 & 0.10 & 0.06 & --0.04 & yes \\
Ge  & Li$_{15}$Ge$_4$ & $I\overline{4}3d$ & mp-1777 & 0.22 & 0.23 & 0.14 & -0.01 & yes \\
Rh  & LiRh      & $P\overline{6}m2$ & mp-600561    & 0.04 & --0.11 & 0.09 & --0.05 & yes \\
Pd  & Li$_{15}$Pd$_4$ & $I\overline{4}3d$ & mp-1197547 & 0.17 & 0.14 & 0.09 & 0.03 & no \\
Ag  & LiAg      & $Pm\overline{3}m$ & mp-2426      & --0.12 & 0.13 & 0.09 & --0.13 & yes \\
Cd  & LiCd      & $Fd\overline{3}m$ & mp-1437      & --0.21 & 0.03 & 0.07 & --0.07 & yes \\
In  & Li$_{13}$In$_3$ & $Fd\overline{3}m$ & mp-510430 & 0.11 & 0.05 & 0.10 & 0.01 & no \\
Sn  & Li$_{17}$Sn$_4$ & $F\overline{4}3m$ & mp-573471 & 0.20 & 0.17 & 0.12 & 0.03 & no \\
Sb  & Li$_3$Sb  & $Fm\overline{3}m$ & mp-2074      & 0.30 & 0.22 & 0.10 & 0.08 & no \\
Pt  & Li$_2$Pt  & $P6/mmm$          & mp-2170      & 0.01 & --0.08 & 0.07 & --0.06 & yes \\
Au  & Li$_{15}$Au$_4$ & $I\overline{4}3d$ & mp-567395 & 0.20 & --0.02 & 0.08 & 0.12 & no \\
Hg  & \ce{Li3Hg}        & $P6/mmm$          & mp-1646     & 0.08 & 0.09 & 0.10 & --0.02 & yes \\
Tl  & Li$_3$Tl  & $Fm\overline{3}m$ & mp-7396      & --0.01 & 0.04 & 0.09 & --0.09 & yes \\
Pb  & Li$_{17}$Pb$_4$ & $F\overline{4}3m$ & mp-574275 & 0.17 & 0.10 & 0.11 & 0.06 & no \\
Bi  & Li$_3$Bi  & $Fm\overline{3}m$ & mp-23222     & 0.23 & 0.15 & 0.10 & 0.08 & no \\
\hline \hline
\end{tabular*}
\end{table}

\section{Effective vacancy increase for the best solutes}\label{sec:eff_vac_increase}

The corresponding increase in effective vacancy concentration was evaluated at room temperature ($T = 300$~K) relative to the intrinsic vacancy concentration in pristine bcc Li, following~\cref{eq:enhancement_factor}. The maximum enhancement factor at maximum solubility calculated thermodynamically ($\eta_{\mathrm{th}}$) is calculated as follows: 

\begin{equation}
    \eta_{\mathrm{th}}  = 1 + z\exp{\left(-\frac{E_{\mathrm{eff}}}{k_{\mathrm{B}}T} \right)},
    \label{eq:vac_conc_eff_increase}
\end{equation}
where $z$ is the coordination number in the system ($z=8$ for bcc structure).

However, because the thermodynamically derived solubility limit and the experimental limit may differ substantially (see~\cref{tab:solubility}) due to many-body interactions and the general dependence on solute concentration $c$ (i.e., $E_{\mathrm{s}} = E_{\mathrm{s}}(c)$ and $E_{\mathrm{b}} = E_{\mathrm{b}}(c)$), we instead used a more accurate approach with a variable solute concentration $c < c_{\mathrm{max}}$. In addition, we omitted the coordination number $z$ (the number of equivalent binding sites around a solute) because the correlation factor between vacancies cannot be reliably estimated from DFT calculations. To maintain a conservative estimate, we set $z = 1$, which reduces the maximum enhancement factor at the solubility limit in~\cref{eq:enhancement_factor} to

\begin{equation}
    \eta = 1 + c \, \exp{\left(\frac{E_{\mathrm{b}}}{k_{\mathrm{B}}T} \right)} = 1 + c\, \tilde{\eta}
    \label{eq:vac_conc_eff_increase_only_eb}
\end{equation}
where $c$ is the manually selected atomic concentration of solute, chosen to align with experimentally achievable values ($c < c_{\mathrm{max}}$) and $\tilde{\eta}= \exp{\left(\frac{E_{\mathrm{b}}}{k_{\mathrm{B}}T}\right)}$ is the Boltzmann factor describing the vacancy-concentration enhancement per solute atom.

In essence, we obtain a simplified version of the Lomer model in the dilute limit (low solute concentrations)~\cite{lomer1958equilibrium} with $z=1$, formulated as follows:

\begin{equation}
    n_{\mathrm{alloy}} = n_{\mathrm{Li}} \left[ 1 - c + zc \, \exp{\left(\frac{E_{\mathrm{b}}}{k_{\mathrm{B}}T} \right)} \right],
\end{equation}
where $n_{\mathrm{Li}}$ is the equilibrium vacancy concentration in pristine Li, $c$ is the atomic fraction of the solute element, and $z=1$ is the coordination number of the host structure assumed in this simplification.

The enhancement factor ($\tilde{\eta}$) for the best solute candidates and the corresponding maximum enhancement factors $\eta_{\mathrm{max}}$ at solubility limit are listed in~\cref{tb:vac_conc_gain}.

\begin{table}[H]
\caption{The most promising solute candidates, along with their binding energies ($E_{\mathrm{b}}$), the Boltzmann factor describing the vacancy-concentration enhancement per solute atom ($\tilde{\eta}$, see \cref{eq:vac_conc_eff_increase_only_eb}), the experimental solubility limit $c_{\mathrm{max}}$, and the vacancy enhancement factor at this limit ($\eta_{\mathrm{max}}$), evaluated at $T = 300$~K and sorted by decreasing $\eta_{\mathrm{max}}$.}
\label{tb:vac_conc_gain}
\centering
\setlength{\tabcolsep}{13pt}
\begin{tabular*}{0.53\columnwidth}{l r r r r}
\hline \hline
Element  & $E_{\mathrm{b}}$, meV &  $\tilde{\eta}$ & $c_{\mathrm{max}}$, at.\% & $\eta_{\mathrm{max}}$ \\ \hline
Ag      &  129  &   146.9  &   9.0   &  14.2  \\
Ge      &  226  &  6260.8  &   0.1   &   7.3  \\
Cd      &   72  &    16.2  &  18.0   &   3.9  \\
Mg      &   10  &     1.5  &  75.5   &   2.1  \\
Hg      &   98  &    44.3  &   1.0   &   1.4  \\
Ga      &  100  &    47.9  &   0.1   &   1.0  \\
Rh      &   93  &    36.5  &   0.1   &   1.0  \\
Tl      &   89  &    31.3  &   0.1   &   1.0  \\
Pt      &   67  &    13.4  &   0.1   &   1.0  \\
\hline \hline
\end{tabular*}
\end{table}

\section{Effect of alloying on critical current density}

\subsection{Critical current density from vacancy concentration in Li anode}\label{sec:ccd_estimations_si}

In Li/SE interface, the transport limit is due to the slow stripping and low stripping current~\cite{kasemchainan2019critical, lu2021critical}. The voids are formed due to the vacancy supply to the interface, as they cannot be effectively taken away from the interface. These findings are also consistent with studies by Krauskopf~\textit{et al.}~\cite{krauskopf2019toward, krauskopf2020fast} of Li/LLZO interfaces. The natural critical current density (CCD) of the Li/LLZO interface without externally applied pressure is $J_0 \simeq 280~\mu\mathrm{A~cm}^{-2}$~\cite{flatscher2020natural}. The reference values are calculated as averaged from the experimental works and were compiled in the work by Burov~\textit{et al.}~\cite{burov2026machinelearningassistedmolecular}.

During stripping, the flux of Li atoms removed from the interface equals to the anode side should match the flux of vacancies from SE side, otherwise voids and pores will form at the interface from the anode side. The flux of vacancies from the interface to the bulk of Li anode equals:

\begin{equation}
J_{\mathrm{v}} = e \, D_{\mathrm{Li}} \, \frac{\Delta C_{\mathrm{v}}^{\mathrm{vol}}}{L_{\mathrm{Li}}},
\end{equation}
where  $e = 1.602\times 10^{-19}\,\mathrm{C}$ is the elementary charge,  $D_{\mathrm{Li}} = 3\times 10^{-4}\,\mathrm{cm}^2/\mathrm{s}$ is the vacancy diffusion coefficient in bcc Li~\cite{mali19886li,messer1975nuclear,lodding1970isotope_li_diffusion},  $\Delta C_{\mathrm{v}}^{\mathrm{vol}}$ is the volumetric vacancy concentration difference between bulk Li and interface,  $L_{\mathrm{Li}}$ is the characteristic diffusion length in Li.

At the natural CCD, the interface is nearly depleted of vacancies, so 

\begin{equation*}
\Delta C_{\mathrm{v}}^{\mathrm{vol}} \approx C_{\mathrm{v}}^{\mathrm{Li}} = \rho_{\mathrm{Li}} \times 2.0 \times 10^{-7},
\end{equation*}
where $\rho_{\mathrm{Li}} \approx 4.68\times 10^{22}\,\mathrm{cm}^{-3}$ being the atomic density of bcc Li. The fraction of vacancies ($2.0\times 10^{-7}$) were derived from the work by Frank~\textit{et al.}~\cite{frank1996first}.

If we increase the effective bulk vacancy concentration by a factor of $N$, the available concentration gradient increases by the same factor and the maximum CCD will increase proportionally, giving

\begin{equation*}
J_{\mathrm{Li}}(N) = N \, J_0 = N \times 280~\mu\mathrm{A\,cm}^{-2},
\end{equation*}
where $J_0 = 280~\mu\mathrm{A~cm}^{-2}$ is the natural critical current density of the Li/LLZO interface without externally applied pressure~\cite{flatscher2020natural}, and $N$ is the vacancy enhancement factor.

This linear scaling holds as long as the Li metal remains the bottleneck and transport through the SE (Ga-LLZO) does not become the limiting phase, which is the upper bound of the alloying limit. The current density cannot exceed the maximum Li flux that the Ga-LLZO electrolyte can sustain. Using Fick’s law for the electrolyte:

\begin{equation}
J_{\mathrm{LLZO}}^{\max} = e \, D_{\mathrm{LLZO}} \, \frac{C_{\mathrm{v}}^{\mathrm{LLZO}} \, n_{\mathrm{LLZO}}}{L_{\mathrm{LLZO}}},
\end{equation}
where  $D_{\mathrm{LLZO}} \simeq 1.5\times 10^{-8}~\mathrm{cm}^2~\mathrm{s}^{-1}$~\cite{chen2020microstructural,mishra2025stabilization,kanai2021low,tao2023preparation},  $C_{\mathrm{v}}^{\mathrm{LLZO}} = 0.576$ (fraction of vacant Li sites in Ga-LLZO),  $n_{\mathrm{LLZO}}$ is the Li atomic density in LLZO. For Ga-LLZO, $\rho \approx 5.34~\mathrm{g\,cm}^{-3}$~\cite{tao2023preparation} gives $n_{\mathrm{LLZO}} \approx 2.68\times 10^{22}~\mathrm{cm}^{-3}$,  $L_{\mathrm{LLZO}}$ is the LLZO thickness (diffusion length in the electrolyte).

To compare this with the natural bulk Li limit, form the ratio:

\begin{equation*}
\frac{J_{\mathrm{LLZO}}^{\max}}{J_0}
=
\frac{D_{\mathrm{LLZO}}\,C_{\mathrm{v}}^{\mathrm{LLZO}}\,n_{\mathrm{LLZO}}\,/\,L_{\mathrm{LLZO}}}
     {D_{\mathrm{Li}}\,C_{\mathrm{v}}^{\mathrm{Li}}\,n_{\mathrm{Li}}\,/\,L_{\mathrm{Li}}}
\end{equation*}

Assuming the diffusion lengths are comparable ($L_{\mathrm{Li}} \approx L_{\mathrm{LLZO}}$), this simplifies to

\begin{equation*}
\frac{J_{\mathrm{LLZO}}^{\max}}{J_0}
\approx
\frac{(1.5\times10^{-8})(0.576)(2.68\times10^{22})}
     {(3\times10^{-4})(2.0\times10^{-7})(4.68\times10^{22})}
\approx 83.
\end{equation*}

Therefore,

\begin{equation*}
J_{\mathrm{LLZO}}^{\max} \approx 83 \times 280\ \mu\mathrm{A\,cm}^{-2}
= 23.2\ \mathrm{mA~cm}^{-2}
\end{equation*}

Combining both limits. The actual critical current density as a function of $N$ is the smaller of the two limiting fluxes:

\begin{equation*}
J_{\mathrm{crit}}(N) = \min \bigl( N \times 280 \ \mu\mathrm{A~cm}^{-2},\; J_{\mathrm{LLZO}}^{\max} \bigr)
\end{equation*}

Using our estimated cap $J_{\mathrm{LLZO}}^{\max} \approx 22.1\,\mathrm{mA~cm}^{-2}$, we obtain the piecewise result:

\begin{equation}
J_{\mathrm{crit}}(N) \approx 
\begin{cases}
N \times 280\ \mu\mathrm{A~cm}^{-2}, & N \le 83 \\ 
23.2 \ \mathrm{mA~cm}^{-2}, & N > 83
\end{cases}
\label{eq:ccd_gain}
\end{equation}

Hence, the maximum critical current density achievable by increasing the Li vacancy concentration is approximately $J_{\mathrm{crit}}^{\max} \approx 22.1\ \mathrm{mA~cm}^{-2}$ under the equal-length assumption. Increasing $N$ beyond $\sim 83$ brings no further benefit, because the Ga-LLZO electrolyte then becomes the limiting factor.

\subsection{CCD estimations after the alloying}

Knowing the vacancy enhancement factor, the gain in effective vacancy concentration for the alloy can be calculated as: 

The effective vacancy concentration ratio is defined as:
\begin{equation}
    N = \frac{n_{\mathrm{alloy}}}{n_{\mathrm{Li}}} = \eta_{\mathrm{max}},
    \label{eq:gain_vac_conc}
\end{equation}
where $n_{\mathrm{Li}}$ and $n_{\mathrm{alloy}}$ are the effective vacancy concentrations in pristine Li and the Li--$M$ alloy at solubility limit of $X$, respectively; $\eta_{\mathrm{max}}$ is the maximum vacancy enhancement factor at solubility limit $c_{\mathrm{max}}$ (see \cref{eq:vac_conc_eff_increase_only_eb}).

The maximum theoretical capacity was calculated as:

\begin{equation}
    Q_{\mathrm{th}} = \frac{nF}{3.6M(\mathrm{Li}-X)},
    \label{eq:cap_theor}
\end{equation}
where $n$ is the number of electrons transferred per formula unit, $F$ is the Faraday constant (96485~C\,mol$^{-1}$), and $M(\mathrm{Li}-X)$ is the molar mass of the Li--$X$ solid solution.

The maximum CCD for each solute element can be estimated from~\cref{eq:ccd_gain} and~\cref{eq:gain_vac_conc} using the effective vacancy concentration enhancement factor (see~\cref{tb:vac_conc_gain}) and the solubility limit of the element (see~\cref{tab:solubility}). The results are compiled in~\cref{tab:ccd_gain_all}.

% pristine Li metal anode has specific capacity of 3860 mAh/g

\begin{table}[H]
\centering
\setlength{\tabcolsep}{10pt}
\renewcommand{\arraystretch}{1.15}
\caption{Thermodynamic estimations of critical current density (CCD) values for Li--$X$ alloys at the maximum solubility limit of each solute ($c_{\mathrm{max}}$, see~\cref{tab:solubility}). The table lists the Boltzmann factor $\tilde{\eta} = \exp(E_{\mathrm{b}}/k_{\mathrm{B}}T)$ at $T = 300$~K and the vacancy enhancement factor at this limit, $\eta_{\mathrm{max}} = 1 + (c_{\mathrm{max}}/100)\,\tilde{\eta}$ (see~\cref{tb:vac_conc_gain}), where $c_{\mathrm{max}}/100$ is the atomic fraction. $M$ is the molar mass of the solute. CCD values are $0.280 \times \eta_{\mathrm{max}}$~mA~cm$^{-2}$. Both capacities use~\cref{eq:cap_theor} with one electron per Li atom, $n = 1 - c$, and $M(\mathrm{Li}{-}X) = (1-c)M_{\mathrm{Li}} + cM$. $Q(c_{\mathrm{max}})$ is evaluated at the solubility limit (for elemental Li, $c = 0$). $Q_{\mathrm{th}}$ is evaluated at the lower concentration where the estimated CCD equals 1~mA~cm$^{-2}$, and is omitted when the solubility limit does not reach that current.}
\label{tab:ccd_gain_all}
\begin{tabular}{lrrrrcrr}
\hline \hline
Element & $c_{\mathrm{max}}$ (at.\%) & $\tilde{\eta}$ & $\eta_{\mathrm{max}}$ & CCD (mA~cm$^{-2}$) & $M$ (g/mol) & $Q(c_{\mathrm{max}})$ (mAh/g) & $Q_{\mathrm{th}}$ (mAh/g) \\
\hline
Li  & 100.0 &    1.0 &   1.0 & 0.280 &   6.94 & 3862 &  --- \\
Rh  &   0.1 &   36.5 &   1.0 & 0.280 & 102.91 & 3805 &  --- \\
Pt  &   0.1 &   13.4 &   1.0 & 0.280 & 195.08 & 3756 &  --- \\
Tl  &   0.1 &   31.3 &   1.0 & 0.280 & 204.38 & 3751 &  --- \\
Ga  &   0.1 &   47.9 &   1.0 & 0.280 &  69.72 & 3823 &  --- \\
Hg  &   1.0 &   44.3 &   1.4 & 0.392 & 200.59 & 2989 &  --- \\
Mg  &  75.5 &    1.5 &   2.1 & 0.588 &  24.31 &  327 &  --- \\
Cd  &  18.0 &   16.2 &   3.9 & 1.092 & 112.41 &  848 &  952 \\
Ge  &   0.1 & 6260.8 &   7.3 & 2.044 &  72.63 & 3822 & 3845 \\
Ag  &   9.0 &  146.9 &  14.2 & 3.976 & 107.87 & 1522 & 3024 \\
\hline \hline
\end{tabular}
\end{table}

\section{DFT-NEB calculations}

\subsection{Li migration in bcc Li}

\begin{figure}[H]
\centering
\includegraphics[width=0.7\columnwidth]{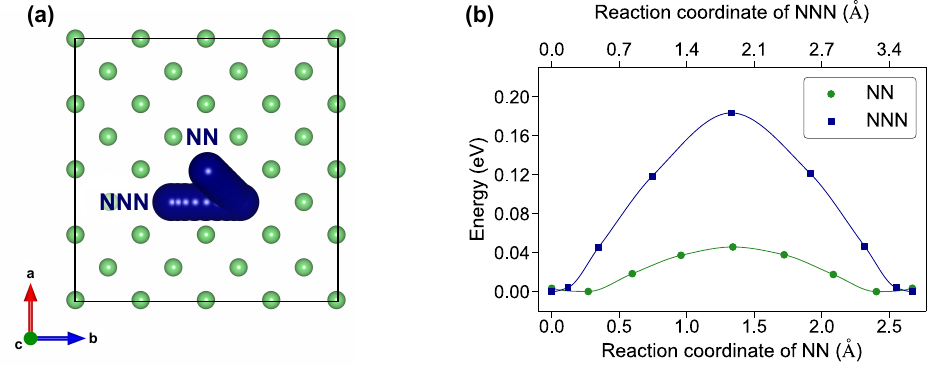}
\caption{DFT-NEB calculated migration pathways of Li in bcc-Li for nearest-neighbor (NN) and next-nearest-neighbor (NNN) jumps. (a) Schematic of the migration pathways, and (b) corresponding energy profiles.}
\label{fig:neb_li_both}
\end{figure}

Using DFT-based nudged elastic band (NEB) calculations, we computed the migration barriers for nearest-neighbor (NN) and next-nearest-neighbor (NNN) Li hops, as shown in~\cref{fig:neb_li_both}. The calculated NN barrier is 46~meV, while the NNN barrier is 183~meV. These values are in good agreement with previously reported literature values of 35~meV and 210~meV, respectively~\cite{behara2024fundamental}.

\subsection{Solute element migration}

\begin{figure}[H]
\centering
\includegraphics[width=0.9\columnwidth]{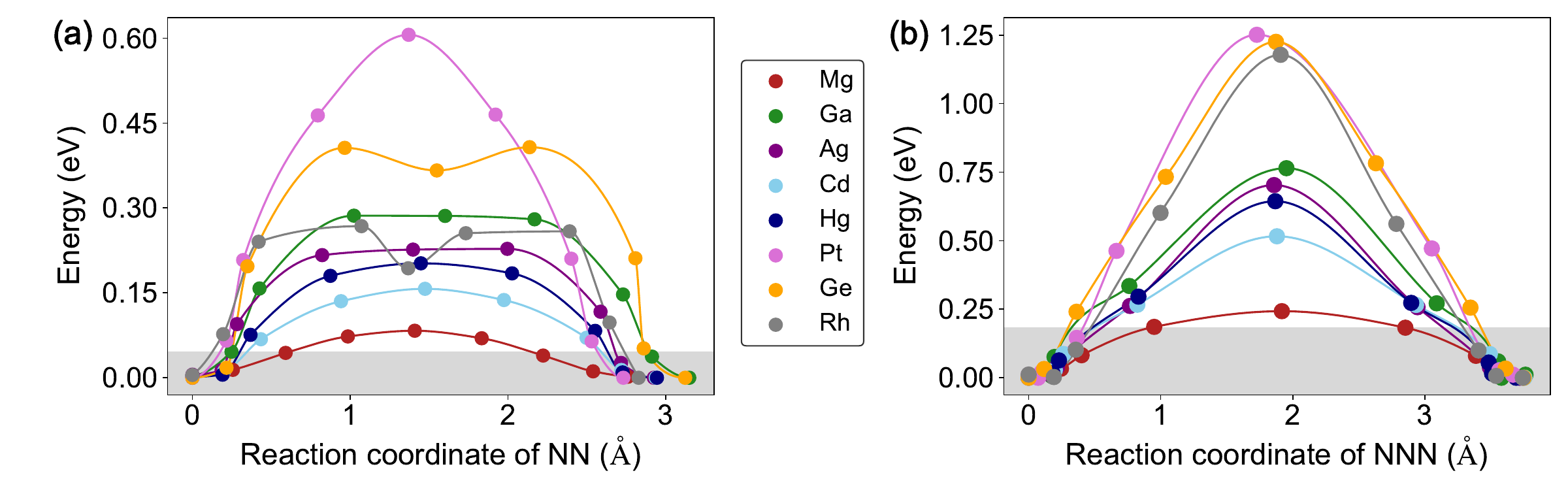}
\caption{DFT-NEB calculated migration pathways of solutes in $\ce{Li127\it{X}}$, where $X$ is the solute, for nearest-neighbor (NN) and next-nearest-neighbor (NNN) jumps. (a) NN energy profile, and (b) NNN energy profile. Gray shaded areas on subplots are migration barriers in bcc Li for NN and NNN pathways, respectively.}
\label{fig:solutes_NN_NNN}
\end{figure}

\subsection{Percolation pathway of solute--vacancy complex}

\begin{figure}[H]
\centering
\includegraphics[width=0.99\columnwidth]{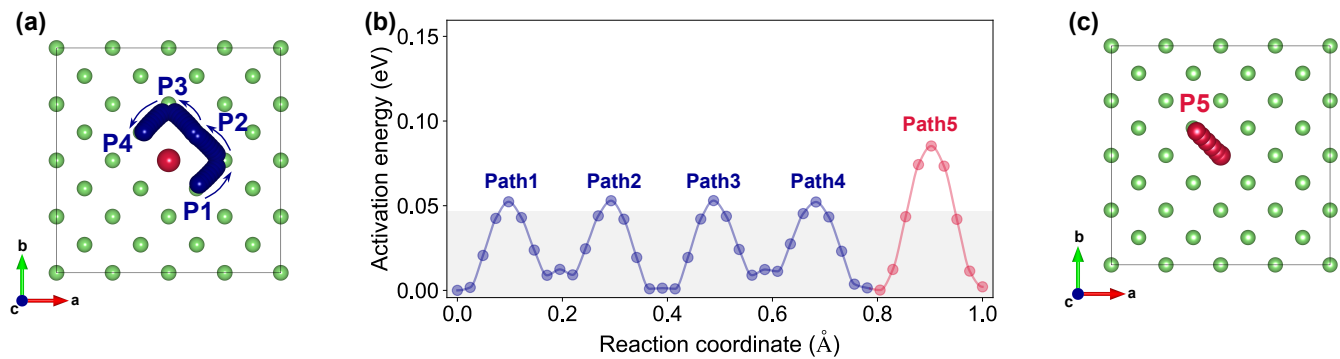}
\caption{(a) Single-ion Li migration in the Li$_{127}$Mg structure, with Li atoms depicted in blue. (b) DFT--NEB energy profile for single-ion percolation along the [$\overline{1}$10] direction, showing Li migration paths (Path1--Path4, blue) and the Mg migration path (Path5, red). (c) Single-ion Mg migration in the Li$_{127}$Mg structure, with Mg atoms depicted in red.}
\label{fig:neb_all_mg}
\end{figure}

\subsection{Vacancy enhancement versus trapping under Ag alloying}\label{sec:ag_vacancy_trapping}

The effect of Ag alloying on the effective Li vacancy diffusivity in bcc Li was analyzed using DFT-NEB calculations. The results reveal a competition between local vacancy enrichment near Ag and reduced mobility of vacancies trapped at Ag. The key DFT-NEB parameters are summarized in~\cref{tab:ag_doping_summary}.

\begin{table}[!htbp]
\centering
\caption{Summary of DFT-NEB calculated energetics for Li diffusion in a bcc Li supercell containing a substitutional Ag solute.}
\setlength{\tabcolsep}{12pt}
\label{tab:ag_doping_summary}
\begin{tabular}{l c l}
\hline\hline
Property & Value (eV) & Implication \\
\hline
Li vacancy formation energy & $0.50$ & High formation energy; low intrinsic vacancy concentration \\
Ag solution energy ($E_{\mathrm{s}}$) & $-0.12$ & Favorable Ag incorporation \\
Ag--vacancy binding energy ($E_{\mathrm{b}}$) & $0.13$ & Vacancies trapped near Ag \\
Li migration barrier (pristine) & $0.04$ & Fast intrinsic Li transport \\
Li migration barrier (near Ag) & $0.04$ & Local Li hops preserved \\
Vacancy--Ag escape barrier ($E_{\mathrm{esc}}$) & $0.12$ & Extra barrier to leave the Ag neighborhood \\
Ag migration barrier & $0.22$ & Ag nearly immobile relative to Li (vacancy-mediated path) \\
\hline\hline
\end{tabular}
\end{table}

\subsubsection{Local vacancy enrichment near Ag}

Because $E_{\mathrm{s}}<0$ for Ag, the dilute-solution Boltzmann factor based on $E_{\mathrm{eff}}=E_{\mathrm{s}}-E_{\mathrm{b}}$ is not applicable (see above). The local vacancy enrichment is therefore estimated from the binding energy alone:
\begin{equation}
\eta_{\mathrm{b}} = \exp\left(\frac{E_{\mathrm{b}}}{k_{\mathrm{B}} T}\right)
\label{eq:enhancement_factor_sum}
\end{equation}
At room temperature ($T = 300$~K, $k_{\mathrm{B}}T \approx 0.02585$~eV),
\begin{equation}
\eta_{\mathrm{b}} = \exp\left(\frac{0.13}{0.02585}\right) \simeq \exp(5) \simeq 146.9
\label{eq:enhancement_value}
\end{equation}
This enrichment is confined to the first coordination shell of Ag. In bcc Li the 1NN coordination is $z=8$; many-body solute--vacancy interactions may modify the local energetics at higher Ag concentrations.

\subsubsection{1NN site fraction and mean vacancy enhancement}

For an Ag concentration $c_{\mathrm{Ag}}$ (at.\%), the fraction of lattice sites in Ag 1NN shells is estimated as
\begin{equation}
f_{\mathrm{nn}} = \frac{z\, c_{\mathrm{Ag}}}{100} = \frac{8 c_{\text{Ag}}}{100},
\label{eq:volume_fraction}
\end{equation}
assuming non-overlapping shells (valid while $f_{\mathrm{nn}}\lesssim 1$ or $c_{\mathrm{Ag}}\lesssim 12.5$~at.\%). The corresponding spatially averaged vacancy-concentration enhancement relative to pristine Li is the following:
\begin{equation}
\eta_{\mathrm{avg}} = 1 + (\eta_{\mathrm{b}} - 1)\, f_{\mathrm{nn}}
\label{eq:effective_enhancement}
\end{equation}

\subsubsection{Diffusivity ratio with vacancy escape}

Local enrichment increases the vacancy population near Ag, but long-range transport requires vacancies to leave the trap. With escape barrier $E_{\mathrm{esc}}=0.12$~eV and pristine Li migration barrier $E_{\mathrm{a}}^{\mathrm{Li}}=0.04$~eV, the excess barrier is $E_{\mathrm{esc}}-E_{\mathrm{a}}^{\mathrm{Li}}=0.08$~eV. We adopt a two-region estimate in which pristine sites retain the intrinsic hop rate, while Ag 1NN sites contribute with the local enrichment $\eta_{\mathrm{b}}$ reduced by the escape penalty:
\begin{equation}
r_D = (1 - f_{\mathrm{nn}}) + \eta_{\mathrm{b}}\, f_{\mathrm{nn}}\, \exp\left(-\frac{E_{\mathrm{esc}} - E_{\mathrm{a}}^{\mathrm{Li}}}{k_{\mathrm{B}} T}\right),
\label{eq:net_enhancement}
\end{equation}
where $r_D$ is the estimated diffusivity ratio relative to pristine bcc Li. Equivalently, with $\eta_{\mathrm{b}}=146.9$:

\begin{equation}
r_D(c_{\mathrm{Ag}}) = 1 - f_{\mathrm{nn}} + 146.9\, f_{\mathrm{nn}}\, \exp\left(-\frac{0.08}{k_{\mathrm{B}} T}\right),
\quad f_{\mathrm{nn}}=\frac{8 c_{\mathrm{Ag}}}{100}
\label{eq:net_scaling}
\end{equation}

At $T=300$~K the escape factor is $\exp(-0.08/0.02585)\simeq 0.045$, so each Ag-influenced site contributes a net factor $\eta_{\mathrm{b}}\times 0.045\simeq 6.7$ relative to a pristine site, which is still larger than unity, despite trapping.

\subsubsection{Concentration dependence}

\cref{tab:ag_enhancement_values} lists $f_{\mathrm{nn}}$, $\eta_{\mathrm{avg}}$, and $r_D$ for representative Ag concentrations at $T=300$~K. Within this dilute, non-overlapping-shell model, $r_D>1$ for all concentrations: the local vacancy enrichment outweighs the escape penalty on Ag 1NN sites, while the remaining crystal retains pristine mobility. The ratio increases approximately linearly with $c_{\mathrm{Ag}}$, reaching $r_D\simeq 5.6$ at 10~at.\% Ag. Overlap of 1NN shells, solute--solute interactions, and correlation effects are neglected and would reduce these estimates at high alloying concentrations.

\begin{table}[!htbp]
\setlength{\tabcolsep}{10pt}
\centering
\caption{Ag alloying estimates for bcc Li at $T = 300$~K ($\eta_{\mathrm{b}}=146.9$, $z=8$).}
\label{tab:ag_enhancement_values}
\begin{tabular}{l c c c}
\hline\hline
Ag concentration (at.\%) & $f_{\mathrm{nn}}$ & $\eta_{\mathrm{avg}}$ & $r_D$ \\
\hline
0.5 & 0.040 & 6.9 & 1.23 \\
1.0 & 0.080 & 12.8 & 1.46 \\
2.0 & 0.160 & 24.7 & 1.92 \\
3.0 & 0.240 & 36.5 & 2.38 \\
4.0 & 0.320 & 48.4 & 2.84 \\
5.0 & 0.400 & 60.2 & 3.30 \\
6.0 & 0.480 & 72.0 & 3.76 \\
7.0 & 0.560 & 83.9 & 4.22 \\
8.0 & 0.640 & 95.7 & 4.68 \\
9.0 & 0.720 & 107.6 & 5.14 \\
10.0 & 0.800 & 119.4 & 5.60 \\
\hline\hline
\end{tabular}
\end{table}

\subsubsection{Implications and limits of the vacancy-mediated model}

With $z=8$ and $\eta_{\mathrm{b}}=146.9$, the two-region estimate predicts enhanced vacancy-mediated Li transport for dilute Ag alloying at room temperature ($r_D>1$), scaling roughly linearly with $c_{\mathrm{Ag}}$ up to the onset of shell overlap. Along the vacancy-mediated path, Ag remains nearly immobile: its migration barrier (0.22~eV) is much higher than that of Li (0.04~eV). Spin-lattice relaxation (SLR) measurements, however, indicate enhanced Ag diffusion in Li-rich solid solutions and suggest that Ag can diffuse faster than Li~\cite{korblein1985diffusion}, in apparent conflict with the vacancy-mediated NEB barriers. Interstitial and multi-hop pathways have been proposed and may lower the Ag activation energy by about a factor of two~\cite{korblein1985diffusion,thomas2024thermodynamic}. Resolving this discrepancy motivates MLMD simulations that include cooperative hops, lattice vibrations, finite-temperature effects, and alternative migration channels.

\subsection{Effect of cooperative multi-ion diffusion}\label{sec:neb_coop_effects}

The cooperative migration of several atoms can reduce the activation energy of diffusion, as previously shown by He~\textit{et al.}~\cite{He2017}. To model the simplest cooperative process, we considered two-stage jumps of the Li--solute pair: Li$\xleftrightarrow{\mathrm{Path 1}}$Vac and Cd$\xleftrightarrow{\mathrm{Path 2}}$Vac, as well as the corresponding process in which both jumps occur simultaneously.

\begin{figure}[H]
\centering
\includegraphics[width=0.99\columnwidth]{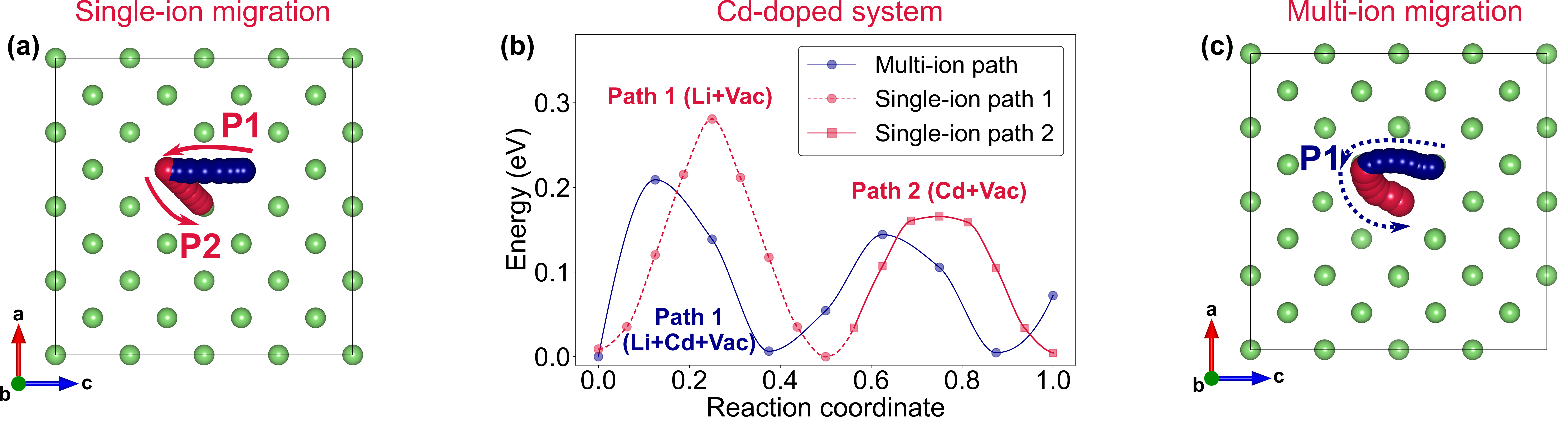}
\caption{(a) Single-ion migration of Li and Cd in the \ce{Li127Cd} structure:  Li $\xleftrightarrow{\text{Path 1}}$ Vac (blue atoms) and 
Cd $\xleftrightarrow{\text{Path 2}}$ Vac (red atoms). (b) Energy profiles of the single-ion and multi-ion processes:  the red dashed line represents single-ion Path~1, the red solid line represents single-ion Path~2, and the blue solid line represents the multi-ion path.
(c) Multi-ion migration of Li and Cd, in which both jumps occur simultaneously (blue corresponds to Li atoms, red to Cd atoms).}
\label{fig:neb_coop_cd}
\end{figure}

We calculated single-ion (\cref{fig:neb_coop_cd}.a) and cooperative (\cref{fig:neb_coop_cd}.b) diffusion of the solute--vacancy complex in the $\ce{Li127Cd}$ system. In the single-ion process, a Li atom first migrates to the vacant site (path 1, red dashed line), followed by a Cd atom migrating to the newly created vacancy (previously occupied by Li), as illustrated by path 2 (red solid line). In the multi-ion process, these two jumps occur simultaneously. As a result, the trajectories of Li and Cd become slightly curved, and the activation energy of this cooperative process is reduced to 209~meV from 281~meV for the single-ion process.

\begin{figure}[H]
\centering
\includegraphics[width=0.65\columnwidth]{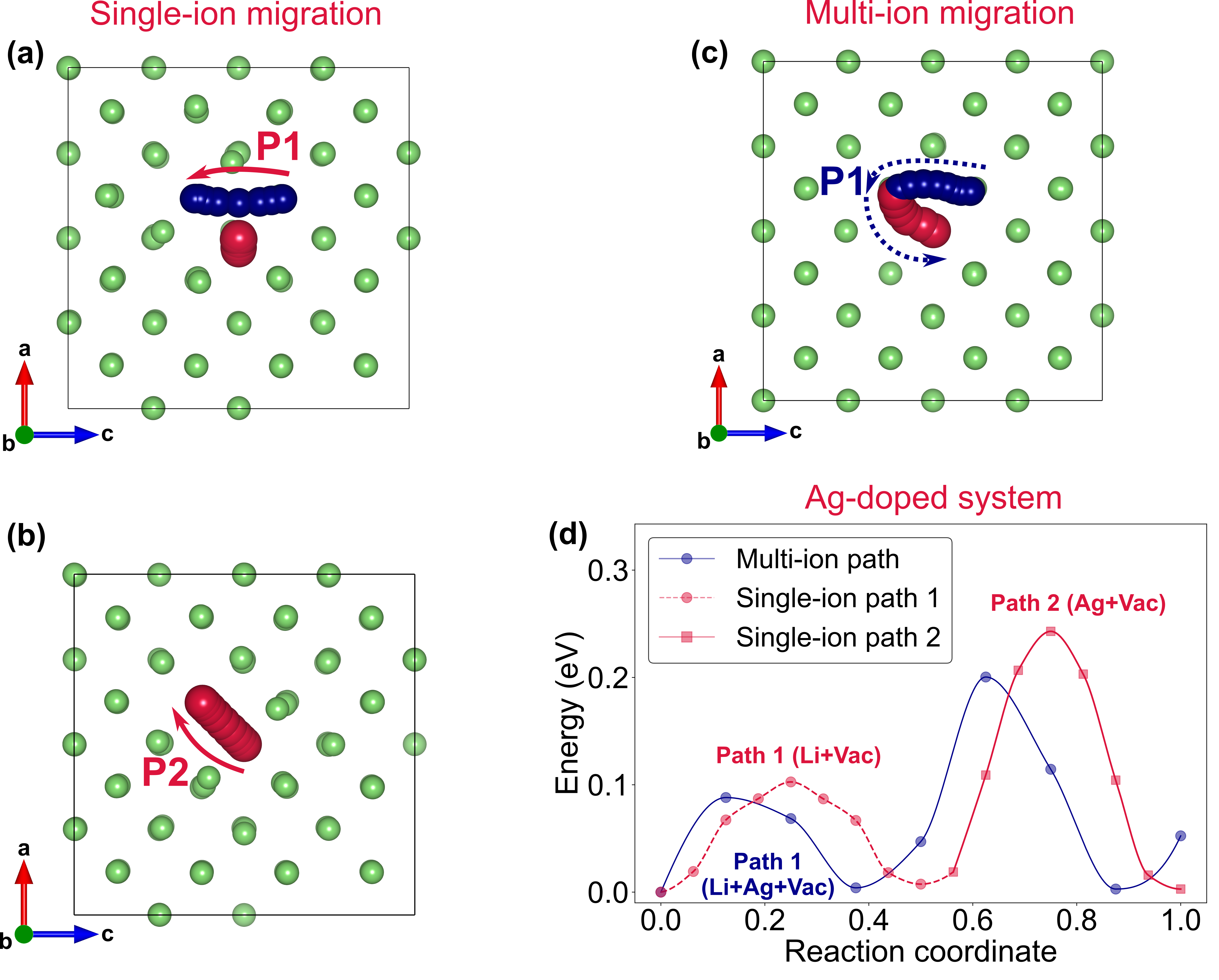}
\caption{Single-ion and multi-ion migration pathways in the Li$_{127}$Ag structure. (a) Single-ion migration of Li via Path~1 (Li$\xleftrightarrow{\mathrm{Path 1}}$Vac). (b) Single-ion migration of Ag via Path~2 (Ag$\xleftrightarrow{\mathrm{Path 2}}$Vac). (c) Multi-ion migration, where both Li and Ag jumps occur simultaneously (Li atoms in blue, Ag atoms in red). (d) Energy profiles: the red dashed line corresponds to single-ion Path~1, the red solid line to single-ion Path~2, and the blue solid line to the multi-ion migration pathway.}
\label{fig:neb_coop_ag}
\end{figure}

\section{Validation of Hume--Rothery rules}

The Hume--Rothery (H--R) rules govern the conditions under which an element can dissolve in a metal to form a solid solution~\cite{hume1934freezing}.  They are formulated in  \cref{eq:hr_rules_1,eq:hr_rules_2,eq:hr_rules_3,eq:hr_rules_4}. In this section, we extend the number of considered elements to 65, including As, Rb, Sr, Tc, Cs, Ba, Eu, Gd, Tb, Dy, Ho, Er, Tm, Yb, and Lu, which are considered only in this section.

\begin{figure}[H]
\centering
\includegraphics[width=0.99\columnwidth]{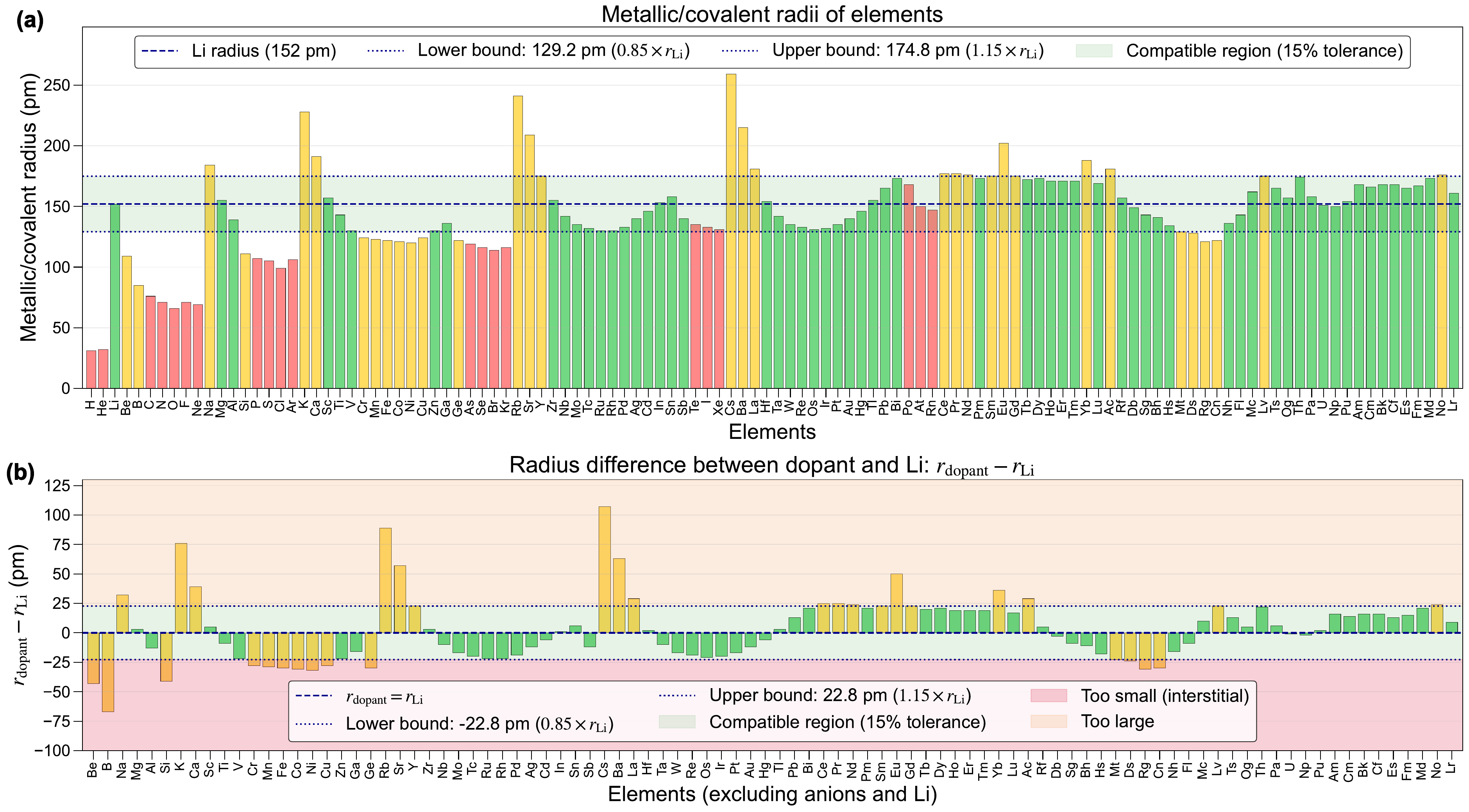}
\caption{(a) Radii of species (metallic for metals, covalent for non-metals) ordered by atomic number. The green shaded region indicates the compatibility zone ($\Delta r_i < 15\%$). Bars are color-coded by radius deviation: green (compatible), red (too small), orange (too large). (b) Radius difference $r_{\mathrm{solute}} - r_{\mathrm{Li}}$, with shaded regions marking compatible (green), too small (red), and too large (orange) zones. Green bars represent compatible solutes ($\Delta r_i < 15\%$), while orange bars represent all incompatible solutes (both too small and too large), simplifying the visualization of overall pass/fail status.  }
\label{fig:metallic_radii_analysis}
\end{figure}

To apply the first Hume--Rothery rule, we use metallic radii for solutes that possess a metallic phase (appropriate for solid solutions with Li metal) and covalent radii otherwise.~\cref{fig:metallic_radii_analysis} displays all radii and their differences from Li. The reference values were taken from the work by Cordero~\textit{et al.}~\cite{cordero2008covalent} and scaled in necessary, for example $r_{\mathrm{CN=8}}=0.97\,r_{\mathrm{CN=12}}$. The size rule is traditionally employed as the first screening criterion for solid solutions. Here, we validate its effectiveness within a DFT-driven screening funnel and examine whether the classical $\Delta r_i < 15\%$ criterion requires modification for accurate prediction of Li-based solid solutions.

The fourth rule has no standard quantitative metric, so we validate it by determining the optimal electronegativity difference between Li and the solute. For this purpose, we use Pauling's electronegativity scale~\cite{pauling1932nature}. We take the experimental phase diagrams (see~\cref{sec:exp_diagrams}) as the ground truth for this validation.

First, we divided the solutes into four categories, using the experimental data from~\cref{tab:solubility}: ``extensive solubility'', when the solid-solution region exceeds 1\%; ``limited solubility'', when it is approximately 1\%; ``minor solubility'', when it is less than 1\%; and ``insoluble'', when the solute does not form a solid solution with Li.

In our validation, we prioritized precision over recall, $F$-score, receiver operating characteristic (ROC) curve, or other metrics, as the cost of DFT calculations is negligible compared to the risk of missing a solid-solution candidate that could have industrial applications.

\begin{figure}[H]
\centering
\includegraphics[width=0.50\columnwidth]{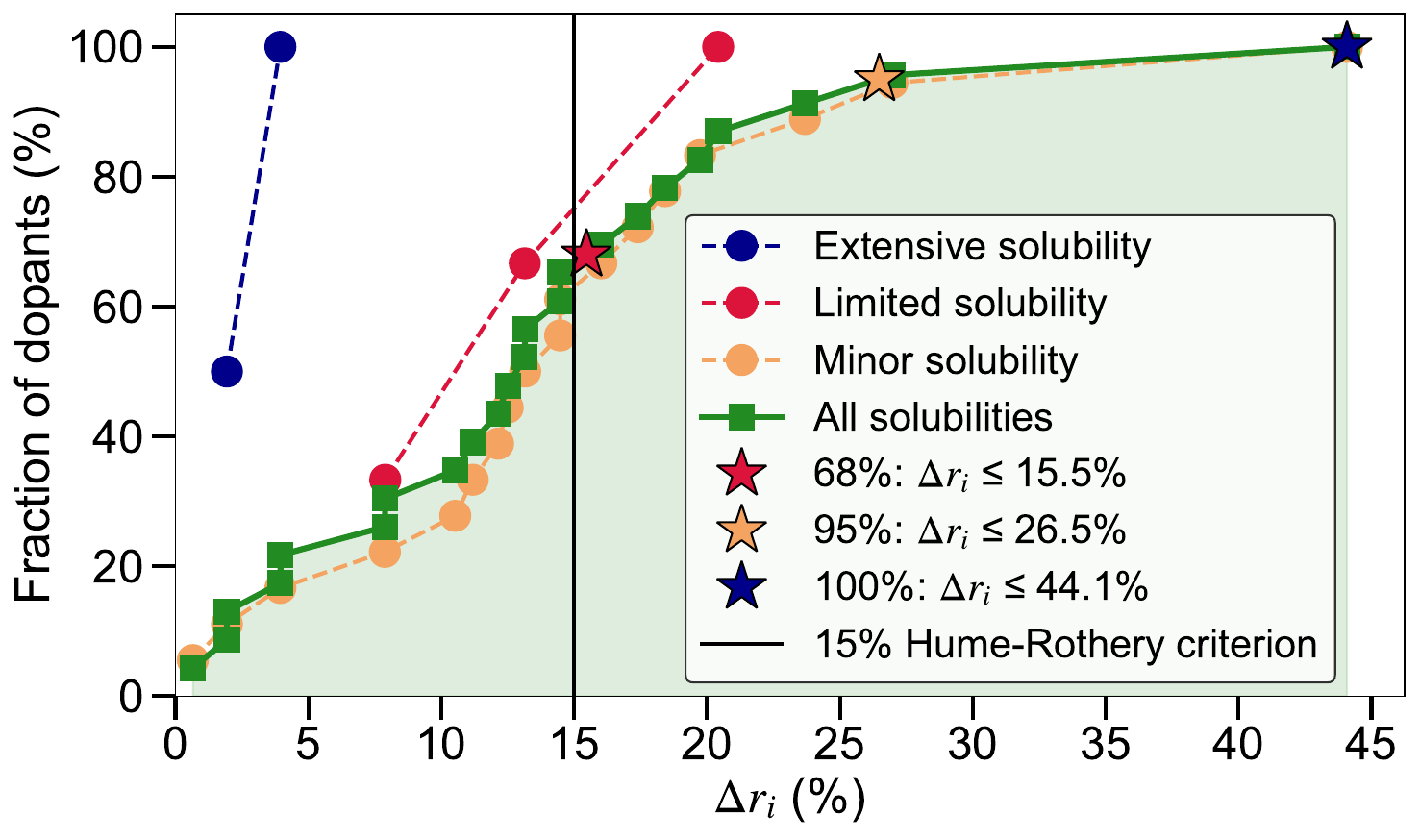}
\caption{Validation of the first Hume--Rothery rule, related to the difference in host--solute metallic radii. Blue denotes extensive solubility ($>1\%$), red denotes limited solubility ($\sim 1\%$), orange denotes minor solubility ($<1\%$), and green denotes the combined region. Star symbols mark $\sigma$ (65\% precision), $2\sigma$ (95\% precision), and the ideal prediction (100\% precision).}
\label{fig:hr_rule_1}
\end{figure}

The validation of the first H--R rule is visualized in~\cref{fig:hr_rule_1}. The data show that only solutes with extensive solubility are correctly predicted by this rule and satisfy the criterion $\Delta r_i \leq 15\%$. Some solutes with smaller solubility regions are underpredicted with this criterion. To predict all solutes across all solubility regimes, the criterion should be increased to $\Delta r_{\mathrm{opt}} \leq 44.1\%$.

\begin{figure}[H]
\centering
\includegraphics[width=0.55\columnwidth]{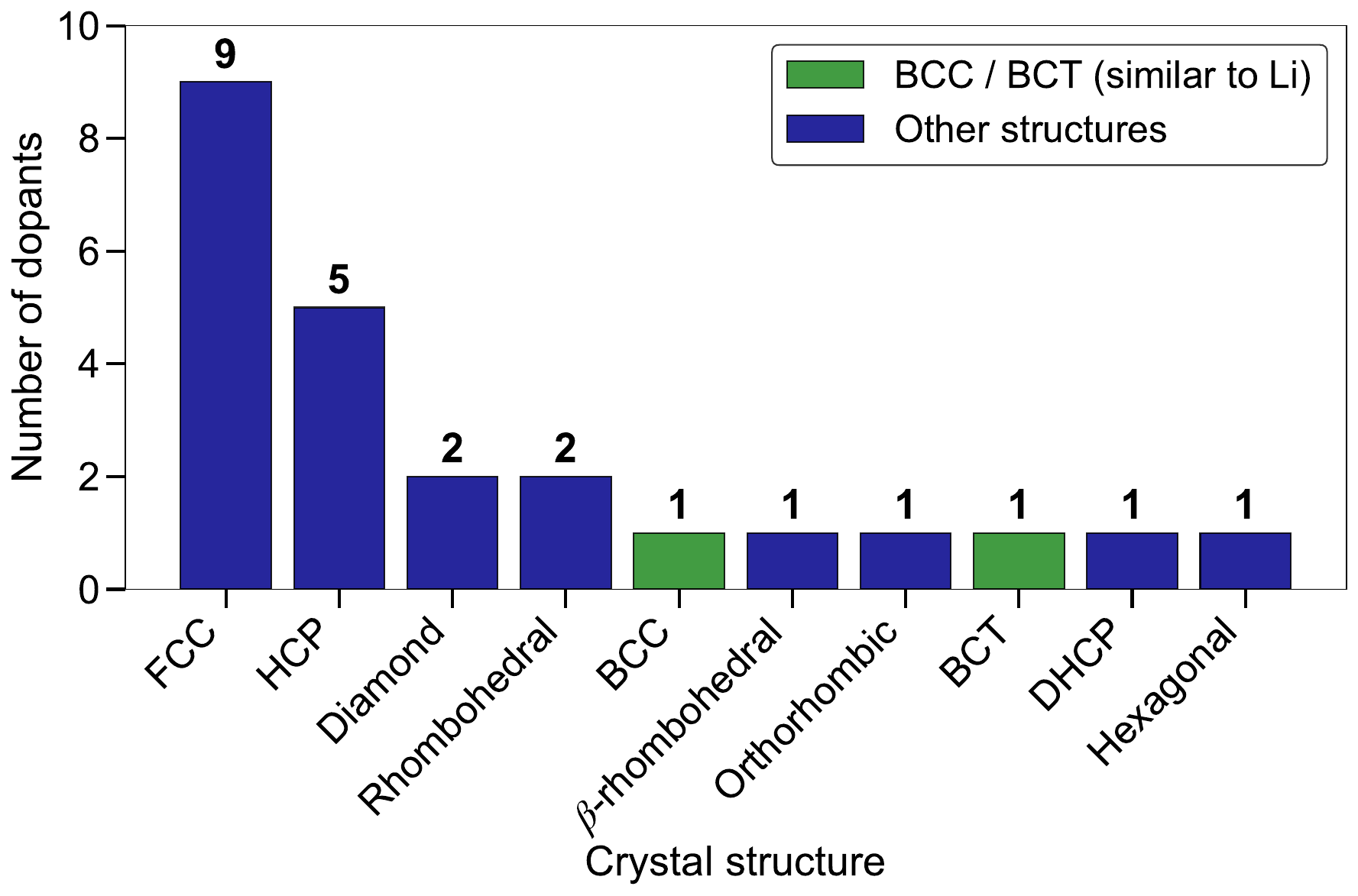}
\caption{Crystal structures of solutes considered in this work. We restrict our study to body-centered cubic (bcc) and body-centered tetragonal (bct) lattices, which are structurally similar to the Li metal host.  }
\label{fig:hr_rule_2}
\end{figure}

Next, we checked the second H--R rule for crystal lattices (\cref{eq:hr_rules_2}) with the results shown in~\cref{fig:hr_rule_2}. Over all lattices, we can consider only body-centered tetragonal (bct) and body-centered cubic (bcc) as the same. In contrast, face-centered cubic (fcc), hexagonal-closed packed (hcp), rhombohedral, $\beta$-rhombohedral, orthorhombic, diamond and double hexagonal close-packed (dhcp) lattices are considered as a different. As a result, only three (Na, In, Sn) out of twenty one (14\% in total) of solid solutions with Li could be predicted with the second rule. All other solutes would be discarded from the further screening stages.

\begin{figure}[H]
\centering
\includegraphics[width=0.75\columnwidth]{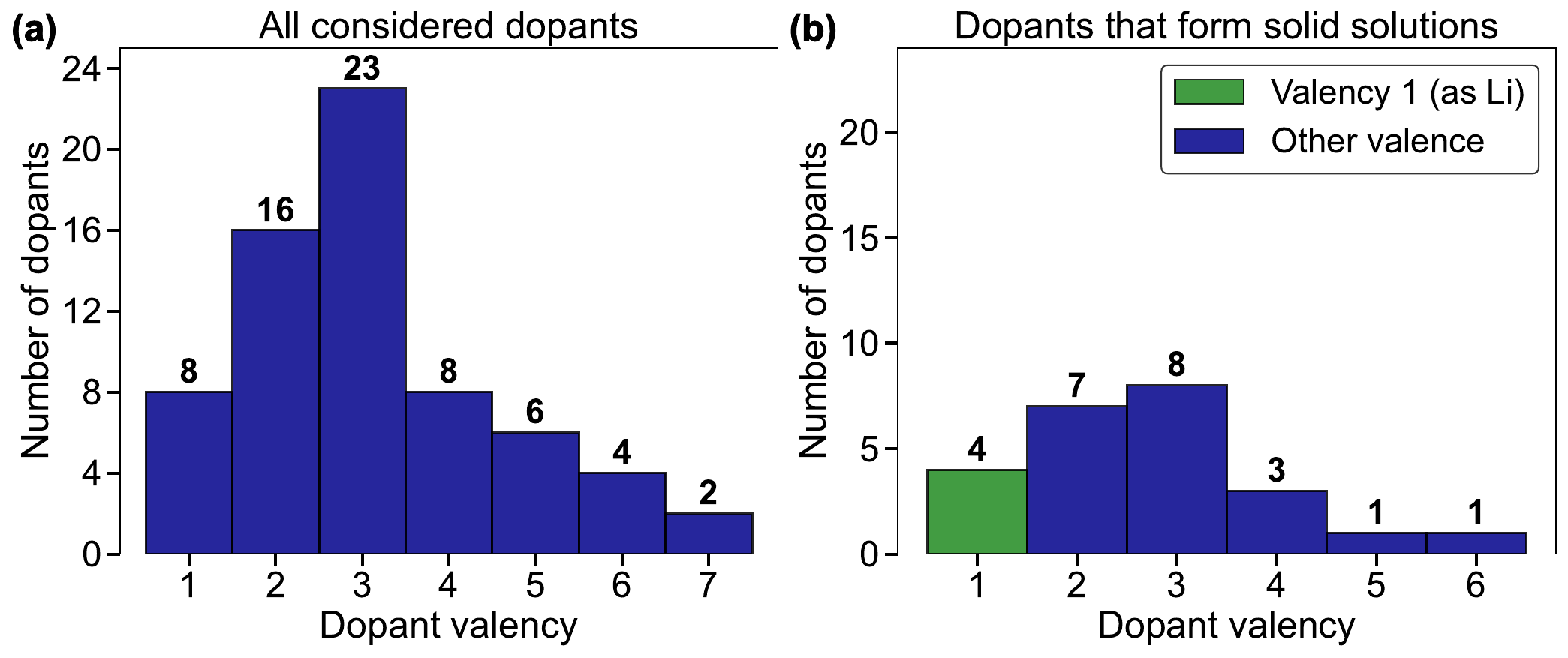}
\caption{Validation of the third Hume--Rothery rule. (a) Valency distribution for all solutes considered. (b) Valency distribution for solutes that form solid solutions with bcc Li.}
\label{fig:hr_rule_3}
\end{figure}

The H--R valency rule (\cref{eq:hr_rules_3}) was developed mainly for transition and noble metals and performs poorly for Li metal (see~\cref{fig:hr_rule_3}). Across an extended list of 65 solute elements studied, only nine share lithium’s valency. Restricting to solutes that form solid solutions with bcc Li, just four of 21 match Li’s valency. Consequently, only 19\% of the solutes satisfy the H--R valency rule. 

In general, we cannot modify this rule or the previous one for the bcc Li system. The new modified rules would be too broad, as they would fail to filter out unwanted solutes without also eliminating desirable ones.

\section{Crystal Orbital Hamilton Population Analysis}\label{sec:cohp_si}

\subsection{Applicability of COHP analysis to Li solid solutions}\label{sec:cohp_justified}

Li solid solutions exhibit metallic long-range bonding; however, COHP calculations are inherently limited to a finite effective environment. The developers of the COHP method demonstrated its applicability to metallic systems, as shown previously by Deringer~\textit{et al.}~\cite{deringer2011crystal} for the Na system, where it accurately represented the metallic bonding character.

\begin{figure}[H]
\centering
\includegraphics[width=0.45\columnwidth]{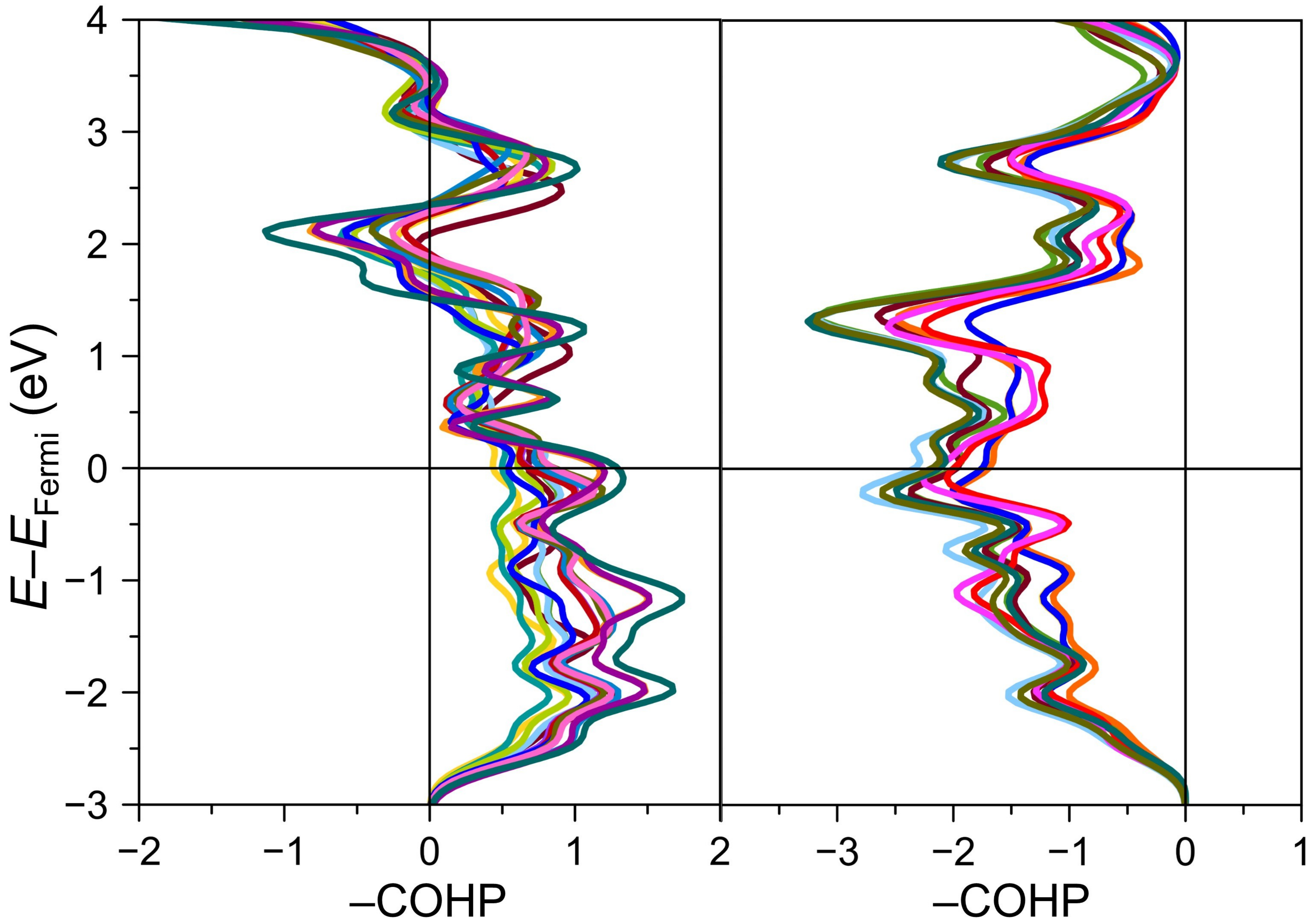}
\caption{Calculated COHPs for Li--Li bonds in the Li--Si system, both with and without a vacancy at the 1NN position (Li:Si+$\mathrm{V_{1NN}}$). Left: COHPs of bonds in the cubic environment. Right: COHPs of bonds in the octahedral environment. }
\label{fig:cohp_conv}
\end{figure}

Nevertheless, convergence tests were performed to verify this capability for lithium, with the results presented in~\cref{fig:cohp_conv}. Here, we compared the COHPs of various Li--Li bonds in the Li--Si system. Hereinafter, we use the notation $V_{i}^{j}$ to denote the location of the vacancy and the environment within which the COHP and ICOHP were calculated. We use $i={\mathrm{1NN,~2NN,~near,~sep}}$ to denote the position of the vacancy at nearest-neighbor (1NN) and next-nearest-neighbor (2NN) sites. ``Near'' denotes the position with the highest solute--vacancy binding energy and the most probable vacancy location (see $E_{\mathrm{b}}^{\mathrm{refined}}$ in~\cref{eq:eff_for_2nn}), while ``sep'' denotes the isolated $\mathrm{Li}-X$ complex, in which the vacancy and solute are separated by 11.91~\AA. $j={\mathrm{cub,~oct}}$ denotes the cubic and octahedral environments within which the COHP and ICOHP were calculated (see~\cref{fig:cohp_main}). The obtained data show that various bonds in the Li bulk crystal exhibit identical --COHP profiles, confirming that COHP analysis can be reliably applied to Li-based materials. To ensure accuracy in the bond analysis, we employed a 128-atom Li supercell.

\begin{figure}[H]
\centering
\includegraphics[width=0.9\columnwidth]{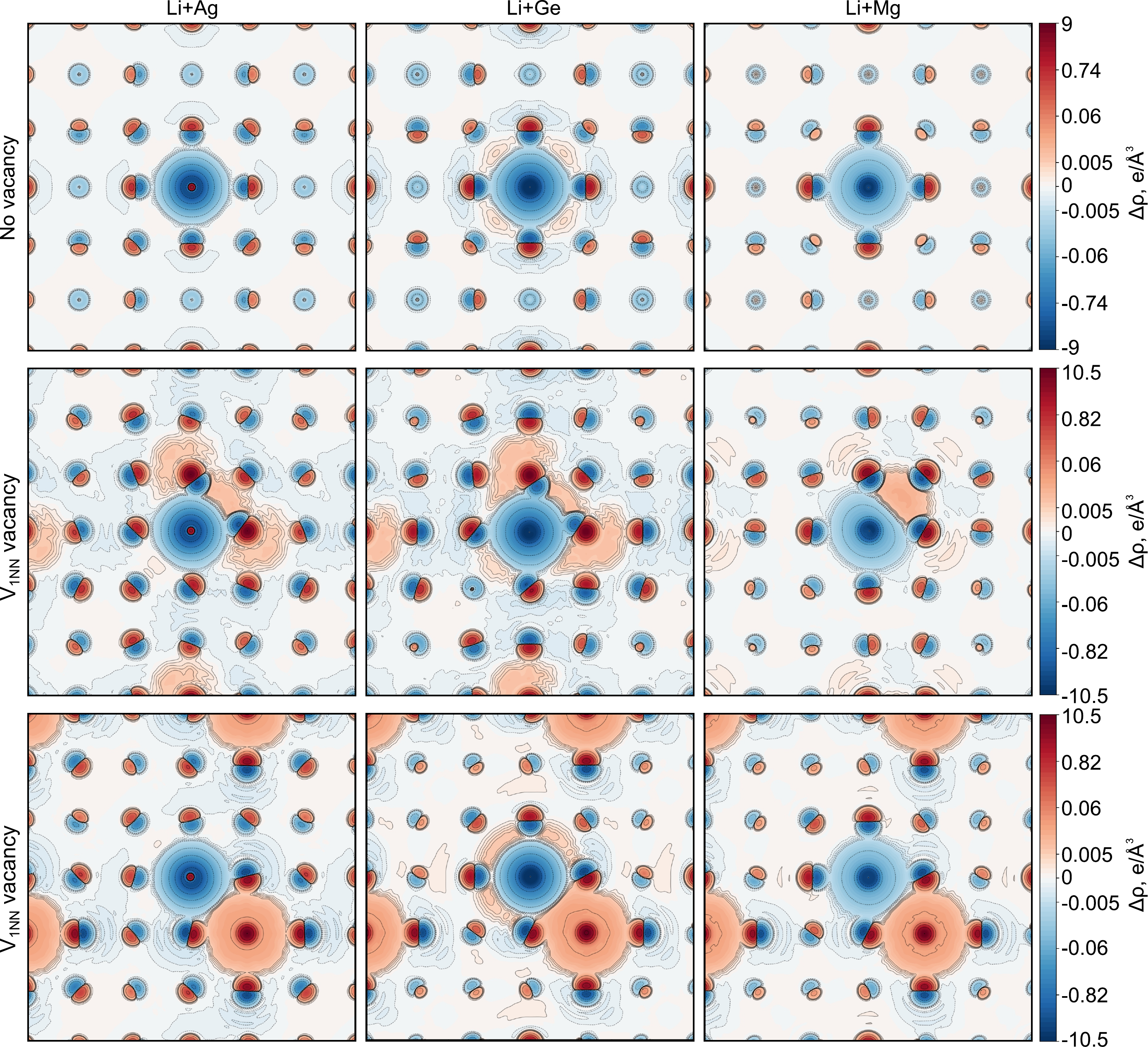}
\caption{Charge-density difference ($\Delta \rho$, $\mathrm{e\, \AA}^{-3}$) for supercells containing solute elements: Ag, Ge, and Mg. Columns compare the three hosts; rows show the lattice with no vacancy and with $\mathrm{V_{1NN}}$ and $\mathrm{V_{2NN}}$ vacancies. Red and blue mark charge accumulation and depletion.}
\label{fig:charge_diff}
\end{figure}

Furthermore, we calculated the charge difference between solute-containing and solute-free structures, with the results shown in \cref{fig:charge_diff}. The results indicate that the most prominent charge difference occurs around the solute element and the removed Li atom ($\mathrm{V_{1NN}}$ and $\mathrm{V_{2NN}}$ vacancies). The levels of accumulated and depleted electron density are the most prominent, whereas the changes on other atoms and positions are less prominent and should be considered as second-order contributions to the COHP method.

\subsection{Calculated COHPs for Li--$\boldsymbol{X}$ systems}\label{sec:cohp_calculated}

\begin{figure}[H]
\centering
\includegraphics[width=1.0\columnwidth]{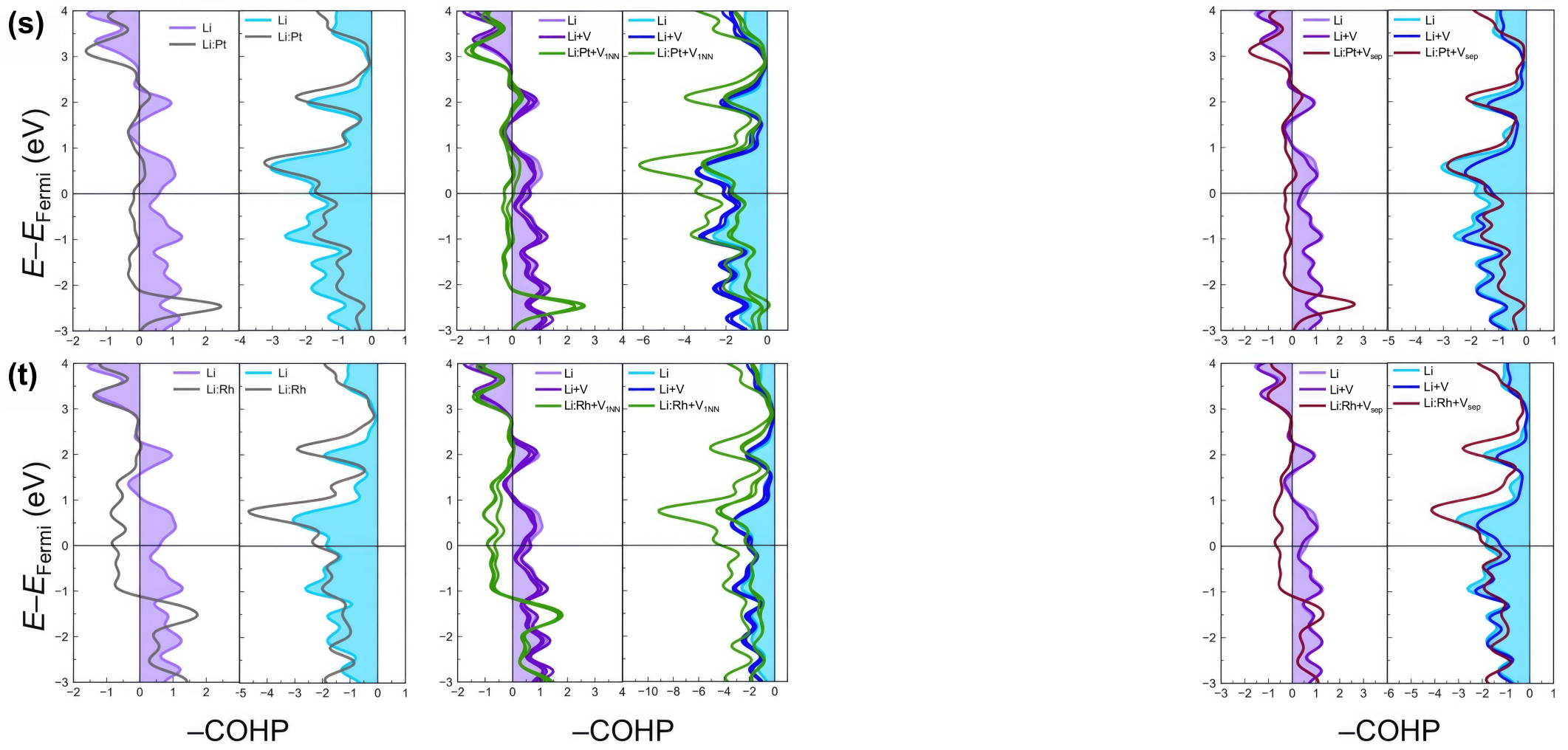}
\caption{--COHP plots for Li--solute bonds in cubic (blue) and octahedral (violet) Li environments. Each panel corresponds to a different solute atom: (a)~Ag, (b)~Cd, (c)~Ga, (d)~Ge, (e)~Hg, (f)~Mg, (g)~Al, (h)~Au, (i)~Bi, (j)~Ca, (k)~Cu, (l)~In, (m)~Ir, (n)~Pd, (o)~Si, (p)~Tl, (q)~Zn, (r)~Pb, (s)~Pt, and (t)~Rh. The four columns correspond, from left to right, to the vacancy-free system (gray) and systems with vacancies at the $\mathrm{V_{1NN}}$ (green), $\mathrm{V_{2NN}}$ (turquoise), and $\mathrm{V_{sep}}$ (crimson) positions. For reference, --COHP curves for pristine Li with vacancies are shown in dark blue and dark purple.}
\label{fig:llli}
\end{figure}

The COHPs for Li--$X$ systems are shown in~\cref{fig:llli}. We considered four types of structures and the corresponding rows: (1) vacancy-free structures with a single Li atom substituted by the solute element $X$; (2) structures with the solute element and a vacancy at nearest-neighbor positions ($\mathrm{V_{1NN}}$, 2.98~\AA); (3) structures with the solute element and a vacancy at next-nearest-neighbor positions ($\mathrm{V_{2NN}}$, 3.41~\AA); and (4) structures with the solute element and an isolated vacancy ($\mathrm{V_{sep}}$, 11.91~\AA). Some $\mathrm{V_{1NN}}$ and $\mathrm{V_{2NN}}$ configurations are missing, since we generally selected the configuration with the maximum binding energy among them (see $E_{\mathrm{b}}^{\mathrm{refined}}$ in \cref{eq:eff_for_2nn}).

\begin{figure}[H]
\centering
\includegraphics[width=1.0\columnwidth]{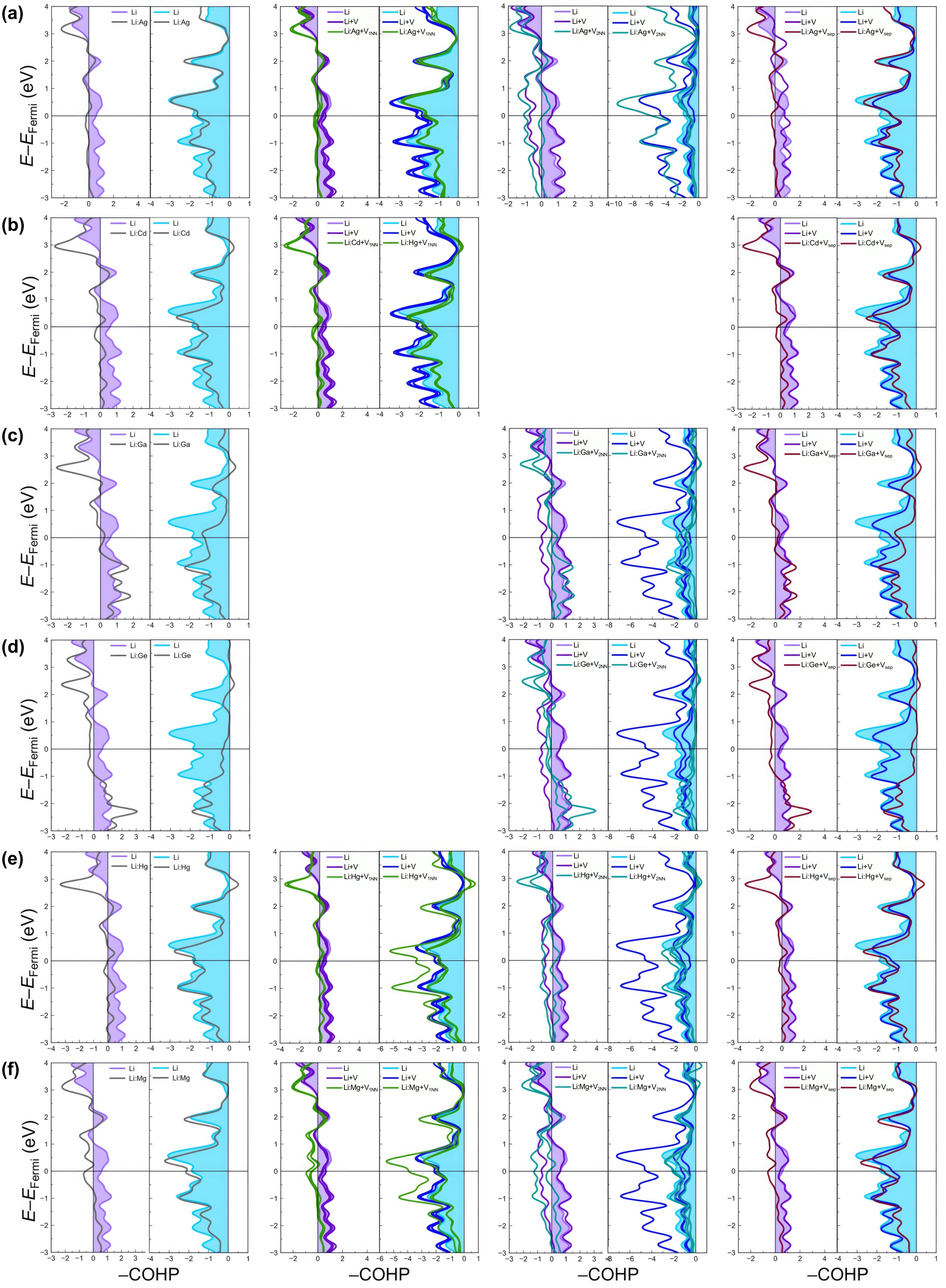}
\label{fig:lli1}
\end{figure}
\begin{figure}[H]
\centering
\includegraphics[width=1.0\columnwidth]{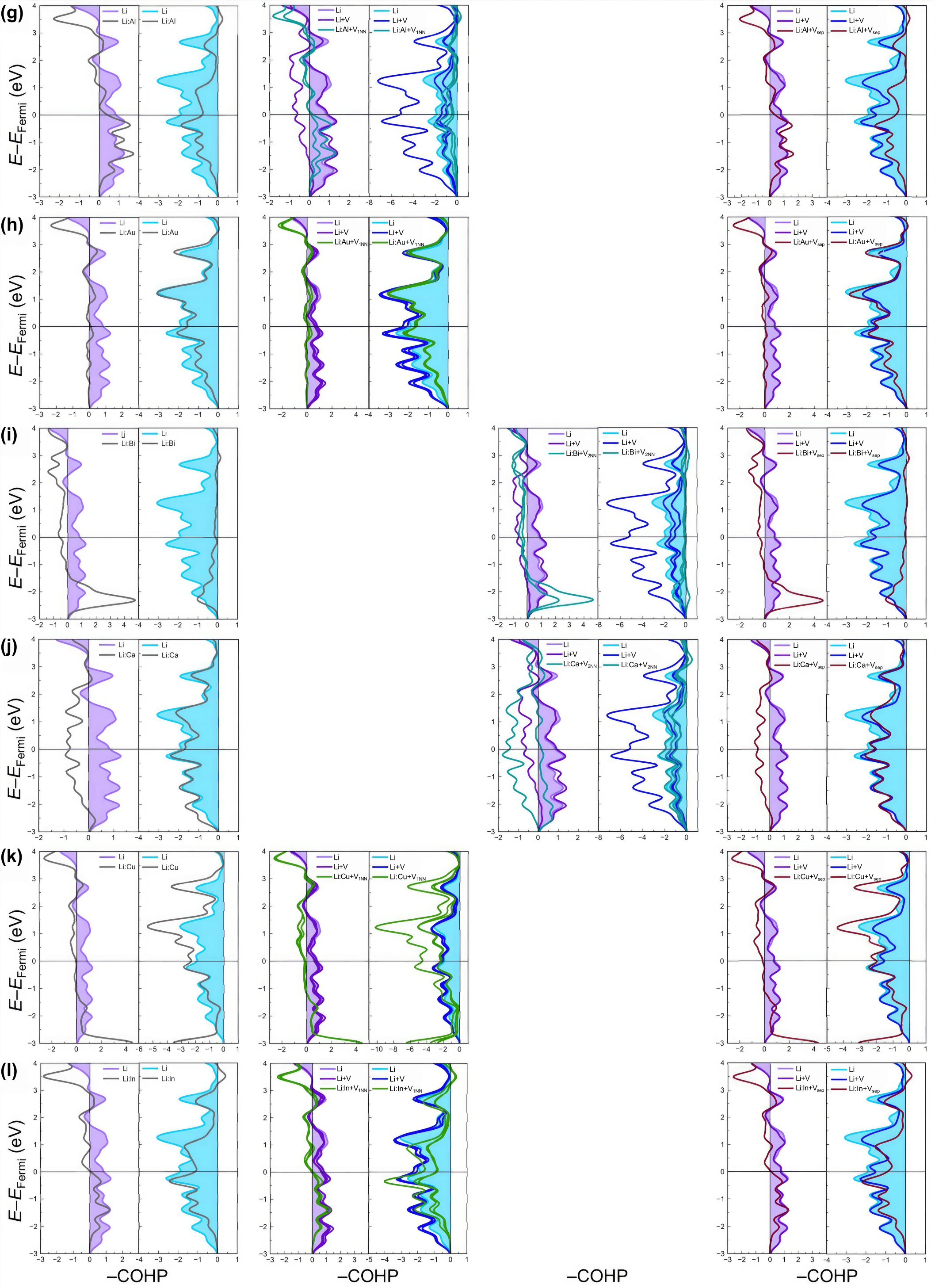}
\label{fig:lli2}
\end{figure}
\begin{figure}[H]
\centering
\includegraphics[width=1.0\columnwidth]{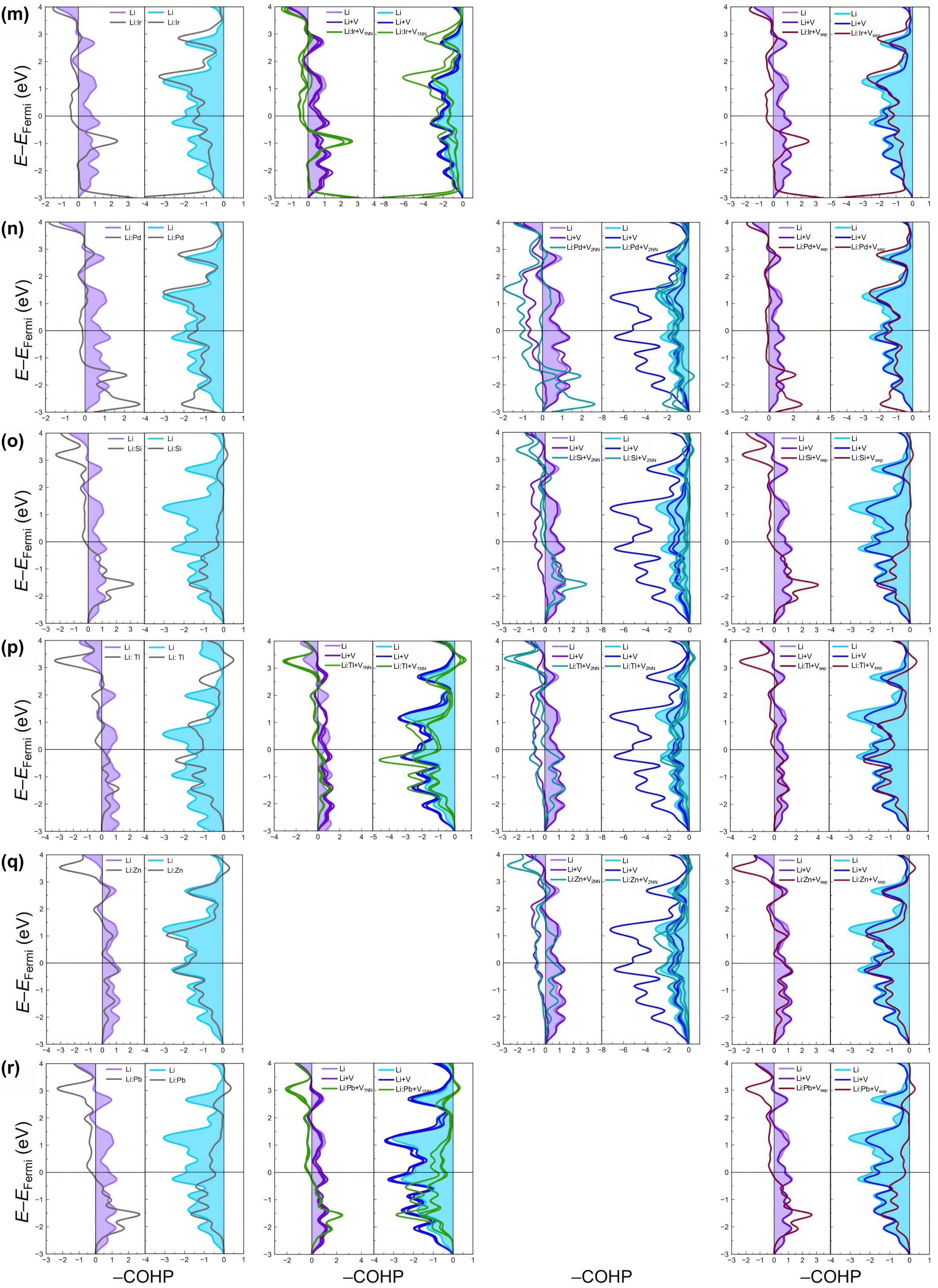}
\label{fig:lli3}
\end{figure}

\subsection{Calculated ICOHPs for Li--$\boldsymbol{X}$ systems}\label{sec:icohp_calculated}

Since introducing a vacancy into a structure may alter different bonds and their corresponding COHPs/ICOHPs within the same environment to varying degrees, we treated them separately in the main COHP plot (see~\cref{fig:cohp_main}). To analyze the nature of the different curves in the COHP plots, we considered them individually. An example of the different bonds and their ICOHPs in the Ag--Li system is provided in~\cref{tab:icohp_ag_li_var}.

However, to find correlations between ICOHPs and thermodynamic quantities such as binding and solution energies, we averaged the ICOHPs over all bonds and used the average values of the bond lengths and corresponding ICOHPs. The compiled data are provided in~\cref{tab:icohp}.

\begin{table}[H]
\centering
\setlength{\tabcolsep}{10pt}
\caption{ICOHP values for Ag--Li bonds in octahedral (O) and cubic (C) coordination environments. The vacancy is located in the cubic environment as the nearest neighbor. Bonds O1--O3 share identical bond lengths and ICOHP values, whereas O4--O6 exhibit elongated distances with correspondingly reduced ICOHP magnitudes. The same trend is observed for C1--C4 and C5--C7.}
\label{tab:icohp_ag_li_var}
\begin{tabular}{lcc}
\hline\hline
Bond length (\AA) & ICOHP (eV) & Atom number \\
\hline
\multicolumn{3}{c}{Octahedral environment} \\
\hline
3.46 & --2.98 & O1 \\
3.46 & --2.99 & O2 \\
3.46 & --2.98 & O3 \\
3.00 & --2.49 & O4 \\
3.00 & --2.49 & O5 \\
3.00 & --2.49 & O6 \\
\hline
\multicolumn{3}{c}{Cubic environment} \\
\hline
2.95 & --2.55 & C1 \\
2.95 & --2.56 & C2 \\
2.95 & --2.56 & C3 \\
2.82 & --2.69 & C4 \\
2.82 & --5.03 & C5 \\
2.82 & --5.00 & C6 \\
2.82 & --5.01 & C7 \\
\hline\hline
\end{tabular}
\end{table}

\begin{table}[H]
    \centering
    \caption{Bond lengths (\AA), ICOHPs, and vacancy-induced changes in ICOHP ($\Delta$ICOHP, in eV per bond) for Li--solute bonds across the four considered configurations: the vacancy-free alloyed system (Li:alloying) and the alloyed systems with a vacancy at the nearest-neighbor ($\mathrm{V_{1NN}}$), next-nearest-neighbor ($\mathrm{V_{2NN}}$), and separated ($\mathrm{V_{sep}}$) positions. Negative $\Delta$ICOHP values (bold) indicate bond strengthening upon vacancy formation, whereas positive values indicate weakening. Dashes denote configurations that were not considered.}
    \label{tab:icohp}
    \begin{tabular}{@{}lccccccccccc@{}}
        \hline\hline
        \multirow{2}{*}{\textbf{Solute element}} & \multicolumn{2}{c}{\textbf{Li:alloying}} & \multicolumn{3}{c}{\textbf{Alloying + $\boldsymbol{\mathrm{V_{1NN}}}$}} & \multicolumn{3}{c}{\textbf{Alloying + $\boldsymbol{\mathrm{V_{2NN}}}$}} & \multicolumn{3}{c}{\textbf{Alloying + $\boldsymbol{\mathrm{V_{sep}}}$}} \\
        \cmidrule(lr){2-3} \cmidrule(lr){4-6} \cmidrule(lr){7-9} \cmidrule(lr){10-12}
        & Bond length & ICOHP & Bond length & ICOHP & $\Delta$ICOHP & Bond length & ICOHP & $\Delta$ICOHP & Bond length & ICOHP & $\Delta$ICOHP \\
        \midrule
        pristine Li & 2.98 & \textbf{--1.85} & 3.02 & --1.24 & --1.85 & 3.02 & --1.24 & 0.61 & 3.01 & --1.14 & 0.71 \\
        Ag & 2.86 & --3.61 & 2.87 & --3.63 & 0.17 & 2.87 & --3.63 & \textbf{--0.02} & 2.88 & --3.42 & 0.20 \\
        Al & 2.88 & 0.42 & 2.89 & --0.24 & --- & 2.89 & --0.24 & --0.66 & 2.90 & 0.17 & --0.25 \\
        Au & 2.78 & --3.90 & --- & --- & \textbf{0.08} & --- & --- & --- & 2.80 & --4.07 & --0.17 \\
        Bi & 2.95 & --0.88 & 2.95 & --1.40 & --- & 2.95 & --1.40 & \textbf{--0.51} & 2.98 & --1.39 & --0.51 \\
        Ca & 3.24 & --2.15 & 3.36 & --2.00 & --- & 3.36 & --2.00 & 0.16 & 3.33 & --1.84 & 0.32 \\
        Cd & 2.94 & --1.50 & --- & --- & \textbf{0.02} & --- & --- & --- & 2.96 & --1.52 & --0.02 \\
        Cu & 2.75 & --5.56 & --- & --- & \textbf{0.26} & --- & --- & --- & 2.78 & --5.00 & 0.56 \\
        Ga & 2.82 & 0.28 & 2.85 & -0.32 & --- & 2.85 & --0.32 & \textbf{--0.60} & 2.84 & 0.20 & --0.07 \\
        Ge & 2.74 & 0.68 & 2.77 & 0.04 & --- & 2.77 & 0.04 & \textbf{--0.65} & 2.76 & 0.41 & --0.28 \\
        Hg & 2.87 & --2.41 & 2.89 & --2.29 & \textbf{0.36} & 2.89 & --2.29 & 0.12 & 2.90 & --2.52 & --0.11 \\
        In & 2.92 & --0.04 & 2.94 & --0.29 & \textbf{0.02} & 2.94 & --0.29 & --0.25 & 2.95 & --0.15 & --0.11 \\
        Ir & 2.62 & --5.29 & --- & --- & 0.14 & --- & --- & --- & 2.64 & --5.09 & 0.19 \\
        Mg & 2.99 & --2.20 & 3.02 & --1.72 & \textbf{0.43} & 3.02 & --1.72 & 0.48 & 3.03 & --1.69 & 0.51 \\
        Pb & 2.93 & --0.39 & --- & --- & \textbf{0.56} & --- & --- & --- & 2.97 & --0.07 & 0.32 \\
        Pd & 2.72 & --4.50 & 2.75 & --4.26 & --- & 2.75 & --4.26 & \textbf{0.25} & 2.73 & --4.48 & 0.03 \\
        Pt & 2.66 & --4.96 & --- & --- & \textbf{0.16} & --- & --- & --- & 2.69 & --7.56 & --2.60 \\
        Rh & 2.66 & --5.79 & --- & --- & \textbf{0.07} & --- & --- & --- & 2.68 & --5.59 & 0.20 \\
        Si & 2.73 & 0.60 & 2.75 & --0.09 & --- & 2.75 & --0.09 & --0.69 & 2.73 & 0.40 & --0.20 \\
        Zn & 2.83 & --1.50 & 2.88 & --1.62 & --- & 2.88 & --1.62 & \textbf{--0.12} & 2.87 & --0.87 & 0.63 \\
        Tl & 2.93 & --4.52 & 2.96 & --3.46 & \textbf{1.50} & 2.96 & --3.46 & \textbf{1.06} & 2.95 & --3.87 & \textbf{0.66} \\
        \hline\hline
    \end{tabular}
\end{table}

\subsection{Regression of thermodynamic properties on ICOHP descriptors}\label{sec:icohp_regression}

\begin{figure}[H]
\centering
\includegraphics[width=0.9\columnwidth]{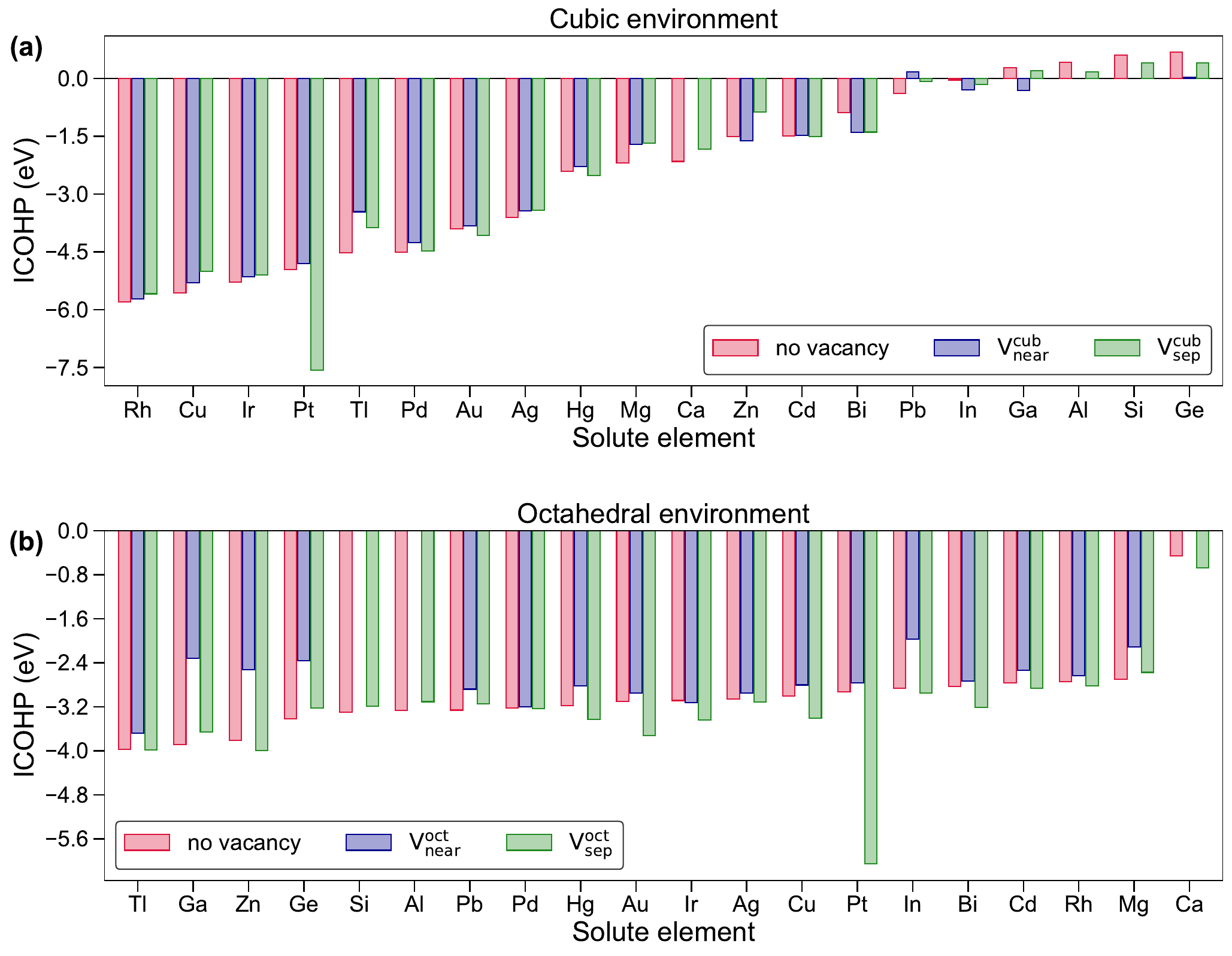}
\caption{Calculated ICOHP values for various solute elements, calculated for the two environments: (a) cubic and (b) octahedral. Red denotes systems with one Li atom replaced by a solute element, blue denotes a vacancy near the solute element ($\mathrm{V_{1NN}}$ or $\mathrm{V_{2NN}}$, maximizing the solute--vacancy binding energy $E_{\mathrm{b}}^{\mathrm{refined}}$ in \cref{eq:eff_for_2nn}), and green denotes the separated solute--vacancy complex with a distance of 11.91~\AA\ between them. }
\label{fig:icohp_bar_solute_near_separated}
\end{figure}

The calculated ICOHPs for the best solute elements, along with several others, are shown in~\cref{fig:icohp_bar_solute_near_separated}.

\begin{figure}[H]
\centering
\includegraphics[width=0.7\columnwidth]{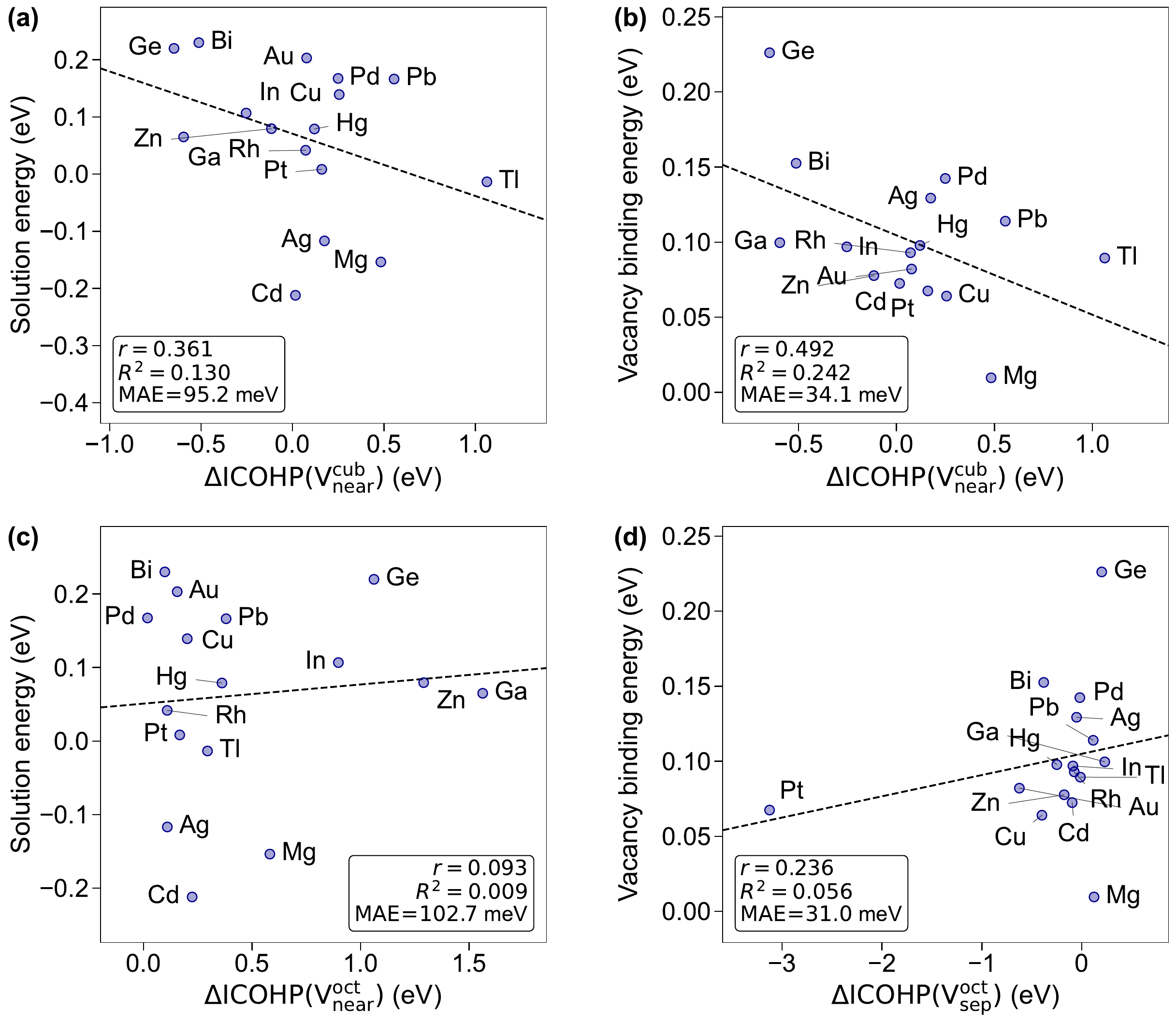}
\caption{Correlation between the linear fits of single descriptors (ICOHP or $\Delta \mathrm{ICOHP}$) and thermodynamic properties in different environments. (a),(c) Solution energies ($E_{\mathrm{s}}$) in cubic and octahedral environments, respectively; (b),(d) solute--vacancy binding energies ($E_{\mathrm{b}}$) in cubic and octahedral environments, respectively. Coefficients of determination ($R^2$), Pearson correlation coefficients ($r$), and mean absolute errors (MAE) are provided for each fit. }
\label{fig:icohp_delta_vs_energies}
\end{figure}

First, we calculated prediction metrics, namely the coefficient of determination ($R^2$), the Pearson correlation coefficient ($r$), and the mean absolute error (MAE), for single descriptors such as ICOHP(no vacancy) or the difference between the calculated ICOHP and that of pristine Li without solute element or vacancy. For instance, $\Delta \mathrm{ICOHP(} \mathrm{V_{sep}})=\mathrm{ICOHP(} \mathrm{V_{sep}}) - \mathrm{ICOHP(} \mathrm{no~vacancy})$. Using these descriptors, we constructed linear regression models between thermodynamic properties and ICOHP descriptors. The results are compiled in the first part of \cref{tab:icohp-prediction-metrics} and visualized in \cref{fig:icohp_delta_vs_energies}. However, such simple descriptors show no correlation with thermodynamic properties such as binding and solution energies. The strongest correlation is between $E_{\mathrm{s}}$ and $\Delta \mathrm{ICOHP} \mathrm{(V_{near}^{\mathrm{cub}}})$, with $R^2=0.13$ and $r=0.361$, which indicates no correlation. Similarly, the strongest correlation for $E_{\mathrm{b}}$ is with $\Delta \mathrm{ICOHP} \mathrm{(V_{sep}})$, with $R^2=0.24$ and $r=0.263$, which also indicates no correlation.

\begin{table*}[ht]
  \centering
  \caption{Prediction metrics for the solution energy $E_{\mathrm{s}}$ and the vacancy binding energy $E_{\mathrm{b}}$, using various ICOHP descriptors. Single descriptors are one-feature linear models. Complex predictors are two-feature models; pairs that span the same descriptor space are listed once. $R^2$ is the coefficient of determination, $r$ is the Pearson correlation of the prediction with the target, and MAE is the mean absolute error in meV. Bold marks the best value in that block and column.}
  \label{tab:icohp-prediction-metrics}
  \begin{tabular*}{\textwidth}{@{\extracolsep{\fill}}l*{12}{c}@{}}
    \hline \hline
    & \multicolumn{6}{c}{Cubic environment} & \multicolumn{6}{c}{Octahedral environment} \\
    \cmidrule(lr){2-7}\cmidrule(lr){8-13}
    & \multicolumn{3}{c}{$E_{\mathrm{s}}$} & \multicolumn{3}{c}{$E_{\mathrm{b}}$}
    & \multicolumn{3}{c}{$E_{\mathrm{s}}$} & \multicolumn{3}{c}{$E_{\mathrm{b}}$} \\
    \cmidrule(lr){2-4}\cmidrule(lr){5-7}\cmidrule(lr){8-10}\cmidrule(lr){11-13}
    Set of descriptors & $R^2$ & $r$ & MAE & $R^2$ & $r$ & MAE & $R^2$ & $r$ & MAE & $R^2$ & $r$ & MAE \\
    \midrule
    \multicolumn{13}{l}{\emph{Single descriptors}} \\
    \addlinespace[2pt]
    $\mathrm{ICOHP(no~vacancy)}$ & 0.026 & 0.161 & 98.7 & 0.138 & 0.371 & 31.8 & \textbf{0.041} & \textbf{0.203} & 105.8 & 0.047 & 0.217 & 33.5 \\
    $\mathrm{ICOHP}(\mathrm{V}_{\mathrm{near}})$ & 0.009 & 0.097 & 99.9 & 0.091 & 0.301 & 32.0 & 0.009 & 0.095 & 104.9 & 0.012 & 0.111 & 32.6 \\
    $\mathrm{ICOHP}(\mathrm{V}_{\mathrm{sep}})$ & 0.015 & 0.122 & 98.3 & 0.098 & 0.312 & \textbf{31.0} & 0.007 & 0.084 & 104.3 & 0.015 & 0.121 & 31.3 \\
    $\Delta\mathrm{ICOHP}(\mathrm{V}_{\mathrm{near}})$ & \textbf{0.130} & \textbf{0.361} & \textbf{95.2} & \textbf{0.242} & \textbf{0.492} & 34.1 & 0.009 & 0.093 & 102.7 & 0.008 & 0.090 & 33.2 \\
    $\Delta\mathrm{ICOHP}(\mathrm{V}_{\mathrm{sep}})$ & 0.007 & 0.081 & 104.5 & 0.008 & 0.092 & 32.5 & 0.000 & 0.021 & \textbf{102.4} & \textbf{0.056} & \textbf{0.236} & \textbf{31.0} \\ \hline
    \addlinespace[6pt]
    \multicolumn{13}{l}{\emph{Complex descriptors}} \\
    \addlinespace[2pt]
    $\mathrm{ICOHP(no~vacancy)} \,\&\, \Delta\mathrm{ICOHP}(\mathrm{V}_{\mathrm{near}})$ & \textbf{0.133} & \textbf{0.365} & 96.3 & \textbf{0.254} & \textbf{0.504} & 33.1 & \textbf{0.042} & \textbf{0.205} & 105.4 & 0.049 & 0.221 & 33.5 \\
    $\mathrm{ICOHP(no~vacancy)} \,\&\, \Delta\mathrm{ICOHP}(\mathrm{V}_{\mathrm{sep}})$ & 0.035 & 0.186 & 100.8 & 0.152 & 0.390 & 32.4 & \textbf{0.042} & \textbf{0.205} & 106.7 & \textbf{0.083} & \textbf{0.289} & 31.3 \\
    $\mathrm{ICOHP}(\mathrm{V}_{\mathrm{near}}) \,\&\, \Delta\mathrm{ICOHP}(\mathrm{V}_{\mathrm{sep}})$ & 0.019 & 0.137 & 102.1 & 0.109 & 0.331 & \textbf{32.3} & 0.010 & 0.100 & 104.0 & 0.075 & 0.273 & \textbf{30.4} \\
    $\mathrm{ICOHP}(\mathrm{V}_{\mathrm{sep}}) \,\&\, \Delta\mathrm{ICOHP}(\mathrm{V}_{\mathrm{near}})$ & 0.132 & 0.363 & 96.1 & 0.253 & 0.503 & 33.0 & 0.016 & 0.126 & 105.1 & 0.023 & 0.151 & 32.0 \\
    $\Delta\mathrm{ICOHP}(\mathrm{V}_{\mathrm{near}}) \,\&\, \Delta\mathrm{ICOHP}(\mathrm{V}_{\mathrm{sep}})$ & 0.130 & 0.361 & \textbf{94.9} & 0.244 & 0.494 & 34.0 & 0.009 & 0.093 & \textbf{102.9} & 0.056 & 0.237 & 31.2 \\
    \hline \hline
  \end{tabular*}
\end{table*}

Next, we employed more complex descriptors involving several calculated ICOHPs and evaluated their prediction metrics. The best-performing models are compiled in the second part of \cref{tab:icohp-prediction-metrics}. The top models for the binding and solution energies, predicted using ICOHPs in cubic and octahedral environments, are as follows:

\begin{align}
E_{\mathrm{s}}^{\mathrm{cub}}
  &= 0.061
   - 0.004\,\mathrm{ICOHP}
   - 0.121\,\Delta\mathrm{ICOHP}(\mathrm{V}_{\mathrm{near}}), \\
E_{\mathrm{b}}^{\mathrm{cub}}
  &= 0.111
   + 0.003\,\mathrm{ICOHP}(\mathrm{V}_{\mathrm{near}})
   - 0.048\,\Delta\mathrm{ICOHP}(\mathrm{V}_{\mathrm{near}}), \\
E_{\mathrm{s}}^{\mathrm{oct}}
  &= -0.155
   - 0.068\,\mathrm{ICOHP}
   - 0.005\,\Delta\mathrm{ICOHP}(\mathrm{V}_{\mathrm{sep}}), \\
E_{\mathrm{b}}^{\mathrm{oct}}
  &= 0.042
   - 0.020\,\mathrm{ICOHP}
   + 0.012\,\Delta\mathrm{ICOHP}(\mathrm{V}_{\mathrm{sep}}),
   \label{eq:icohp_regressions}
\end{align}
where $\mathrm{V}_{i}$ is the position of the vacancy at the corresponding site and $E_{i}^{j}$ is the thermodynamic property $i=\mathrm{s,~b}$ (solution and binding energies) in environment $j=\mathrm{cub,~oct}$.

\begin{figure}[H]
\centering
\includegraphics[width=0.75\columnwidth]{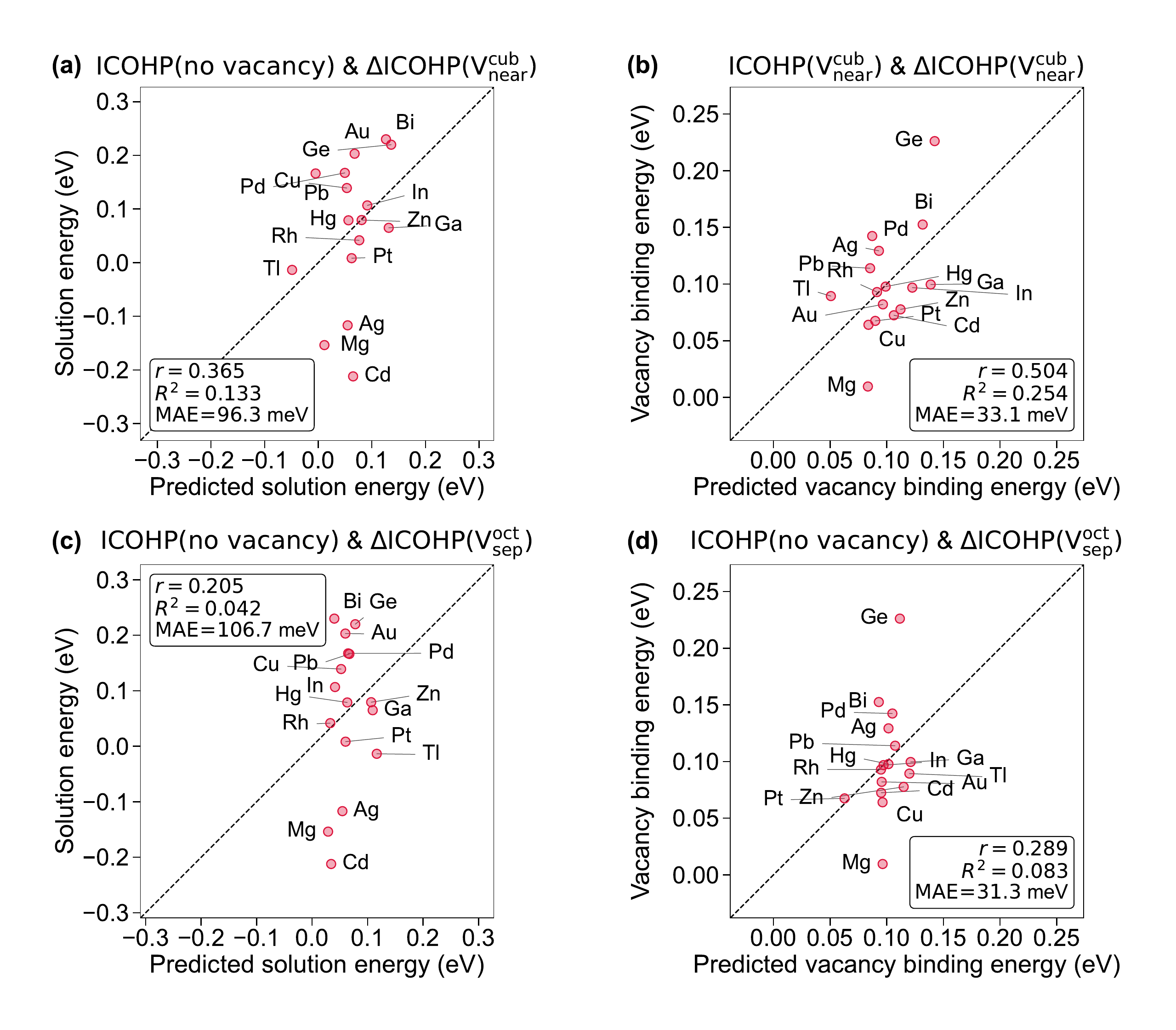}
\caption{Correlation between the linear fits of complex combinations of descriptors (ICOHPs and $\Delta \mathrm{ICOHPs}$) and thermodynamic properties in different environments. (a),(c) Solution energies ($E_{\mathrm{s}}$) in cubic and octahedral environments, respectively; (b),(d) solute--vacancy binding energies ($E_{\mathrm{b}}$) in cubic and octahedral environments, respectively. Coefficients of determination ($R^2$), Pearson correlation coefficients ($r$), and mean absolute errors (MAE) are provided for each fit. }
\label{fig:icohp_energy_best_correlation}
\end{figure}

The performance of the best models with complex descriptors is shown in~\cref{fig:icohp_energy_best_correlation}. For both thermodynamic properties, ICOHPs in the cubic environment yield better performance. The best descriptors for solution energies are ICOHP with only the solute element present (ICOHP(no vacancy)) and the difference between the ICOHPs of the near vacancy position and pristine Li ($\Delta \mathrm{ICOHP(V_{near})}$). The corresponding metrics are $R^2=0.133$ and $r=0.365$. The best descriptors for binding energies are the ICOHP of the vacancy near the solute element (ICOHP($\mathrm{V_{near}}$)) and the difference between the ICOHPs of the near vacancy position and pristine Li ($\Delta \mathrm{ICOHP(V_{near})}$). The corresponding metrics are $R^2=0.254$ and $r=0.504$, which indicate moderate correlation.

\section{Machine-learning models to predict thermodynamic properties}\label{sec:ML_surrogate_models}

\subsection{Feature significance analysis}\label{sec:feature_significance}

To interpret model predictions, we evaluated feature importance for three target properties across the set of $N=19$ top-performing solute candidates: the solution energy ($E_{\mathrm{s}}$), the solute--vacancy binding energy ($E_{\mathrm{b}}$), and the effective vacancy formation energy ($E_{\mathrm{eff}}$). Each candidate was encoded using a numerical descriptor vector comprising three feature categories: (i) MatMiner~\cite{ward2018matminer} descriptors and structural features derived from the solute's room-temperature phase; (ii) Materials Project properties of the these phases (energy per atom, energy above the convex hull, etc.) and the solute chemical potential ($\mu_{\mathrm{solute}}$); and (iii) Hume--Rothery rule features defined in~\cref{eq:hr_rules_1,eq:hr_rules_2,eq:hr_rules_3,eq:hr_rules_4}.

For each target, a random-forest regressor ($n_{\mathrm{estimators}}=600$), ($\mathrm{max_{features}}=\sqrt{p}$)) was trained after median imputation of missing entries. Feature relevance was quantified by permutation importance with respect to the coefficient of determination ($R^{2}$) with $n_{\mathrm{repeats}}=40$. Model quality on the same sample was monitored by three-fold cross-validated ($R^{2}$).

To reduce redundancy in the feature space, we applied a correlation-based filtering procedure. For any pair of features with absolute Pearson correlation exceeding $|r| > 0.99$, we retained only the feature with the higher permutation importance and discarded the other. This step was particularly relevant for MatMiner descriptors, which often include highly correlated summary statistics (e.g., maximum, minimum, mean, and median values) that yielded similar importance scores; in such cases, we kept only the single feature with the highest permutation importance among the correlated group. Remaining features were labeled significant when the mean permutation importance exceeded both its standard deviation over permutations and an absolute cutoff of ($10^{-4}$). The resulting significant subsets were used for ranking plots, pairwise feature-correlation matrices, and Pearson correlations against the three targets. 

\begin{figure}[H]
\centering
\includegraphics[width=0.75\columnwidth]{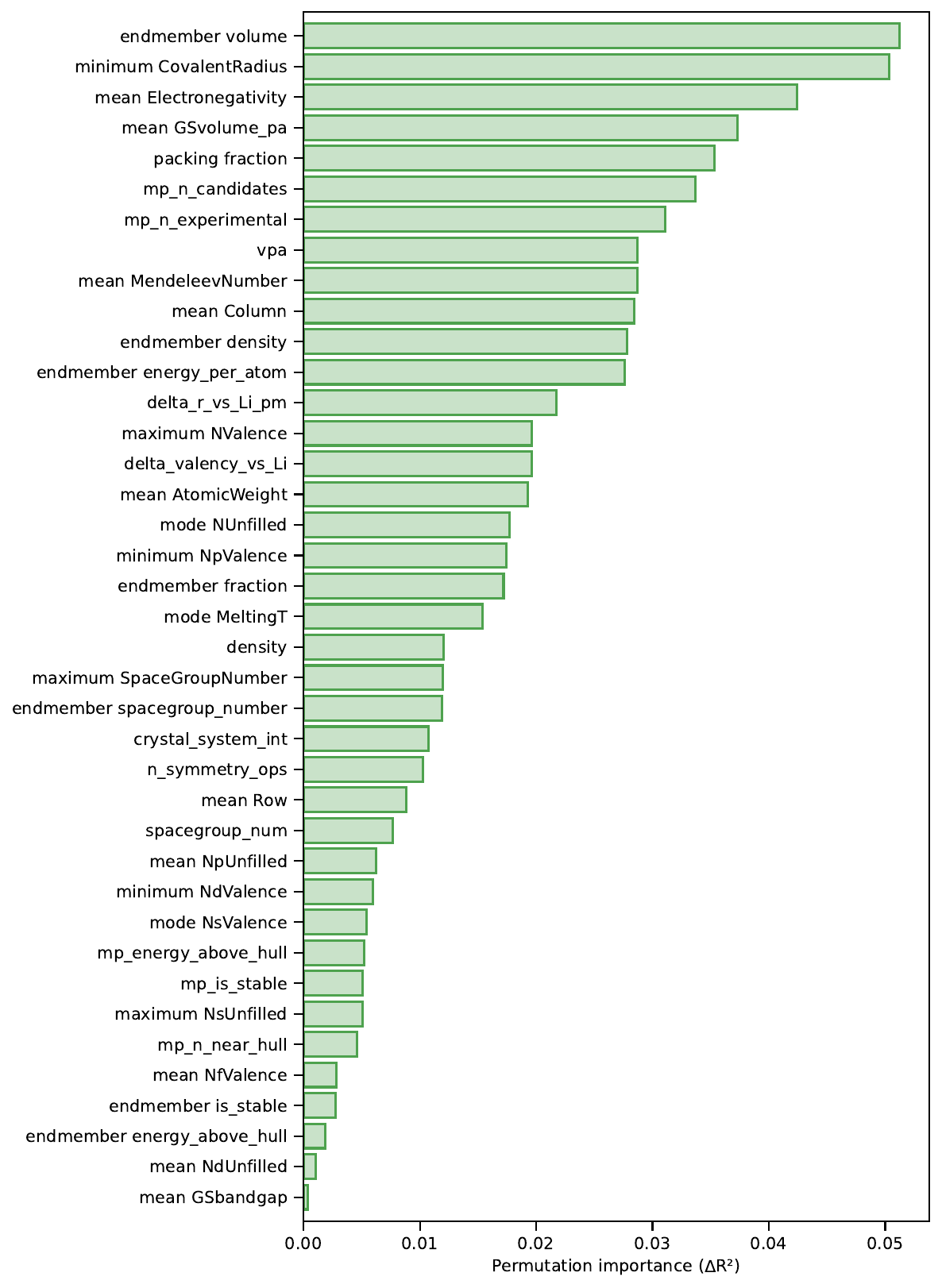}
\caption{Feature importance ranking for solution energy ($E_{\mathrm{s}}$) prediction, derived from random forest permutation importance scores. The label ``end-member'' refers to the crystal structure of the corresponding end-member phase as listed in~\cref{tab:chemical_potentials_all}, whereas all other tags indicate the room-temperature crystal structures of the respective solutes.   }
\label{fig:feature_importance_sol}
\end{figure}

\begin{figure}[H]
\centering
\includegraphics[width=0.99\columnwidth]{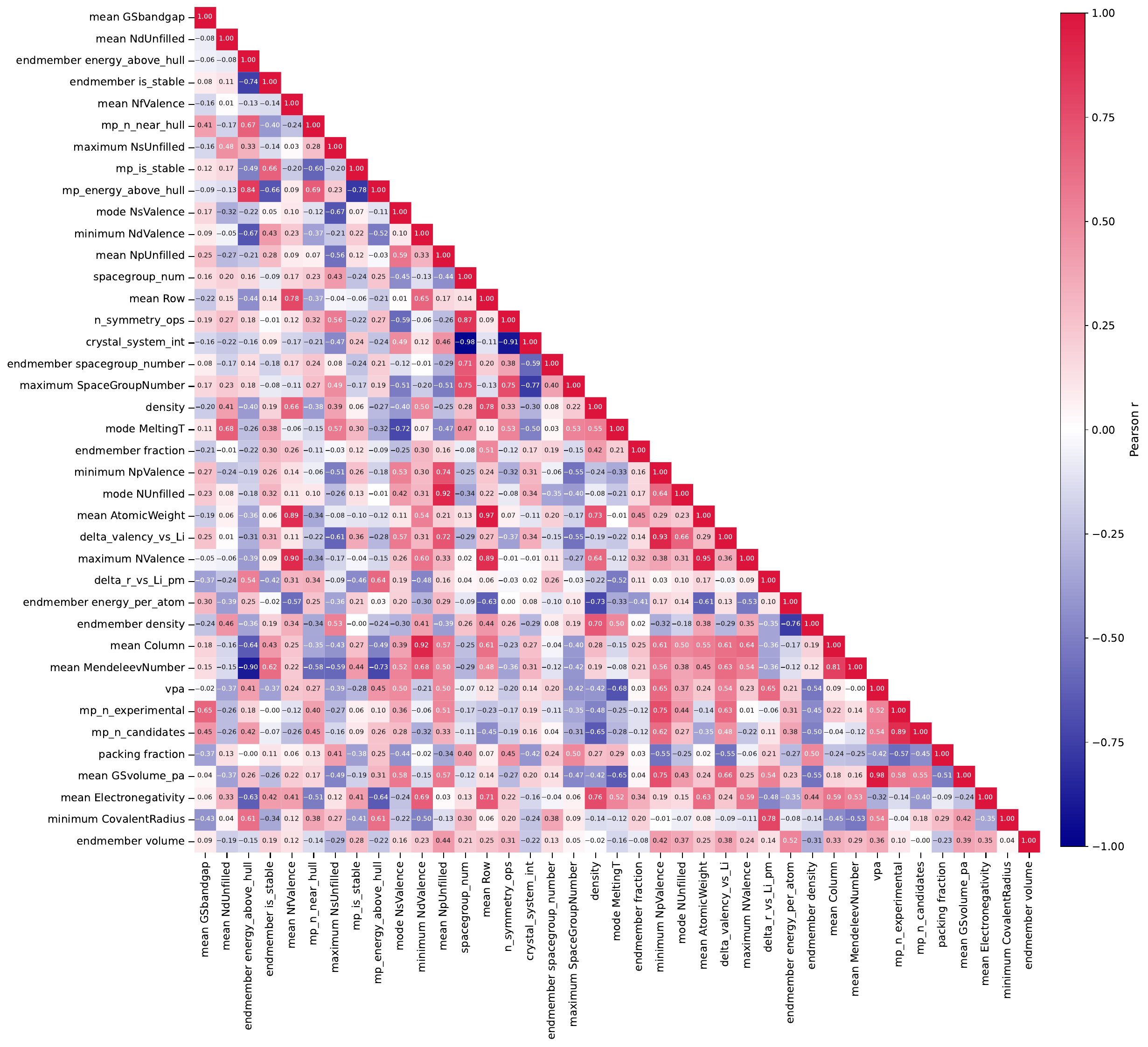}
\caption{Pearson correlation coefficients between the most significant features and the solution energy ($E_{\mathrm{s}}$). The ``end-member'' label denotes the crystal structure of the corresponding end-member phase in~\cref{tab:chemical_potentials_all}, whereas all other tags indicate the room-temperature crystal structures of the respective solutes.}
\label{fig:corr_solution}
\end{figure}

\begin{figure}[H]
\centering
\includegraphics[width=0.75\columnwidth]{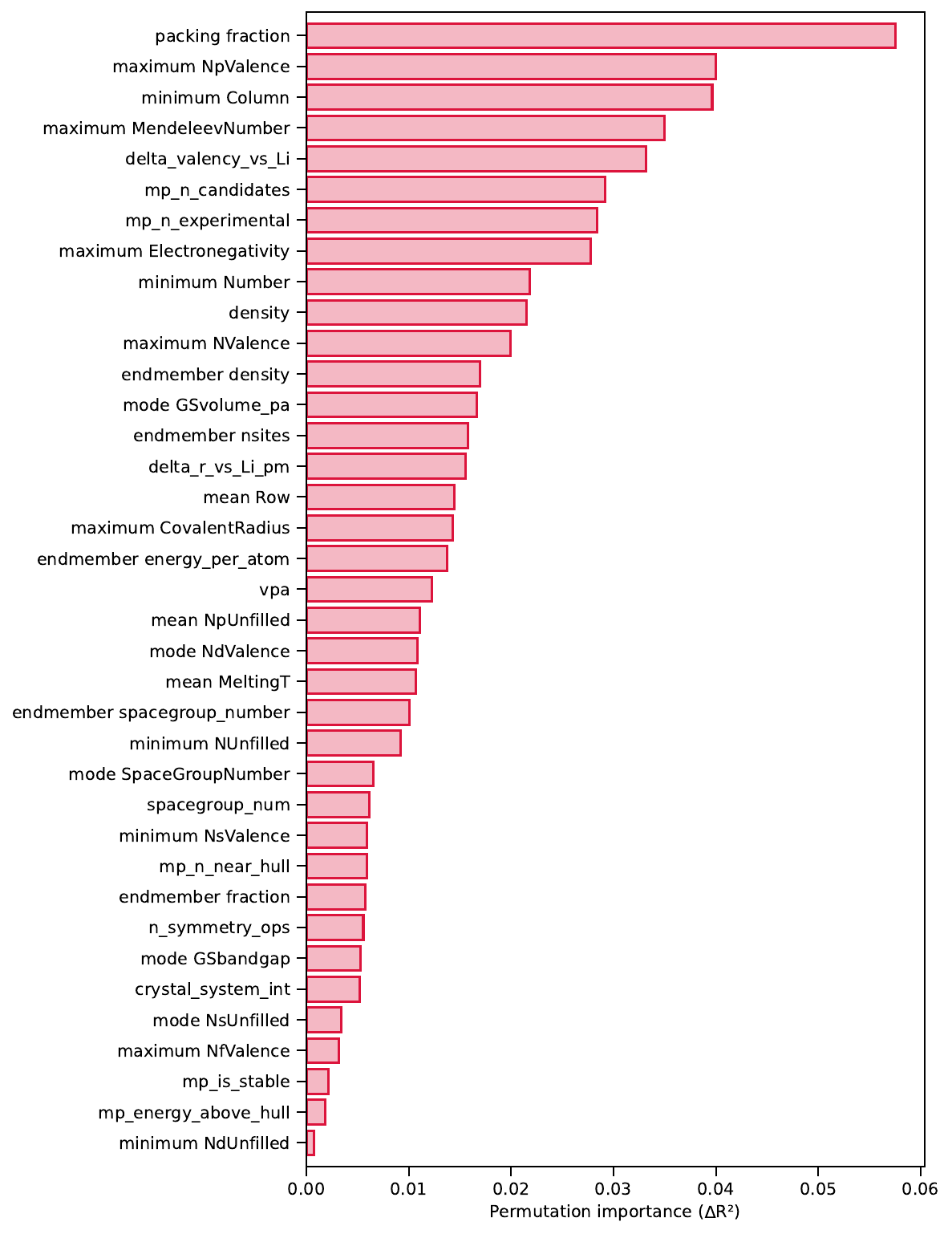}
\caption{Feature importance ranking for solute--vacancy binding energy ($E_{\mathrm{b}}$) prediction, derived from random forest permutation importance scores. The label ``end-member'' refers to the crystal structure of the corresponding end-member phase as listed in~\cref{tab:chemical_potentials_all}, whereas all other tags indicate the room-temperature crystal structures of the respective solutes.}
\label{fig:feature_importance_bind}
\end{figure}

\begin{figure}[H]
\centering
\includegraphics[width=0.99\columnwidth]{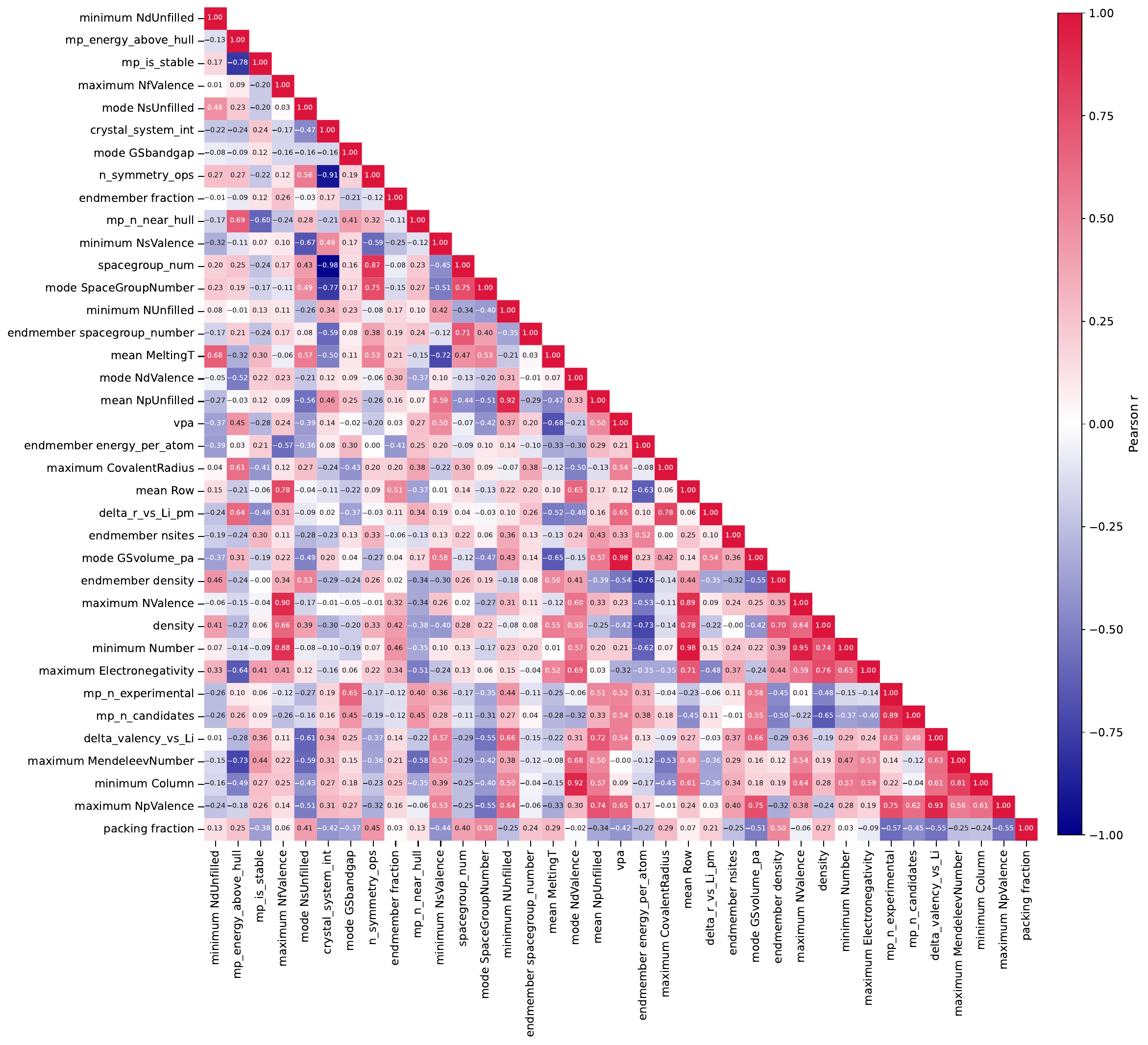}
\caption{Pearson correlation coefficients between the most significant features and the solute--vacancy binding energy ($E_{\mathrm{b}}$). The ``end-member'' label denotes the crystal structure of the corresponding end-member phase in~\cref{tab:chemical_potentials_all}, whereas all other tags indicate the room-temperature crystal structures of the respective solutes.}
\label{fig:corr_bind}
\end{figure}

\begin{figure}[H]
\centering
\includegraphics[width=0.75\columnwidth]{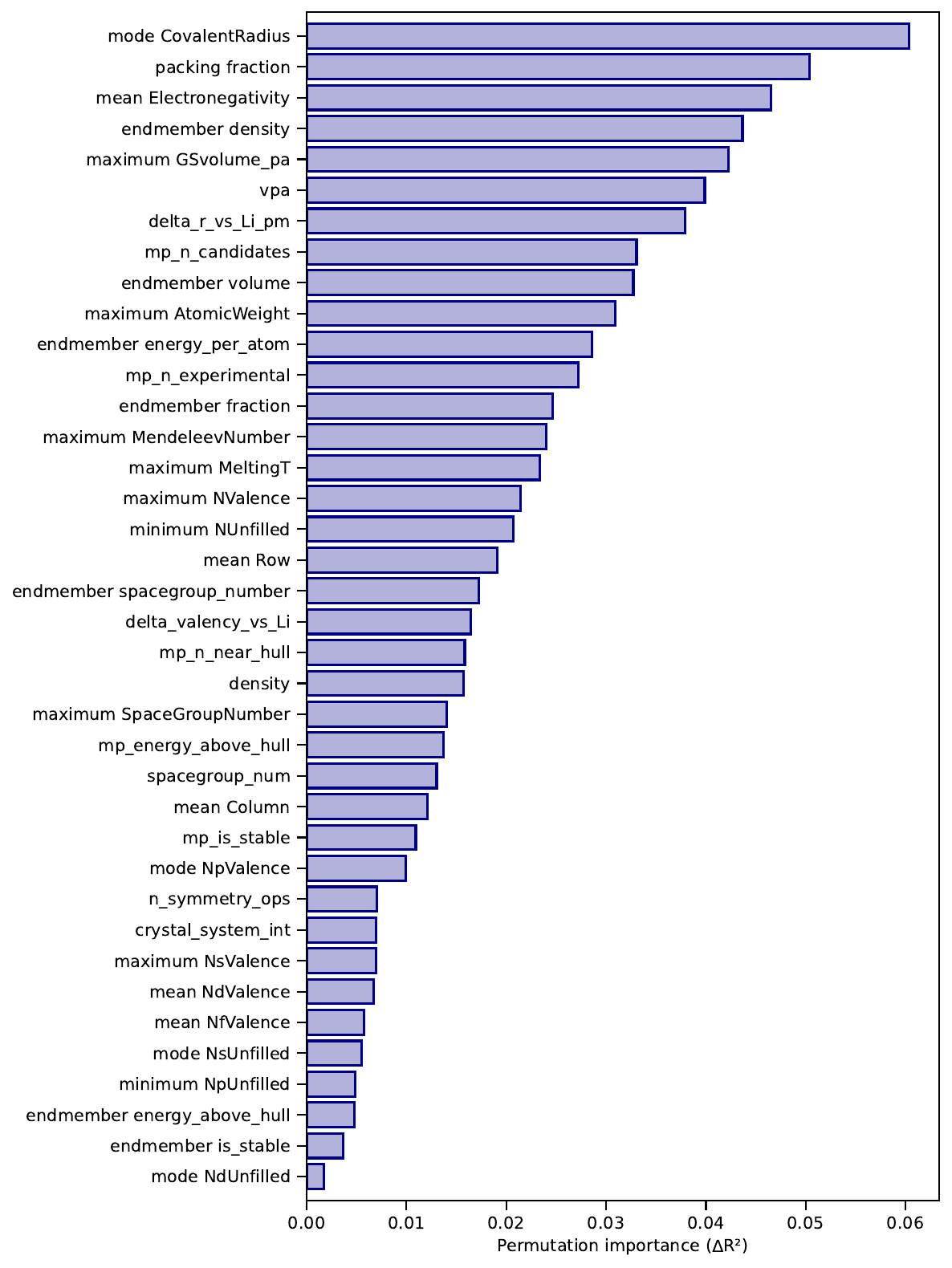}
\caption{Feature importance ranking for vacancy effective energy ($E_{\mathrm{eff}}$) prediction, derived from random forest permutation importance scores. The label ``end-member'' refers to the crystal structure of the corresponding end-member phase as listed in~\cref{tab:chemical_potentials_all}, whereas all other tags indicate the room-temperature crystal structures of the respective solutes.}
\label{fig:feature_importance_sol_min_bind}
\end{figure}

\begin{figure}[H]
\centering
\includegraphics[width=0.99\columnwidth]{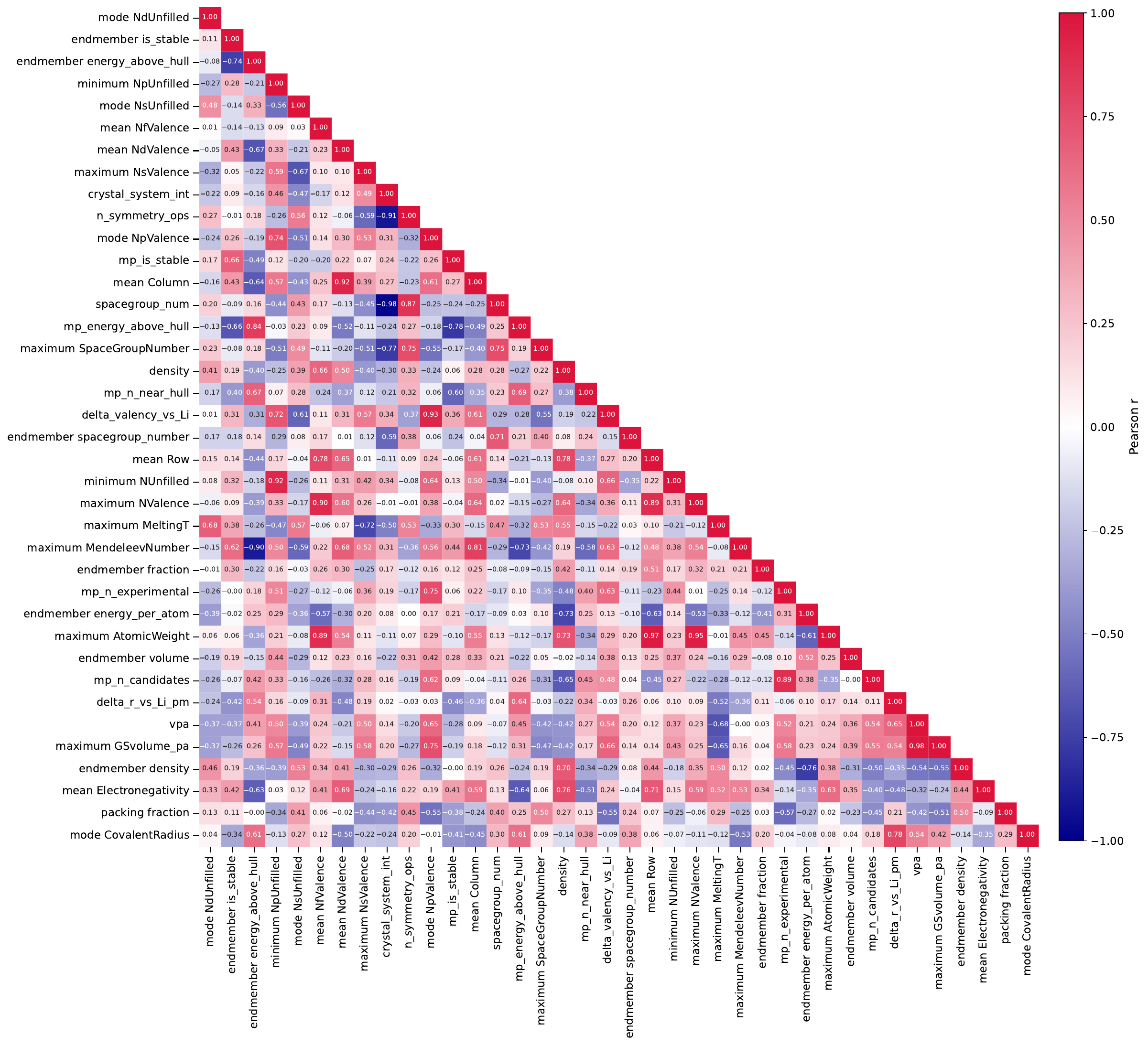}
\caption{Pearson correlation coefficients between the most significant features and the vacancy effective energy ($E_{\mathrm{eff}}$). The ``end-member'' label denotes the crystal structure of the corresponding end-member phase in~\cref{tab:chemical_potentials_all}, whereas all other tags indicate the room-temperature crystal structures of the respective solutes.}
\label{fig:corr_eff_en}
\end{figure}

For each target property, we present both feature importance rankings and prediction parity plots: $E_{\mathrm{s}}$ in~\cref{fig:feature_importance_sol,fig:corr_solution}, $E_{\mathrm{b}}$ in~\cref{fig:feature_importance_bind,fig:corr_bind}, and $E_{\mathrm{eff}}$ in~\cref{fig:feature_importance_sol_min_bind,fig:corr_eff_en}. Across all three target properties, features derived from the end-member phases consistently exhibited higher importance than those from the room-temperature phases of the solutes. Furthermore, the Hume--Rothery rule descriptors achieved at best only a top 5 ranking in importance, falling substantially below the MatMiner-generated features. These findings suggest that classical descriptors are insufficient for identifying suitable solute candidates in Li-based solid solutions, and that more sophisticated feature representations are necessary to accurately predict the target properties.

Here, we obtained nearly the same feature importance for all solute candidates and the best ones. Therefore, we used the same set of features for both type of predictions.

\begin{figure}[H]
\centering
\includegraphics[width=0.58\columnwidth]{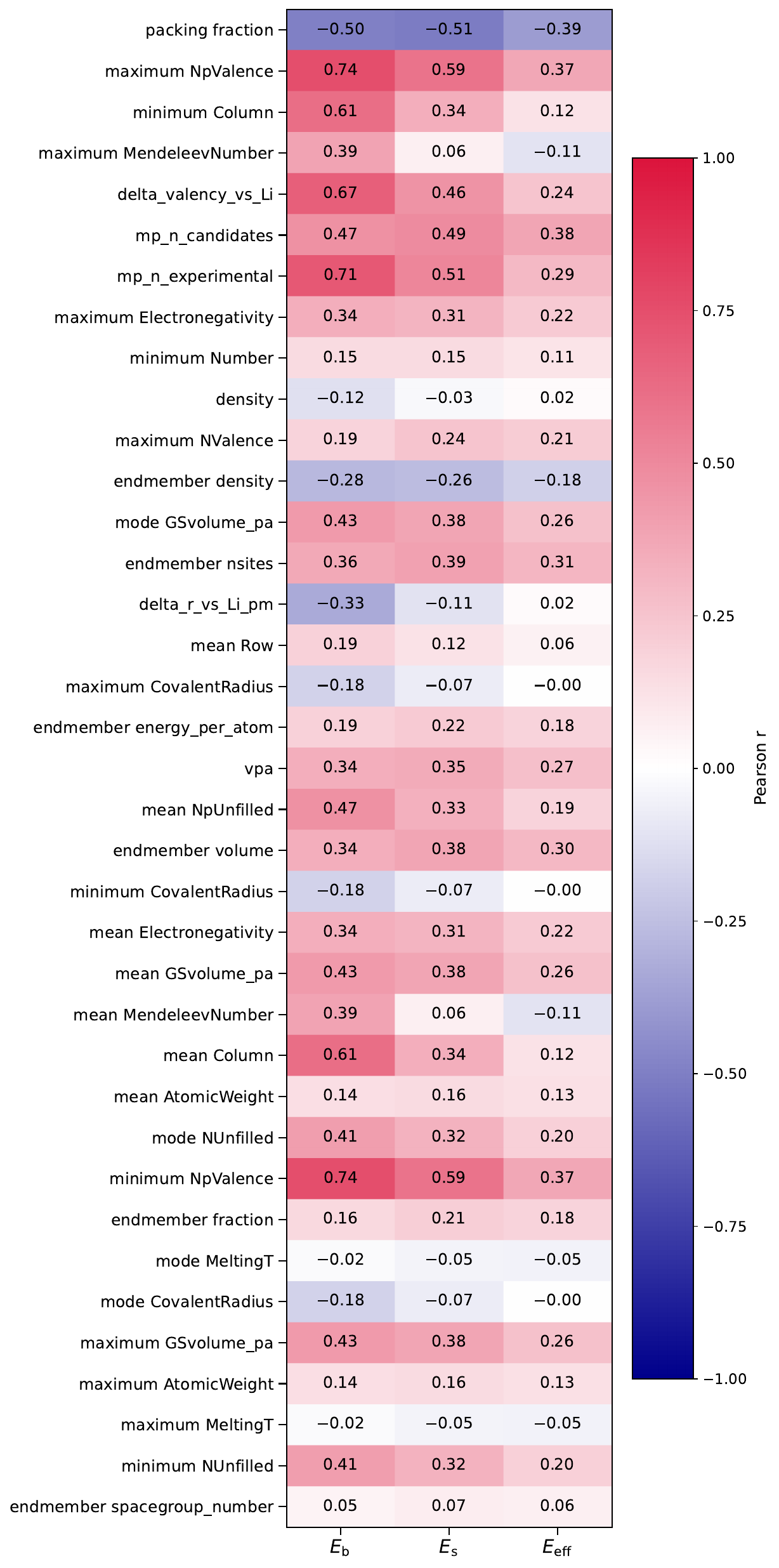}
\caption{Pearson correlation coefficients ($r$) between the top-ranked unique features and three key energetic properties: solution energies ($E_{\mathrm{s}}$), solute--vacancy binding energies ($E_{\mathrm{b}}$), and vacancy effective energies ($E_{\mathrm{eff}}$). Red and blue hues denote positive and negative correlations, respectively. The ``end-member'' label denotes the crystal structure of the corresponding end-member phase in~\cref{tab:chemical_potentials_all}, whereas all other tags indicate the room-temperature crystal structures of the respective solutes.}
\label{fig:pearson_correlation_all}
\end{figure}

Next, we examined the pairwise correlations between all significant features and the three target properties, with the results visualized in~\cref{fig:pearson_correlation_all}. Overall, no consistent correlation pattern emerges across all target properties. For example, the covalent radius of the end-member exhibits a positive correlation with $E_{\mathrm{s}}$ ($r = 0.37$), but shows negative correlations with $E_{\mathrm{b}}$ ($r = -0.05$) and $E_{\mathrm{eff}}$ ($r = -0.24$). This suggests that a unified feature set for all targets is suboptimal, and that separate feature sets should be selected for each target property.

\subsection{Regression models}\label{sec:ml_models_best}

Using the significant features identified for each target property, we selected the optimal machine learning model by minimizing the cross-validation root-mean-square error (RMSE) for the best 19 solute element candidates (opt-only hold-out: 9 train / 10 test; 17 features after filtering). The evaluated models spanned several families: linear regularized regressors (Ridge, Lasso, ElasticNet), Support Vector Regression (SVR), ensemble methods (Random Forest and GradientBoosting), and optimized gradient-boosting frameworks (XGBoost and CatBoost). All implementations were based on Scikit-learn~\cite{pedregosa2011scikit} and the respective native libraries~\cite{chen2016xgboost,prokhorenkova2018catboost}. Hyperparameter optimization was performed via randomized search, and the best-performing model for each target was determined by the lowest cross-validation RMSE.

\begin{figure}[H]
\centering
\includegraphics[width=0.99\columnwidth]{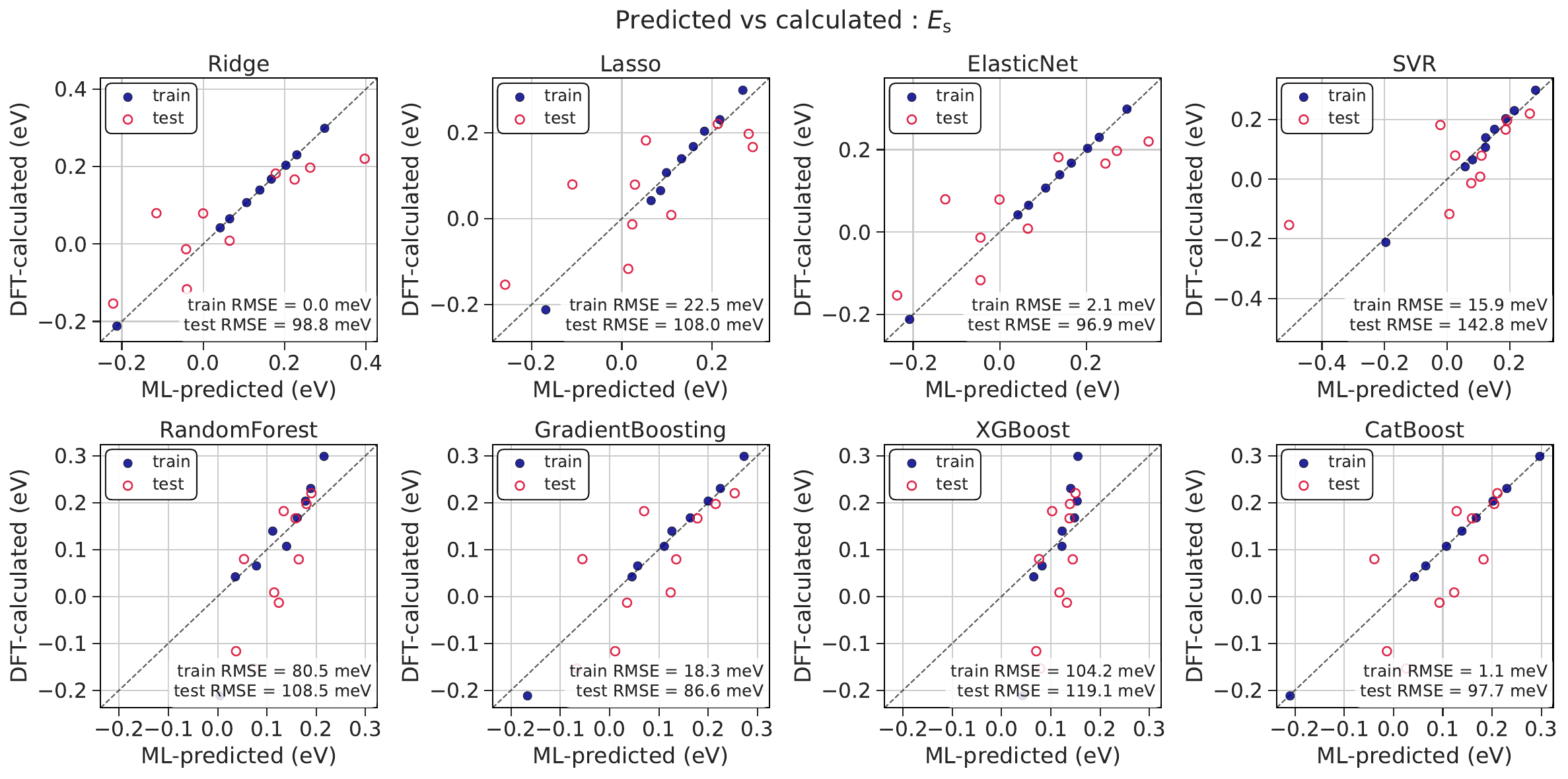}
\caption{Prediction performance of the machine learning models for solution energies ($E_{\mathrm{s}}$). Training set predictions are shown as blue filled circles, whereas test set predictions are represented by red open circles. The root-mean-square errors (RMSE) for both the training and test sets are provided alongside the data.}
\label{fig:ml_models_all_solution}
\end{figure}

The prediction performance of the evaluated models for $E_{\mathrm{s}}$ is visualized in~\cref{fig:ml_models_all_solution}, and the corresponding error metrics are summarized in~\cref{tab:ml_solution}. CatBoost emerged as the best-performing model by cross-validation RMSE, with a hold-out test RMSE of 98~meV.

\begin{table}[H]
\centering
\setlength{\tabcolsep}{8pt}
\renewcommand{\arraystretch}{1.15}
\caption{Hold-out performance of machine learning regression models for $E_{\mathrm{s}}$ predictions on the training and test sets. All errors are reported in meV. The highest $R^{2}$ value and the lowest error for each column are indicated in bold.}
\label{tab:ml_solution}
\begin{tabular}{lcccccccc}
\hline\hline
Model & \multicolumn{4}{c}{Training set} & \multicolumn{4}{c}{Test set} \\
\cline{2-5}\cline{6-9}
 & $R^{2}$ & MaxAE & MAE & RMSE & $R^{2}$ & MaxAE & MAE & RMSE \\
\hline
Ridge & \textbf{1.000} & \textbf{0.0} & \textbf{0.0} & \textbf{0.0} & 0.375 & 194.4 & 80.8 & 98.8 \\
Lasso & 0.974 & 43.7 & 19.5 & 22.5 & 0.252 & 188.9 & 95.5 & 108.0 \\
ElasticNet & 1.000 & 4.2 & 1.5 & 2.1 & 0.398 & 205.1 & 85.0 & 96.9 \\
SVR & 0.987 & 16.2 & 15.8 & 15.9 & --0.308 & 351.3 & 102.1 & 142.8 \\
RandomForest & 0.663 & 217.0 & 50.1 & 80.5 & 0.245 & 228.4 & 84.1 & 108.5 \\
GradientBoosting & 0.983 & 45.4 & 12.4 & 18.3 & \textbf{0.519} & \textbf{134.6} & \textbf{74.6} & \textbf{86.6} \\
XGBoost & 0.435 & 254.5 & 70.0 & 104.2 & 0.091 & 232.2 & 97.9 & 119.1 \\
CatBoost & 1.000 & 2.1 & 0.9 & 1.1 & 0.388 & 179.8 & 80.4 & 97.7 \\
\hline\hline
\end{tabular}
\end{table}

\begin{figure}[H]
\centering
\includegraphics[width=0.99\columnwidth]{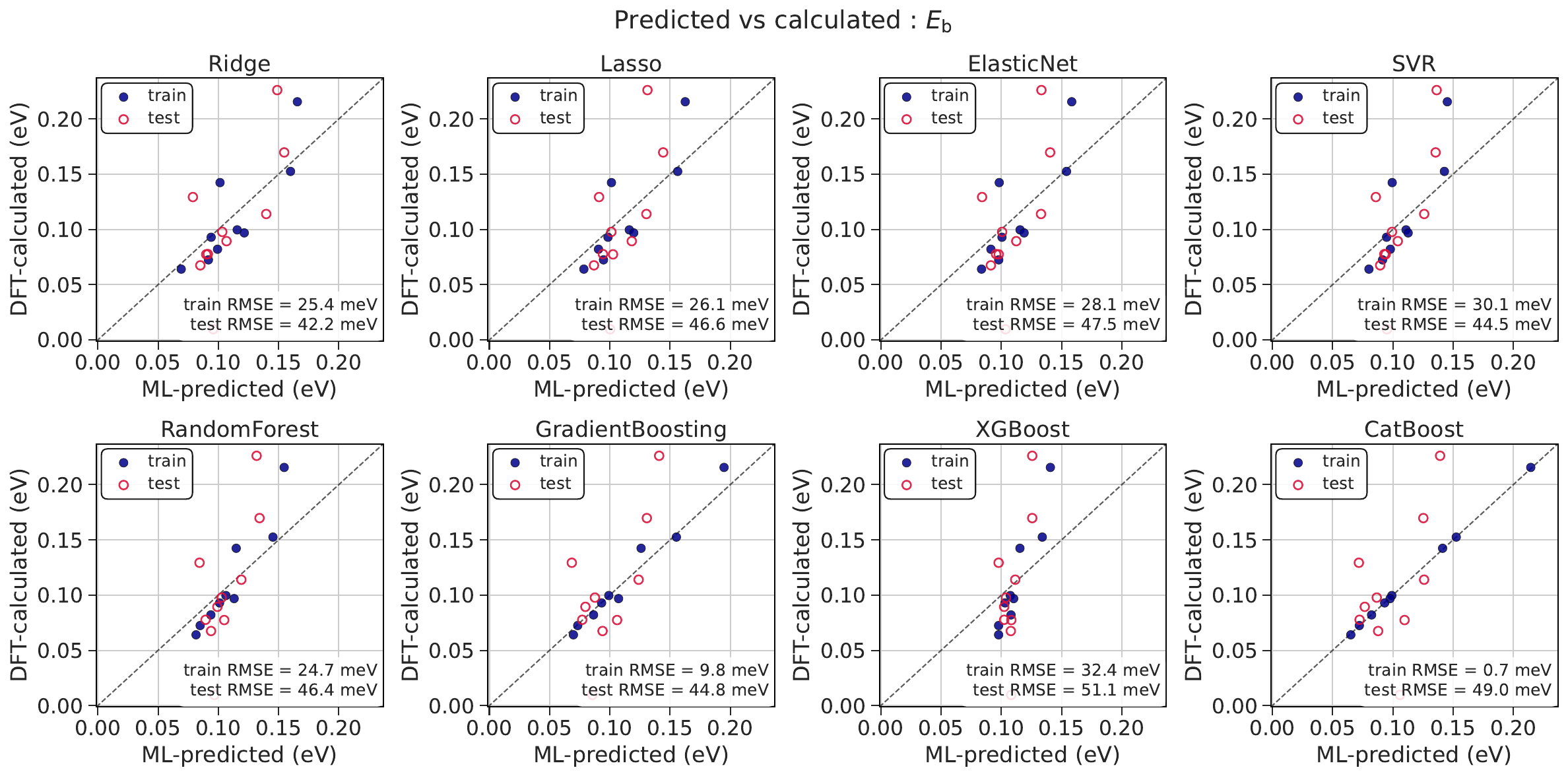}
\caption{Prediction performance of the machine learning models for solute--vacancy binding energies ($E_{\mathrm{b}}$). Training set predictions are shown as blue filled circles, whereas test set predictions are represented by red open circles. The root-mean-square errors (RMSE) for both the training and test sets are provided alongside the data.}
\label{fig:ml_models_all_binding}
\end{figure}

The prediction performance of the evaluated models for $E_{\mathrm{b}}$ is visualized in~\cref{fig:ml_models_all_binding}, and the corresponding error metrics are summarized in~\cref{tab:ml_binding}. Ridge regression emerged as the best-performing model by cross-validation RMSE, with a hold-out test RMSE of 42~meV.

\begin{table}[H]
\centering
\setlength{\tabcolsep}{8pt}
\renewcommand{\arraystretch}{1.15}
\caption{Hold-out performance of machine learning regression models for $E_{\mathrm{b}}$ predictions on the training and test sets. All errors are reported in meV. The highest $R^{2}$ value and the lowest error for each column are indicated in bold.}
\label{tab:ml_binding}
\begin{tabular}{lcccccccc}
\hline\hline
Model & \multicolumn{4}{c}{Training set} & \multicolumn{4}{c}{Test set} \\
\cline{2-5}\cline{6-9}
 & $R^{2}$ & MaxAE & MAE & RMSE & $R^{2}$ & MaxAE & MAE & RMSE \\
\hline
Ridge & 0.691 & 50.0 & 20.2 & 25.4 & \textbf{0.441} & 86.1 & \textbf{32.1} & \textbf{42.2} \\
Lasso & 0.674 & 53.0 & 20.9 & 26.1 & 0.317 & 94.9 & 35.9 & 46.6 \\
ElasticNet & 0.621 & 57.1 & 22.5 & 28.1 & 0.292 & 93.8 & 36.8 & 47.5 \\
SVR & 0.567 & 70.4 & 22.6 & 30.1 & 0.378 & 89.8 & 33.5 & 44.5 \\
RandomForest & 0.708 & 60.9 & 18.7 & 24.7 & 0.324 & 94.3 & 34.7 & 46.4 \\
GradientBoosting & 0.954 & 20.8 & 6.9 & 9.8 & 0.370 & \textbf{85.2} & 34.6 & 44.8 \\
XGBoost & 0.496 & 74.9 & 26.3 & 32.4 & 0.179 & 100.5 & 39.2 & 51.1 \\
CatBoost & \textbf{1.000} & \textbf{1.1} & \textbf{0.6} & \textbf{0.7} & 0.246 & 96.2 & 37.9 & 49.0 \\
\hline\hline
\end{tabular}
\end{table}

\begin{figure}[H]
\centering
\includegraphics[width=0.99\columnwidth]{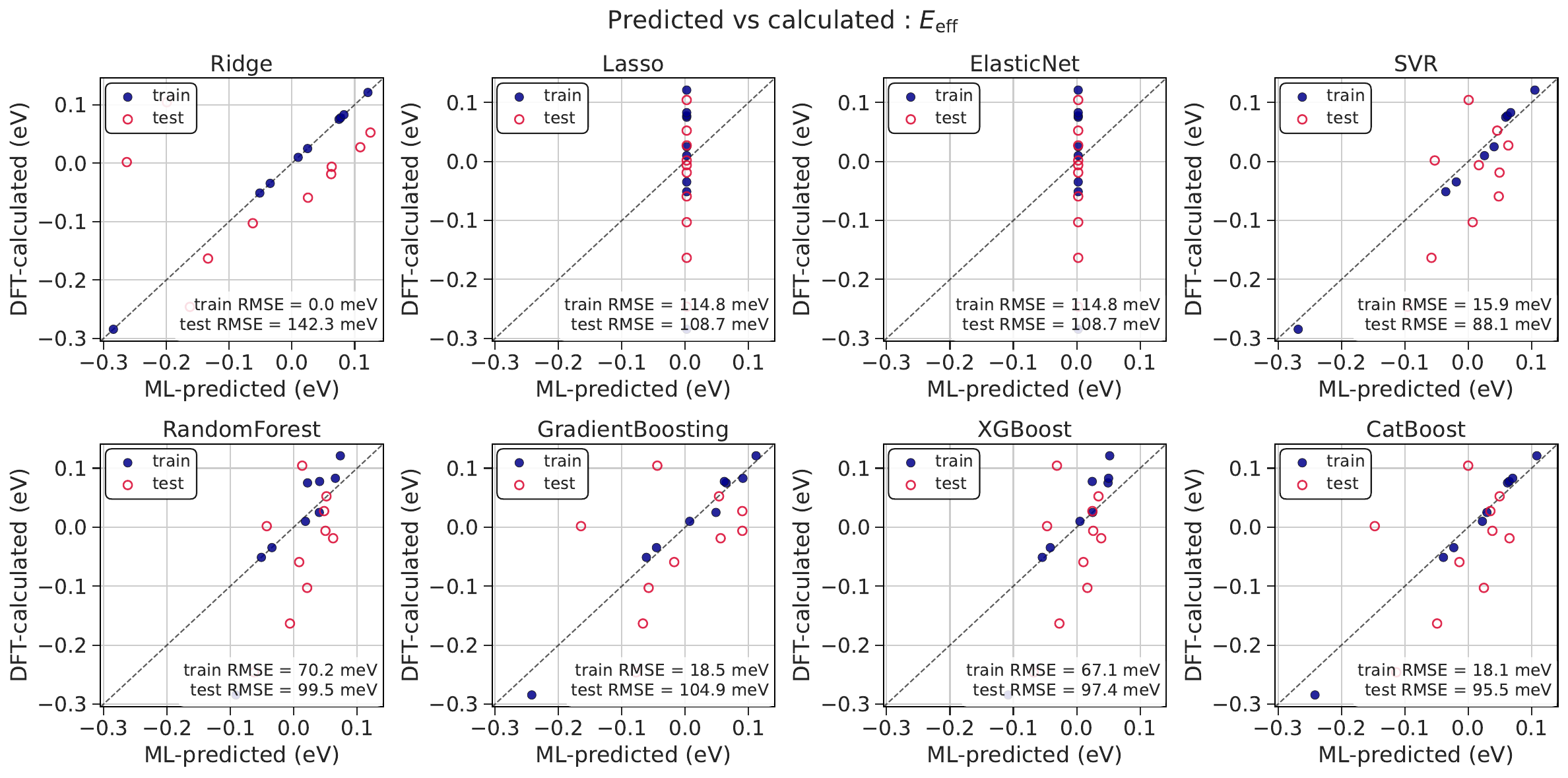}
\caption{Prediction performance of the machine learning models for vacancy effective energies ($E_{\mathrm{eff}}$). Training set predictions are shown as blue filled circles, whereas test set predictions are represented by red open circles. The root-mean-square errors (RMSE) for both the training and test sets are provided alongside the data.}
\label{fig:ml_models_all_effective}
\end{figure}

The prediction performance of the evaluated models for $E_{\mathrm{eff}}$ is visualized in~\cref{fig:ml_models_all_effective}, and the corresponding error metrics are summarized in~\cref{tab:ml_solution_minus_binding}. SVR emerged as the best-performing direct model by cross-validation RMSE, with a hold-out test RMSE of 88~meV.

\begin{table}[H]
\centering
\setlength{\tabcolsep}{8pt}
\renewcommand{\arraystretch}{1.15}
\caption{Hold-out performance of machine learning regression models for $E_{\mathrm{eff}}$ predictions on the training and test sets. All errors are reported in meV. The highest $R^{2}$ value and the lowest error for each column are indicated in bold.}
\label{tab:ml_solution_minus_binding}
\begin{tabular}{lcccccccc}
\hline\hline
Model & \multicolumn{4}{c}{Training set} & \multicolumn{4}{c}{Test set} \\
\cline{2-5}\cline{6-9}
 & $R^{2}$ & MaxAE & MAE & RMSE & $R^{2}$ & MaxAE & MAE & RMSE \\
\hline
Ridge & \textbf{1.000} & \textbf{0.0} & \textbf{0.0} & \textbf{0.0} & --1.039 & 303.5 & 111.2 & 142.3 \\
Lasso & 0.000 & 286.8 & 83.9 & 114.8 & --0.190 & 248.5 & 78.8 & 108.7 \\
ElasticNet & 0.000 & 286.8 & 83.9 & 114.8 & --0.190 & 248.5 & 78.8 & 108.7 \\
SVR & 0.981 & 16.2 & 15.9 & 15.9 & \textbf{0.218} & 150.5 & \textbf{76.5} & \textbf{88.1} \\
RandomForest & 0.626 & 193.3 & 41.4 & 70.2 & 0.003 & 183.8 & 82.7 & 99.5 \\
GradientBoosting & 0.974 & 42.4 & 14.8 & 18.5 & --0.108 & 168.6 & 90.2 & 104.9 \\
XGBoost & 0.659 & 176.4 & 41.3 & 67.1 & 0.044 & 178.3 & 79.8 & 97.4 \\
CatBoost & 0.975 & 42.7 & 15.0 & 18.1 & 0.081 & \textbf{149.2} & 81.2 & 95.5 \\
\hline\hline
\end{tabular}
\end{table}

\begin{figure}[H]
\centering
\includegraphics[width=0.33\columnwidth]{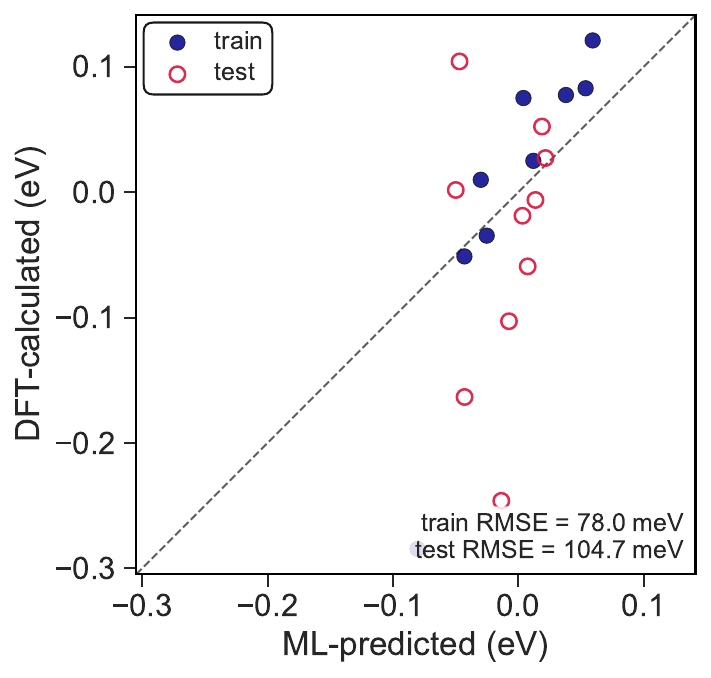}
\caption{The direct ML-prediction of $E_{\mathrm{eff}}$ prediction. Training set predictions are shown as blue filled circles, whereas test set predictions are represented by red open circles.}
\label{fig:ml_best_eff_direct}
\end{figure}

We next evaluated an indirect approach for predicting $E_{\mathrm{eff}}$ by combining the best-performing models for $E_{\mathrm{s}}$ (CatBoost) and $E_{\mathrm{b}}$ (Ridge). As visualized in~\cref{fig:ml_part_main}.c, this indirect method yielded a test RMSE of 71~meV, better than the 88~meV test RMSE obtained via direct SVR prediction (see~\cref{fig:ml_best_eff_direct}). Based on this comparison, we adopted the indirect prediction as the superior approach for $E_{\mathrm{eff}}$.

\subsection{Metrics of the best ML models}

\begin{table}[H]
\centering
\setlength{\tabcolsep}{8pt}
\renewcommand{\arraystretch}{1.15}
\caption{Performance metrics for the best-performing machine-learning regression models on training and test sets for each target property: $E_{\mathrm{s}}$, $E_{\mathrm{b}}$, and $E_{\mathrm{eff}}$. For $E_{\mathrm{eff}}$, we report results for two prediction strategies: (i) direct ML prediction and (ii) indirect estimation derived from the optimal models for $E_{\mathrm{s}}$ and $E_{\mathrm{b}}$ via the funnel relationship (\cref{eq:funnel_rule_3}). All errors are reported in meV.}
\label{tab:ml_best_models_metrics}
\begin{tabular}{llrrrrrrrr}
\hline\hline
Target property & Model & \multicolumn{4}{c}{Training set} & \multicolumn{4}{c}{Test set} \\
\cline{3-6}\cline{7-10}
   &   &  $R^{2}$ & MaxAE & MAE & RMSE & $R^{2}$ & MaxAE & MAE & RMSE \\
\hline
$E_{\mathrm{s}}$ & CatBoost & 1.000 & 2.1 & 0.9 & 1.1 & 0.388 & 179.8 & 80.4 & 97.7 \\
$E_{\mathrm{b}}$ & Ridge Regression & 0.691 & 50.0 & 20.2 & 25.4 & 0.441 & 86.1 & 32.1 & 42.2 \\
$E_{\mathrm{eff}}$ (direct) & SVR & 0.981 & 16.2 & 15.9 & 15.9 & 0.218 & 150.5 & 76.5 & 88.1 \\
$E_{\mathrm{eff}}$ (derived) & CatBoost \& Ridge & 0.846 & 48.3 & 20.0 & 25.1 & --0.213 & 98.1 & 66.4 & 70.9 \\
\hline\hline
\end{tabular}
\end{table}

\section{Reference experimental phase diagrams}\label{sec:exp_diagrams}

\begin{table}[H] 
\setlength{\tabcolsep}{10pt}
\caption{The experimental maximum equilibrium solid solubility in bcc Li, compiled from the data of Goodall~\textit{et al.}~\cite{Goodall2019} and reference phase diagrams (see Figures below), is presented alongside solubility predictions derived from a Boltzmann-type model based on calculated solution energies.} 
\label{tab:solubility} 
\centering 
\begin{tabular}{lcccc} 
\hline \hline
Element & $E_{\mathrm{s}},$~eV & $c_{\max},$~at.\% & $T,$~K & $c_{\mathrm{pred}},$~at.\% \\ \hline 
Na & 0.23 & $<0.1$ & 444 & 0.55 \\ 
Mg & --0.15 & 75.5 & 865 & 100.00 \\
Al & 0.13 & 0 & --- & --- \\ 
Si & 0.33 & $<0.1$ & 454 & 0.08 \\
Ca & 0.35 & 2.2 & 503 & 0.07 \\ 
Sc & 0.84 & $<0.1$ & 453 & 0.0002 \\ 
Ni & 0.44 & 0 & --- & --- \\ 
Cu & 0.13 & $<0.1$ & 453 & 0.36 \\ 
Zn & 0.09 & 1.5 & 434 & 1.31 \\ 
Ga & 0.08 & 0.1 & 454 & 1.43 \\ 
Ge & 0.26 & $<0.1$ & 454 & 0.42 \\ 
Se & 0.94 & 0 & --- & --- \\
Rh & 0.06 & $<0.1$ & 454 & 1.79 \\ 
Pd & 0.19 & $<0.1$ & 418 & 0.63 \\ 
Ag & --0.12 & 9.0 & 418 & 100.00 \\ 
Cd & --0.19 & 18.0 & 487 & 100.00 \\ 
In & 0.12 & 0.2 & 429 & 0.42 \\ 
Sn & 0.22 & $<0.1$ & 454 & 0.32 \\
Sb & 0.32 & 0 & --- & --- \\ 
Ir & 0.35 & 0.1 & 454 & 0.07 \\ 
Pt & 0.06 & $<0.1$ & 442 & 1.17 \\ 
Au & 0.22 & 0.7 & 449 & 0.33 \\ 
Hg & 0.07 & 1.0 & 432 & 0.95 \\ 
Tl & 0.02 & $<0.1$ & 454 & 2.55 \\
Pb & 0.18 & $<0.1$ & 454 & 0.52 \\ 
Bi & 0.23 & $<0.1$ & 454 & 0.32 \\ 
\hline \hline 
\end{tabular}
\end{table}

% Z = 3
\begin{figure}[H]
\centering
\includegraphics[width=0.65\columnwidth]{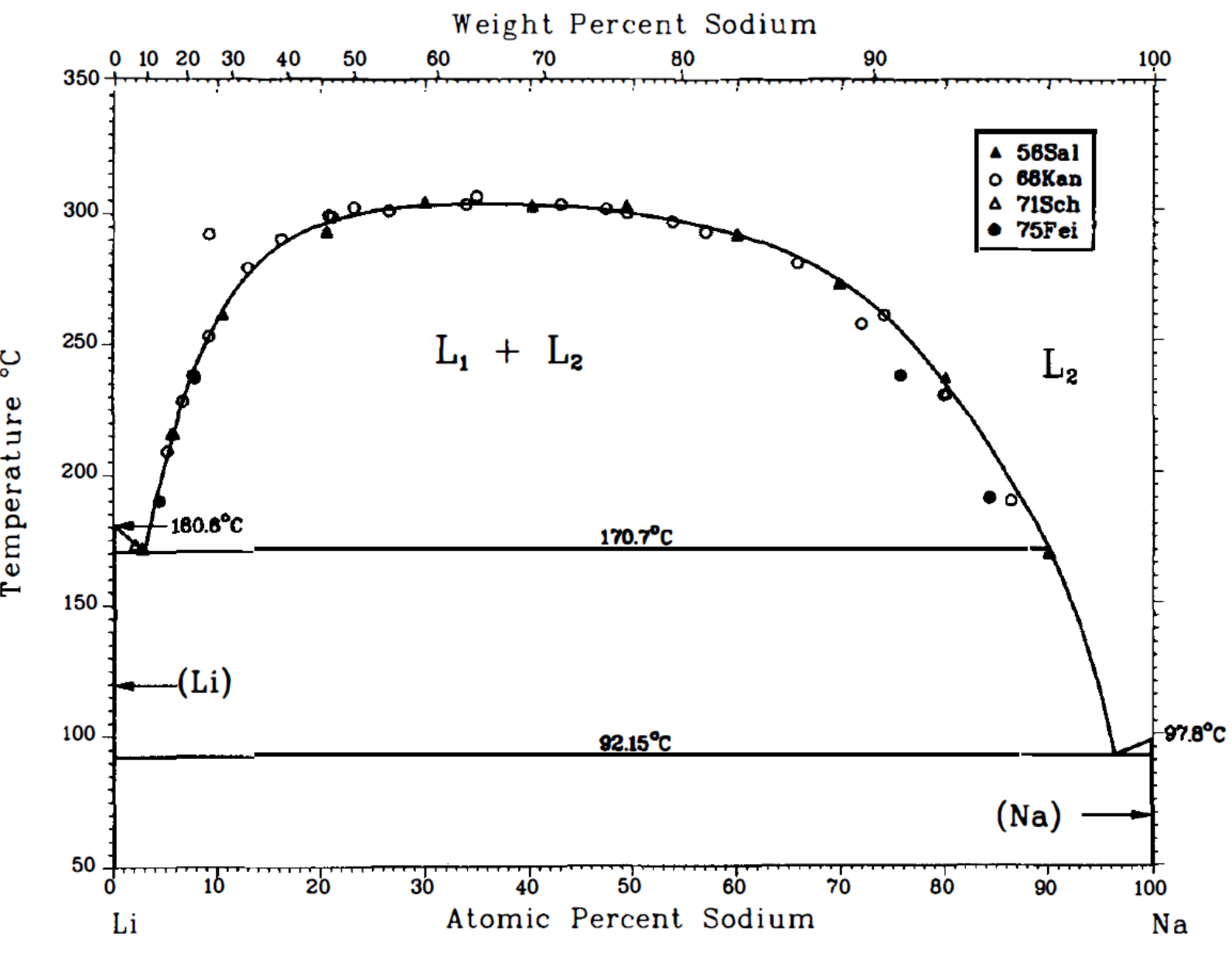}
\caption{Li--Na phase diagram. Here, Na forms a solid solution with Li in a narrow region of concentrations. Our reference phase is Na (mp-127). Data is from the work by Bale~\textit{et al.}~\cite{bale1989li}.}
\label{fig:phase_diagram_na}
\end{figure}

% Z = 12
\begin{figure}[H]
\centering
\includegraphics[width=0.47\columnwidth]{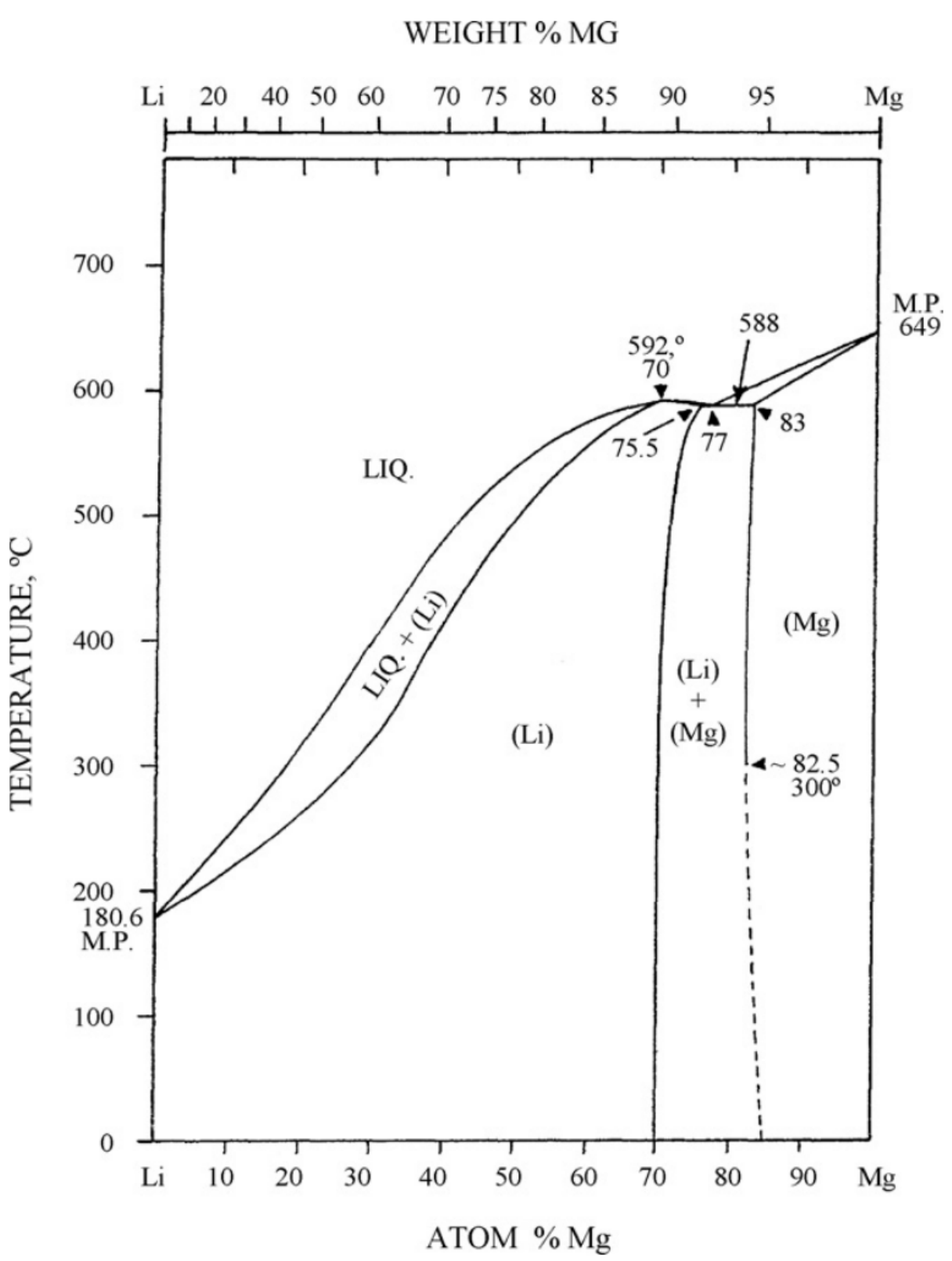}
\caption{Li--Mg phase diagram. Here, Mg forms a solid solution with Li in a wide region of concentrations. Our reference phase is Mg (mp-153). Data is from the work by Guidotti~\textit{et al.}~\cite{guidotti2008thermally}.}
\label{fig:phase_diagram_mg}
\end{figure}

% Z = 13
\begin{figure}[H]
\centering
\includegraphics[width=0.5\columnwidth]{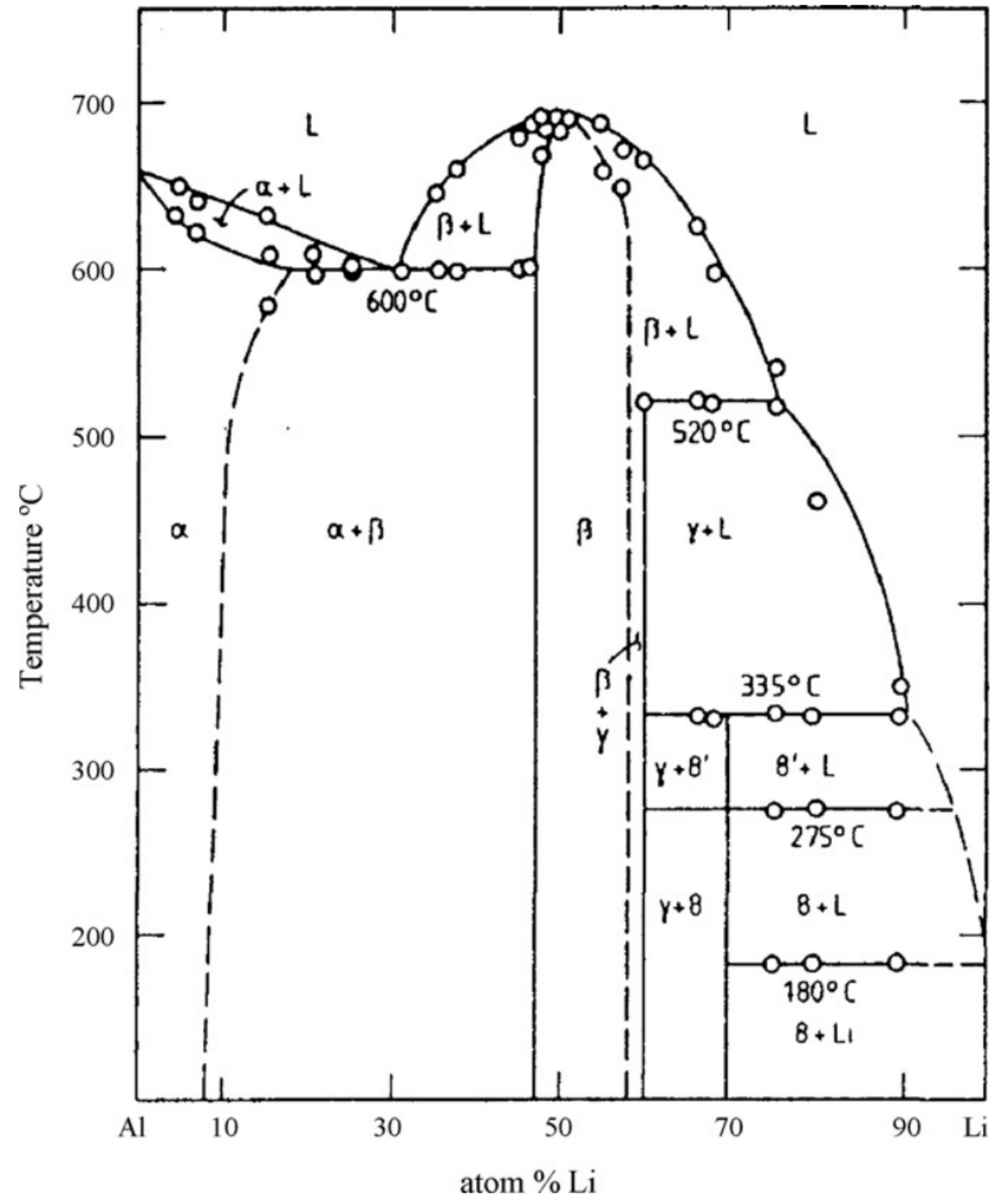}
\caption{Li--Al phase diagram. Here, Al does not form a solid solution with Li. Our reference phase is $\ce{Li9Al4}$ (mp-568404). Data is from the work by Guidotti~\textit{et al.}~\cite{guidotti2008thermally}.}
\label{fig:phase_diagram_al}
\end{figure}

% Z = 14
\begin{figure}[H]
\centering
\includegraphics[width=0.5\columnwidth]{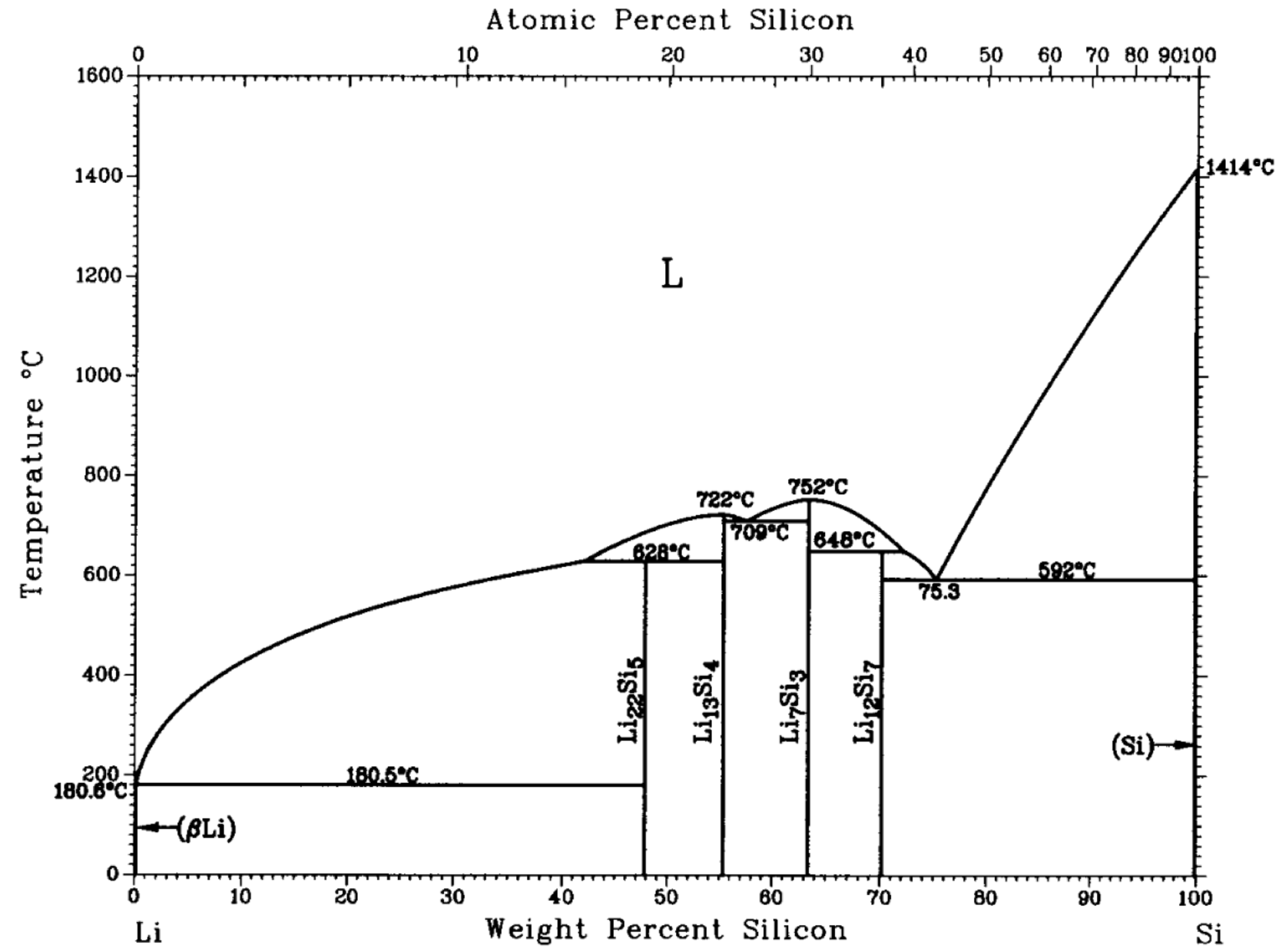}
\caption{Li--Si phase diagram. Here, Si forms a solid solution with Li in a narrow region of concentrations. Our reference phase is $\ce{Li21Si5}$ (mp-29720). Data is from the work by Okamoto~\textit{et al.}~\cite{okamoto1990si}.}
\label{fig:phase_diagram_si}
\end{figure}

% Z = 20
\begin{figure}[H]
\centering
\includegraphics[width=0.7\columnwidth]{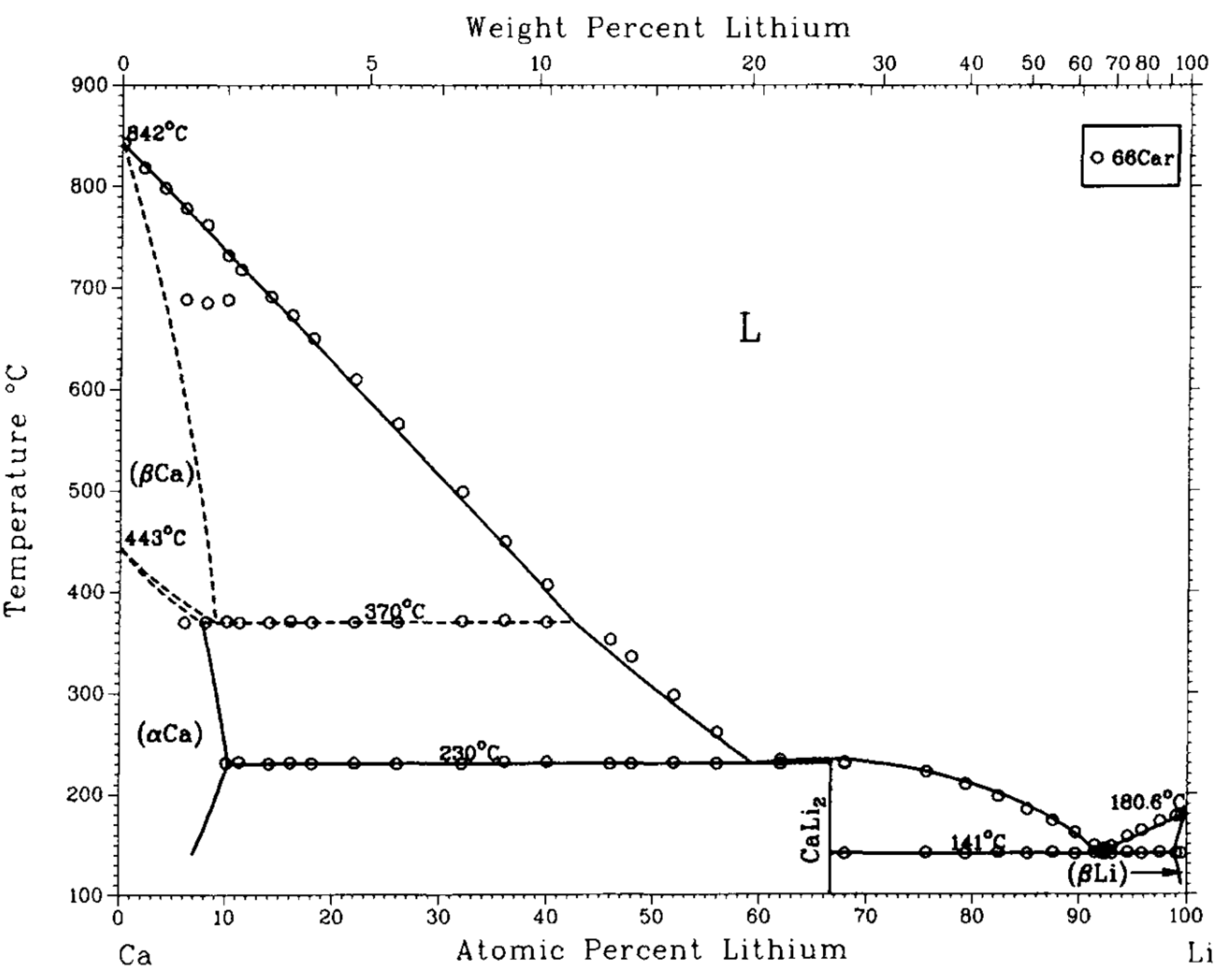}
\caption{Li--Ca phase diagram. Here, Ca forms a solid solution with Li in a narrow region of concentrations. Our reference phase is $\ce{Li2Ca}$ (mp-11644). Data is from the work by Bale~\textit{et al.}~\cite{bale1987ca}.}
\label{fig:phase_diagram_ca}
\end{figure}

% Z = 28
\begin{figure}[H]
\centering
\includegraphics[width=0.5\columnwidth]{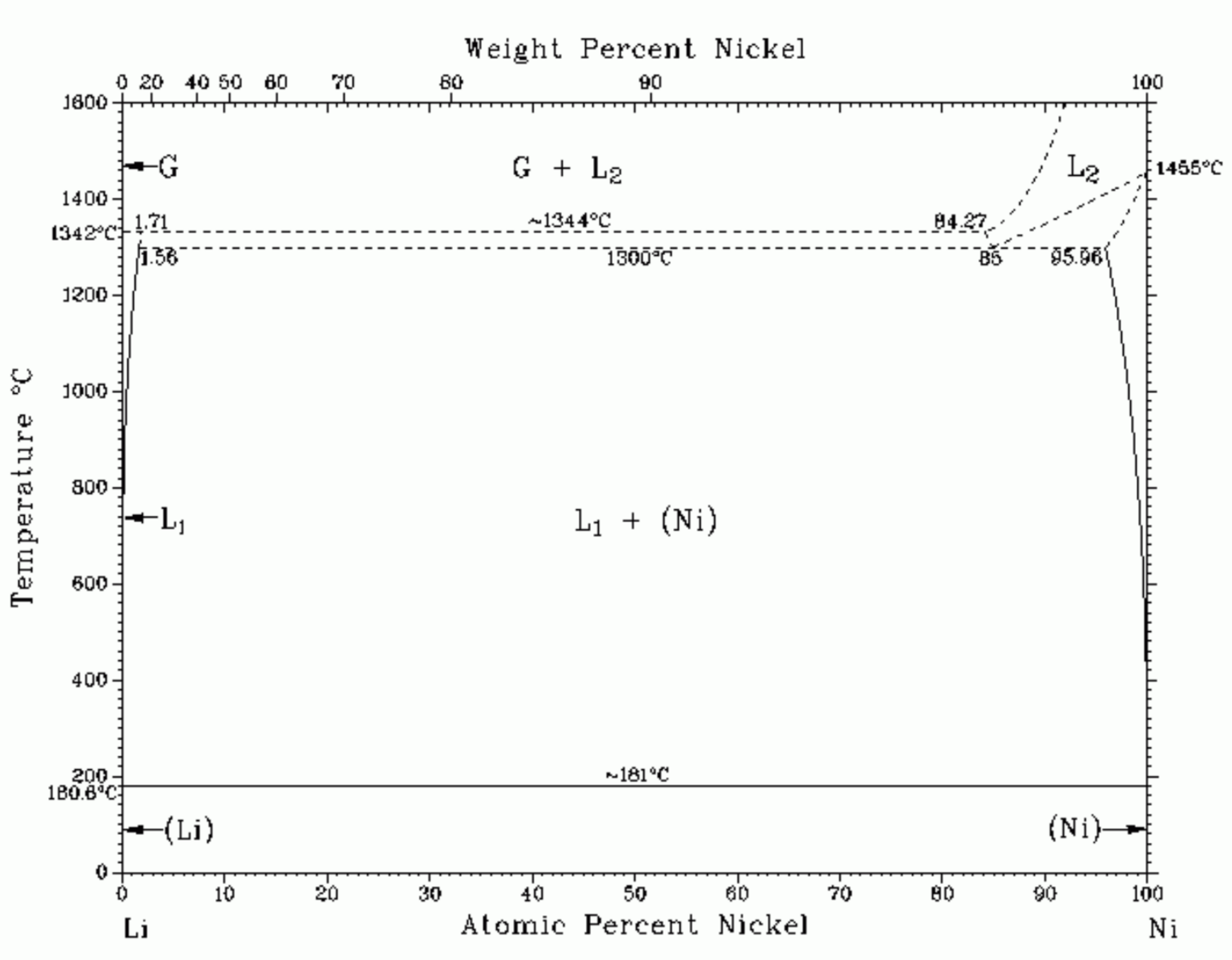}
\caption{Li--Ni phase diagram. Here, Ni does not form a solid solution with Li. Our reference phase is $\ce{Ni}$ (mp-23). Data is from the work by Massalski~\textit{et al.}~\cite{massalski1986binary}.}
\label{fig:phase_diagram_ni}
\end{figure}

% Z = 29
\begin{figure}[H]
\centering
\includegraphics[width=0.7\columnwidth]{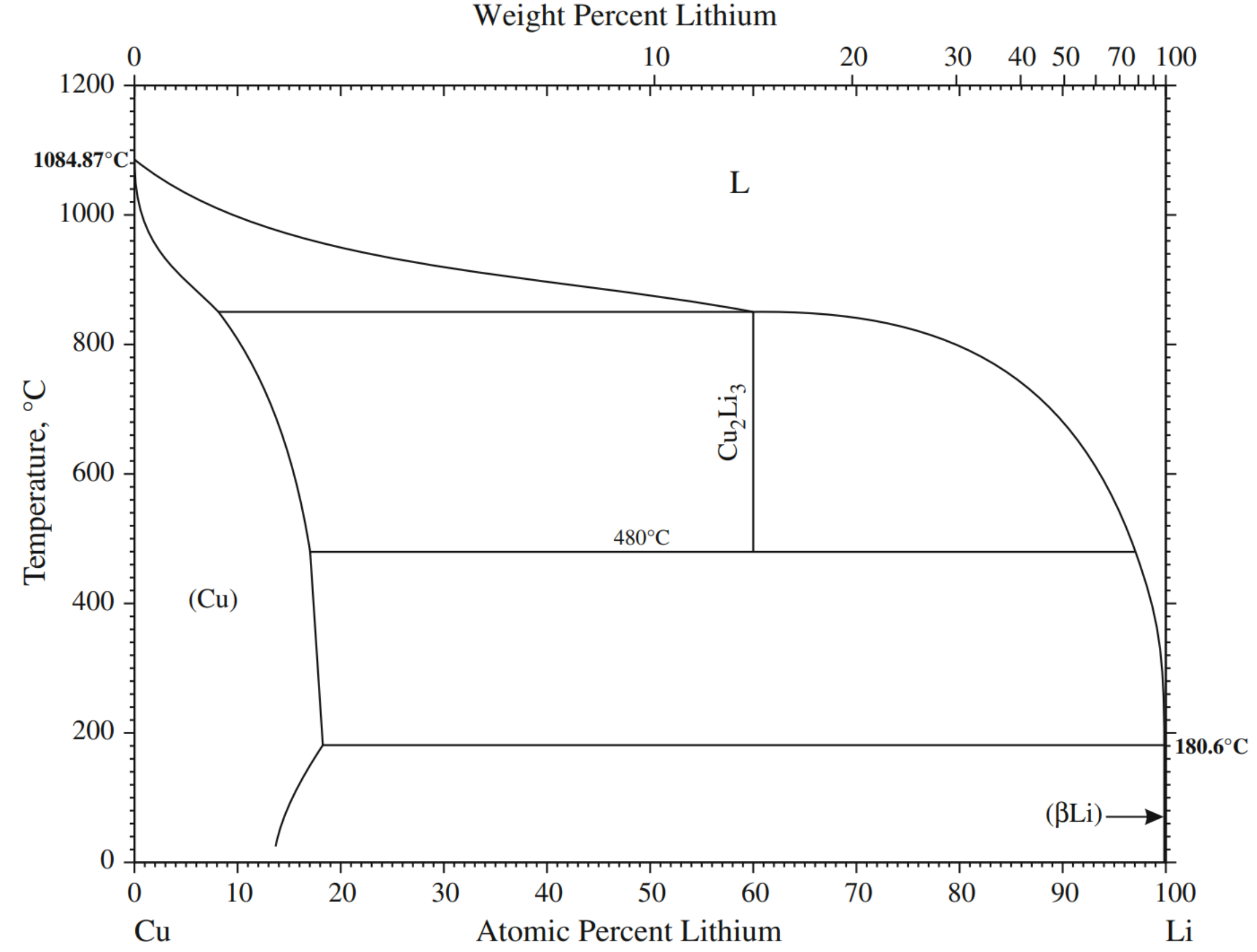}
\caption{Li--Cu phase diagram. Here, Cu forms a solid solution with Li in an extremely narrow range of concentrations. Our reference phase is $\ce{Cu}$ (mp-30). Data is from the work by Okamoto~\textit{et al.}~\cite{okamoto2011cu}.}
\label{fig:phase_diagram_cu}
\end{figure}

% Z = 30
\begin{figure}[H]
\centering
\includegraphics[width=0.7\columnwidth]{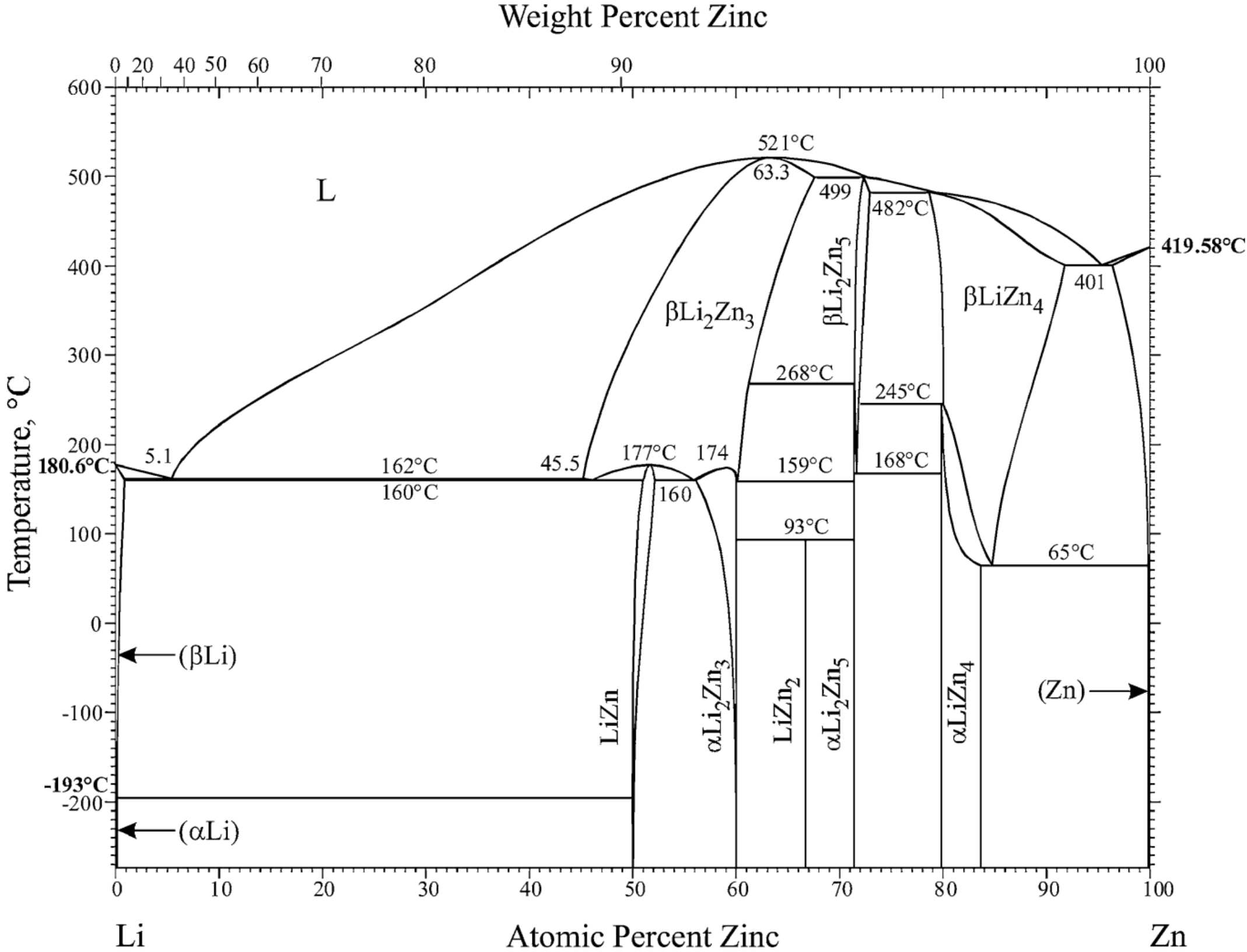}
\caption{Li--Zn phase diagram. Here, Zn forms a solid solution with Li in a narrow region of concentrations. Our reference phase is $\ce{LiZn}$ (mp-1934). Data is from the work by Okamoto~\textit{et al.}~\cite{okamoto2012zn}.}
\label{fig:phase_diagram_zn}
\end{figure}

% Z = 31
\begin{figure}[H]
\centering
\includegraphics[width=0.7\columnwidth]{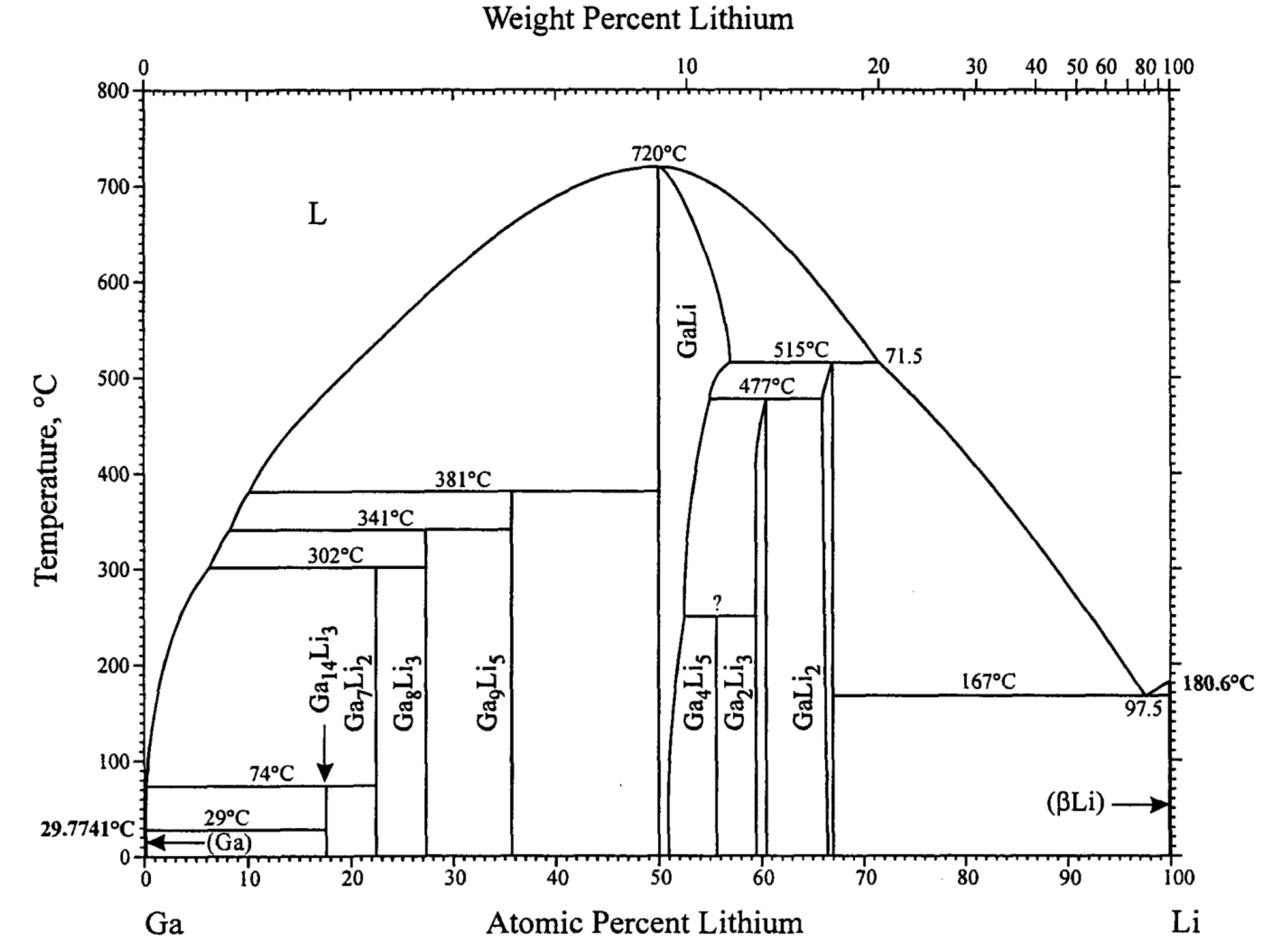}
\caption{Li--Ga phase diagram. Here, Ga forms a solid solution with Li in a narrow range of concentrations. Our reference phase is $\ce{Li2Ga}$ (mp-29210). Data is from the work by Okamoto~\textit{et al.}~\cite{okamoto2006ga}.}
\label{fig:phase_diagram_ga}
\end{figure}

% Z = 32
\begin{figure}[H]
\centering
\includegraphics[width=0.7\columnwidth]{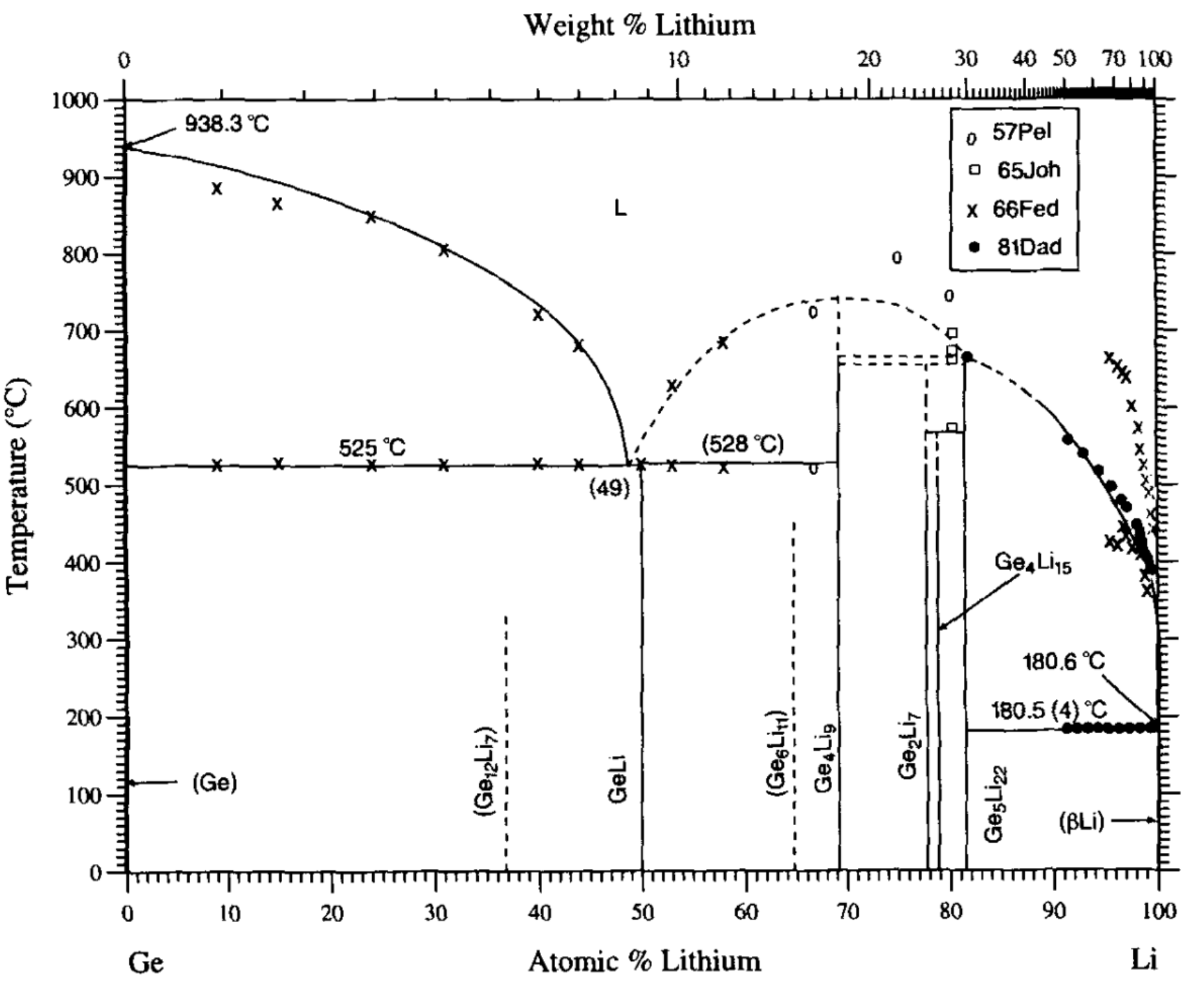}
\caption{Li--Ge phase diagram. Here, Ge forms a solid solution with Li in an extremely narrow range of concentrations. Our reference phase is $\ce{Li15Ge4}$ (mp-1777). Data is from the work by Sangster~\textit{et al.}~\cite{sangster1997ge}.}
\label{fig:phase_diagram_ge}
\end{figure}

% Z = 34
\begin{figure}[H]
\centering
\includegraphics[width=0.7\columnwidth]{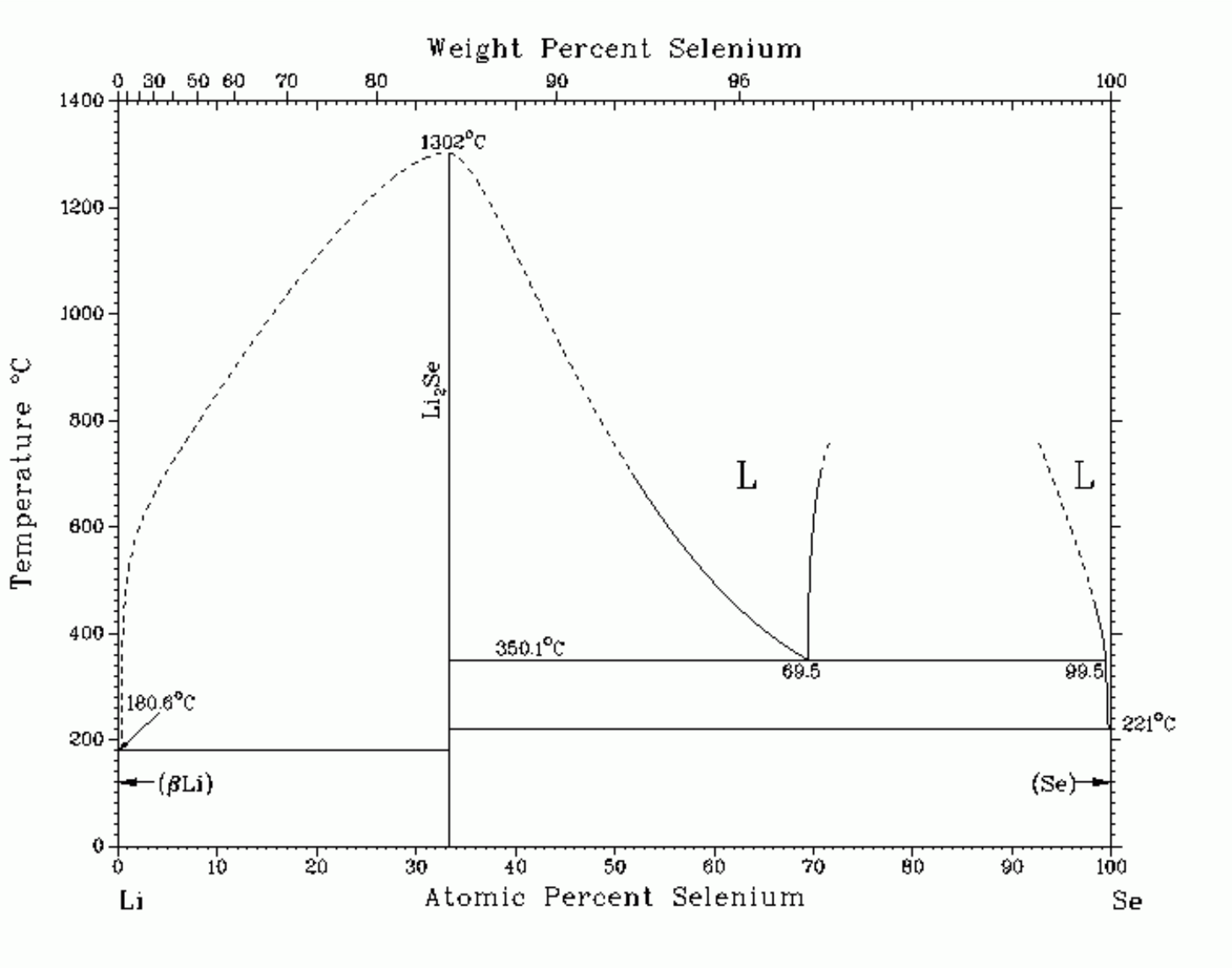}
\caption{Li--Se phase diagram. Here, Se forms a solid solution with Li in a narrow range of concentrations. Data is from the work by Cunningham~\textit{et al.}~\cite{cunningham1971phase_se}.}
\label{fig:phase_diagram_se}
\end{figure}

% Z = 45
\begin{figure}[H]
\centering
\includegraphics[width=0.7\columnwidth]{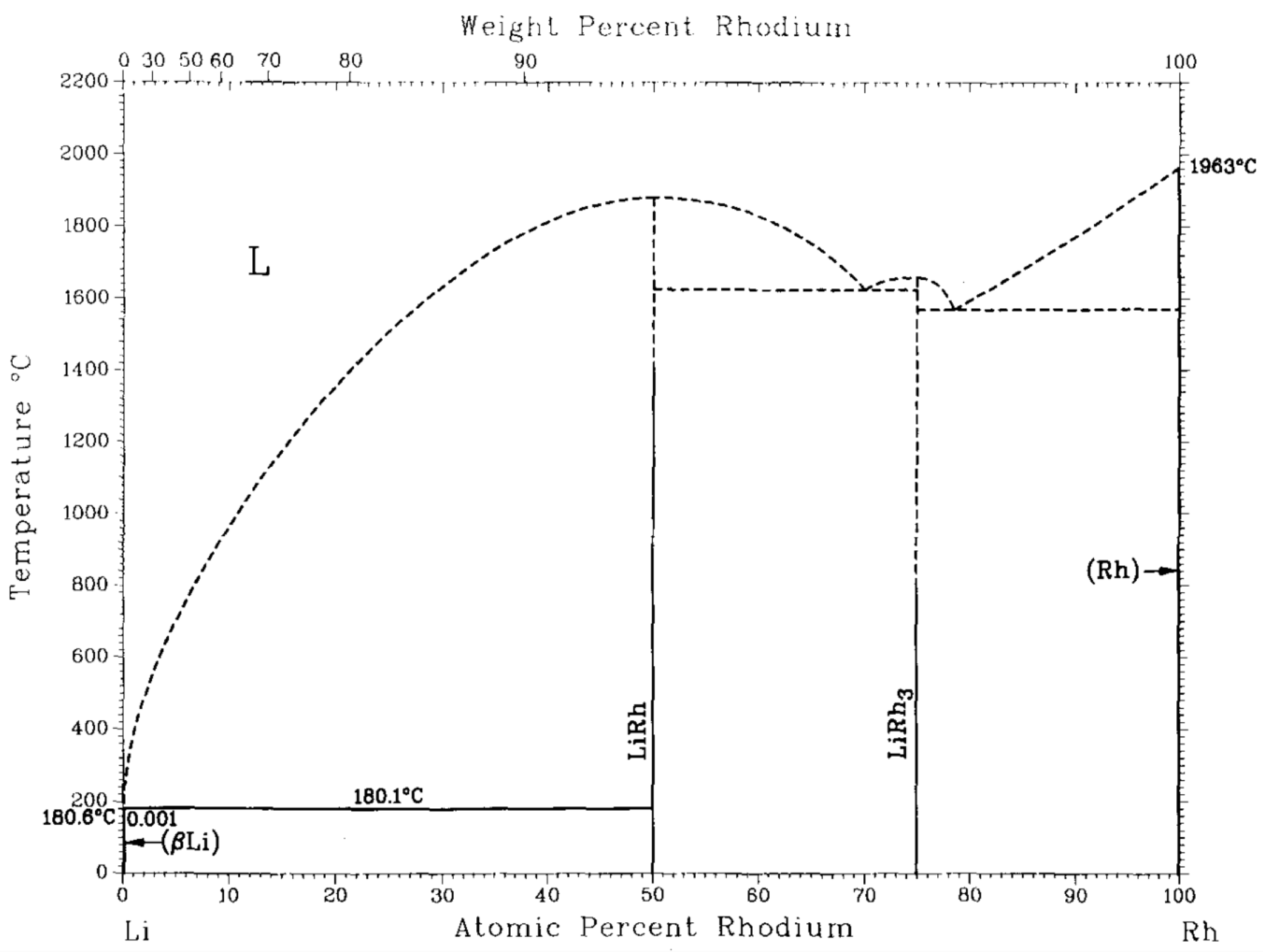}
\caption{Li--Rh phase diagram. Here, Rh forms a solid solution with Li in an extremely narrow region of concentrations. Our reference phase is $\ce{LiRh}$ (mp-600561). Data is from the work by Sangster~\textit{et al.}~\cite{sangster1991rh}.}
\label{fig:phase_diagram_rh}
\end{figure}

% Z = 46
\begin{figure}[H]
\centering
\includegraphics[width=0.7\columnwidth]{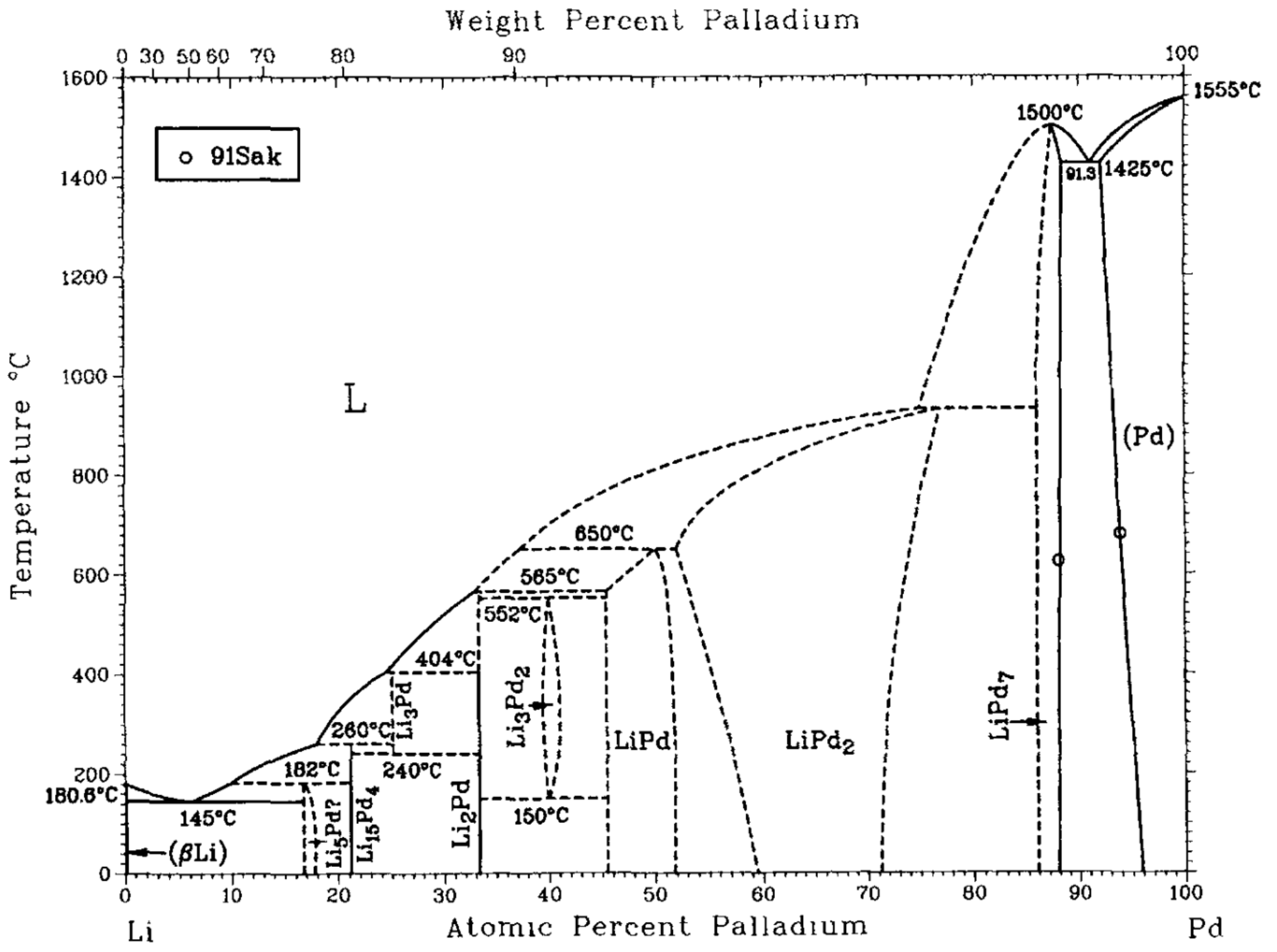}
\caption{Li--Pd phase diagram. Here, Pd forms a solid solution with Li in a narrow region of concentrations. Our reference phase is $\ce{Li15Pd4}$ (mp-1197547). Data is from the work by Okamoto~\textit{et al.}~\cite{okamoto1993pd}.}
\label{fig:phase_diagram_pd}
\end{figure}

% Z = 47
\begin{figure}[H]
\centering
\includegraphics[width=0.7\columnwidth]{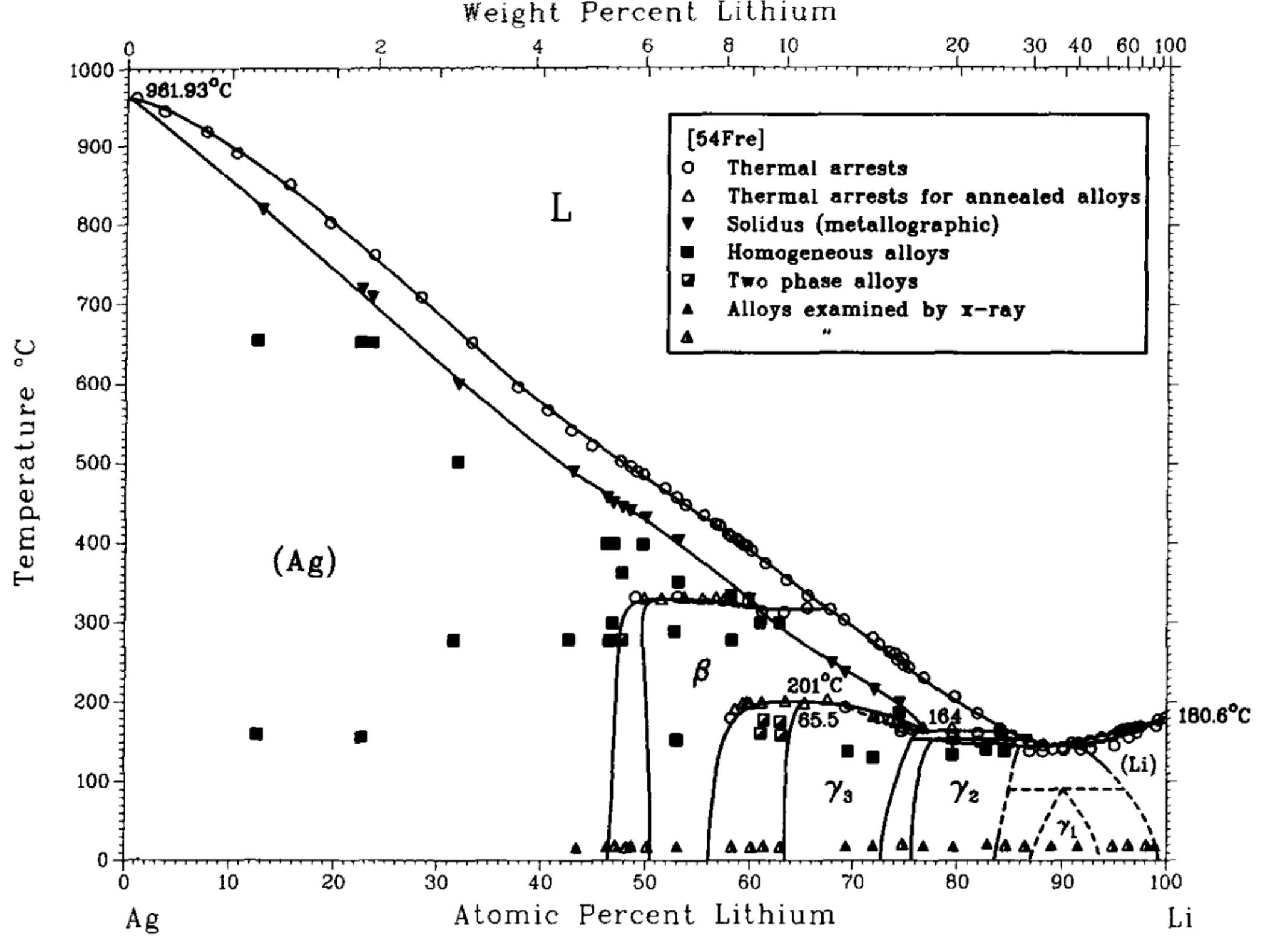}
\caption{Li--Ag phase diagram. Here, Ag forms a solid solution with Li in a narrow region of concentrations. Our reference phase is $\ce{LiAg}$ (mp-2426). Data is from the work by Pelton~\textit{et al.}~\cite{pelton1986ag}.}
\label{fig:phase_diagram_ag}
\end{figure}

% Z = 48
\begin{figure}[H]
\centering
\includegraphics[width=0.7\columnwidth]{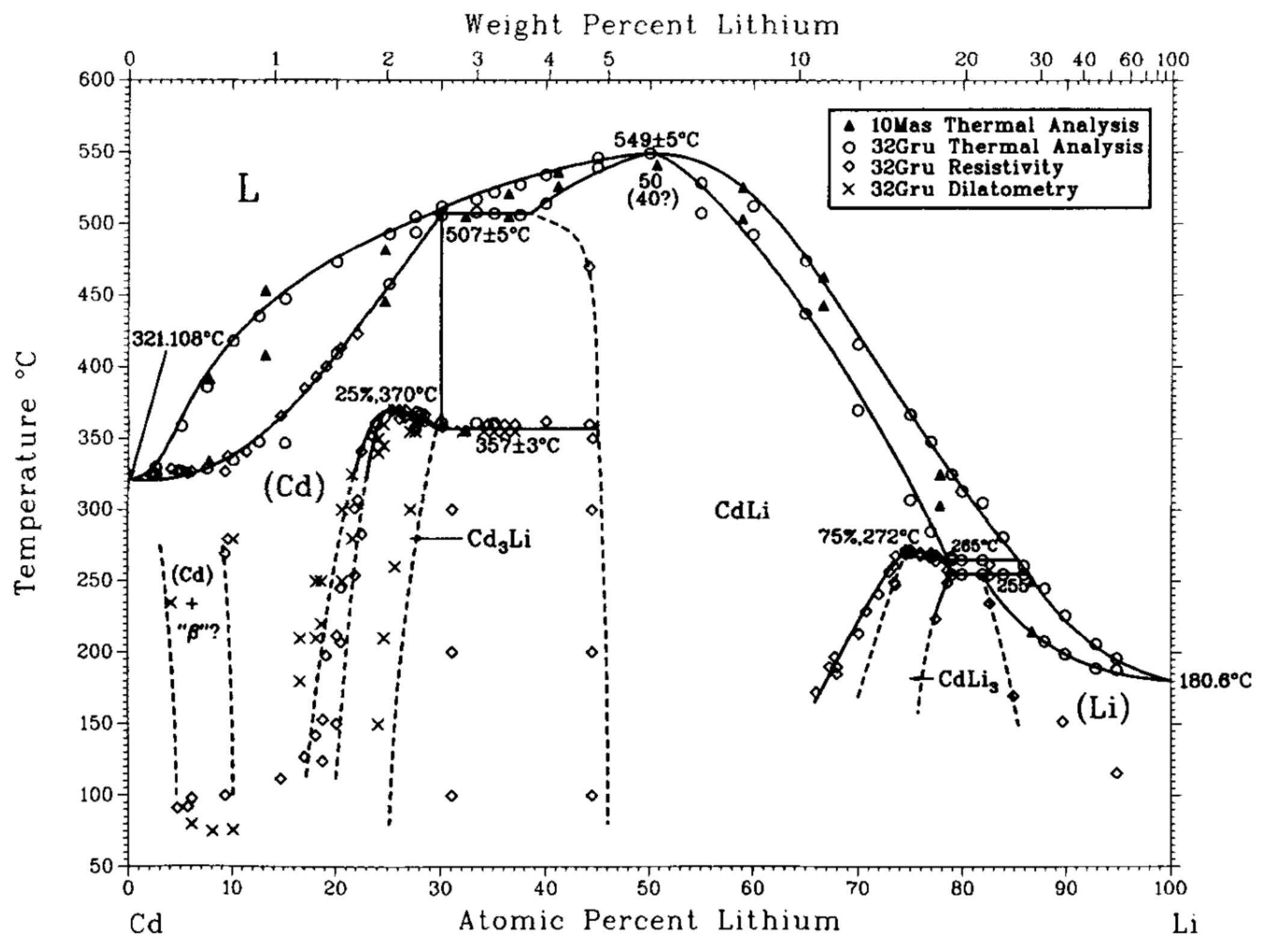}
\caption{Li--Cd phase diagram. Here, Cd forms a solid solution with Li in a wide region of concentrations. Our reference phase is $\ce{LiCd}$ (mp-1437). Data is from the work by Pelton~\textit{et al.}~\cite{pelton1988cd}.}
\label{fig:phase_diagram_cd}
\end{figure}

% Z = 49
\begin{figure}[H]
\centering
\includegraphics[width=0.7\columnwidth]{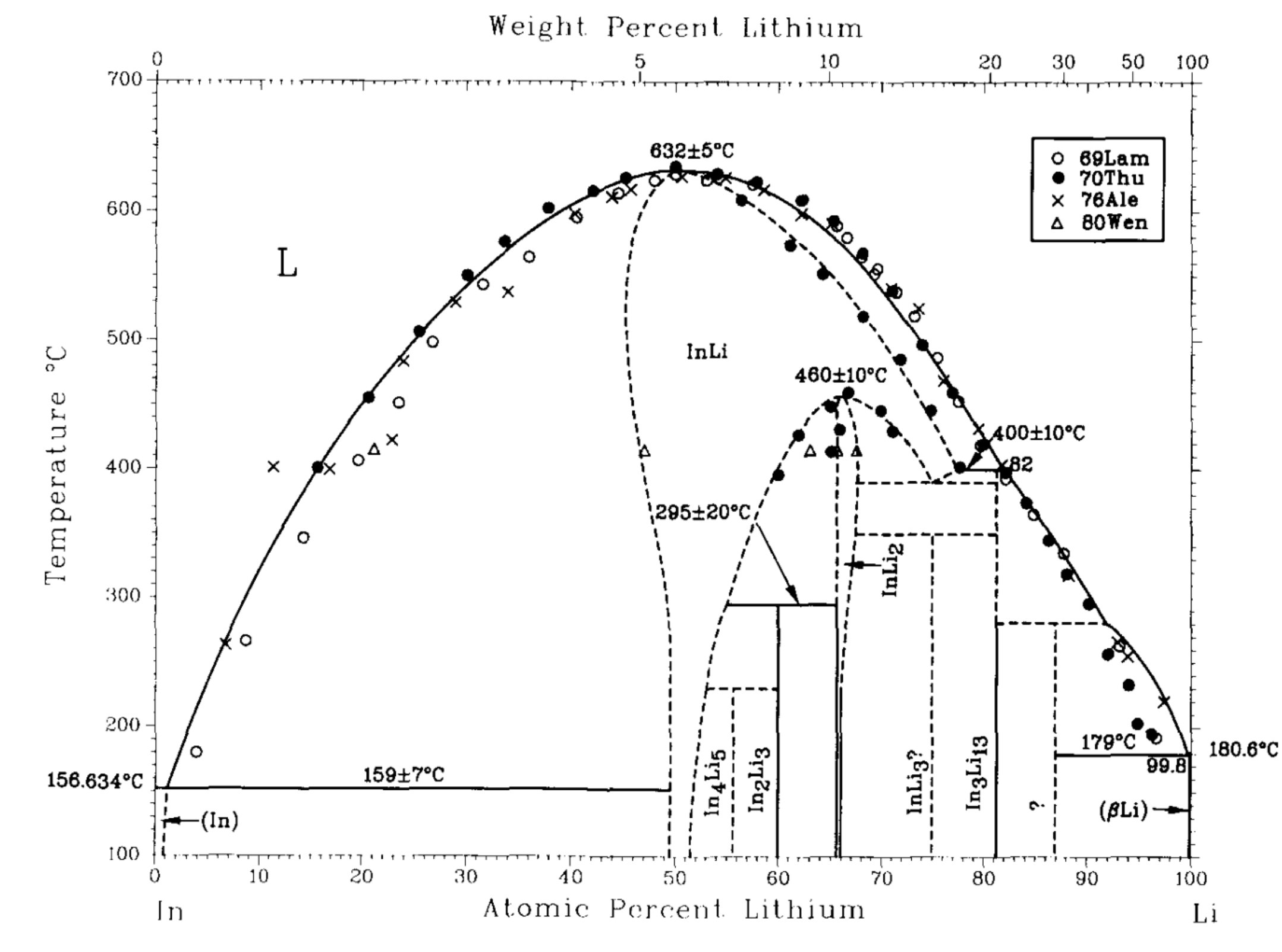}
\caption{Li--In phase diagram. Here, In forms a solid solution with Li in a narrow region of concentrations. Our reference phase is $\ce{Li13In3}$ (mp-510430). Data is from the work by Sangster~\textit{et al.}~\cite{sangster1991In}.}
\label{fig:phase_diagram_in}
\end{figure}

% Z = 50

% Z = 51
\begin{figure}[H]
\centering
\includegraphics[width=0.5\columnwidth]{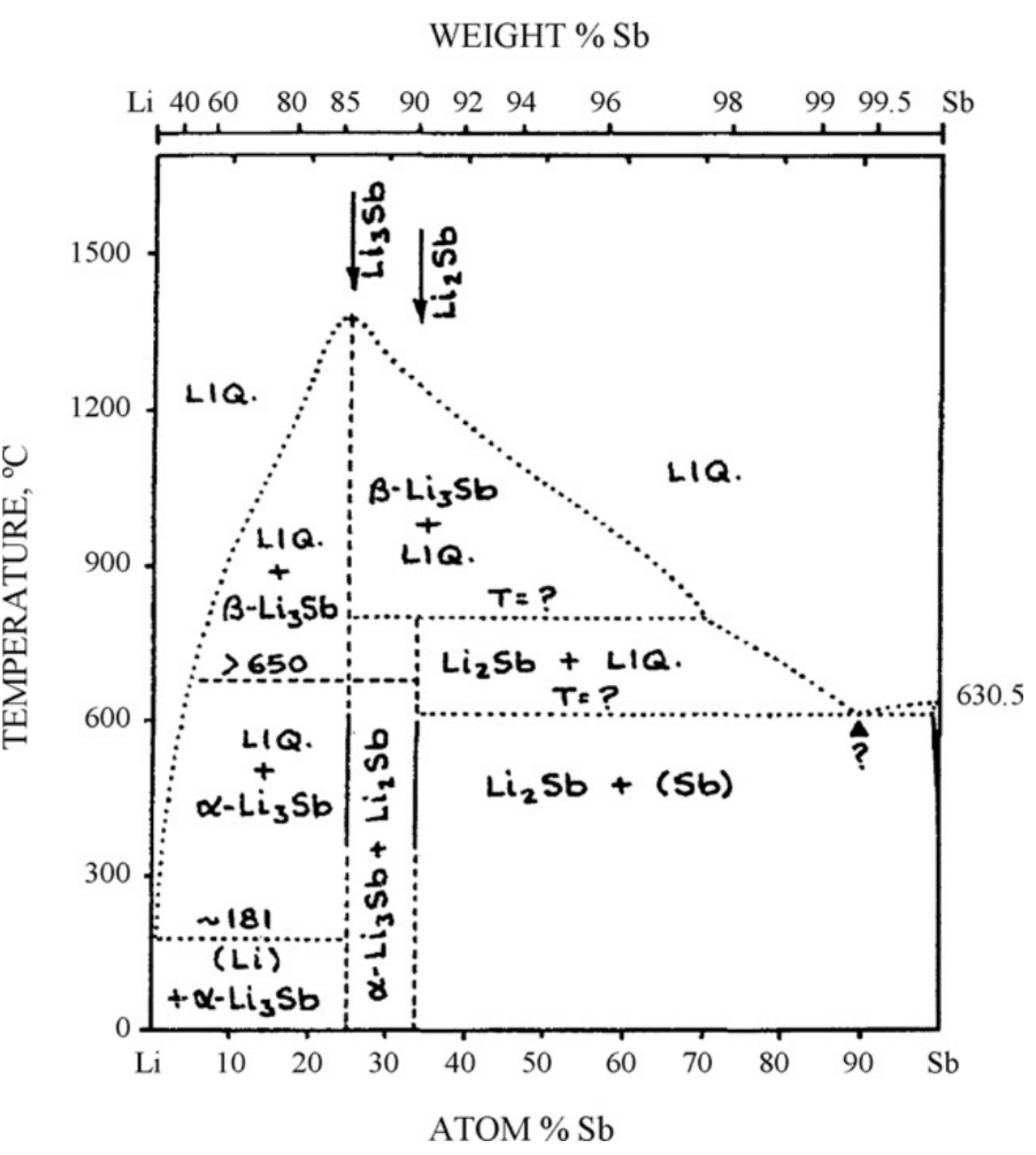}
\caption{Li--Sb phase diagram. Here, Sb does not form a solid solution with Li. Our reference phase is $\ce{Li3Sb}$ (mp-2074). Data is from the work by Guidotti~\textit{et al.}~\cite{guidotti2008thermally}.}
\label{fig:phase_diagram_sb}
\end{figure}

% Z = 52
\begin{figure}[H]
\centering
\includegraphics[width=0.7\columnwidth]{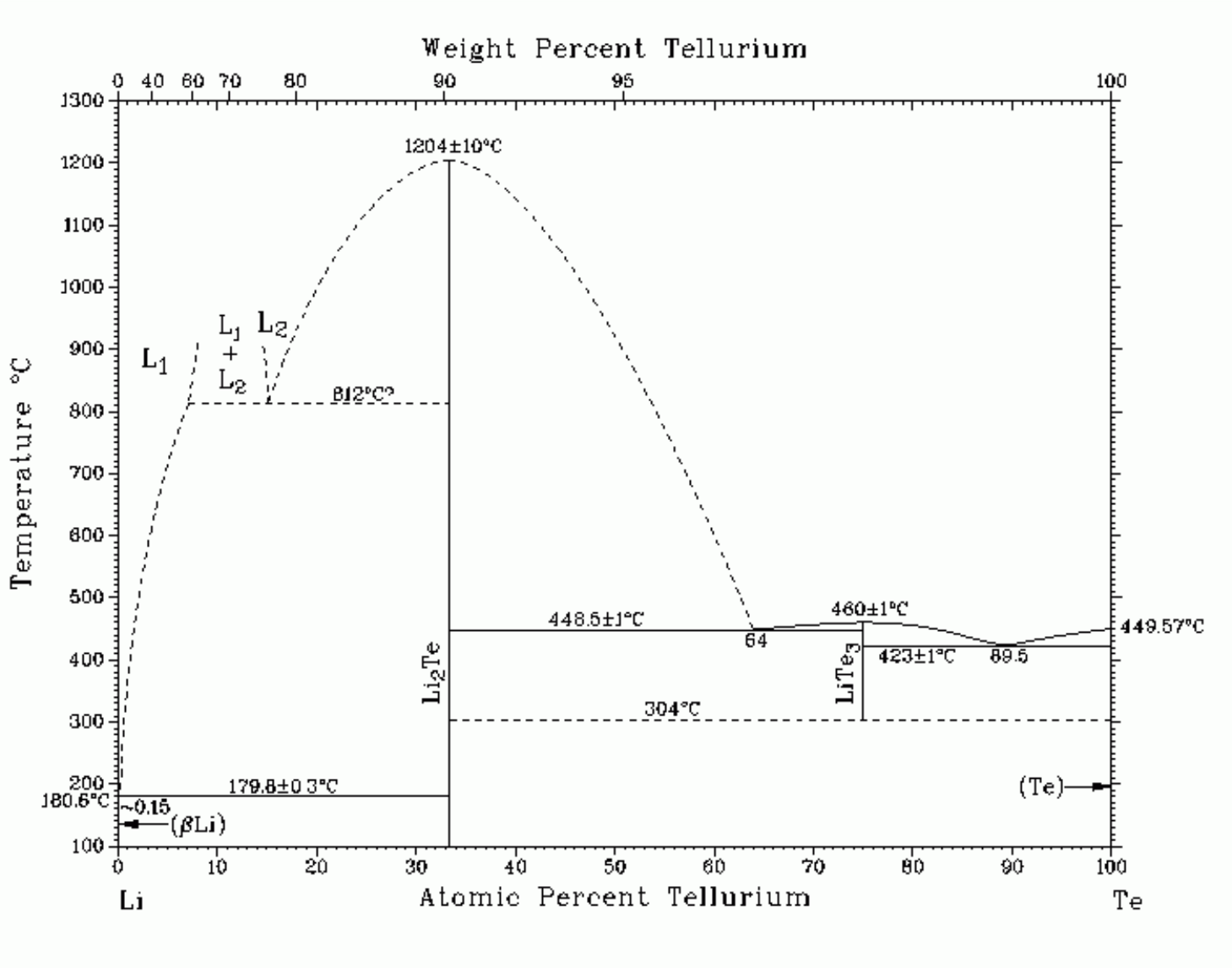}
\caption{Li--Te phase diagram. Here, Te forms a solid solution with Li in a narrow range of concentrations. Data is from the work by Cunningham~\textit{et al.}~\cite{cunningham1973phase_te}.}
\label{fig:phase_diagram_te}
\end{figure}

% Z = 77
\begin{figure}[H]
\centering
\includegraphics[width=0.7\columnwidth]{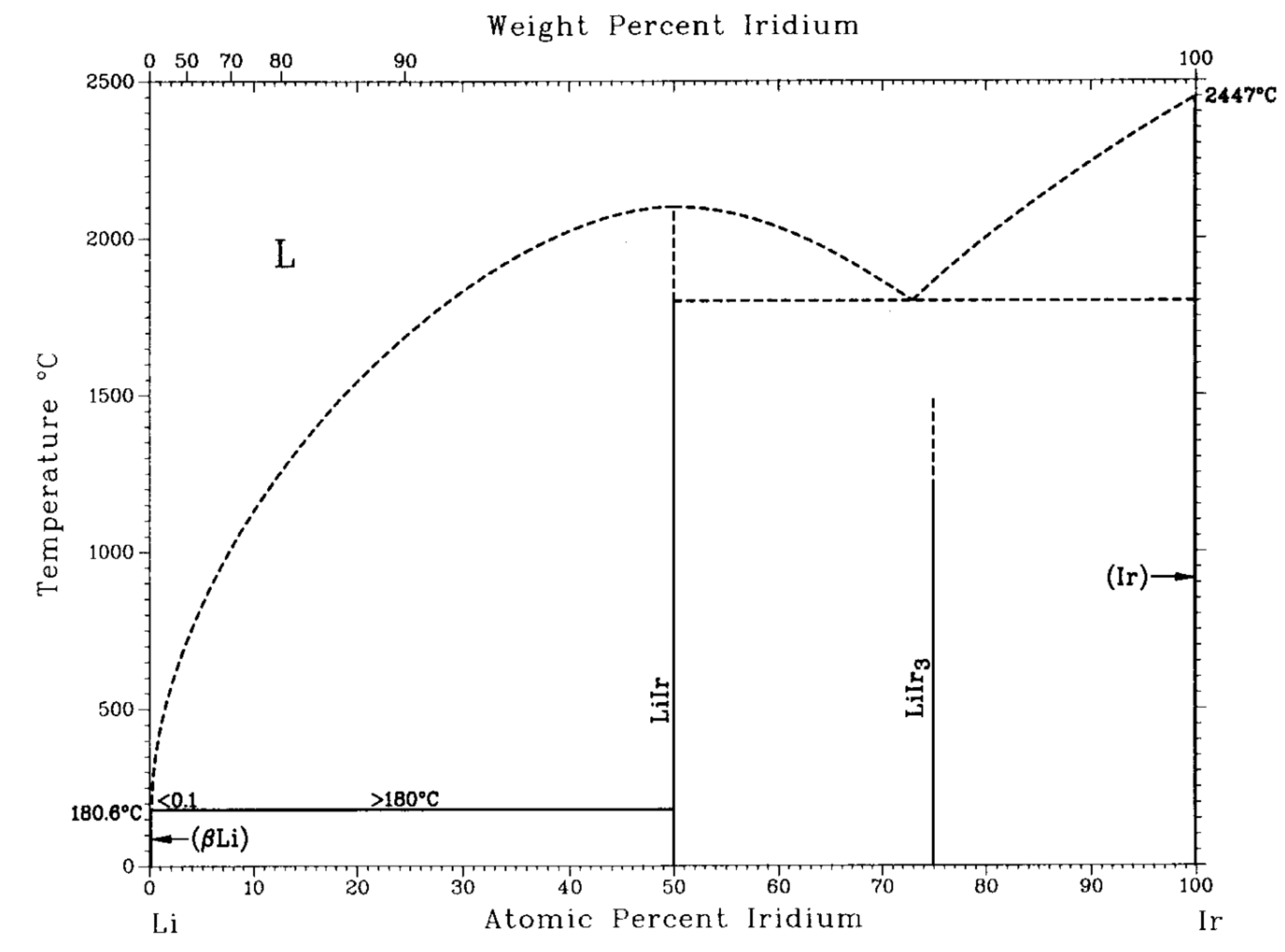}
\caption{Li--Ir phase diagram. Here, Ir forms a solid solution with Li in a narrow range of concentrations. Our reference phase is $\ce{LiIr}$ (mp-279). Data is from the work by Sangster~\textit{et al.}~\cite{sangster1992ir}.}
\label{fig:phase_diagram_ir}
\end{figure}

% Z = 78
\begin{figure}[H]
\centering
\includegraphics[width=0.7\columnwidth]{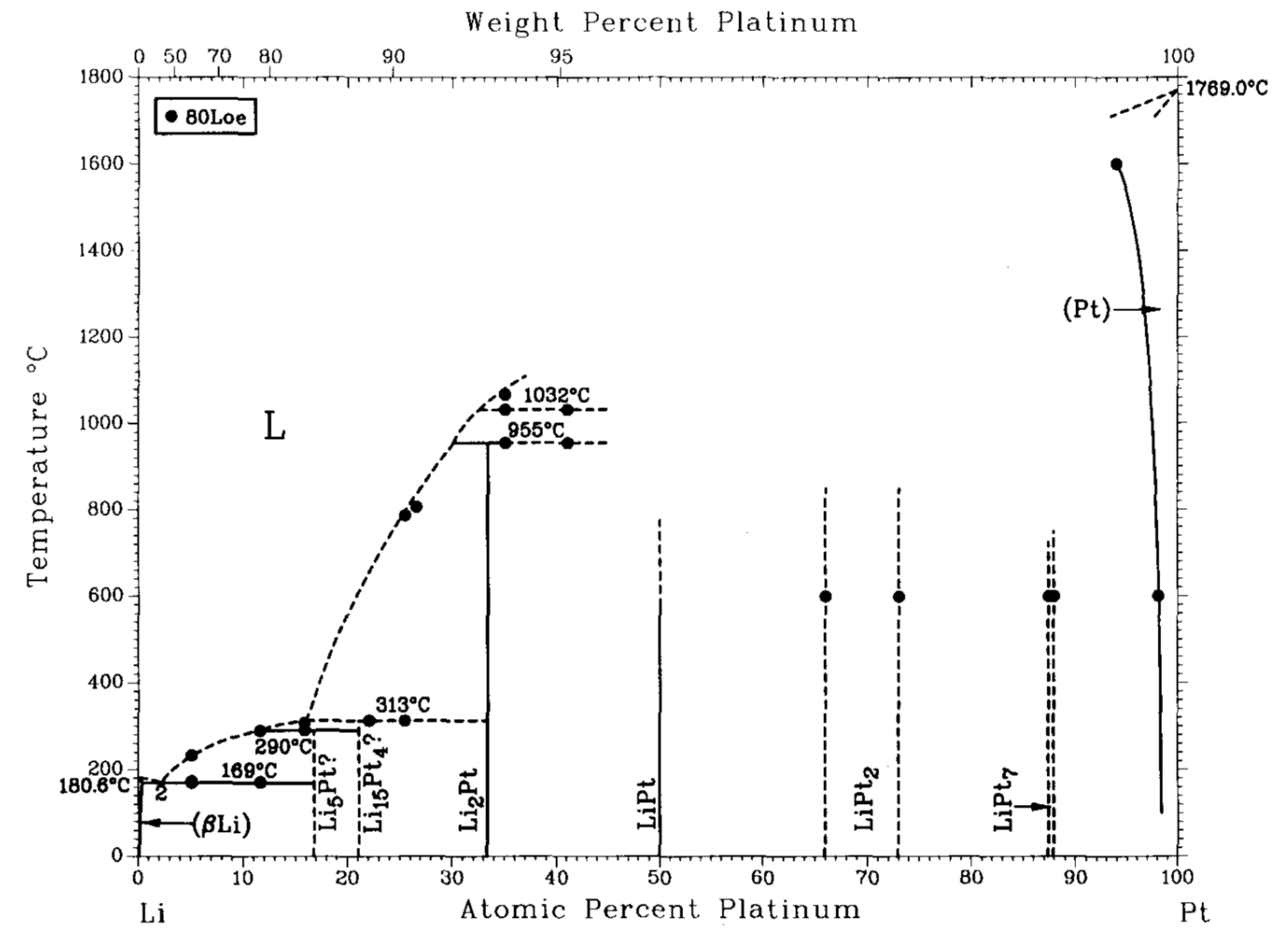}
\caption{Li--Pt phase diagram. Here, Pt forms a solid solution with Li in a narrow range of concentrations. Our reference phase is $\ce{Li2Pt}$ (mp-2170). Data is from the work by Sangster~\textit{et al.}~\cite{sangster1991pt}.}
\label{fig:phase_diagram_pt}
\end{figure}

% Z = 79
\begin{figure}[H]
\centering
\includegraphics[width=0.7\columnwidth]{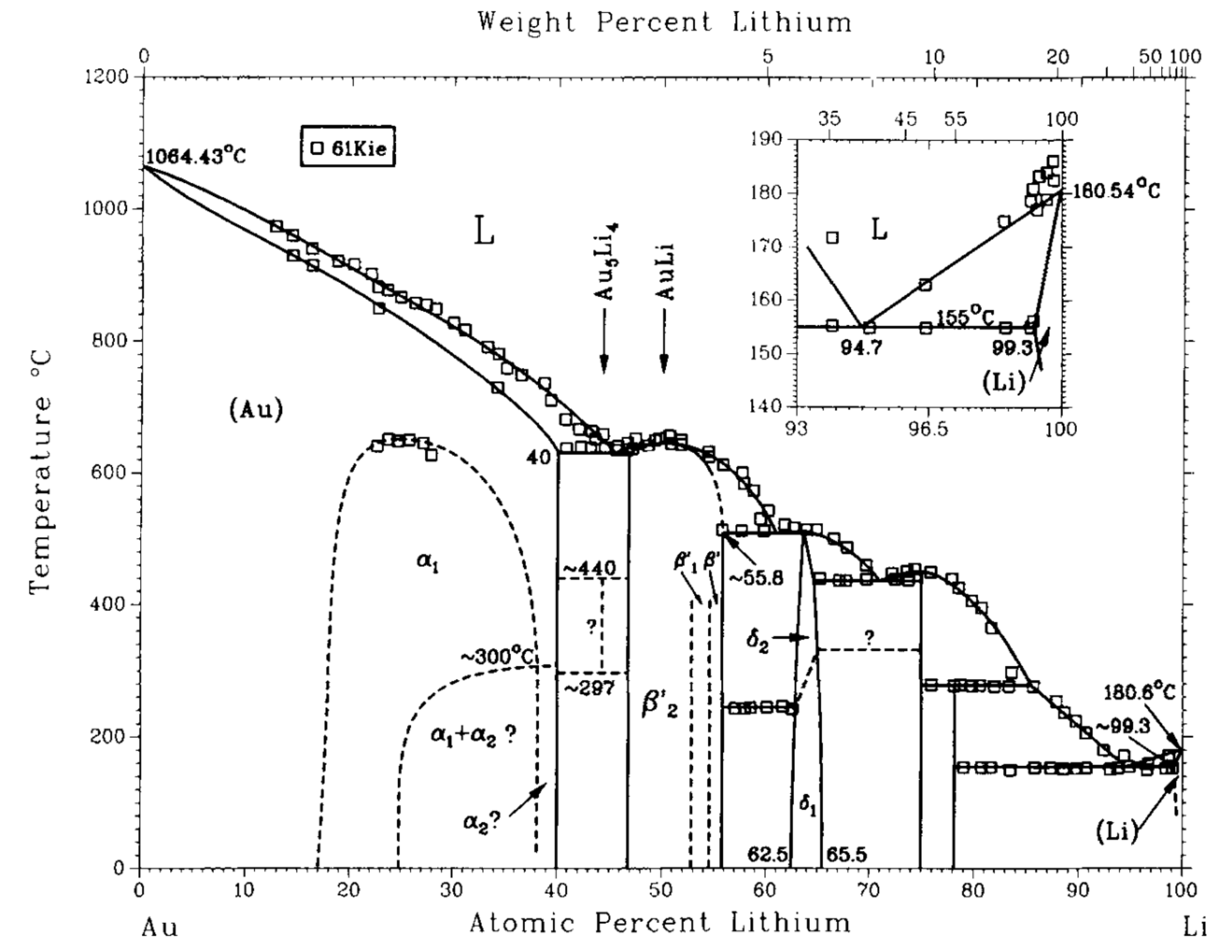}
\caption{Li--Au phase diagram. Here, Au forms a solid solution with Li in a narrow region of concentrations. Our reference phase is $\ce{Li15Au4}$ (mp-567395). Data is from the work by Pelton~\textit{et al.}~\cite{pelton1986au}.}
\label{fig:phase_diagram_au}
\end{figure}

% Z = 80

% Z = 81
\begin{figure}[H]
\centering
\includegraphics[width=0.7\columnwidth]{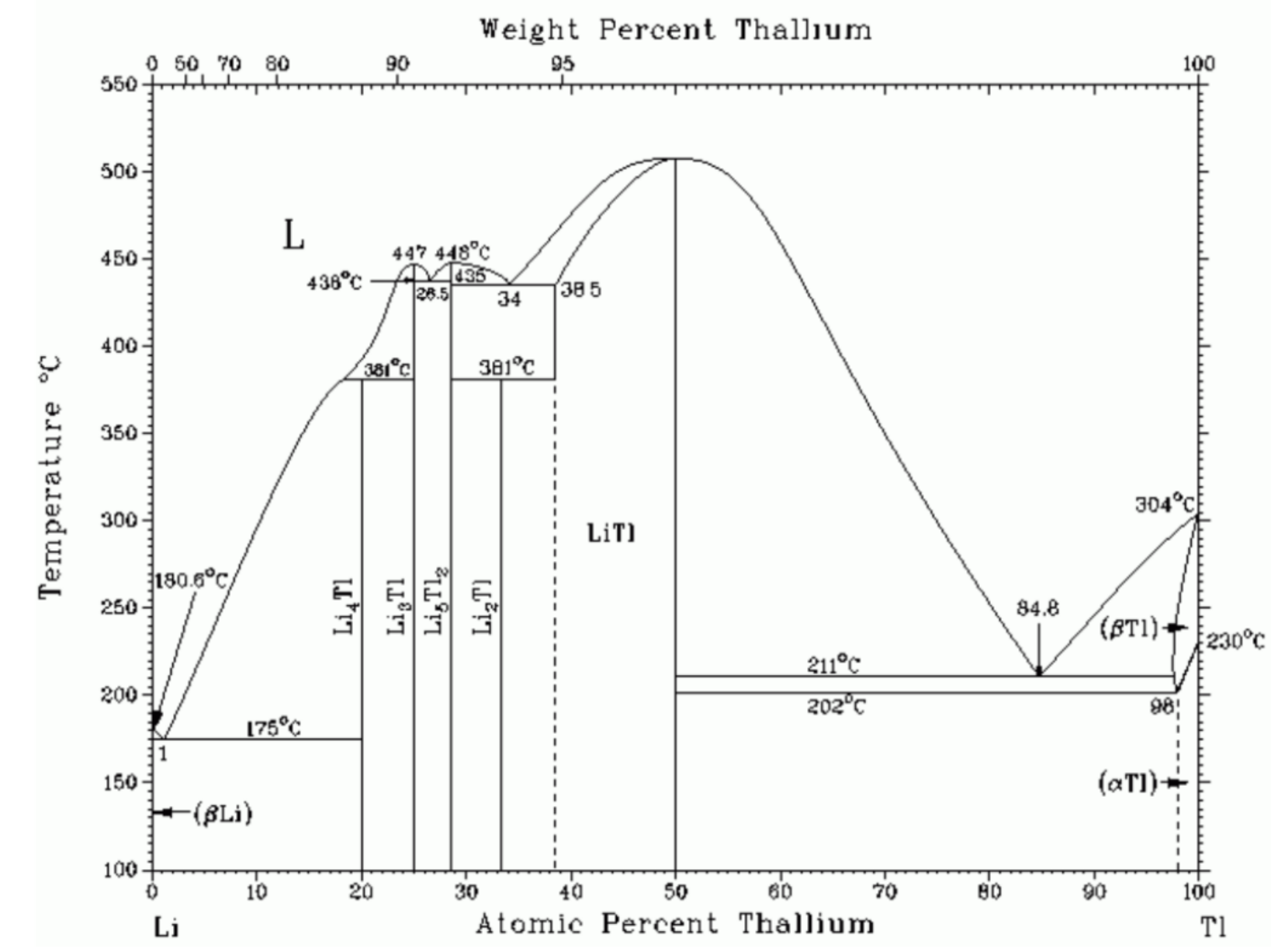}
\caption{Li--Tl phase diagram. Here, Tl forms a solid solution with Li in a narrow range of concentrations. Our reference phase is $\ce{Li3Tl}$ (mp-7396). Data is from the work by Grube~\textit{et al.}~\cite{grube1934elektrische}.}
\label{fig:phase_diagram_tl}
\end{figure}

% Z = 82

% Z = 83
\begin{figure}[H]
\centering
\includegraphics[width=0.6\columnwidth]{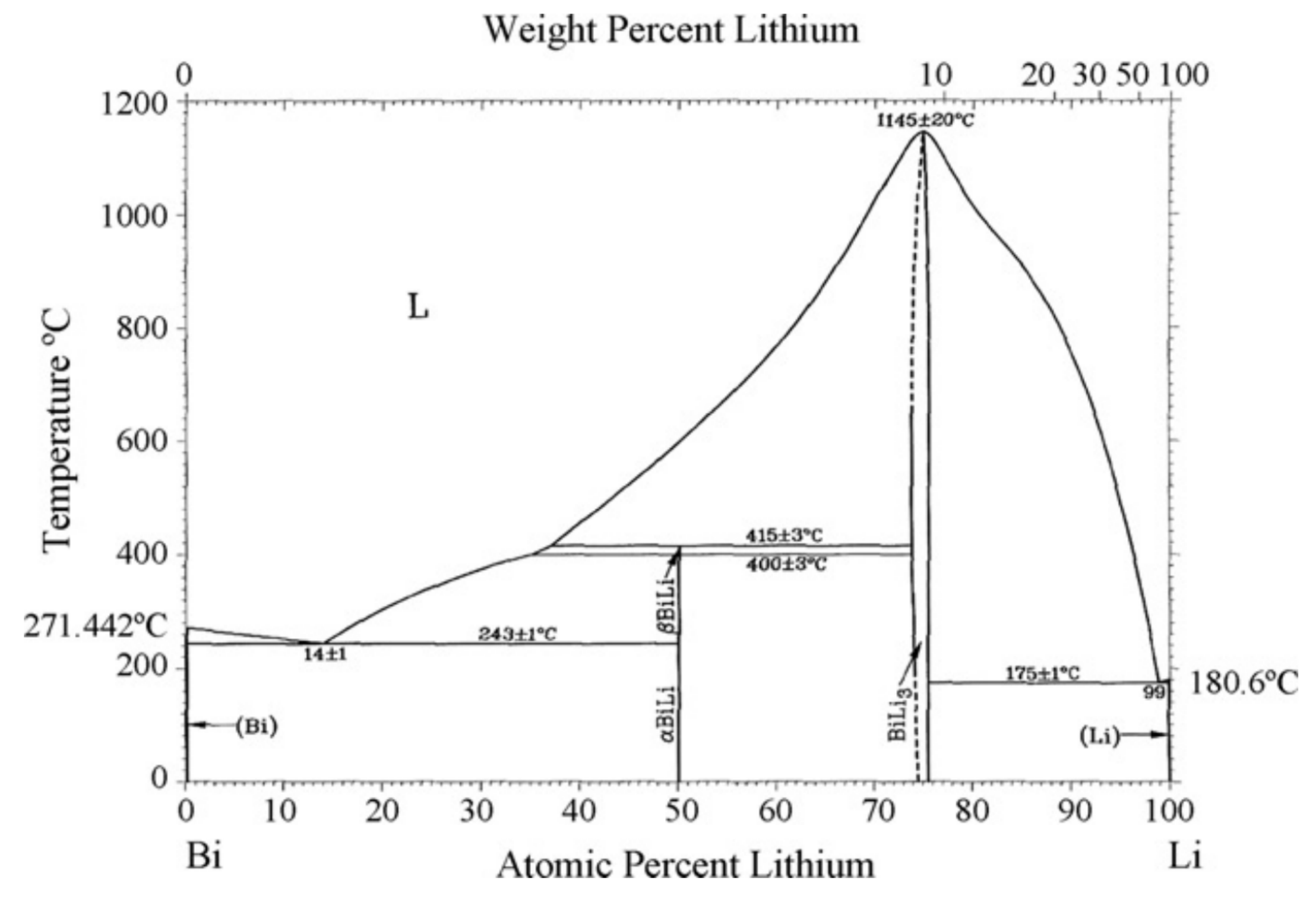}
\caption{Li--Bi phase diagram. Here, Bi forms a solid solution with Li in a narrow range of concentrations. Our reference phase is $\ce{Li3Bi}$ (mp-23222). Data is from the work by Guidotti~\textit{et al.}~\cite{guidotti2008thermally}.}
\label{fig:phase_diagram_bi}
\end{figure}

\def\bibsection{\section*{Supplementary References}} 
\putbib[lib]
\end{bibunit}

\end{document}